\documentclass[prb, twocolumn, 
superscriptaddress,
showpacs,preprintnumbers,
notitlepage,amsmath,amssymb
]{revtex4-1}

\usepackage{xcolor}
\usepackage{times} 
\usepackage{mathtools}

\DeclarePairedDelimiter\floor{\lfloor}{\rfloor}

\usepackage{graphicx}

\usepackage[colorlinks=true,citecolor=blue]{hyperref}
\usepackage{amsmath,latexsym,amssymb,bm,color,graphicx,multirow,array,wrapfig,caption,scalerel,subcaption,datetime}
\usepackage[section]{placeins}

\newcounter{Num}
\newcommand{\FigName}{Fig_}
\newcommand{\NewFigName}[1]{\renewcommand{\FigName}{#1}\tikzsetexternalprefix{\FigName}\setcounter{Num}{0}}

\newcommand{\al}{\alpha}
\newcommand{\be}{\beta}

\newcommand{\s}{\sigma}

\newcommand{\Z}{\mathbb Z}
\newcommand{\eq}[1]{Eq.~(\ref{#1})}

\usepackage{tikz,ifthen}
\usetikzlibrary{external}
\usetikzlibrary{calc}
\usetikzlibrary{decorations.markings}
\usetikzlibrary{shapes.geometric}
\tikzset{->-/.style={decoration={
  markings,
  mark=at position #1 with {\arrow{>}}},postaction={decorate}}
}

\begin{document}

\title{Towards a complete classification of non-chiral topological phases in 2D fermion systems}
 
\author{Jing-Ren Zhou}
\affiliation{Department of Physics, The Chinese University of Hong Kong, Shatin, New Territories, Hong Kong, China}

\author{Qing-Rui Wang}
\affiliation{Yau Mathematical Sciences Center, Tsinghua University, Haidian, Beijing, China}

\author{Zheng-Cheng Gu}
\email{zcgu@phy.cuhk.edu.hk}
\affiliation{Department of Physics, The Chinese University of Hong Kong, Shatin, New Territories, Hong Kong, China}

\date{\today}

\begin{abstract}
In recent years, fermionic topological phases of quantum matter has attracted a lot of attention. In a pioneer work by Gu, Wang and Wen, the concept of equivalence classes of fermionic local unitary(FLU) transformations was proposed to systematically understand non-chiral topological phases in 2D fermion systems and an incomplete classification was obtained. On the other hand, the physical picture of fermion condensation and its corresponding super pivotal categories give rise to a generic mathematical framework to describe fermionic topological phases of quantum matter. In particular, it has been pointed out that in the fermionic string-net models of certain fermionic topological phases, there exists the so-called q-type strings which have no analogues in bosonic theories. In this paper, we generalize the Gu, Wang and Wen construction to include those fermionic topological phases with q-type strings. We argue that all non-chiral fermionic topological phases in 2+1D are characterized by a set of tensors $(N^{ij}_{k},F^{ij}_{k},F^{ijm,\alpha\beta}_{kln,\chi\delta},n_{i},d_{i})$, which satisfy a set of nonlinear algebraic equations parameterized by phase factors $\Xi^{ijm,\alpha\beta}_{kl}$ and $\Xi^{ij}_{kln,\chi\delta}$. Moreover, consistency conditions among algebraic equations give rise to additional constraints on these phase factors which allow us to construct a topological invariant partition for an arbitrary triangulation of 3D spin manifold. Finally, several examples with q-type strings are discussed, including the Fermionic topological phase from Tambara-Yamagami category for $\mathbb{Z}_{2N}$, which can be regarded as the $\mathbb{Z}_{2N}$ parafermion generalization of Ising fermionic topological phase.
\end{abstract}

\maketitle

{
  \hypersetup{linkcolor=black}
  \tableofcontents
}

\section{Introduction}

\subsection{The goal of this paper}
Since the discovery of fractional quantum Hall effect (FQHE)\cite{Tsui1982}, it has been realized that these peculiar quantum matters can be described by a new type of order---topological order\cite{Wen90}. The topological order of FQHE can be characterized by its precise quantization of the Hall conductance, fractionalized charge and fractionalized statistics carried by elementary excitations\cite{Laughlin1983}. Mathematically, it is well known that topological order in 2D bosonic systems can be systematically described and classified by the advanced mathematical theory –-- unitary modular tensor category (UMTC)\cite{Kitaev2006}. On the other hand, it has also been realized that the patterns of long-range entanglement\cite{Chen2010} gives rise to an essential physical picture to understand various topological phases. In particular, the equivalence classes of local unitary(LU) transformations\cite{Chen2010} allows us to construct fixed point wavefunctions to classify all non-chiral topological phases in 2D bosonic systems\cite{Kit03,Levin2005}.

Nevertheless, the UMTC framework can not be applied to fermion systems directly. Topological phases in interacting fermion systems are strictly richer
than bosonic systems due to the Fock space structure of fermionic Hilbert space. In addition to the well known FQHE states which are known as chiral topological phases, many new examples of non-chiral topological phases are constructed for 2D fermion systems\cite{FNSWW04,fto,Gu2015,Lan2016}. Interestingly, it has been shown that a fermionic generalization of Pentagon relation is necessary for understanding topological phases in 2D fermion systems. Very recently, the physical picture of fermion condensation and its corresponding super pivotal categories\cite{fc2019} give rise to a generic mathematical framework to derive the fermionic Pentagon relation\cite{Gu2015} and understand the underlying physics for almost all non-chiral topological phases in 2D fermion systems. Most surprisingly, it has been pointed out that there are two distinct types of objects in the resulting super fusion categories, and the so-called q-type objects have no analogues in bosonic theories\cite{fc2019}. Nevertheless, it is still unclear how to understand the algebraic relations generated by fermion condensation\cite{wan2017,fc2019,lou2021} from the patterns of long-range entanglement for 2D fermion systems.

In this paper, we aim at generalizing the equivalence classes of fermionic LU(fLU) transformation framework to construct and classify all non-chiral topological phases, including those cases with q-type objects in 2D fermion systems. Then we will try to understand the deep relationship between fermion condensation picture and the equivalence classes of fLU transformations. Below we will briefly review the precise meaning of fermionic topological phases and fLU transformations.    

\subsection{Gapped quantum liquids}
The classification of gapped quantum phases is in general beyond the Landau symmetry breaking paradigm. For bosonic systems, we define that two gapped quantum systems belong to the same equivalence class if they are connected by a sequence of LU transformations without closing the energy gap, and the LU transformations are generated by a finite-time evolution of local Hermitian operators \cite{Hastings05,Bravyi06,Bravyi10,Chen2010}:
\begin{equation}
\vert \Phi \rangle \sim \vert \Phi' \rangle 
\text{      iff }
\vert \Phi' \rangle =
\mathcal{T}e^{i\int d\tau\widetilde{H}(\tau)}
\vert \Phi \rangle,
\end{equation}
where $\mathcal{T}$ is the time-ordering operator
and $\widetilde{H}(\tau)=\sum_{i}O_{i}(\tau)$ is a summation of local Hermitian operators. Under such a equivalence relation, the trivial phase is connected to the direct-product state, and other nontrivial phases are long-range entangled and called topologically ordered phases. 

In discrete spacetime, e.g. on a lattice, the LU transformations can be expressed by a finite depth quantum circuit, generated by piece-wise local unitary operators $U_{pwl}=\prod_{i}U(i)$, where $\{U(i)\}$ is a set of unitary operators acting on non-overlapping regions. A quantum circuit with depth $M$ is given by: $U^M_{circ}=U^{(1)}_{pwl}U^{(2)}_{pwl}...U^{(M)}_{pwl}$. Thus the discrete version of the equivalence relation is written as:
\begin{equation}
\vert \Phi \rangle \sim \vert \Phi' \rangle 
\text{      iff }
\vert \Phi' \rangle =
U^M_{circ}
\vert \Phi \rangle.
\end{equation}

More precisely, in this paper we only consider a subset of gapped quantum phases, namely the gapped quantum liquid (GQL)\cite{Zeng2015} which can be defined on arbitrary lattice geometry. In addition, we are also allowed to remove or add degree of freedoms into the systems. Thus, the equivalence classes should be redefined as the generalized local unitary (gLU) transformations satisfying $U_g^\dagger U_g = P$ and $U_gU_g^\dagger= P'$, where $P$ and $P'$ are two projectors. In particular, the action of $P$ does not change the state $\vert \Phi \rangle$.
In such a way, some of the quantum gapped phases cannot be included, e.g. the fracton topological order\cite{FC1,FC2,FC3}.

\subsection{Fermionic gapped quantum liquids and its classification}

In fermionic systems, the underset degrees of freedom are fermions and the total Hilberst space is Fock space instead of a simple tensor product of local Hilbert space. Hence we should redefine the LU transformations as the fermionic LU (fLU) transformations\cite{Gu2015}:
\begin{equation}
\vert \Phi \rangle \sim \vert \Phi' \rangle 
\text{      iff }
\vert \Phi' \rangle =
\mathcal{T}e^{i\int dg\widetilde{H}_{f}(g)}
\vert \Phi \rangle,
\end{equation}
which can also be discretized as the fermionic quantum circuits, where the local fermionic Hamiltonian $\widetilde{H}_{f}(g)=\sum_{i}\mathcal{O}_{i}(g)$ is a summation of pseudo-local bosonic operators $\mathcal{O}_{i}(g)$. Here $\mathcal{O}_{i}(g)$ is a product of even number of local fermionic operators (due to the conservation of fermion parity) and any number of local bosonic operators. It is called "pseudo-local" as it is local for a fermion system in a sense that the fLU transformations acting on different local regions, but are non-local due to the global anti-commutation relation of the fermion creation or annihilation operators. 
Similarly, the fLU transformations can also be redefined as the generalized fLU(gfLU) transformations for fermionic GQL(fQGL).
Thus, the fermionic topological orders are classified by the equivalence classes of gfLU transformations $\tilde U_g$, which are projective unitary operators. Up to some unitary transformations, $\tilde U_g$ is a Hermitian
projection operator:
\begin{align}
 \label{UUp}
\tilde U_g &=U_1P_gU_2, & P_g^2&=P_g, & P_g^\dagger &=P_g,
\nonumber\\
U_1^\dagger U_1&=1, & U_2^\dag U_2 &=1.
\end{align}
We will call such a gfLU transformation a
primitive gfLU transformation. A generic  gfLU
transformation is a product of several  primitive gfLU
transformations which may contain several hermitian
projectors and unitary transformations, for example,
$\tilde U_g =U_1 P_g U_2 P^\prime_g U_3$. We note that $\tilde U_g$ contains only even numbers of fermionic operators (i.e. $\tilde U_g$ is a pseudo-local bosonic operator).
We also regard the inverse of $\tilde U_g$, $\tilde U_g^\dagger$, as a gfLU
transformation.  An fLU transformation is viewed as a special
case of gfLU transformations.  Clearly $\tilde U_g^\dagger \tilde U_g=\tilde P$ and $\tilde U_g
\tilde U_g^\dag=\tilde P'$ are two Hermitian projectors.

Similar to bosonic systems, $\tilde U_g$ can  generate a wavefunction renormalization which allows us to connect the same fGQL state defined on different lattice geometry with different degrees of freedoms. In this paper, by constructing the most generic fixed point wavefucntions from $\tilde U_g$, we argue that all non-chiral fermionic topological phases in 2D fermion systems are characterized by a set of tensors $(N^{ij}_{k},F^{ij}_{k},F^{ijm,\alpha\beta}_{kln,\chi\delta},n_{i},d_{i})$, which are data in fermionic string-net models that satisfy a set of nonlinear algebraic equations parameterized by phase factors $\Xi^{ijm,\alpha\beta}_{kl}$ and $\Xi^{ij}_{kln,\chi\delta}$. In particular, in order to to include those fermionic topological phases with q-type objects, the tensor $F^{ijm,\alpha\beta}_{kln,\chi\delta}$ must be a gfLU transformation instead of the usual fLU transformation in Gu, Wang and Wen's construction. In such a way, we reveal the origin of q-type objects and naturally explain why they do not have analogues in bosonic theories from quantum information perspective. Moreover, consistency conditions among algebraic equations give rise to additional constraints on these phase factors which allow us to construct a topological invariant partition for an arbitrary triangulation of 3D spin manifold.

The rest of the paper is organized as follows: In section \ref{fixwave}, we construct the most general fixed-point wavefunction for non-chiral fermionic topological orders in 2D.  
Then we derive the conditions for all wavefunction renormalization moves with the inclusion of q-type strings, i.e., the conditions on $F$-move, $O$-move, $Y$-move, $H$-move, and dual $F$/$H$-move. Thus we obtain a set of most general algebraic equations in section \ref{Summary}.
In section \ref{pf}, we explicit construct the topological invariant partition function for an arbitrary triangulation of 3D spin manifold. We find that the relations among phase factors for constructing the partition function can be obtained from the fermionic Pentagon equation and four projective unitary conditions for $F$-move. These relations match with the results from fermion condensation theory\cite{fc2019}, as illustrated in section \ref{fcm}. In section \ref{ex}, several examples with q-type strings are studied, including the fermionic topological phase from Tambara-Yamagami category for $\mathbb{Z}_{2N}$, which can be regarded as the $\mathbb{Z}_{2N}$ parafermion generalization of Ising fermionic topological phase. Finally, we summarized this work in section \ref{con}. 

In Appendix \ref{AppendixA}, we review some basic concepts in super pivotal category introduced in Ref.\onlinecite{fc2019}. Appendix \ref{fc} introduces the explict steps to do fermion condensation. We apply the fermion condensation scheme to derive several equivalence relations on fixed-point states with q-type strings, and derive all fermionic $F$-symbols for the four examples from their corresponding bosonic theory. In Appendix \ref{Amove}, we define a special sequence of moves, whose equivalence relation gives the phase factor $\Delta^{mji,\alpha\delta}_{nl}$, which is involved in the relations among phase factors needed for constructing the partition function. Appendix \ref{check23} is a proof that all possible 2-3 moves induced by time-ordering are consistent with the four projective unitary conditions as well as the relations among the corresponding phase factors.





\section{Wavefunction renormalization for generic non-chiral topological phases in 2D fermion systems}
\label{fixwave}
\subsection{Fixed-point wavefunctions on a graph}
Since the wave-function renormalization may change the
lattice structure, we will consider quantum state defined on a
generic trivalent graph G with a branching structure such that each vertex has two incoming or one incoming edges.
Similar to the construction of string-net model for bosonic systems, we assume each edge has $N + 1$ states, labeled by $i = 0, . . . ,N$. Each
vertex also has physical states. The string fusion rules and the local fermion parity are both encoded in the vertex states $\alpha=1,...,N^{ij}_{k}$ or $\beta=1,...,N_{ij}^k$, where $N^{ij}_{k}(N_{ij}^k)$ is the number of fusion states with two incoming (outgoing) strings $i,j$ and one outgoing(incoming) string $k$, graphically represented as $\begin{matrix}\includegraphics[scale=.35]{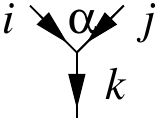}\end{matrix}$ or $\begin{matrix}\includegraphics[scale=.35]{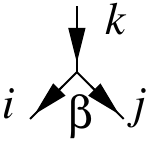}\end{matrix}$. Generally, we have
\begin{equation}
N^{ij}_{k}=B^{ij}_{k}+F^{ij}_{k},
\end{equation}
where $B^{ij}_{k}$ is the number of bosonic fusion states,  and $F^{ij}_{k}$ is the number of fermionic fusion states (a local fermion excitation is involved), represented as a solid dot $\begin{matrix}\includegraphics[scale=.35]{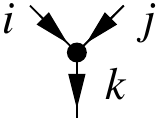}\end{matrix}$. We introduce a number $s(\alpha)$ to indicate the vertex states is bosonic or fermionic: $s(\alpha)=0$ if the state $\alpha$ is bosonic and $s(\alpha)=1$ if it is fermionic.  

In this paper, we will assume that
\begin{align}
 N^{ij}_k  =N_{ij}^k ,\ \ \ \
 B^{ij}_k  =B_{ij}^k ,\ \ \ \
 F^{ij}_k  =F_{ij}^k ,
\end{align}
as required by unitarity. Our fixed-point state is
a superposition of those basis states
\begin{align}
\label{Gstate}
|\psi_\text{fix}\rangle=\sum_\text{all conf.}
\psi_\text{fix}\left(
\vcenter{\hbox{\includegraphics[scale=.3]{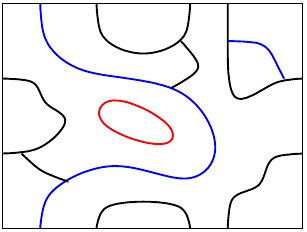}}}
\right)
\left
|
\vcenter{\hbox{\includegraphics[scale=.3]{strnet}}}
\right \rangle.
\end{align}

In the bosonic string-net models, there is a very strong assumption
that the above graphic states on two graphs are the same if
the two graphs have the same topology.  However, since different
vertices and edges are distinct and a generic graph
state does not have such a topological invariance.  
Similar as the construction in Ref. \onlinecite{Gu2015}, here we will consider
vertex-labeled graphs (v-graphs) where each vertex is
assigned an index $\underline\alpha$.  Two v-graphs are said to be
topologically the same if one graph can be continuously
deformed into the other in such a way that vertex labelings
of the two graphs matches.  

\subsection{The structure of fixed-point wavefunctions}

Firstly, we need to divide the state on each edge into m-type and q-type strings. 
When all strings $i,j,k$ are m-type, generally $B^{ij}_{k}$ is not equal to $F^{ij}_{k}$, however, when there is a q-type string involved in the fusion (at least two strings in $i,j,k$ are q-type), we must have
$B^{ij}_{k}=F^{ij}_{k}$ (The physical reason of such an assumption will be explained below). Thus we can introduce the function $B(\alpha)=1,...,B^{ij}_{k}$ to extract the bosonic fusion state of $\alpha$:
\begin{equation}
B(\alpha)
=
\left\{\begin{array}{l}
\alpha
\text{, \ \ if }s(\alpha)=0\\ 
\alpha-B^{ij}_{k}
\text{, \ \ if }s(\alpha)=1
\end{array}\right.
\end{equation}
We note that $B(\alpha)$ is only defined when q-type strings are involved in $i,j,k$. Here we introduce the notation $\cdot f$ to denote the changing of fermion parity without changing the corresponding bosonic state, i.e., $B(\alpha\cdot f)=B(\alpha)$. 

Now let us consider the fixed-point wavefunctions on a patch $\psi_\text{fix}\begin{pmatrix}
\includegraphics[scale=.35]{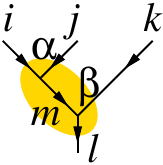}\end{pmatrix}$ , where the boundary string states $i,j,k,l$ are fixed, while yellow shaded ellipse means that the inner fusion states $\alpha,\beta$ and the inner string state $m$ may vary. (More precisely, $\psi_\text{fix}\begin{pmatrix}
\includegraphics[scale=.35]{F1gAA}\end{pmatrix}$ should be regarded as function $\phi_{ijkl,\Gamma}(\alpha,\beta,m)$ where the indices
on the other part of the graph are summarized by $\Gamma$.)
All such fixed-point wavefunctions(as functions of $\alpha,\beta,m$) form a linear space called the support space $V^{ijk}_{l}$, whose dimension is called the support dimension $D^{ijk}_{l}$.

For the fixed-point wavefunctions $\psi_\text{fix}\begin{pmatrix}
\includegraphics[scale=.35]{F1gAA}\end{pmatrix}$, the number of inner states $\{\alpha,\beta,m\}$ is $N^{ijk}_{l}=\sum_{m}N^{ij}_{m}N^{mk}_{l}$. Specially, if the inner string $m$ is a q-type string, the support space $V^{ijk}_{l}$ should mod out the following equivalence relations generated by string $m$, in fermion parity-even and odd sectors respectively:
\begin{align}
\psi_{\text{fix}}\begin{pmatrix} \includegraphics[scale=.40]{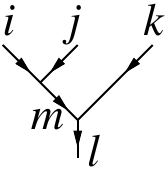} 
\end{pmatrix}\sim
\psi_{\text{fix}}\begin{pmatrix} \includegraphics[scale=.40]{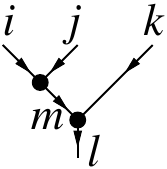} 
\end{pmatrix},
\label{ef1}
\end{align}
\begin{align}
\psi_{\text{fix}}\begin{pmatrix} \includegraphics[scale=.40]{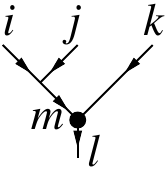} 
\end{pmatrix}\sim
\psi_{\text{fix}}\begin{pmatrix} \includegraphics[scale=.40]{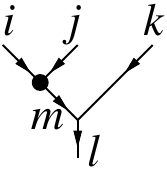} 
\end{pmatrix},
\label{ef2}
\end{align}
which can be altogether denoted as:
\begin{align}
\psi_{\text{fix}}\begin{pmatrix} \includegraphics[scale=.40]{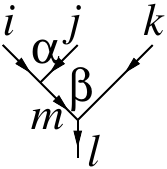} 
\end{pmatrix}\sim
\psi_{\text{fix}}\begin{pmatrix} \includegraphics[scale=.40]{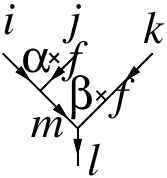} 
\end{pmatrix},
\label{ef12}
\end{align}
where $f$ denotes a transparent (local) fermion excitation, and $\times f$ means changing the fermion parity on a fusion state via attaching a transparent fermion, which does not have to preserve the original bosonic state in general, i.e. $B(\alpha\times f)$ generally may not be the same as $B(\alpha)$. However, we require $(\alpha\times f)\times f=\alpha$ as attaching a pair of transparent fermions should not affect the fusion state on each vertex. Here the equivalence relation $\sim$ is up to a phase. Physically, the first relation means that a pair of transparent fermions can be created or annihilated on q-type strings (that's why we must require $B^{ij}_{k}=F^{ij}_{k}$ once q-type string is involved in the fusion), and the second relation means that a local fermion excitation can slide along q-type strings freely.
Thus, we can assign a number $n_{i}$ to each string, with $n_{i}=1$ for a m-type string and $n_{i}=2$ for a q-type string. 
Mathematically,the number $n_{i}$ can be regarded as the dimension of the endomorphism algebra of string $i$, as explained in Appendix \ref{AppendixA}. As a result, the support dimension $D^{ijk}_{l}$ is generally equal to or less than the number of inner states $N^{ijk}_{l}$, and
\begin{equation}
D^{ijk}_{l}=\underset{m}{\sum} \frac{N^{ij}_{m}N^{mk}_{l}}{n_{m}}.
\end{equation}

Similarly, the support space of the fixed-point wavefunctions on $\psi_\text{fix}\begin{pmatrix}
\includegraphics[scale=.35]{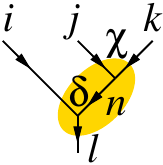}\end{pmatrix}$ should also mod out the following equivalence relation if $n$ is a q-type string:
\begin{align}
\psi_{\text{fix}}\begin{pmatrix} \includegraphics[scale=.40]{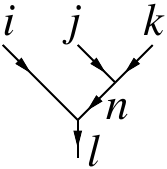} 
\end{pmatrix}\sim
\psi_{\text{fix}}\begin{pmatrix} \includegraphics[scale=.40]{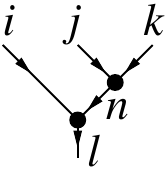} 
\end{pmatrix},
\label{ef3}
\end{align}
\begin{align}
\psi_{\text{fix}}\begin{pmatrix} \includegraphics[scale=.40]{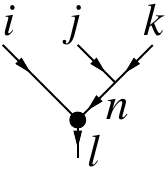} 
\end{pmatrix}\sim
\psi_{\text{fix}}\begin{pmatrix} \includegraphics[scale=.40]{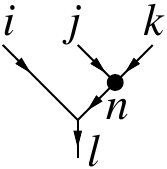} 
\end{pmatrix},
\label{ef4}
\end{align}
which can be in together denoted as:
\begin{align}
\psi_{\text{fix}}\begin{pmatrix} \includegraphics[scale=.40]{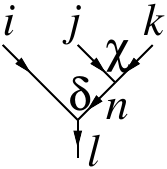} 
\end{pmatrix}\sim
\psi_{\text{fix}}\begin{pmatrix} \includegraphics[scale=.40]{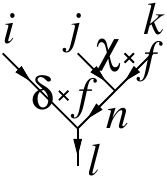} 
\end{pmatrix},
\label{ef34}
\end{align}
where similarly the changing of fermion parity $\times f$ here does not have to preserve the original bosonic fusion state. The support dimension here is
\begin{equation}
D^{ijk}_{l}=\underset{n}{\sum} \frac{N^{in}_{l}N^{jk}_{n}}{n_{n}}.
\end{equation}


\subsection{$F$-move}

The first type of wavefunction renormalization is the $F$-move, which is a gfLU transformation between the two fixed-point wavefunctions $\psi_\text{fix}\begin{pmatrix}
\includegraphics[scale=.35]{F1gAA}\end{pmatrix}$ and $\psi_\text{fix}\begin{pmatrix}
\includegraphics[scale=.35]{F2gAA}\end{pmatrix}$ (We assume that the two fixed point wavefucntions are the same for the other parts of the graph). 
 
Apparently, the support dimensions on both sides are equal:
\begin{equation}
\underset{m}{\sum} \frac{N^{ij}_{m}N^{mk}_{l}}{n_{m}}
=\underset{n}{\sum} \frac{N^{in}_{l}N^{jk}_{n}}{n_{n}}.
\label{F2}
\end{equation}
Since the fermion parity-odd sector and the parity-even sector are independent, this relation can be further split as
\begin{equation}
\underset{m}{\sum} \frac{B^{ij}_{m}B^{mk}_{l}+F^{ij}_{m}F^{mk}_{l}}{n_{m}}
=\underset{n}{\sum} \frac{B^{in}_{l}B^{jk}_{n}+F^{in}_{l}F^{jk}_{n}}{n_{n}},
\label{F3}
\end{equation}
\begin{equation}
\underset{m}{\sum} \frac{B^{ij}_{m}F^{mk}_{l}+F^{ij}_{m}B^{mk}_{l}}{n_{m}}
=\underset{n}{\sum} \frac{B^{in}_{l}F^{jk}_{n}+F^{in}_{l}B^{jk}_{n}}{n_{n}}.
\label{F4}
\end{equation}

In fact, for fermionic systems, the total Hilbert space is a Fock space, and we need to specify the ordering of the fermionic states in the fixed point wavefunctions. For example, $\psi_\text{fix}\begin{pmatrix}
\includegraphics[scale=.35]{F1gAA}\end{pmatrix}$ actually means $\psi_\text{fix}^{\underline{\alpha}\underline{\beta},...}\begin{pmatrix}
\includegraphics[scale=.35]{F1gAA}\end{pmatrix}$ where the fermionic state on vertex $\underline{\beta}$ is always created before the fermionic state on vertex $\underline{\alpha}$. An elegant way to count the ordering of fermionic states is to introduce
the Majorana numbers $\theta_{\underline{\alpha}},\theta_{\underline{\beta}},...$, where $\underline{\alpha},\underline{\beta},...$ denote the vertices carrying the fusion state $\alpha,\beta,...$. The Majorana numbers satisfy:
\begin{align}
& \theta_{\underline{\alpha}}^2=1,\ \ \ \
\theta_{\underline{\alpha}}\theta_{\underline{\beta}}= -\theta_{\underline{\beta}} \theta_{\underline{\alpha}} \text{ for any }\alpha\neq \beta ,
\nonumber\\
& \theta_{\underline{\alpha}}^\dagger = \theta_{\underline{\alpha}},\ \ \ \ \
(\theta_{\underline{\alpha}}...\theta_{\underline{\beta}})^\dagger =
\theta_{\underline{\beta}}...\theta_{\underline{\alpha}}.
\label{gnr}
\end{align}
Thus, we can define the ordering independent fixed-point wavefunctions $\Psi_{\text{fix}}$ by attaching Majorana numbers on $\psi_{\text{fix}}^{\underline{\alpha}\underline{\beta},...}$.
\begin{align}
\Psi_{\text{fix}}\begin{pmatrix} \includegraphics[scale=.40]{F1g} 
\end{pmatrix}
=
\theta^{s(\alpha)}_{\underline{\alpha}}
\theta^{s(\beta)}_{\underline{\beta}}
\psi_{\text{fix}}^{\underline{\alpha}\underline{\beta},...}\begin{pmatrix} \includegraphics[scale=.40]{F1g} 
\end{pmatrix}
\label{gn1}
\end{align}
Similarly, for the patch $\psi_\text{fix}\begin{pmatrix}
\includegraphics[scale=.35]{F2gAA}\end{pmatrix}$, we can also define:
\begin{align}
\Psi_{\text{fix}}\begin{pmatrix} \includegraphics[scale=.40]{F2g1} 
\end{pmatrix}
=
\theta^{s(\chi)}_{\underline{\chi}}
\theta^{s(\delta)}_{\underline{\delta}}
\psi_{\text{fix}}^{\underline{\chi}\underline{\delta},...}\begin{pmatrix} \includegraphics[scale=.40]{F2g1} 
\end{pmatrix}
.
\label{gn2}
\end{align}
where in $\psi_{\text{fix}}^{\underline{\chi}\underline{\delta},...}$ the fermionic state on vertex $\underline{\delta}$ is always created before the fermionic state on vertex $\underline{\chi}$. 

On the other hand, since the fermion ordering in $\psi_\text{fix}^{\underline{\alpha}\underline{\beta},...}\begin{pmatrix}
\includegraphics[scale=.35]{F1gAA}\end{pmatrix}$ and $\psi_\text{fix}^{\underline{\chi}\underline{\delta},...}\begin{pmatrix}
\includegraphics[scale=.35]{F2gAA}\end{pmatrix}$ will be naturally induced by the branching structure of the graph, below we will also omit the subscript ${\underline{\alpha}\underline{\beta},...}$ and ${\underline{\chi}\underline{\delta},...}$ throughout the whole paper without confusion. However, one should always keep in mind that $\Psi_{\text{fix}}$ is the Majorana number valued ordering independent wavefunction while $\psi_{\text{fix}}$ is the ordering dependent fixed-point wavefunctions.


Similar to the Gu, Wang and Wen construciton, we can introduce a Majorana number valued $\mathcal{F}$-move without specifying the ordering of fermions on vertices of both patches:
\begin{align}
\Psi_\text{fix}\begin{pmatrix} \includegraphics[scale=.40]{F1g} \end{pmatrix}
\simeq
\sum_{n\chi\delta}
\mathcal{F}^{ijm,\alpha\beta}_{kln,\chi\delta}
\Psi_\text{fix}
\begin{pmatrix} \includegraphics[scale=.40]{F2g1} \end{pmatrix},
\label{F1}
\end{align}
where 
\begin{equation}
\mathcal{F}^{ijm,\alpha\beta}_{kln,\chi\delta} =
\theta^{s(\alpha)}_{\underline{\alpha}}
\theta^{s(\beta)}_{\underline{\beta}}
\theta^{s(\delta)}_{\underline{\delta}}
\theta^{s(\chi)}_{\underline{\chi}}
F^{ijm,\alpha\beta}_{kln,\chi\delta},
\label{F9}
\end{equation}
which follows our Majorana number convention in Eq.(\ref{gn1}) and Eq.(\ref{gn2}). The $\mathcal{F}$-move is non-zero only when all the fusion states are non-vanishing and the fermion parity is conserved: $s(\alpha)+s(\beta)+s(\chi)+s(\delta)=0 \text{ mod } 2$. Or in other words,
\begin{align}
F^{ijm,\alpha\beta}_{kln,\chi\delta}=0 \text{ when }
N^{ij}_{m}<1 \text{ or } N^{mk}_{l}<1 \text{ or }
N^{jk}_{n}<1
\nonumber\\
 \text{ or } N^{in}_{l}<1,
\text{or } s(\alpha)+s(\beta)+s(\chi)+s(\delta)=1 \text{ mod } 2.
\label{F5}
\end{align}
Here the complex number valued $F$-symbol is defined according to the special fermion ordering scheme discussed above. The unique advantage of the Majorana number valued $\mathcal{F}$-move is that the anti-commuting nature of fermion creating/annhilation operators is naturally encoded in such a gFLU and we do not need to worry about the fermion ordering problem when considering a sequence of gFLU transformations. 

However, very different from the Gu, Wang and Wen construciton where $\mathcal{F}$-move is assumed to be unitary, here if $m$ is q-type,  we can only fix the target space up to a certain superposition of two equivalent states in the 1-dimensional projective space: $\Psi_\text{fix}\begin{pmatrix}
\includegraphics[scale=.35]{F1g}\end{pmatrix}$ and $\Psi_\text{fix}\begin{pmatrix}
\includegraphics[scale=.35]{F1g4fff}\end{pmatrix}$. Therefore, depending on whether $m$ is an m-type or q-type string, the $\mathcal{F}$-move can be unitary or projective unitary. In particular, when $m$ is q-type, the projective unitary condition should map to both of the two equivalent states, and we require:





\begin{widetext}

\begin{equation}
\underset{n\chi\delta}{\sum} F^{ijm',\alpha'\beta'}_{kln,\chi\delta} (F^{ijm,\alpha\beta}_{kln,\chi\delta})^{*}=
\left\{\begin{array}{l}
\delta_{mm'}\delta_{\alpha\alpha'}\delta_{\beta\beta'}
\text{, \ \ if }m\text{ is m-type
}\\ 
\frac{1}{n_{m}} (\delta_{mm'}\delta_{\alpha\alpha'}\delta_{\beta\beta'}+
\Xi^{ijm,\alpha\beta}_{kl}
\delta_{mm'}\delta_{(\alpha\times f)\alpha'}\delta_{(\beta\times f)\beta'})
\text{,  \ \ if }m\text{ is q-type}
\end{array}\right.
\label{fprojc1}
\end{equation} 
We note that the Majorana numbers cancel out due to relation in Eq.(\ref{gnr}), and we can write down the projective unitary condition for the complex valued $F$-moves without Majorana numbers. Here $\Xi^{ijm,\alpha\beta}_{kl}$ is a phase factor satisfying $(\Xi^{ijm,\alpha\beta}_{kl})^*=\Xi^{ijm,(\alpha\times f)(\beta\times f)}_{kl}$. It depends on strings $i,j,k,l,m$ and fusion states $\alpha,\beta$.  We should note that $B(\alpha\times f )$ and $B(\beta\times f)$ do not have to be the same as $B(\alpha)$ and $B(\beta)$ in general. But the explicit corresponding bosonic fusion state of $\alpha\times f$ can be determined by $\alpha$ and the three strings $i,j,k$ attached to it. Similarly, $B(\beta\times f)$ can be determined by $\beta$ and strings $m,k,l$.  If $m$ is q-type, 
this projective unitary condition can be viewed as the following projective map:

\begin{align}
\frac{1}{2}\psi_{\text{fix}}\begin{pmatrix} \includegraphics[scale=.40]{F1g} 
\end{pmatrix}+
\frac{(\Xi^{ijm,\alpha\beta}_{kl})^*}{2}
\psi_{\text{fix}}\begin{pmatrix} \includegraphics[scale=.40]{F1g4fff} 
\end{pmatrix}
\rightarrow
\psi_{\text{fix}}\begin{pmatrix} \includegraphics[scale=.40]{F1g} 
\end{pmatrix}
,
\label{PJ1}
\end{align}
\begin{align}
\frac{\Xi^{ijm,\alpha\beta}_{kl}}{2}
\psi_{\text{fix}}\begin{pmatrix} \includegraphics[scale=.40]{F1g} 
\end{pmatrix}+
\frac{1}{2}\psi_{\text{fix}}\begin{pmatrix} \includegraphics[scale=.40]{F1g4fff} 
\end{pmatrix}
\rightarrow
\psi_{\text{fix}}\begin{pmatrix} \includegraphics[scale=.40]{F1g4fff} 
\end{pmatrix}
,
\label{PJ2}
\end{align}

\end{widetext}

If we view both $\psi_\text{fix}\begin{pmatrix}
\includegraphics[scale=.35]{F1gAA}\end{pmatrix}$ and $\psi_\text{fix}\begin{pmatrix}
\includegraphics[scale=.35]{F2gAA}\end{pmatrix}$ as column basis vectors in each support space,
the above expression can also be rewritten in matrix form:
\begin{equation}
P=F^{ijk}_{l}
(F^{ijk}_{l})^{\dagger},
\label{Fmatrixproj}
\end{equation}
where $P$ is a projective matrix with the following form:
\begin{equation}
P
 =
\left(
\begin{array}{cc}
\frac{1}{2} & \frac{(\Xi^{ijm,\alpha\beta}_{kl})^*}{2}\\ 
 \frac{\Xi^{ijm,\alpha\beta}_{kl}}{2} & \frac{1}{2}%
\end{array}%
\right),
\end{equation}
Apparently, it satisfies $P^2= P$.

From Eq.(\ref{PJ1}) and Eq.(\ref{PJ2}), we see that the phase factor $\Xi^{ijm,\alpha\beta}_{kl}$ is actually the phase difference between the two equivalent states:
\begin{align}
\psi_{\text{fix}}\begin{pmatrix} \includegraphics[scale=.40]{F1g4fff} 
\end{pmatrix}
=
\Xi^{ijm,\alpha\beta}_{kl}
\psi_{\text{fix}}\begin{pmatrix} \includegraphics[scale=.40]{F1g} 
\end{pmatrix}
,
\nonumber\\
\text{\ \ \ \ \ \ \ \ \ \ \ \ \ \ \ \ \ \ \ \ \ \ \ \ \ \ \ \ \ \ \ \ \ \ \ \ \ \ \ \ 
if }m\text{ is q-type.}
\end{align}

According to the $F$-move (with the aforementioned fermionic state ordering convention), we have:
\begin{align} 
\psi_{\text{fix}}\begin{pmatrix} \includegraphics[scale=.35]{F1g4fff} 
\end{pmatrix}\simeq\sum_{n\chi\delta}F^{ijm,(\alpha\times f)(\beta\times f)}_{kln,\chi\delta}\psi_{\text{fix}}\begin{pmatrix} \includegraphics[scale=.35]{F2g1} 
\end{pmatrix}
\end{align}
and
\begin{align}
\psi_{\text{fix}}\begin{pmatrix} \includegraphics[scale=.35]{F1g} 
\end{pmatrix}\simeq 
\sum_{n\chi\delta}F^{ijm,\alpha\beta}_{kln,\chi\delta}\psi_{\text{fix}}\begin{pmatrix} \includegraphics[scale=.35]{F2g1} 
\end{pmatrix}.
\end{align} 
Comparing each term with fixed $n,\chi,\delta$, we immediately obtain a relation between the $F$-moves on two equivalent states: 
\begin{align}
F^{ijm,(\alpha\times f)(\beta\times f)}_{kln,\chi\delta}
=
\Xi^{ijm,\alpha\beta}_{kln}
F^{ijm,\alpha\beta}_{kln,\chi\delta}   
\text{, \ \ if }m\text{ is q-type.}
\label{Fequiup}
\end{align}





If we reverse the initial space and target  space, 
we can obtain the inverse fermionic $\mathcal{F}$-move:
\begin{align}
\Psi_\text{fix}\begin{pmatrix} \includegraphics[scale=.40]{F2g1} \end{pmatrix}
\simeq
\sum_{m\alpha\beta}
(\mathcal{F}^{ijm,\alpha\beta}_{kln,\chi\delta})^*
\Psi_\text{fix}
\begin{pmatrix} \includegraphics[scale=.40]{F1g} \end{pmatrix}.
\end{align} 

If $n$ is q-type, we can only fix the target space up to certain superposition of the two equivalent states in Eq.(\ref{ef3}) or Eq.(\ref{ef4}), and there will be another independent projective unitary condition for $F$-move (Similarly the Majorana numbers cancel out so that we can write down the relation for $F$-move):
\begin{widetext}

\begin{equation}
\underset{m\alpha\beta}{\sum} (F^{ijm,\alpha\beta}_{kln',\chi'\delta'})^{*} F^{ijm,\alpha\beta}_{kln,\chi\delta} =
\left\{\begin{array}{l}
\delta_{nn'}\delta_{\chi\chi'}\delta_{\delta\delta'}
\text{, \ \ if }n\text{ is m-type
}\\ 
\frac{1}{n_{n}} (\delta_{nn'}\delta_{\chi\chi'}\delta_{\delta\delta'}
+
\Xi^{ij}_{kln,\chi\delta}
\delta_{nn'}\delta_{(\chi\times f)\chi'}\delta_{(\delta\times f)\delta'})
\text{, \ \ if }n\text{ is q-type}
\end{array}\right.
\label{fprojc2}
\end{equation}
where $\Xi^{ij}_{kln,\chi\delta}$ is another phase factor satisfying $(\Xi^{ij}_{kln,\chi\delta})^*=\Xi^{ij}_{kln,(\chi\times f)(\delta\times f)}$. It depends on strings $i,j,k,l,n$ and fusion states $\chi,\delta$.  $B(\chi\times f )$ and $B(\delta\times f)$ do not have to be the same as $B(\chi)$ and $B(\delta)$ respectively in general.  If $n$ is q-type, 
this projective unitary condition can be viewed as the following projective map:
\begin{align}
\frac{1}{2}\psi_{\text{fix}}\begin{pmatrix} 
\includegraphics[scale=.40]{F2g1} 
\end{pmatrix}
+
\frac{(\Xi^{ij}_{kln,\chi\delta})^*}{2}
\psi_{\text{fix}}\begin{pmatrix} \includegraphics[scale=.40]{F2g4fff} 
\end{pmatrix}
\rightarrow
\psi_{\text{fix}}\begin{pmatrix} \includegraphics[scale=.40]{F2g1} 
\end{pmatrix}
,
\end{align}
\begin{align}
\frac{\Xi^{ij}_{kln,\chi\delta}}{2}
\psi_{\text{fix}}\begin{pmatrix} \includegraphics[scale=.40]{F2g1} 
\end{pmatrix}+
\frac{1}{2}\psi_{\text{fix}}\begin{pmatrix} \includegraphics[scale=.40]{F2g4fff} 
\end{pmatrix}
\rightarrow
\psi_{\text{fix}}\begin{pmatrix} \includegraphics[scale=.40]{F2g4fff} 
\end{pmatrix}
,
\end{align}
\end{widetext}
In terms of matrix form, we have:
\begin{equation}
P'=
(F^{ijk}_{l})^{\dagger}
F^{ijk}_{l},
\label{Fmatrixproj}
\end{equation}
where the projective matrix takes the form:
\begin{equation}
P'
=
\left(
\begin{array}{cc}
\frac{1}{2} & \frac{(\Xi^{ij}_{kln,\chi\delta})^*}{2} \\ 
\frac{\Xi^{ij}_{kln,\chi\delta}}{2} & \frac{1}{2}%
\end{array}%
\right),
\end{equation}
which also satisfies $(P')^2= P'$.





The phase factor $\Xi^{ij}_{kln,\chi\delta}$ is actually the phase difference between these two equivalent states:
\begin{align}
\psi_{\text{fix}}\begin{pmatrix} \includegraphics[scale=.40]{F2g4fff} 
\end{pmatrix}
=
\Xi^{ij}_{kln,\chi\delta}
\psi_{\text{fix}}\begin{pmatrix} \includegraphics[scale=.40]{F2g} 
\end{pmatrix}
,
\nonumber\\
\text{\ \ \ \ \ \ \ \ \ \ \ \ \ \ \ \ \ \ \ \ \ \ \ \ \ \ \ \ \ \ \ \ \ \ \ \ \ \ \ \ 
if }n\text{ is q-type.}
\end{align}
from which we can obtain another relation between the $F$-moves on two equivalent states:
\begin{align}
(F^{ijm,\alpha\beta}_{kln,(\chi\times f)(\delta\times f)})^*
=
\Xi^{ij}_{kln,\chi\delta}
(F^{ijm,\alpha\beta}_{kln,\chi\delta})^*
\text{, \ \ if }n\text{ is q-type}
.
\label{Fequidown}
\end{align}


\subsection{Fermionic pentagon equation}
Similar to the Gu, Wang and Wen construction, if we apply the gFLU transformations on a bigger patch of the graph, certain consistent condition is required.
The so-called fermionic pentagon equation is essentially a consistency relation on two paths connecting two fixed point wavefunctions:

\begin{widetext}

\begin{align}
\Psi_\text{fix} 
\begin{pmatrix} 
\includegraphics[scale=.40]{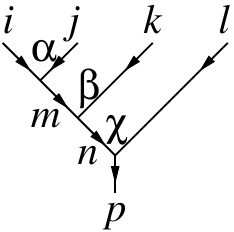}
\end{pmatrix}
&\simeq\sum_{t\eta\psi}
\mathcal{F}^{ijm,\alpha\beta}_{knt,\eta\psi}
\Psi_\text{fix} 
\begin{pmatrix}
\includegraphics[scale=.40]{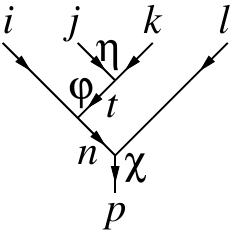}
\end{pmatrix}
\simeq\sum_{t\eta\psi;s\kappa\gamma}
\mathcal{F}^{ijm,\alpha\beta}_{knt,\eta\psi}
\mathcal{F}^{itn,\psi\chi}_{lps,\kappa\gamma}
\Psi_\text{fix} 
\begin{pmatrix}
\includegraphics[scale=.40]{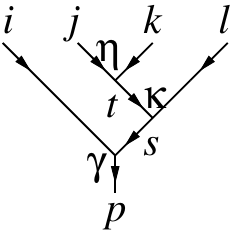}
\end{pmatrix}
\nonumber\\
&\simeq\sum_{t\eta\kappa;\psi;s\kappa\gamma;q\delta\phi}
\mathcal{F}^{ijm,\alpha\beta}_{knt,\eta\psi}
\mathcal{F}^{itn,\psi\chi}_{lps,\kappa\gamma}
\mathcal{F}^{jkt,\eta\kappa}_{lsq,\delta\phi}
\Psi_\text{fix} 
\begin{pmatrix}
\includegraphics[scale=.40]{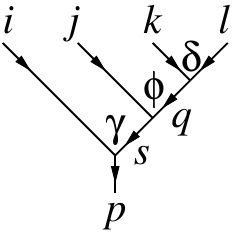}
\end{pmatrix},
\end{align}
\begin{align}
\Psi_\text{fix} 
\begin{pmatrix}
\includegraphics[scale=.40]{pent1g} 
\end{pmatrix}
&\simeq
\sum_{q\delta\epsilon}
\mathcal{F}^{mkn,\beta\chi}_{lpq,\delta\epsilon}
\Psi_\text{fix} 
\begin{pmatrix}
\includegraphics[scale=.40]{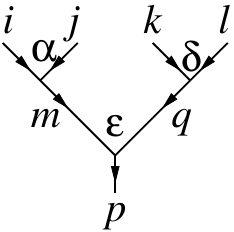} 
\end{pmatrix}
\simeq\sum_{q\delta\epsilon;s\phi\gamma}
\mathcal{F}^{mkn,\beta\chi}_{lpq,\delta\epsilon}
\mathcal{F}^{ijm,\alpha\epsilon}_{qps,\phi\gamma}
\Psi_\text{fix} 
\begin{pmatrix}
\includegraphics[scale=.40]{pent3g} 
\end{pmatrix},
\end{align}
which leads to:
\begin{equation}
\underset{\epsilon}{\sum} 
\mathcal{F}^{mkn,\beta\chi}_{lpq,\delta\epsilon} 
\mathcal{F}^{ijm,\alpha\epsilon}_{qps,\phi\gamma}
\simeq
\underset{t\eta\psi\kappa}{\sum} 
\mathcal{F}^{ijm,\alpha\beta}_{knt,\eta\psi} 
\mathcal{F}^{itn,\psi\chi}_{lps,\kappa\gamma} 
\mathcal{F}^{jkt,\eta\kappa}_{lsq,\delta\phi}
.
\label{ffpenta}
\end{equation}

By eliminating the Majorana numbers and canceling out the constant phase factors via a proper phase shift of the $F$-symbol, we can use a constant phase factor to change $\simeq$ into $=$:
\begin{equation}
\underset{\epsilon}{\sum} 
F^{mkn,\beta\chi}_{lpq,\delta\epsilon} 
F^{ijm,\alpha\epsilon}_{qps,\phi\gamma}
=
(-1)^{s(\alpha)s(\delta)}
\underset{t\eta\psi\kappa}{\sum} 
F^{ijm,\alpha\beta}_{knt,\eta\psi} 
F^{itn,\psi\chi}_{lps,\kappa\gamma} 
F^{jkt,\eta\kappa}_{lsq,\delta\phi}
.
\label{fpenta}
\end{equation}

\end{widetext}

\subsection{$O$-move}

The second type of wavefunction renormalization is the $O$-move, 
graphically expressed as:

\begin{align}
\psi_\text{fix}\begin{pmatrix} \includegraphics[scale=.40]{ioip} \end{pmatrix}
\simeq
O^{ij,\alpha\beta}_{k}
\psi_\text{fix}
\begin{pmatrix} \includegraphics[scale=.40]{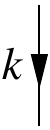} \end{pmatrix}.
\end{align}


We only permit parity-even $O$-move, i.e., 
\begin{equation}
O^{ij,\alpha\beta}_{k}=0 \text{ when }
N^{ij}_{k}<1 \text{ or } 
s(\alpha)+s(\beta)=1 \text{ mod } 2.
\label{O2}
\end{equation} 
The support space of $\psi_\text{fix}\begin{pmatrix}
\includegraphics[scale=.35]{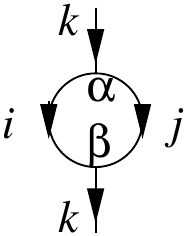}\end{pmatrix}$ should also mod out the following equivalence relation when $k$ is q-type:
\begin{align}
\psi_{\text{fix}}\begin{pmatrix} \includegraphics[scale=.40]{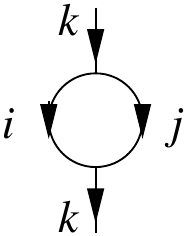} 
\end{pmatrix}
\sim
\psi_{\text{fix}}\begin{pmatrix} \includegraphics[scale=.40]{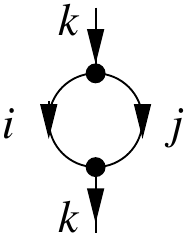} 
\end{pmatrix}.
\label{eo1}
\end{align}

We still use the convention to assign Majorana numbers from top to bottom and define the fermin ordering independent wavefunction as:
\begin{align}
\Psi_{\text{fix}}\begin{pmatrix} \includegraphics[scale=.40]{O1} 
\end{pmatrix}
=
\theta^{s(\alpha)}_{\underline{\alpha}}\theta^{s(\beta)}_{\underline{\beta}}
\psi_{\text{fix}}\begin{pmatrix} \includegraphics[scale=.40]{O1} 
\end{pmatrix}
\label{go}.
\end{align}
and rewrite the ordering independent $\mathcal{O}$-move as:
\begin{align}
\Psi_\text{fix}\begin{pmatrix} \includegraphics[scale=.40]{ioip} \end{pmatrix}
\simeq
\mathcal{O}^{ij,\alpha\beta}_{k}
\Psi_\text{fix}
\begin{pmatrix} \includegraphics[scale=.40]{kline} \end{pmatrix}.
\end{align}
where the fermionic $\mathcal{O}$-move is defined as:
\begin{equation}
\mathcal{O}^{ij,\alpha\beta}_{k} =
\theta^{s(\alpha)}_{\underline{\alpha}}\theta^{s(\beta)}_{\underline{\beta}}
O^{ij,\alpha\beta}_{k},
\label{O4}
\end{equation}

However, the $O$-move  itself is not a gFLU transformation in general, as in the patch $\psi_\text{fix}\begin{pmatrix}
\includegraphics[scale=.35]{iOip}\end{pmatrix}$, fermion parity-odd states actually exist when $k$ is q-type, but we only permit parity-even $O$-move. Therefore, we should define a three-vertices $\widetilde{O}$-move as a gFLU transformation, which includes the following six different cases:
\begin{align}
\psi_\text{fix}\begin{pmatrix} \includegraphics[scale=.40]{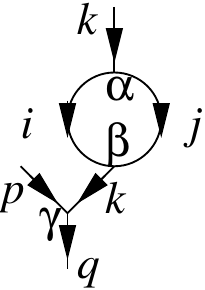} \end{pmatrix}
\simeq
\widetilde{O_1}^{ij,\alpha\beta\gamma}_{kpq,\lambda}
\psi_\text{fix}
\begin{pmatrix} \includegraphics[scale=.40]{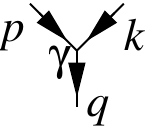} \end{pmatrix},
\end{align}
\begin{align}
\psi_\text{fix}\begin{pmatrix} \includegraphics[scale=.40]{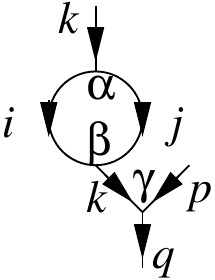} \end{pmatrix}
\simeq
\widetilde{O_2}^{ij,\alpha\beta\gamma}_{kpq,\lambda}
\psi_\text{fix}
\begin{pmatrix} \includegraphics[scale=.40]{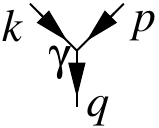} \end{pmatrix},
\end{align}
\begin{align}
\psi_\text{fix}\begin{pmatrix} \includegraphics[scale=.40]{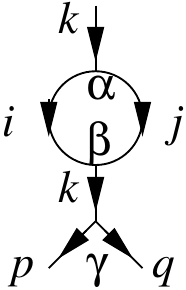} \end{pmatrix}
\simeq
\widetilde{O_3}^{ij,\alpha\beta\gamma}_{kpq,\lambda}
\psi_\text{fix}
\begin{pmatrix} \includegraphics[scale=.40]{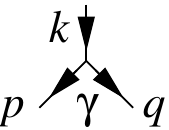} \end{pmatrix},
\end{align}
\begin{align}
\psi_\text{fix}\begin{pmatrix} \includegraphics[scale=.40]{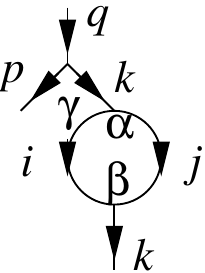} \end{pmatrix}
\simeq
\widetilde{O_4}^{ij,\alpha\beta\gamma}_{kpq,\lambda}
\psi_\text{fix}
\begin{pmatrix} \includegraphics[scale=.40]{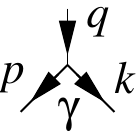} \end{pmatrix},
\end{align}
\begin{align}
\psi_\text{fix}\begin{pmatrix} \includegraphics[scale=.40]{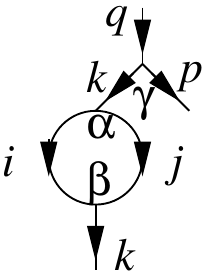} \end{pmatrix}
\simeq
\widetilde{O_5}^{ij,\alpha\beta\gamma}_{kpq,\lambda}
\psi_\text{fix}
\begin{pmatrix} \includegraphics[scale=.40]{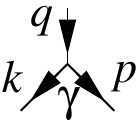} \end{pmatrix},
\end{align}
\begin{align}
\psi_\text{fix}\begin{pmatrix} \includegraphics[scale=.40]{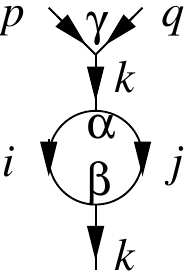} \end{pmatrix}
\simeq
\widetilde{O_6}^{ij,\alpha\beta\gamma}_{kpq,\lambda}
\psi_\text{fix}
\begin{pmatrix} \includegraphics[scale=.40]{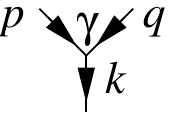} \end{pmatrix},
\end{align}
where in all cases the state $\lambda$ is related to $\gamma$ by:
\begin{equation}
\lambda=
\left\{\begin{array}{l}
\gamma
\text{, \ \ if }s(\alpha)+s(\beta)=0
\\ 
\gamma\times f
\text{, \ \ if }s(\alpha)+s(\beta)=1  \text{ and }k \text{ is q-type}
\end{array}\right.
.
\end{equation}
Different from the original $O$-move, when $k$ is q-type, our newly defined $\widetilde{O}$-move permits parity-odd sector, i.e.  $s(\alpha)+s(\beta)=1$. In this case, the additional fermion is moved to the third vertex, as a fermion can slide on a q-type string. This is the reason why we have $\lambda=\gamma\times f$ when $s(\alpha)+s(\beta)=1  \text{ and }k \text{ is q-type}$. 

We define the fermionic three-vertex $\widetilde{O}$-move as
\begin{equation}
\widetilde{\mathcal{O}_a}^{ij,\alpha\beta\gamma}_{kpq,\lambda} =
\theta^{s(\alpha)}_{\underline{\alpha}}
\theta^{s(\beta)}_{\underline{\beta}}
\theta^{s(\gamma)}_{\underline{\gamma}}
\theta^{s(\lambda)}_{\underline{\gamma}}
\widetilde{O_a}^{ij,\alpha\beta\gamma}_{kpq,\lambda}
\text{, \ \ for }a=1,2,3,
\end{equation}
and
\begin{equation}
\widetilde{\mathcal{O}_a}^{ij,\gamma\alpha\beta}_{kpq,\lambda} =
\theta^{s(\gamma)}_{\underline{\gamma}}
\theta^{s(\alpha)}_{\underline{\alpha}}
\theta^{s(\beta)}_{\underline{\beta}}
\theta^{s(\lambda)}_{\underline{\gamma}}
\widetilde{O_a}^{ij,\gamma\alpha\beta}_{kpq,\lambda}
\text{, \ \ for }a=4,5,6.
\end{equation}

Since the $\widetilde{\mathcal{O}}$-move is a gFLU transformation, after cancelling out the Majorana numbers, it must satisfy:
\begin{equation}
\underset{ij\alpha\beta}{\sum} 
\widetilde{O_a}^{ij,\alpha\beta\gamma}_{kpq,\lambda} 
(\widetilde{O_a}^{ij,\alpha\beta\gamma}_{kpq,\lambda})^*
=
1
\text{, \ \ for }a=1,2,3,4,5,6.
\label{UO}
\end{equation}
From Eq.(\ref{UO}), depending on whether $k$ is m-type or q-type, we have the following conditions for $O$-move:

(1) When $k$ is m-type,
 \begin{equation}
\underset{ij\alpha\beta}{\sum} 
O^{ij,\alpha\beta}_{k} (O^{ij,\alpha\beta}_{k})^{*}
=
1,
\label{O3m}
\end{equation}
This is because $\widetilde{O}$ is identical to $O$-move when $k$ is m-type. 

(2) When $k$ is q-type, we divide Eq.(\ref{UO}) in parity-even and odd sectors:
\begin{align}
&\underset{ij\alpha\beta}{\sum} 
\widetilde{O_a}^{ij,\alpha\beta\gamma}_{kpq,\lambda} 
(\widetilde{O_a}^{ij,\alpha\beta\gamma}_{kpq,\lambda})^*
\delta_{s(\alpha)s(\beta)}\nonumber\\
&+
\underset{ij\alpha\beta}{\sum} 
\widetilde{O_a}^{ij,\alpha\beta\gamma}_{kpq,\lambda} 
(\widetilde{O_a}^{ij,\alpha\beta\gamma}_{kpq,\lambda})^*
\delta_{s(\alpha)(s(\beta)+1)}
=1.
\end{align}
For parity-even sector $s(\alpha)=s(\beta)$, the three-vertex $\widetilde{O}$-move just equals to the  corresponding $O$-move. While for parity-odd sector $s(\alpha)=s(\beta)+1$, the three-vertex $\widetilde{O}$-move differs from the corresponding $O$-move by sliding a fermion or creating two fermions on a q-type string, which can at most cause a phase difference(see more detailed computations for all $\widetilde{O}$-move below). Thus we can replace the three-vertex $\widetilde{O}$-move by the original $O$-move, where in the parity-odd sector the general phase factor difference cancels out, we finally get:
 \begin{equation}
 2
\underset{ij\alpha\beta}{\sum} 
O^{ij,\alpha\beta}_{k} (O^{ij,\alpha\beta}_{k})^{*}
=
1.
\label{O3q}
\end{equation}

Combining Eq.(\ref{O3m}) and Eq.(\ref{O3q}), the original $O$-move satisfies:
\begin{equation}
n_{k}
\underset{ij\alpha\beta}{\sum} 
O^{ij,\alpha\beta}_{k} (O^{ij,\alpha\beta}_{k})^{*}
=
1,
\label{O3}
\end{equation}
We stress that since the parity-odd states actually exist when $k$ is q-type, there is also a equivalence relation:
\begin{align}
\psi_{\text{fix}}\begin{pmatrix} \includegraphics[scale=.40]{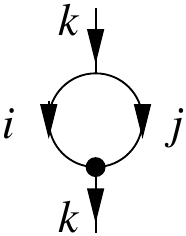} 
\end{pmatrix}\sim
\psi_{\text{fix}}\begin{pmatrix} \includegraphics[scale=.40]{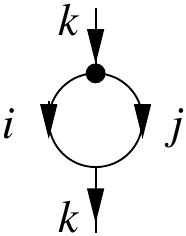} 
\end{pmatrix},
\label{eo2}
\end{align}
Physically, as a fermion can slide freely on a q-type string, we can move the fermion outside the patch and still apply the parity-even $O$-move. Such a scheme may only cause a phase factor difference, which is exactly achieved by our newly defined three-vertex $\widetilde{O}$-move.

We define $\Lambda^{ij,\alpha\beta}_{k}$ as the phase difference of the two equivalent states:
\begin{align}
\psi_{\text{fix}}\begin{pmatrix} \includegraphics[scale=.40]{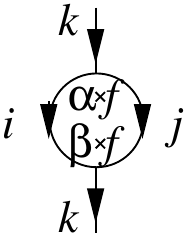} 
\end{pmatrix}
=
\Lambda^{ij,\alpha\beta}_{k}
\psi_{\text{fix}}\begin{pmatrix} \includegraphics[scale=.40]{ioip} 
\end{pmatrix}
,
\nonumber\\
\text{\ \ \ \ \ \ \ \ \ \ \ \ \ \ \ \ \ \ \ \ \ \ \ \ \ \ \ \ \ \ \ \ \ \ \ \ \ \ \ \ \ \ \ \ \ \ \ \ 
if }k\text{ is q-type,}
\end{align}
from which we have a relation between the $O$-moves of two equivalent states:
\begin{align}
O^{ij,(\alpha\times f)(\beta\times f)}_{k}
=
\Lambda^{ij,\alpha\beta}_{k}
O^{ij,\alpha\beta}_{k}   
\text{, \ \ if }k\text{ is q-type}
,
\label{eo}
\end{align}
where the phase factor $\Lambda^{ij,\alpha\beta}_{k}$ generally depends on strings $i,j,k$ and fusion states $\alpha,\beta$. The phase factor has the property $(\Lambda^{ij,\alpha\beta}_{k})^*=\Lambda^{ij,(\alpha\times f)(\beta\times f)}_{k}$.

\subsection{$Y$-move}

The third type of wavefunction renormalization is the $Y$-move, which is a completeness condition relating $\psi_\text{fix}\begin{pmatrix}
\includegraphics[scale=.35]{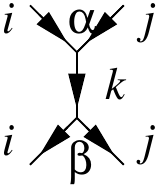}\end{pmatrix}$ to $\psi_\text{fix}\begin{pmatrix}
\includegraphics[scale=.35]{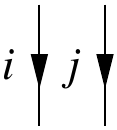}\end{pmatrix}$. Similarly when $k$ is q-type, the support space of $\psi_\text{fix}\begin{pmatrix}
\includegraphics[scale=.35]{Y}\end{pmatrix}$ should mod out the following equivalence relations:
\begin{align}
\psi_{\text{fix}}\begin{pmatrix} \includegraphics[scale=.40]{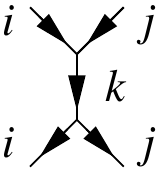} 
\end{pmatrix}\sim
\psi_{\text{fix}}\begin{pmatrix} \includegraphics[scale=.40]{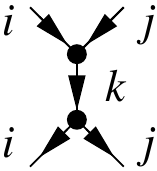} 
\end{pmatrix}.
\label{ey1}
\end{align}

As the $Y$-move exists as a completeness condition, we can always assume that in the above two equivalent states, the changing of fermion parity never changes the bosonic fusion states. In this paper, we denote a changing of fermion parity that \textit{may} change the bosonic state as $\times f$, and denote a changing of fermion parity that never changes the bosonic state as $\cdot f$ (The definition of $\cdot f$ is enclosed in the definition of $\times f$). 

When $k$ is m-type, the completeness condition is graphically expressed as:
\begin{align}
\sum_{k\alpha\beta}
Y^{ij}_{k,\alpha\beta}
\psi_\text{fix}\begin{pmatrix} \includegraphics[scale=.40]{Y} \end{pmatrix}
\simeq
\psi_\text{fix}
\begin{pmatrix} \includegraphics[scale=.40]{ijline} \end{pmatrix}.
\label{ym}
\end{align}
Specially, when $k$ is q-type, it is written as:
\begin{align}
&\sum_{kB(\alpha)B(\beta)}
Y^{ij}_{k,\alpha\beta}
(
\psi_\text{fix}\begin{pmatrix} \includegraphics[scale=.40]{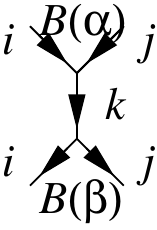} \end{pmatrix}
+
(\widetilde{\Lambda}^{ij,\alpha\beta}_{k})^*
\psi_\text{fix}\begin{pmatrix} \includegraphics[scale=.40]{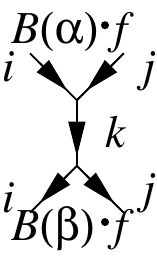} \end{pmatrix}
)
\nonumber\\
&\ \ \ \ \ \ \ \ \ \ \ \ \ \ \ \ \ \ \ \ \ \ \ \ \ \ \ \ \ \ \ \ \ \ \ \ \ \ \ \ \ \ \ \ \ \ \ \ \ \ \ \ \ 
\simeq
\psi_\text{fix}
\begin{pmatrix} \includegraphics[scale=.40]{ijline} \end{pmatrix},
\label{yq}
\end{align}
where the weight coefficient $Y^{ij}_{k,\alpha\beta}$ should count for the pair of two equivalent states, and the summation is over all bosonic states of $\alpha$ and $\beta$. $\widetilde{\Lambda}^{ij,\alpha\beta}_{k}$ is defined as the phase difference of the two equivalent states:
\begin{align}
\psi_{\text{fix}}\begin{pmatrix} \includegraphics[scale=.40]{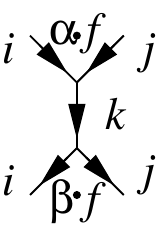} 
\end{pmatrix}
=
\widetilde{\Lambda}^{ij,\alpha\beta}_{k}
\psi_{\text{fix}}\begin{pmatrix} \includegraphics[scale=.40]{Y} 
\end{pmatrix}
,
\nonumber\\
\text{\ \ \ \ \ \ \ \ \ \ \ \ \ \ \ \ \ \ \ \ \ \ \ \ \ \ \ \ \ \ \ \ \ \ \ \ \ \ \ \ \ \ \ \ \ \ \ \ 
if }k\text{ is q-type,}
\label{yy1}
\end{align}
from which we have a relation between the $Y$-moves of two equivalent states:
\begin{align}
Y^{ij}_{k,(\alpha\cdot f)(\beta\cdot f)}
=
(\widetilde{\Lambda}^{ij}_{k,\alpha\beta})^*
Y^{ij}_{k,\alpha\beta}
\text{, \ \ if }k\text{ is q-type}
,
\label{ey}
\end{align}
where $\widetilde{\Lambda}^{ij}_{k,\alpha\beta}$ is also a phase factor satisfying $(\widetilde{\Lambda}^{ij}_{k,\alpha\beta})^*=\widetilde{\Lambda}^{ij}_{k,(\alpha\cdot f)(\beta\cdot f)}$.

However, since we have Eq.(\ref{ey}), $Y^{ij}_{k,\alpha\beta}(\widetilde{\Lambda}^{ij,\alpha\beta}_{k})^*$ can be rewritten as $Y^{ij}_{k,(\alpha\cdot f)(\beta\cdot f)}$, and Eq.(\ref{yq}) can be still written in the form of Eq.(\ref{ym}). But we should keep in mind that when $k$ is q-type, the two equivalents states are always paired in counting weights in the completeness condition.

Similar to $\mathcal{O}$-move, we can also define
the fermionic $\mathcal{Y}$-move as:
\begin{equation}
\mathcal{Y}^{ij}_{k,\alpha\beta} =
\theta^{s(\beta)}_{\underline{\beta}}
\theta^{s(\alpha)}_{\underline{\alpha}}
Y^{ij}_{k,\alpha\beta}.
\label{Y2}
\end{equation}

\subsection{A gauge freedom and a relation between $O$-move and $Y$-move}

There is a gauge freedom in the bosonic states in the support space $V^{ij}_{k}$, i.e. we can do the following transformation on the fixed-point wavefunctions:
\begin{align}
\Psi_\text{fix}\begin{pmatrix} \includegraphics[scale=.40]{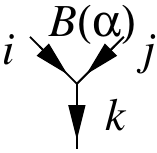} \end{pmatrix}
\simeq
\sum_{B(\beta)}
u^{ij,B(\alpha)}_{k,B(\beta)}
\Psi_\text{fix}
\begin{pmatrix} \includegraphics[scale=.40]{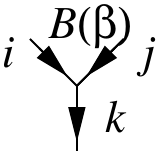} \end{pmatrix},
\end{align}
where $u^{ij}_{k}$ is a unitary matrix.

Therefore, since we only permit parity-even $O$-move, i.e. the fermion parity on the two vertices are always the same, we can make a gauge choice on the $O$-move such that the basis choices in the fusion space $V^{ij}_{k}$ and the splitting space $V^{k}_{ij}$ are always the same, i.e.
\begin{equation}
O^{ij,\alpha\beta}_{k}
=
O^{ij,\alpha}_{k}\delta_{\alpha\beta}
\label{Ogauge}
\end{equation}
Under such a gauge, Eq.(\ref{eo}) is written as:
\begin{align}
O^{ij,(\alpha\times f)}_{k}
=
\Lambda^{ij,\alpha}_{k}
O^{ij,\alpha}_{k}
\text{, \ \ if }k\text{ is q-type.}
\label{org}
\end{align}

Similarly, we can make the same gauge choice on $Y$-move:
\begin{equation}
Y^{ij}_{k,\alpha\beta}
=
Y^{ij}_{k,\alpha}\delta_{\alpha\beta},
\label{Ygauge}
\end{equation}
and Eq.(\ref{ey}) can also be simplified as:
\begin{align}
Y^{ij}_{k,(\alpha\cdot f)}
=
(\widetilde{\Lambda}^{ij}_{k,\alpha})^*
Y^{ij}_{k,\alpha}   
\text{, \ \ if }k\text{ is q-type.}
\label{yrg}
\end{align}

\begin{widetext}

There is a relation between some ordering-independent $\mathcal{O}$-moves and a $\mathcal{Y}$-move. We discuss in two cases. Depending on $k$ is m-type or q-type, we have:

(1) If $k$ is m-type, 

\begin{align}
\theta^{s(\alpha)}_{\underline{\alpha}}
\theta^{s(\alpha)}_{\underline{\lambda}}
O^{ij,\alpha}_{k}
\Psi_\text{fix}\begin{pmatrix} \includegraphics[scale=.40]{kline} \end{pmatrix}
&\simeq
\Psi_\text{fix}
\begin{pmatrix} \includegraphics[scale=.40]{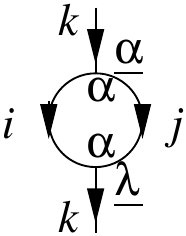} \end{pmatrix}
\simeq
\sum_{k'\beta}
\mathcal{Y}^{ij}_{k',\beta}
\Psi_\text{fix}
\begin{pmatrix} \includegraphics[scale=.40]{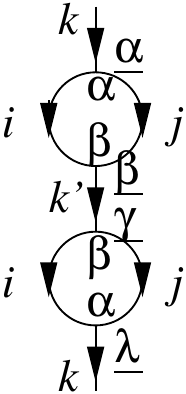} \end{pmatrix}
\simeq
\mathcal{Y}^{ij}_{k,\alpha}
\Psi_\text{fix}
\begin{pmatrix} \includegraphics[scale=.40]{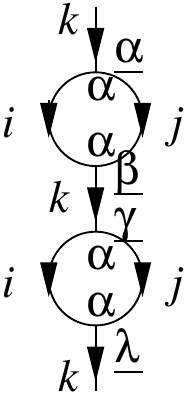} \end{pmatrix}
\nonumber\\
&\simeq
\theta^{s(\alpha)}_{\underline{\gamma}}
\theta^{s(\alpha)}_{\underline{\beta}}
Y^{ij}_{k,\alpha}
\theta^{s(\alpha)}_{\underline{\alpha}}
\theta^{s(\alpha)}_{\underline{\beta}}
O^{ij,\alpha}_{k}
\theta^{s(\alpha)}_{\underline{\gamma}}
\theta^{s(\alpha)}_{\underline{\lambda}}
O^{ij,\alpha}_{k}
\Psi_\text{fix}
\begin{pmatrix} \includegraphics[scale=.40]{kline} \end{pmatrix}
\simeq
\theta^{s(\alpha)}_{\underline{\alpha}}
\theta^{s(\alpha)}_{\underline{\lambda}}
Y^{ij}_{k,\alpha}
O^{ij,\alpha}_{k}
O^{ij,\alpha}_{k}
\Psi_\text{fix}
\begin{pmatrix} \includegraphics[scale=.40]{kline} \end{pmatrix}
,
\end{align}
we have:
\begin{equation}
1
=
Y^{ij}_{k,\alpha}O^{ij,\alpha}_{k},
\end{equation}
where we can choose the constant phase of $Y^{ij}_{k,\alpha}$ such that $\simeq$ is replaced by $=$.

(2) If $k$ is q-type,

\begin{align}
&\ \ \ \ \ \ \ \ \ \ \ \ \ \ 
\theta^{s(\alpha)}_{\underline{\alpha}}
\theta^{s(\alpha)}_{\underline{\lambda}}
O^{ij,\alpha}_{k}
\Psi_\text{fix}\begin{pmatrix} \includegraphics[scale=.40]{kline} \end{pmatrix}
\simeq
\Psi_\text{fix}
\begin{pmatrix} \includegraphics[scale=.40]{OY1} \end{pmatrix}
\nonumber\\
&\simeq
\sum_{k'\beta}
\mathcal{Y}^{ij}_{k',\beta}
\Psi_\text{fix}
\begin{pmatrix} \includegraphics[scale=.40]{OY2} \end{pmatrix}
\simeq
\mathcal{Y}^{ij}_{k,\alpha}
\Psi_\text{fix}
\begin{pmatrix} \includegraphics[scale=.40]{OY3} \end{pmatrix}
+
\mathcal{Y}^{ij}_{k,(\alpha\cdot f)}
\Psi_\text{fix}
\begin{pmatrix} \includegraphics[scale=.40]{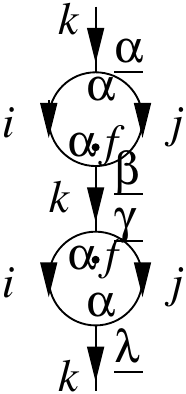} \end{pmatrix}
\nonumber\\
&\simeq
\theta^{s(\alpha)}_{\underline{\gamma}}
\theta^{s(\alpha)}_{\underline{\beta}}
Y^{ij}_{k,\alpha}
\theta^{s(\alpha)}_{\underline{\alpha}}
\theta^{s(\alpha)}_{\underline{\beta}}
O^{ij,\alpha}_{k}
\Psi_\text{fix}
\begin{pmatrix} \includegraphics[scale=.40]{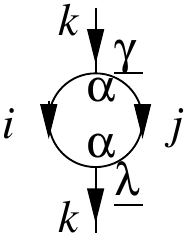} \end{pmatrix}
+
\theta^{s(\alpha\cdot f)}_{\underline{\gamma}}
\theta^{s(\alpha\cdot f)}_{\underline{\beta}}
Y^{ij}_{k,(\alpha\cdot f)}
\theta^{s(\alpha)}_{\underline{\alpha}}
\theta^{s(\alpha\times f)}_{\underline{\beta}}
\theta^{s(\alpha\times f)}_{\underline{\gamma}}
\theta^{s(\alpha)}_{\underline{\gamma}}
\widetilde{O_3}^{ij,\alpha(\alpha\cdot f)(\alpha\cdot f)}_{kij,\alpha}
\Psi_\text{fix}
\begin{pmatrix} \includegraphics[scale=.40]{OY6} \end{pmatrix}
\nonumber\\
&\simeq
\theta^{s(\alpha)}_{\underline{\alpha}}
\theta^{s(\alpha)}_{\underline{\gamma}}
Y^{ij}_{k,\alpha}
O^{ij,\alpha}_{k}
\theta^{s(\alpha)}_{\underline{\gamma}}
\theta^{s(\alpha)}_{\underline{\lambda}}
O^{ij,\alpha}_{k}
\Psi_\text{fix}
\begin{pmatrix} \includegraphics[scale=.40]{kline} \end{pmatrix}
+
\theta^{s(\alpha)}_{\underline{\alpha}}
\theta^{s(\alpha)}_{\underline{\gamma}}
Y^{ij}_{k,(\alpha\cdot f)}
\widetilde{O_3}^{ij,\alpha(\alpha\cdot f)(\alpha\cdot f)}_{kij,\alpha}
\theta^{s(\alpha)}_{\underline{\gamma}}
\theta^{s(\alpha)}_{\underline{\lambda}}
O^{ij,\alpha}_{k}
\Psi_\text{fix}
\begin{pmatrix} \includegraphics[scale=.40]{kline} \end{pmatrix}
\nonumber\\
&\simeq
2
\theta^{s(\alpha)}_{\underline{\alpha}}
\theta^{s(\alpha)}_{\underline{\lambda}}
Y^{ij}_{k,\alpha}
O^{ij,\alpha}_{k}
O^{ij,\alpha}_{k}
\Psi_\text{fix}
\begin{pmatrix} \includegraphics[scale=.40]{kline} \end{pmatrix}
,
\end{align}


where $\widetilde{O_3}^{ij,\alpha(\alpha\cdot f)(\alpha\cdot f)}_{kij,\alpha}$ is related to $O^{ij,\alpha}_{k}$ by:
\begin{align}
\widetilde{O_3}^{ij,\alpha(\alpha\cdot f)(\alpha\cdot f)}_{kij,\alpha}
\psi_\text{fix}
\begin{pmatrix} \includegraphics[scale=.40]{OY6} \end{pmatrix}
\simeq
\psi_\text{fix}\begin{pmatrix} \includegraphics[scale=.40]{OY4} \end{pmatrix}
=
\widetilde{\Lambda}^{ij}_{k,\alpha}
\psi_\text{fix}\begin{pmatrix} \includegraphics[scale=.40]{OY3} \end{pmatrix}
\simeq
\widetilde{\Lambda}^{ij}_{k,\alpha}
O^{ij,\alpha}_{k}
\psi_\text{fix}\begin{pmatrix} \includegraphics[scale=.40]{OY6} \end{pmatrix}
.
\label{deriO3}
\end{align}

\end{widetext}
We note that when we derive the relation on equivalent states, we can consider the fixed-point wavefunctions without Majorana numbers $\psi_\text{fix}$.  But when we derive relations among different renormalization moves, we should always consider fermionic fixed-point wavefuntions $\Psi_\text{fix}$ attached with Majorana numbers. 

From Eq.(\ref{yrg}) and Eq.(\ref{deriO3}), we have $Y^{ij}_{k,\alpha}
O^{ij,\alpha}_{k}=Y^{ij}_{k,(\alpha\cdot f)}
\widetilde{O_3}^{ij,\alpha(\alpha\cdot f)(\alpha\cdot f)}_{kij,\alpha}$. Therefore we obtain

\begin{equation}
1
=
2Y^{ij}_{k,\alpha}O^{ij,\alpha}_{k}.
\label{1yo}
\end{equation}
where we also choose the convention to eliminate the phase difference on both side. Replacing the $Y$-move and $O$-move by the equivalence relations in Eq.(\ref{org}) and Eq.(\ref{yrg}), we get:
\begin{align}
\widetilde{\Lambda}^{ij}_{k,\alpha}
Y^{ij}_{k,(\alpha\cdot f)}
=
\frac{1}{2(\Lambda^{ij,\alpha}_{k})^*
O^{ij,(\alpha\times f)}_{k}},
\label{fyo}
\end{align}
where we note that generally $\alpha\cdot f$ is different from $\alpha\times f$, as generally the bosonic states can be changed in the equivalence relations of $O$-move.


Combining the two cases that $k$ is m-type or q-type, we obtain:
\begin{equation}
Y^{ij}_{k,\alpha}=\frac{1}{n_{k}O^{ij,\alpha}_{k}}.
\label{OY2}
\end{equation}

In addition, from derivation in Eq.(\ref{deriO3}), we see that $\widetilde{O_1}^{ij,\alpha\beta\gamma}_{kpq,\lambda}$ and $\widetilde{O_6}^{ij,\alpha\beta\gamma}_{kpq,\lambda}$ are related:
\begin{equation}
\widetilde{O_3}^{ij,\alpha\beta\gamma}_{kpq,\lambda}
=
\widetilde{O_6}^{ij,\beta\alpha\gamma}_{kpq,\lambda},
\end{equation}

\subsection{Dual $F$-move and a relation between $O$-move and $F$-move}

We can also define a fermionic dual $\mathcal{F}$-move as the following local projective unitary transformation:
\begin{align}
\Psi_\text{fix}\begin{pmatrix} \includegraphics[scale=.40]{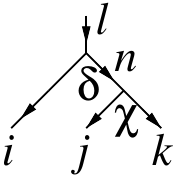} \end{pmatrix}
\simeq
\sum_{m\alpha\beta}
\widetilde{\mathcal{F}}^{ijm,\alpha\beta}_{kln,\chi\delta}
\Psi_\text{fix}
\begin{pmatrix} \includegraphics[scale=.40]{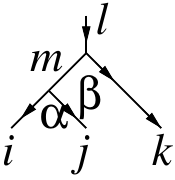} \end{pmatrix},
\label{dualF1}
\end{align}
where
\begin{equation}
\widetilde{\mathcal{F}}^{ijm,\alpha\beta}_{kln,\chi\delta} =
\theta^{s(\delta)}_{\underline{\delta}}
\theta^{s(\chi)}_{\underline{\chi}}
\theta^{s(\alpha)}_{\underline{\alpha}}\theta^{s(\beta)}_{\underline{\beta}}
\widetilde{F}^{ijm,\alpha\beta}_{kln,\chi\delta}.
\label{dualF2}
\end{equation}

When $n$ is q-type, we define $\Xi^{'ij}_{kln,\chi\delta}$ as the phase difference of these two equivalent states:
\begin{align}
\psi_{\text{fix}}\begin{pmatrix} \includegraphics[scale=.40]{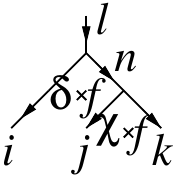} 
\end{pmatrix}
=
\widetilde{\Xi}^{ij}_{kln,\chi\delta}
\psi_{\text{fix}}\begin{pmatrix} \includegraphics[scale=.40]{dualF1} 
\end{pmatrix}
,
\nonumber\\
\text{\ \ \ \ \ \ \ \ \ \ \ \ \ \ \ \ \ \ \ \ \ \ \ \ \ \ \ \ \ \ \ \ \ \ \ \ \ \ \ \ \ \ \ \ \ \ \ \ 
if }n\text{ is q-type,}
\end{align}
from which we have another relation between the dual $F$-moves of two equivalent states:
\begin{align}
\widetilde{F}^{ijm,\alpha\beta}_{kln,(\chi\times f)(\delta\times f)}
=
\widetilde{\Xi}^{ij}_{kln,\chi\delta}
\widetilde{F}^{ijm,\alpha\beta}_{kln,\chi\delta}
\text{, \ \ if }n\text{ is q-type}
.
\label{dualFequidown}
\end{align}
When $m$ is q-type, we define $\widetilde{\Xi}^{ijm,\alpha\beta}_{kln}$ as the phase difference of these two equivalent states:
\begin{align}
\psi_{\text{fix}}\begin{pmatrix} \includegraphics[scale=.40]{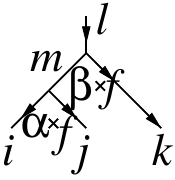} 
\end{pmatrix}
=
\widetilde{\Xi}^{ijm,\alpha\beta}_{kln}
\psi_{\text{fix}}\begin{pmatrix} \includegraphics[scale=.40]{dualF2} 
\end{pmatrix}
,
\nonumber\\
\text{\ \ \ \ \ \ \ \ \ \ \ \ \ \ \ \ \ \ \ \ \ \ \ \ \ \ \ \ \ \ \ \ \ \ \ \ \ \ \ \ \ \ \ \ \ \ \ \ 
if }m\text{ is q-type,}
\end{align}
from which we have a relation between the dual $F$-moves of two equivalent states:
\begin{align}
(\widetilde{F}^{ijm,(\alpha\times f)(\beta\times f)}_{kl,\chi\delta})^*
=
\widetilde{\Xi}^{ijm,\alpha\beta}_{kln}
(\widetilde{F}^{ijm,\alpha\beta}_{kln,\chi\delta})^*
\text{, \ \ if }m\text{ is q-type}.
\label{dualFequiup}
\end{align}

\begin{widetext}

There is a relation between $O$-move, $F$-move and dual $F$-move. On one hand, depending on string $p$ is m-type or q-type, we have:

(1) If $p$ is m-type,

\begin{align}
\Psi_\text{fix}\begin{pmatrix} \includegraphics[scale=.35]{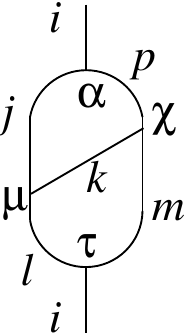} \end{pmatrix}
&\simeq
\sum_{p'\chi'\alpha'}
\mathcal{F}^{jkl,\mu\tau}_{mip',\chi'\alpha'}
\Psi_\text{fix}
\begin{pmatrix} \includegraphics[scale=.35]{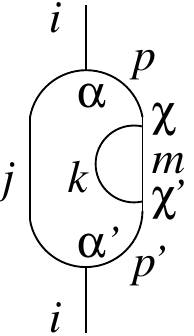} \end{pmatrix}
\simeq
\mathcal{F}^{jkl,\mu\tau}_{mip,\chi\alpha}
\Psi_\text{fix}
\begin{pmatrix} \includegraphics[scale=.35]{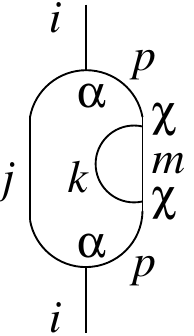} \end{pmatrix}
\nonumber\\
&\simeq
\theta^{s(\mu)}_{\underline{\mu}}
\theta^{s(\tau)}_{\underline{\tau}}
\theta^{s(\alpha)}_{\underline{\alpha'}}
\theta^{s(\chi)}_{\underline{\chi'}}
F^{jkl,\mu\tau}_{mip,\chi\alpha}
\theta^{s(\chi)}_{\underline{\chi}}
\theta^{s(\chi)}_{\underline{\chi'}}
O^{km,\chi}_{p}
\theta^{s(\alpha)}_{\underline{\alpha}}
\theta^{s(\alpha)}_{\underline{\alpha'}}
O^{jp,\alpha}_{i}
\Psi_\text{fix}
\begin{pmatrix} \includegraphics[scale=.4]{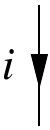} \end{pmatrix}
\nonumber\\
&\simeq
\theta^{s(\alpha)}_{\underline{\alpha}}
\theta^{s(\chi)}_{\underline{\chi}}
\theta^{s(\mu)}_{\underline{\mu}}
\theta^{s(\tau)}_{\underline{\tau}}
F^{jkl,\mu\tau}_{mip,\chi\alpha}
O^{km,\chi}_{p}
O^{jp,\alpha}_{i}
\Psi_\text{fix}
\begin{pmatrix} \includegraphics[scale=.4]{iline} \end{pmatrix}
,
\label{foo1}
\end{align}
where in the second line the Majorana numbers $\theta^{s(\mu)}_{\underline{\mu}}
\theta^{s(\tau)}_{\underline{\tau}}
\theta^{s(\alpha)}_{\underline{\alpha'}}
\theta^{s(\chi)}_{\underline{\chi'}}
\theta^{s(\chi)}_{\underline{\chi}}
\theta^{s(\chi)}_{\underline{\chi'}}
\theta^{s(\alpha)}_{\underline{\alpha}}
\theta^{s(\alpha)}_{\underline{\alpha'}}=
(-1)^{s(\alpha)+s(\chi)}
\theta^{s(\mu)}_{\underline{\mu}}
\theta^{s(\tau)}_{\underline{\tau}}
\theta^{s(\alpha)}_{\underline{\alpha}}
\theta^{s(\chi)}_{\underline{\chi}}=
\theta^{s(\alpha)}_{\underline{\alpha}}
\theta^{s(\chi)}_{\underline{\chi}}
\theta^{s(\mu)}_{\underline{\mu}}
\theta^{s(\tau)}_{\underline{\tau}}
$.

(2) If $p$ is q-type, 

\begin{align}
&\Psi_\text{fix}\begin{pmatrix} \includegraphics[scale=.35]{OF1} \end{pmatrix}
\simeq
\sum_{p'\chi'\alpha'}
\mathcal{F}^{jkl,\mu\tau}_{mip',\chi'\alpha'}
\delta_{pp'}
\Psi_\text{fix}
\begin{pmatrix} \includegraphics[scale=.35]{OF2} \end{pmatrix}
\simeq
\mathcal{F}^{jkl,\mu\tau}_{mip,\chi\alpha}
\Psi_\text{fix}
\begin{pmatrix} \includegraphics[scale=.35]{OF6} \end{pmatrix}
+
\mathcal{F}^{jkl,\mu\tau}_{mip,(\chi\times f)(\alpha\times f)}
\Psi_\text{fix}
\begin{pmatrix} \includegraphics[scale=.35]{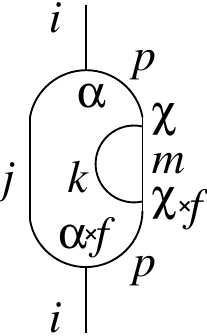} \end{pmatrix}
\nonumber\\
&\simeq
\mathcal{F}^{jkl,\mu\tau}_{mip,\chi\alpha}
\theta^{s(\chi)}_{\underline{\chi}}
\theta^{s(\chi)}_{\underline{\chi'}}
O^{km,\chi}_{p}
\Psi_\text{fix}
\begin{pmatrix} \includegraphics[scale=.35]{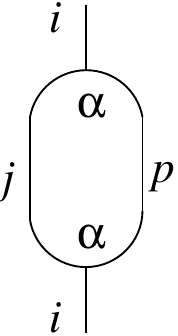} \end{pmatrix}
+
\mathcal{F}^{jkl,\mu\tau}_{mip,(\chi\times f)(\alpha\times f)}
\theta^{s(\chi)}_{\underline{\chi}}
\theta^{s(\chi\times f)}_{\underline{\chi'}}
\theta^{s(\alpha\times f)}_{\underline{\alpha'}}
\theta^{s(\alpha)}_{\underline{\alpha'}}
\widetilde{O_1}^{km,\chi(\chi\times f)(\alpha\times f)}_{pji,\alpha}
\Psi_\text{fix}
\begin{pmatrix} \includegraphics[scale=.35]{OF4} \end{pmatrix}
\nonumber\\
&\simeq
\theta^{s(\mu)}_{\underline{\mu}}
\theta^{s(\tau)}_{\underline{\tau}}
\theta^{s(\alpha)}_{\underline{\alpha'}}
\theta^{s(\chi)}_{\underline{\chi'}}
F^{jkl,\mu\tau}_{mip,\chi\alpha}
\theta^{s(\chi)}_{\underline{\chi}}
\theta^{s(\chi)}_{\underline{\chi'}}
O^{km,\chi}_{p}
\theta^{s(\alpha)}_{\underline{\alpha}}
\theta^{s(\alpha)}_{\underline{\alpha'}}
O^{jp,\alpha}_{i}
\Psi_\text{fix}
\begin{pmatrix} \includegraphics[scale=.4]{iline} \end{pmatrix}
+
\nonumber\\
&\ \ \ \ \ \ 
\theta^{s(\mu)}_{\underline{\mu}}
\theta^{s(\tau)}_{\underline{\tau}}
\theta^{s(\alpha\times f)}_{\underline{\alpha'}}
\theta^{s(\chi\times f)}_{\underline{\chi'}}
F^{jkl,\mu\tau}_{mip,(\chi\times f)(\alpha\times f)}
\theta^{s(\chi)}_{\underline{\chi}}
\theta^{s(\chi\times f)}_{\underline{\chi'}}
\theta^{s(\alpha\times f)}_{\underline{\alpha'}}
\theta^{s(\alpha)}_{\underline{\alpha'}}
\widetilde{O_1}^{km,\chi(\chi\times f)(\alpha\times f)}_{pji,\alpha}
\theta^{s(\alpha)}_{\underline{\alpha}}
\theta^{s(\alpha)}_{\underline{\alpha'}}
O^{jp,\alpha}_{i}
\Psi_\text{fix}
\begin{pmatrix} \includegraphics[scale=.4]{iline} \end{pmatrix}
\nonumber\\
&\simeq
2
\theta^{s(\alpha)}_{\underline{\alpha}}
\theta^{s(\chi)}_{\underline{\chi}}
\theta^{s(\mu)}_{\underline{\mu}}
\theta^{s(\tau)}_{\underline{\tau}}
F^{jkl,\mu\tau}_{mip,\chi\alpha}
O^{km,\chi}_{p}
O^{jp,\alpha}_{i}
\Psi_\text{fix}
\begin{pmatrix} \includegraphics[scale=.4]{iline} \end{pmatrix}
,
\label{foo1q}
\end{align}
where $\widetilde{O_1}^{km,\chi(\chi\times f)(\alpha\times f)}_{pji,\alpha}$ is related to $O^{km,\chi}_{p}$ by:
\begin{align}
\widetilde{O_1}^{km,\chi(\chi\times f)(\alpha\times f)}_{pji,\alpha}
\psi_\text{fix}
\begin{pmatrix} \includegraphics[scale=.40]{OF4} \end{pmatrix}
\simeq
\psi_\text{fix}\begin{pmatrix} \includegraphics[scale=.40]{OF7} \end{pmatrix}
=
\Xi^{jk}_{mip,\chi\alpha}
\psi_\text{fix}\begin{pmatrix} \includegraphics[scale=.40]{OF6} \end{pmatrix}
\simeq
\Xi^{jk}_{mip,\chi\alpha}
O^{km,\chi}_{p}
\psi_\text{fix}\begin{pmatrix} \includegraphics[scale=.40]{OF4} \end{pmatrix}
.
\label{deriO1}
\end{align}

Combining with the case that $p$ is m-type, the general result is given by:
\begin{align}
\Psi_\text{fix}\begin{pmatrix} \includegraphics[scale=.35]{OF1} \end{pmatrix}
\simeq
n_{p}
\theta^{s(\alpha)}_{\underline{\alpha}}
\theta^{s(\chi)}_{\underline{\chi}}
\theta^{s(\mu)}_{\underline{\mu}}
\theta^{s(\tau)}_{\underline{\tau}}
F^{jkl,\mu\tau}_{mip,\chi\alpha}
O^{km,\chi}_{p}
O^{jp,\alpha}_{i}
\Psi_\text{fix}
\begin{pmatrix} \includegraphics[scale=.4]{iline} \end{pmatrix}
,
\label{foo1t}
\end{align}

On the other hand, depending on string $l$ is m-type or q-type, we have:

(1) If $l$ is m-type,

\begin{align}
\Psi_\text{fix}\begin{pmatrix} \includegraphics[scale=.35]{OF1} \end{pmatrix}
&\simeq
\sum_{l'\mu'\tau'}
\widetilde{\mathcal{F}}^{jkl',\mu'\tau'}_{mip,\chi\alpha}
\Psi_\text{fix}
\begin{pmatrix} \includegraphics[scale=.35]{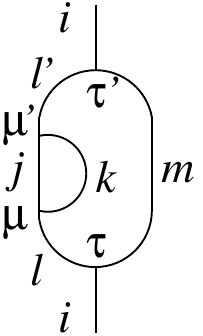} \end{pmatrix}
\simeq
\widetilde{\mathcal{F}}^{jkl,\mu\tau}_{mip,\chi\alpha}
\Psi_\text{fix}
\begin{pmatrix} \includegraphics[scale=.35]{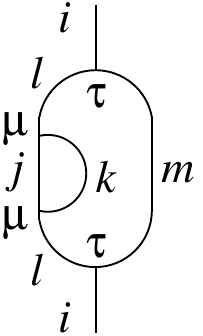} \end{pmatrix}
\nonumber\\
&\simeq
\theta^{s(\alpha)}_{\underline{\alpha}}
\theta^{s(\chi)}_{\underline{\chi}}
\theta^{s(\mu)}_{\underline{\mu'}}
\theta^{s(\tau)}_{\underline{\tau'}}
\widetilde{F}^{jkl,\mu\tau}_{mip,\chi\alpha}
\theta^{s(\mu)}_{\underline{\mu'}}
\theta^{s(\mu)}_{\underline{\mu}}
O^{jk,\mu}_{l}
\theta^{s(\tau)}_{\underline{\tau'}}
\theta^{s(\tau)}_{\underline{\tau}}
O^{lm,\tau}_{i}
\Psi_\text{fix}
\begin{pmatrix} \includegraphics[scale=.4]{iline} \end{pmatrix}
\nonumber\\
&\simeq
\theta^{s(\alpha)}_{\underline{\alpha}}
\theta^{s(\chi)}_{\underline{\chi}}
\theta^{s(\mu)}_{\underline{\mu}}
\theta^{s(\tau)}_{\underline{\tau}}
\widetilde{F}^{jkl,\mu\tau}_{mip,\chi\alpha}
O^{jk,\mu}_{l}
O^{lm,\tau}_{i}
\Psi_\text{fix}
\begin{pmatrix} \includegraphics[scale=.4]{iline} \end{pmatrix}
.
\end{align}

(2) If $l$ is q-type,

\begin{align}
&\Psi_\text{fix}\begin{pmatrix} \includegraphics[scale=.35]{OF1} \end{pmatrix}
\simeq
\sum_{l'\mu'\tau'}
\widetilde{\mathcal{F}}^{jkl',\mu'\tau'}_{mip,\chi\alpha}
\Psi_\text{fix}
\begin{pmatrix} \includegraphics[scale=.35]{OF3} \end{pmatrix}
\simeq
\widetilde{\mathcal{F}}^{jkl,\mu\tau}_{mip,\chi\alpha}
\Psi_\text{fix}
\begin{pmatrix} \includegraphics[scale=.35]{OF9} \end{pmatrix}
+
\widetilde{\mathcal{F}}^{jkl,(\mu\times f)(\tau\times f)}_{mip,\chi\alpha}
\Psi_\text{fix}
\begin{pmatrix} \includegraphics[scale=.35]{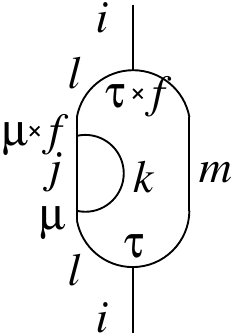} \end{pmatrix}
\nonumber\\
&\simeq
\widetilde{\mathcal{F}}^{jkl,\mu\tau}_{mip,\chi\alpha}
\theta^{s(\mu)}_{\underline{\mu'}}
\theta^{s(\mu)}_{\underline{\mu}}
O^{jk,\mu}_{l}
\Psi_\text{fix}
\begin{pmatrix} \includegraphics[scale=.35]{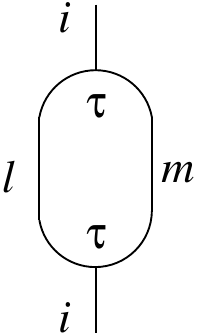} \end{pmatrix}
+
\widetilde{\mathcal{F}}^{jkl,(\mu\times f)(\tau\times f)}_{mip,\chi\alpha}
\theta^{s(\tau\times f)}_{\underline{\tau'}}
\theta^{s(\mu\times f)}_{\underline{\mu'}}
\theta^{s(\mu)}_{\underline{\mu}}
\theta^{s(\tau)}_{\underline{\tau'}}
\widetilde{O_5}^{jk,(\tau\times f)(\mu\times f)\mu}_{lmi,\tau}
\Psi_\text{fix}
\begin{pmatrix} \includegraphics[scale=.35]{OF5} \end{pmatrix}
\nonumber\\
&\simeq
\theta^{s(\alpha)}_{\underline{\alpha}}
\theta^{s(\chi)}_{\underline{\chi}}
\theta^{s(\mu)}_{\underline{\mu'}}
\theta^{s(\tau)}_{\underline{\tau'}}
\widetilde{F}^{jkl,\mu\tau}_{mip,\chi\alpha}
\theta^{s(\mu)}_{\underline{\mu'}}
\theta^{s(\mu)}_{\underline{\mu}}
O^{jk,\mu}_{l}
\theta^{s(\tau)}_{\underline{\tau'}}
\theta^{s(\tau)}_{\underline{\tau}}
O^{lm,\tau}_{i}
\Psi_\text{fix}
\begin{pmatrix} \includegraphics[scale=.4]{iline} \end{pmatrix}
+
\nonumber\\
&\ \ \ \ \ \ 
\theta^{s(\alpha)}_{\underline{\alpha}}
\theta^{s(\chi)}_{\underline{\chi}}
\theta^{s(\mu\times f)}_{\underline{\mu'}}
\theta^{s(\tau\times f)}_{\underline{\tau'}}
\widetilde{F}^{jkl,(\mu\times f)(\tau\times f)}_{mip,\chi\alpha}
\theta^{s(\tau\times f)}_{\underline{\tau'}}
\theta^{s(\mu\times f)}_{\underline{\mu'}}
\theta^{s(\mu)}_{\underline{\mu}}
\theta^{s(\tau)}_{\underline{\tau'}}
\widetilde{O_5}^{jk,(\tau\times f)(\mu\times f)\mu}_{lmi,\tau}
\theta^{s(\tau)}_{\underline{\tau'}}
\theta^{s(\tau)}_{\underline{\tau}}
O^{lm,\tau}_{i}
\Psi_\text{fix}
\begin{pmatrix} \includegraphics[scale=.4]{iline} \end{pmatrix}
\nonumber\\
&\simeq
2
\theta^{s(\alpha)}_{\underline{\alpha}}
\theta^{s(\chi)}_{\underline{\chi}}
\theta^{s(\mu)}_{\underline{\mu}}
\theta^{s(\tau)}_{\underline{\tau}}
\widetilde{F}^{jkl,\mu\tau}_{mip,\chi\alpha}
O^{jk,\mu}_{l}
O^{lm,\tau}_{i}
\Psi_\text{fix}
\begin{pmatrix} \includegraphics[scale=.4]{iline} \end{pmatrix}
,
\end{align}
where $\widetilde{O_5}^{jk,(\tau\times f)(\mu\times f)\mu}_{lmi,\tau}$ is related to $O^{jk,\mu}_{l}$ by:
\begin{align}
\widetilde{O_5}^{jk,(\tau\times f)(\mu\times f)\mu}_{lmi,\tau}
\psi_\text{fix}
\begin{pmatrix} \includegraphics[scale=.40]{OF5} \end{pmatrix}
\simeq
\psi_\text{fix}\begin{pmatrix} \includegraphics[scale=.40]{OF10} \end{pmatrix}
=
\widetilde{\Xi}^{jkl,\mu\tau}_{mi}
\psi_\text{fix}\begin{pmatrix} \includegraphics[scale=.40]{OF9} \end{pmatrix}
\simeq
\widetilde{\Xi}^{jkl,\mu\tau}_{mi}
O^{jk,\mu}_{l}
\psi_\text{fix}\begin{pmatrix} \includegraphics[scale=.40]{OF5} \end{pmatrix}
.
\label{deriO5}
\end{align}

Combining with the case that $p$ is m-type, the general result is given by:
\begin{align}
\Psi_\text{fix}\begin{pmatrix} \includegraphics[scale=.35]{OF1} \end{pmatrix}
\simeq
n_{l}
\theta^{s(\alpha)}_{\underline{\alpha}}
\theta^{s(\chi)}_{\underline{\chi}}
\theta^{s(\mu)}_{\underline{\mu}}
\theta^{s(\tau)}_{\underline{\tau}}
\widetilde{F}^{jkl,\mu\tau}_{mip,\chi\alpha}
O^{jk,\mu}_{l}
O^{lm,\tau}_{i}
\Psi_\text{fix}
\begin{pmatrix} \includegraphics[scale=.4]{iline} \end{pmatrix}
.
\label{foo2t}
\end{align}


Therefore, from Eq.(\ref{foo1t}) and Eq.(\ref{foo2t}), we have the relation:
\begin{align} 
\widetilde{F}^{jkl,\mu\tau}_{mip,\chi\alpha}
=
\frac{n_{p}}{n_{l}} F^{jkl,\mu\tau}_{mip,\chi\alpha}O^{km,\chi}_{p}O^{jp,\alpha}_{i}(O^{lm,\tau}_{i})^{-1}(O^{jk,\mu}_{l})^{-1},
\label{dualff}
\end{align}
from which we obtain two relations on phase factors, if we change the fermion parity on $\mu,\tau$ through Eq.(\ref{Fequiup}), Eq.(\ref{org}) and Eq.(\ref{dualFequiup}), and on $\chi,\alpha$ through Eq.(\ref{Fequidown}), Eq.(\ref{org}) and Eq.(\ref{dualFequidown})  respectively:
\begin{align}
\widetilde{\Xi}^{jkl,\mu\tau}_{mi}=
\Lambda^{lm,\tau}_{i}
\Lambda^{jk,\mu}_{l}
(\Xi^{jkl,\mu\tau}_{mi})^*,
\label{dualFphase}
\end{align}
\begin{align}
\widetilde{\Xi}^{jk}_{mip,\chi\alpha}
=
\Lambda^{km,\chi}_{p}
\Lambda^{jp,\alpha}_{i}
(\Xi^{jk}_{mip,\chi\alpha})^*.
\label{dualFphase2}
\end{align}

We require the dual $F$-move $\widetilde{F}^{jkl,\mu\tau}_{mip,\chi\alpha}$ also to be projective unitary:

\begin{equation}
\underset{p\chi\alpha}{\sum} 
(\widetilde{F}^{jkl,\mu\tau}_{mip,\chi\alpha})^* 
\widetilde{F}^{jkl',\mu'\tau'}_{mip,\chi\alpha}
=
\left\{\begin{array}{l}
\delta_{ll'}\delta_{\mu\mu'}\delta_{\tau\tau'}
\text{, \ \ if }l\text{ is m-type
}\\ 
\frac{1}{n_{l}} (\delta_{ll'}\delta_{\mu\mu'}\delta_{\tau\tau'}+
(\widetilde{\Xi}^{jkl,\mu\tau}_{mi})^*
\delta_{ll'}\delta_{(\mu\times f)\mu'}\delta_{(\tau\times f)\tau'})
\text{, \ \ if }l\text{ is q-type}
\end{array}\right.
\label{dualfp}
\end{equation}
Replacing Eq.(\ref{dualff}) into Eq.(\ref{dualfp}), and by Eq.(\ref{dualFphase}),
\begin{align}
\label{dualF}
&\ \ \ \ \ \ \ 
\underset{p\chi\alpha}{\sum}
(\frac{n_{p}}{n_{l}})^{2}
(F^{jkl,\mu\tau}_{mip,\chi\alpha})^*
F^{jkl',\mu'\tau'}_{mip,\chi\alpha}
\frac{\left| O^{km,\chi}_{p}O^{jp,\alpha}_{i}\right|^{2}}
{(O^{lm,\tau}_{i})^*O^{l'm,\tau'}_{i}
(O^{jk,\mu}_{l'})^*O^{jk,\mu'}_{l}}
\nonumber\\
&=\left\{\begin{array}{l}
\delta_{ll'}\delta_{\mu\mu'}\delta_{\tau\tau'}
\text{, \ \ if }l\text{ is m-type
}\\ 
\frac{1}{n_{l}} (\delta_{ll'}\delta_{\mu\mu'}\delta_{\tau\tau'}+
(\Lambda^{lm,\tau}_{i})^*
(\Lambda^{jk,\mu}_{l})^*
\Xi^{jkl,\mu\tau}_{mi}
\delta_{ll'}\delta_{(\mu\times f)\mu'}\delta_{(\tau\times f)\tau'})
\text{, \ \ if }l\text{ is q-type}
\end{array}\right.
\end{align}


We see that in the above equation, when $l$ is q-type, we have
\begin{equation}
\frac{1}
{(O^{lm,\tau}_{i})^*O^{l'm,\tau'}_{i}
(O^{jk,\mu}_{l'})^*O^{jk,\mu'}_{l}}
=
(\Lambda^{lm,\tau}_{i})^*
(\Lambda^{jk,\mu}_{l})^*
\frac{1}
{\left| O^{lm,\tau}_{i}O^{jk,\mu}_{l}\right|^{2}}
\text{, \ \ when }
\delta_{ll'}\delta_{(\mu\times f)\mu'}\delta_{(\tau\times f)\tau'}=1.
\end{equation}
Then Eq.(\ref{dualF}) can be satisfied by the following ansatz for $O^{ij,\alpha}_{k}$:
\begin{equation}
O^{ij,\alpha}_{k}=
\Phi^{ij,\alpha}_{k}
\sqrt{\frac{d_{i}d_{j}}{n_{i}n_{j}n_{k}D^{2}d_{k}}} \delta^{ij}_{k}
,
\label{ooo}
\end{equation}
where $D^{2}=\underset{i}{\sum} \frac{d^{2}_{i}}{n_{i}}$, and $\Phi^{ij,\alpha}_{k}$ is a general phase factor. And Eq.(\ref{dualF}) reduces to the projective unitary condition for $F$-move in Eq.(\ref{fprojc1}).

\end{widetext}




By Eq.(\ref{OY2}), the $Y$-move has the expression:
\begin{equation}
Y^{ij}_{k,\alpha}=
(\Phi^{ij,\alpha}_{k})^*
\sqrt{\frac{n_{i}n_{j}D^{2}d_{k}}{n_{k}d_{i}d_{j}}} \delta^{ij}_{k}
.
\label{yyy}
\end{equation}
Since we have Eq.(\ref{org}) and Eq.(\ref{yrg}), the phase $\Phi^{ij,\alpha}_{k}$ must satisfy:
\begin{equation}
\Phi^{ij,(\alpha\times f)}_{k}
=
\Lambda^{ij,\alpha}_{k}
\Phi^{ij,\alpha}_{k}
,
\label{p1}
\end{equation}
and
\begin{equation}
\Phi^{ij,(\alpha\cdot f)}_{k}
=
\widetilde{\Lambda}^{ij,\alpha}_{k}
\Phi^{ij,\alpha}_{k}
,
\label{p2}
\end{equation}
where we note that the $\alpha\times f$ in Eq.(\ref{p1}) is determined by the corresponding states in equivalent $O$-moves. Specially, when the fermion parity change in equivalent $O$-moves also does not change the bosonic state, Eq.(\ref{p1}) and Eq.(\ref{p2}) reduce to the same equation. In this special case, we have $\Lambda^{ij,\alpha}_{k}=\widetilde{\Lambda}^{ij,\alpha}_{k}$.

After taking the gauge on $O$-move in Eq.(\ref{Ogauge}), Eq.(\ref{O3}) becomes:
\begin{equation}
n_{k}\underset{ij\alpha}{\sum} O^{ij,\alpha}_{k} (O^{ij,\alpha}_{k})^{*}
=
1.
\end{equation}
Combining with Eq.(\ref{ooo}), we find the quantum dimensions satisfy:
\begin{equation}
\underset{ij}{\sum} \frac{N^{ij}_{k}d_{i}d_{j}}{n_{i}n_{j}}=d_{k}D^{2}.
\end{equation}

From derivations in Eq.(\ref{deriO1}), and Eq.(\ref{deriO5}), we have more relations between different three-vertex $\widetilde{O}$-moves:
\begin{equation}
\widetilde{O_1}^{ij,\alpha\beta\gamma}_{kpq,\lambda}
=
\widetilde{O_4}^{ij,\beta\alpha\gamma}_{kpq,\lambda},
\end{equation}

\begin{equation}
\widetilde{O_2}^{ij,\alpha\beta\gamma}_{kpq,\lambda}
=
\widetilde{O_5}^{ij,\beta\alpha\gamma}_{kpq,\lambda}.
\end{equation}

The other projective unitary condition for dual $F$-move is:

\begin{widetext}

\begin{equation}
\underset{l\mu\tau}{\sum} 
\widetilde{F}^{jkl,\mu\tau}_{mip,\chi\alpha}
(\widetilde{F}^{jkl,\mu\tau}_{mip',\chi'\alpha'})^*
=
\left\{\begin{array}{l}
\delta_{pp'}\delta_{\chi\chi'}\delta_{\alpha\alpha'}
\text{, \ \ if }p\text{ is m-type
}\\ 
\frac{1}{n_{p}} 
(\delta_{pp'}\delta_{\chi\chi'}\delta_{\alpha\alpha'}+
(\widetilde{\Xi}^{jk}_{mip,\chi\alpha})^*
\delta_{pp'}\delta_{(\chi\times f)\chi'}\delta_{(\alpha\times f)\alpha'}
\text{, \ \ if }p\text{ is q-type}
\end{array}\right.
\label{dualfp2}
\end{equation}
Replacing Eq.(\ref{dualff}) into Eq.(\ref{dualfp2}), and by Eq.(\ref{dualFphase2}), we can similarly reduce it to the projective unitary condition in Eq.(\ref{fprojc2}) with the ansatz in Eq.(\ref{ooo}).

\newpage

\end{widetext}

\subsection{$H$-move and an additional constraint between dual $H$-move and $F$-move}

We define the fermionic $\mathcal{H}$-move as the following local projective unitary transformation:
\begin{align}
\Psi_\text{fix}\begin{pmatrix} \includegraphics[scale=.45]{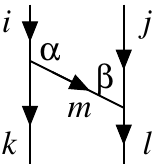} \end{pmatrix}
\simeq
\sum_{n\chi\delta}
\mathcal{H}^{kim,\alpha\beta}_{jln,\chi\delta}
\Psi_\text{fix}
\begin{pmatrix} \includegraphics[scale=.45]{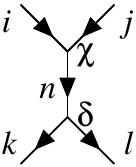} \end{pmatrix},
\label{H}
\end{align}
where
\begin{equation}
\mathcal{H}^{kim,\alpha\beta}_{jln,\chi\delta} =
\theta^{s(\alpha)}_{\underline{\alpha}}
\theta^{s(\beta)}_{\underline{\beta}}
\theta^{s(\delta)}_{\underline{\delta}}
\theta^{s(\chi)}_{\underline{\chi}}
H^{kim,\alpha\beta}_{jln,\chi\delta}.
\end{equation}

Similarly, the fermionic dual $H$-move is defined as
\begin{align}
\Psi_\text{fix}\begin{pmatrix} \includegraphics[scale=.45]{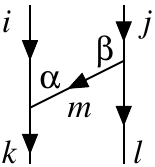} \end{pmatrix}
\simeq
\sum_{n\chi\delta}
\widetilde{\mathcal{H}}^{kim,\alpha\beta}_{jln,\chi\delta}
\Psi_\text{fix}
\begin{pmatrix} \includegraphics[scale=.45]{H1} \end{pmatrix},
\label{dualH1}
\end{align}
where
\begin{equation}
\widetilde{\mathcal{H}}^{kim,\alpha\beta}_{jln,\chi\delta} =
\theta^{s(\beta)}_{\underline{\beta}}
\theta^{s(\alpha)}_{\underline{\alpha}}
\theta^{s(\delta)}_{\underline{\delta}}
\theta^{s(\chi)}_{\underline{\chi}}
\widetilde{H}^{kim,\alpha\beta}_{jln,\chi\delta}.
\label{dualH2}
\end{equation}

Again, when $m$ is a q-type string, there exist the following equivalence relations:
\begin{align}
\psi_{\text{fix}}\begin{pmatrix} \includegraphics[scale=.40]{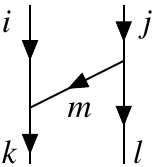} 
\end{pmatrix}\sim
\psi_{\text{fix}}\begin{pmatrix} \includegraphics[scale=.40]{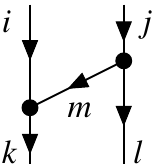} 
\end{pmatrix},
\label{eh1}
\end{align}
\begin{align}
\psi_{\text{fix}}\begin{pmatrix} \includegraphics[scale=.40]{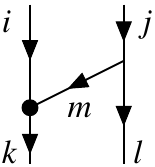} 
\end{pmatrix}\sim
\psi_{\text{fix}}\begin{pmatrix} \includegraphics[scale=.40]{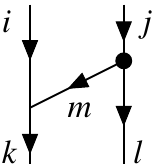} 
\end{pmatrix}.
\label{eh2}
\end{align}
When $n$ is a q-type string, there exist the following equivalence relations:
\begin{align}
\psi_{\text{fix}}\begin{pmatrix} \includegraphics[scale=.40]{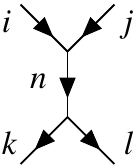} 
\end{pmatrix}\sim
\psi_{\text{fix}}\begin{pmatrix} \includegraphics[scale=.40]{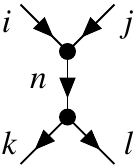} 
\end{pmatrix},
\label{eh3}
\end{align}
\begin{align}
\psi_{\text{fix}}\begin{pmatrix} \includegraphics[scale=.40]{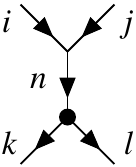} 
\end{pmatrix}\sim
\psi_{\text{fix}}\begin{pmatrix} \includegraphics[scale=.40]{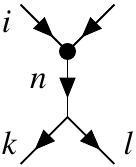} 
\end{pmatrix}.
\label{eh4}
\end{align}

For convenience, below we will show how to derive the dual $H$-move from the combination of $F$, $Y$ and $O$-moves first, and the projective unitarity condition of $H$-move will impose additional conditions on $F$-symbol.  
When $m$ is q-type, we define $\widetilde{\zeta}^{kim,\alpha\beta}_{jl}$ as the phase difference of these two equivalent states:
\begin{align}
\psi_{\text{fix}}\begin{pmatrix} \includegraphics[scale=.40]{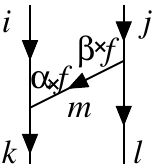} 
\end{pmatrix}
=
\widetilde{\zeta}^{kim,\alpha\beta}_{jl}
\psi_{\text{fix}}\begin{pmatrix} \includegraphics[scale=.40]{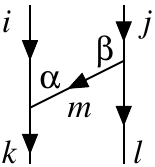} 
\end{pmatrix}
,
\nonumber\\
\text{\ \ \ \ \ \ \ \ \ \ \ \ \ \ \ \ \ \ \ \ \ \ \ \ \ \ \ \ \ \ \ \ \ \ \ \ \ \ \ \ \ \ \ \ \ \ \ \ 
if }m\text{ is q-type,}
\label{dualhe1}
\end{align}
where in general the bosonic states $B(\alpha\times f)$ and $B(\beta\times f)$ may not be the same as $B(\alpha)$ and $B(\beta)$ respectively. Thereby we have a relation between the dual $H$-moves of two equivalent states:
\begin{align}
\widetilde{H}^{kim,(\alpha\times f)(\beta\times f)}_{jln,\chi\delta}
=
\widetilde{\zeta}^{kim,\alpha\beta}_{jl}
\widetilde{H}^{kim,\alpha\beta}_{jln,\chi\delta} 
\text{, \ \ if }m\text{ is q-type}
,
\label{dualhequi1}
\end{align}
where $\widetilde{\zeta}^{kim,\alpha\beta}_{jl}$ satisfies $(\widetilde{\zeta}^{kim,\alpha\beta}_{jl})^*=\widetilde{\zeta}^{kim,(\alpha\times f)(\beta\times f)}_{jl}$.

When $n$ is q-type, we define $\widetilde{\zeta}^{ki}_{jln,\chi\delta}$ as the phase difference between these two equivalent states:
\begin{align}
\psi_{\text{fix}}\begin{pmatrix} \includegraphics[scale=.40]{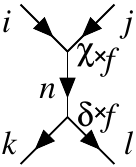} 
\end{pmatrix}
=
\widetilde{\zeta}^{ki}_{jln,\chi\delta}
\psi_{\text{fix}}\begin{pmatrix} \includegraphics[scale=.40]{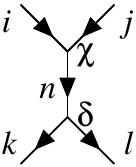} 
\end{pmatrix}
,
\nonumber\\
\text{\ \ \ \ \ \ \ \ \ \ \ \ \ \ \ \ \ \ \ \ \ \ \ \ \ \ \ \ \ \ \ \ \ \ \ \ \ \ \ \ \ \ \ \ \ \ \ \ 
if }n\text{ is q-type,}
\label{dualhh2}
\end{align}
from which we have another relation between the dual $H$-moves of two equivalent states:
\begin{align}
(\widetilde{H}^{kim,\alpha\beta}_{jln,(\chi\times f)(\delta\times f)})^*
=
\widetilde{\zeta}^{ki}_{jln,\chi\delta}
(\widetilde{H}^{kim,\alpha\beta}_{jln,\chi\delta} )^*
\text{, \ \ if }n\text{ is q-type}
,
\label{dualhequi2}
\end{align}
where $\widetilde{\zeta}^{ki}_{jln,\chi\delta}$ satisfies $(\widetilde{\zeta}^{ki}_{jln,\chi\delta})^*=\widetilde{\zeta}^{ki}_{jln,(\chi\times f)(\delta\times f)}$.

\begin{widetext}

There is a relation between the dual $H$-move and $F$-move. Depending on string $j$ is m-type or q-type, we have

(1) If $j$ is m-type,

\begin{align}
\Psi_\text{fix}\begin{pmatrix} \includegraphics[scale=.45]{H3} \end{pmatrix}
&\simeq
\sum_{n\delta}
\mathcal{Y}^{kl}_{n,\delta}
\Psi_\text{fix}
\begin{pmatrix} \includegraphics[scale=.45]{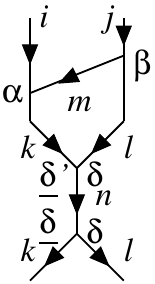} \end{pmatrix}
\simeq
\sum_{n\delta j' \beta'\chi}
\mathcal{Y}^{kl}_{n,\delta}
\mathcal{F}^{imk,\alpha\delta}_{lnj',\beta'\chi}
\Psi_\text{fix}
\begin{pmatrix} \includegraphics[scale=.45]{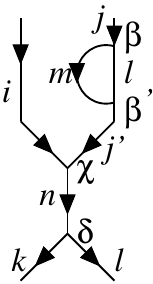} \end{pmatrix}
\simeq
\sum_{n\delta\chi}
\mathcal{Y}^{kl}_{n,\delta}
\mathcal{F}^{imk,\alpha\delta}_{lnj,\beta\chi}
\Psi_\text{fix}
\begin{pmatrix} \includegraphics[scale=.45]{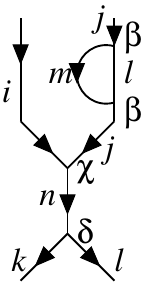} \end{pmatrix}
\nonumber\\
&\simeq
\sum_{n\chi\delta}
\theta^{s(\delta)}_{\underline{\delta}}
\theta^{s(\delta)}_{\underline{\delta'}}
Y^{kl}_{n,\delta}
 \theta^{s(\alpha)}_{\underline{\alpha}}
\theta^{s(\delta)}_{\underline{\delta'}}
 \theta^{s(\chi)}_{\underline{\chi}}
\theta^{s(\beta)}_{\underline{\beta'}}
F^{imk,\alpha\delta}_{lnj,\beta\chi}
\theta^{s(\beta)}_{\underline{\beta}}
\theta^{s(\beta)}_{\underline{\beta'}}
O^{ml,\beta}_{j}
\Psi_\text{fix}
\begin{pmatrix} \includegraphics[scale=.45]{H1} \end{pmatrix}
\nonumber\\
&\simeq
\sum_{n\chi\delta}
\theta^{s(\beta)}_{\underline{\beta}}
\theta^{s(\alpha)}_{\underline{\alpha}}
\theta^{s(\delta)}_{\underline{\delta}}
 \theta^{s(\chi)}_{\underline{\chi}}
Y^{kl}_{n,\delta}
F^{imk,\alpha\delta}_{lnj,\beta\chi}
O^{ml,\beta}_{j}
\Psi_\text{fix}
\begin{pmatrix} \includegraphics[scale=.45]{H1} \end{pmatrix}
,
\label{HF11}
\end{align}

(2) If $j$ is q-type, 

\begin{align}
&\Psi_\text{fix}\begin{pmatrix} \includegraphics[scale=.45]{H3} \end{pmatrix}
\simeq
\sum_{n\delta}
\mathcal{Y}^{kl}_{n,\delta}
\Psi_\text{fix}
\begin{pmatrix} \includegraphics[scale=.45]{H4} \end{pmatrix}
\simeq
\sum_{n\delta j' \beta'\chi}
\mathcal{Y}^{kl}_{n,\delta}
\mathcal{F}^{imk,\alpha\delta}_{lnj',\beta'\chi}
\Psi_\text{fix}
\begin{pmatrix} \includegraphics[scale=.45]{H5} \end{pmatrix}
\nonumber\\
&\simeq
\sum_{n\delta\chi}
\mathcal{Y}^{kl}_{n,\delta}
\mathcal{F}^{imk,\alpha\delta}_{lnj,\beta\chi}
\Psi_\text{fix}
\begin{pmatrix} \includegraphics[scale=.45]{H8} \end{pmatrix}
+
\sum_{n\delta\chi}
\mathcal{Y}^{kl}_{n,\delta}
\mathcal{F}^{imk,\alpha\delta}_{lnj,(\beta\times f)\chi}
\Psi_\text{fix}
\begin{pmatrix} \includegraphics[scale=.45]{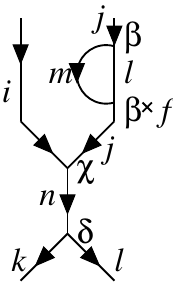} \end{pmatrix}
\nonumber\\
&\simeq
\sum_{n\delta\chi}
\theta^{s(\delta)}_{\underline{\delta}}
\theta^{s(\delta)}_{\underline{\delta'}}
Y^{kl}_{n,\delta}
\theta^{s(\alpha)}_{\underline{\alpha}}
\theta^{s(\delta)}_{\underline{\delta'}}
\theta^{s(\chi)}_{\underline{\chi}}
\theta^{s(\beta)}_{\underline{\beta'}}
F^{imk,\alpha\delta}_{lnj,\beta\chi}
\theta^{s(\beta)}_{\underline{\beta}}
\theta^{s(\beta)}_{\underline{\beta'}}
O^{ml,\beta}_{j}
\Psi_\text{fix}
\begin{pmatrix} \includegraphics[scale=.45]{H1} \end{pmatrix}
+
\nonumber\\
&\ \ \ \ \ \ 
\sum_{n\delta\chi}
\theta^{s(\delta)}_{\underline{\delta}}
\theta^{s(\delta)}_{\underline{\delta'}}
Y^{kl}_{n,\delta}
\theta^{s(\alpha)}_{\underline{\alpha}}
\theta^{s(\delta)}_{\underline{\delta'}}
\theta^{s(\chi)}_{\underline{\chi}}
\theta^{s(\beta\times f)}_{\underline{\beta'}}
F^{imk,\alpha\delta}_{lnj,(\beta\times f)\chi}
\theta^{s(\beta)}_{\underline{\beta}}
\theta^{s(\beta\times f)}_{\underline{\beta'}}
\theta^{s(\chi)}_{\underline{\chi}}
\theta^{s(\chi\times f)}_{\underline{\chi}}
\widetilde{O_1}^{ml,\beta(\beta\times f)\chi}_{jin,(\chi\times f)}
\Psi_\text{fix}
\begin{pmatrix} \includegraphics[scale=.45]{H1} \end{pmatrix}
\nonumber\\
&\simeq
\sum_{n\delta\chi}
\theta^{s(\beta)}_{\underline{\beta}}
\theta^{s(\alpha)}_{\underline{\alpha}}
\theta^{s(\delta)}_{\underline{\delta}}
 \theta^{s(\chi)}_{\underline{\chi}}
Y^{kl}_{n,\delta}
F^{imk,\alpha\delta}_{lnj,\beta\chi}
O^{ml,\beta}_{j}
\Psi_\text{fix}
\begin{pmatrix} \includegraphics[scale=.45]{H1} \end{pmatrix}
+
\sum_{n\delta\chi}
\theta^{s(\beta)}_{\underline{\beta}}
\theta^{s(\alpha)}_{\underline{\alpha}}
\theta^{s(\delta)}_{\underline{\delta}}
 \theta^{s(\chi\times f)}_{\underline{\chi}}
Y^{kl}_{n,\delta}
F^{imk,\alpha\delta}_{lnj,\beta(\chi\times f)}
O^{ml,\beta}_{j}
\Psi_\text{fix}
\begin{pmatrix} \includegraphics[scale=.45]{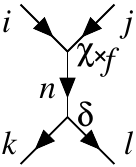} \end{pmatrix}
\nonumber\\
&\simeq
\sum_{n\chi\delta}
2
\theta^{s(\beta)}_{\underline{\beta}}
\theta^{s(\alpha)}_{\underline{\alpha}}
\theta^{s(\delta)}_{\underline{\delta}}
 \theta^{s(\chi)}_{\underline{\chi}}
Y^{kl}_{n,\delta}
F^{imk,\alpha\delta}_{lnj,\beta\chi}
O^{ml,\beta}_{j}
\Psi_\text{fix}
\begin{pmatrix} \includegraphics[scale=.45]{H1} \end{pmatrix}
,
\label{HF12}
\end{align}
where we have
\begin{align}
\sum_{n\chi\delta}
\widetilde{O_1}^{ml,\beta(\beta\times f)\chi}_{jin,(\chi\times f)}
\psi_\text{fix}
\begin{pmatrix} \includegraphics[scale=.40]{H1} \end{pmatrix}
\simeq
\sum_{n\chi\delta}
\psi_\text{fix}\begin{pmatrix} \includegraphics[scale=.40]{H9} \end{pmatrix}
=
\sum_{n\chi\delta}
\Xi^{im}_{lnj,\beta(\chi\times f)}
\psi_\text{fix}\begin{pmatrix} \includegraphics[scale=.40]{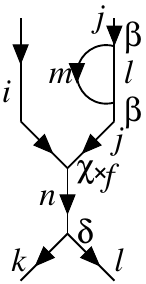} \end{pmatrix}
\simeq
\sum_{n\chi\delta}
\Xi^{im}_{lnj,\beta(\chi\times f)}
O^{ml,\beta}_{j}
\psi_\text{fix}\begin{pmatrix} \includegraphics[scale=.40]{H1f} \end{pmatrix}
.
\end{align}

Combining with the case that $\beta$ is fermionic and the case $j$ is m-type, the relation is written as
\begin{align}
\widetilde{H}^{kim,\alpha\beta}_{jln,\chi\delta}=
n_{j} 
Y^{kl}_{n,\delta}
F^{imk,\alpha\delta}_{lnj,\beta\chi}
O^{ml,\beta}_{j}.
\label{hyfo}
\end{align}

With the ansatz Eq.(\ref{ooo}) and Eq.(\ref{yyy}), we have
\begin{equation}
\widetilde{H}^{kim,\alpha\beta}_{jln,\chi\delta}=
\Phi^{ml,\beta}_{j}
(\Phi^{kl,\delta}_{n})^*
\sqrt{\frac{n_{j}n_{k}d_{m}d_{n}}{n_{m}n_{n}d_{j}d_{k}}}
F^{imk,\alpha\delta}_{lnj,\beta\chi}
\label{HF}
\end{equation}

Similar to $F$-move, we also require the dual $H$-move to be projective unitary:

\begin{equation}
\underset{n\chi\delta}{\sum} 
\widetilde{H}^{kim',\alpha'\beta'}_{jln,\chi\delta} (\widetilde{H}^{kim,\alpha\beta}_{jln,\chi\delta})^{*}
=\left\{\begin{array}{l}
\delta_{mm'}\delta_{\alpha\alpha'}\delta_{\beta\beta'}
\text{, \ \ if }m\text{ is m-type
}\\ 
\frac{1}{n_{m}} (\delta_{mm'}\delta_{\alpha\alpha'}\delta_{\beta\beta'}+
\widetilde{\zeta}^{kim,\alpha\beta}_{jl}
\delta_{mm'}\delta_{(\alpha\times f)\alpha'}\delta_{(\beta\times f)\beta'})
\text{, \ \ if }m\text{ is q-type}
\end{array}\right.
\end{equation}
If $m$ is q-type, and in the fermion parity-even sector for example, this projective unitary condition for dual $H$-move can be viewed as the following projective map:
\begin{align}
\frac{1}{2}\psi_{\text{fix}}\begin{pmatrix} \includegraphics[scale=.40]{dh1} 
\end{pmatrix}+
\frac{(\widetilde{\zeta}^{kim,\alpha\beta}_{jl})^*}{2}\psi_{\text{fix}}\begin{pmatrix} \includegraphics[scale=.40]{dh4fff} 
\end{pmatrix}
\rightarrow
\psi_{\text{fix}}\begin{pmatrix} \includegraphics[scale=.40]{dh1} 
\end{pmatrix}
,
\end{align}
\begin{align}
\frac{\widetilde{\zeta}^{kim,\alpha\beta}_{jl}}{2}
\psi_{\text{fix}}\begin{pmatrix} \includegraphics[scale=.40]{dh1} 
\end{pmatrix}+
\frac{1}{2}\psi_{\text{fix}}\begin{pmatrix} \includegraphics[scale=.40]{dh4fff} 
\end{pmatrix}
\rightarrow
\psi_{\text{fix}}\begin{pmatrix} \includegraphics[scale=.40]{dh4fff} 
\end{pmatrix}
,
\end{align}
In terms of matrix form, we have:
\begin{equation}
\widetilde{P}
=
\left(
\begin{array}{cc}
\frac{1}{2} &  \frac{(\widetilde{\zeta}^{kim,\alpha\beta)^*}_{jl}}{2}\\ 
\frac{\widetilde{\zeta}^{kim,\alpha\beta}_{jl}}{2}& \frac{1}{2}%
\end{array}%
\right),
\end{equation}
which also satisfies $\widetilde{P}^2=\widetilde{P}$.  Relation in Eq.(\ref{hyfo}) induces the following equivalence relation on $F$-move:
\begin{align}
F^{imk,(\alpha\times f)\delta}_{lnj,(\beta\times f)\chi}
=
\frac{
\widetilde{H}^{kim,(\alpha\times f)(\beta\times f)}_{jln,\chi\delta} 
O^{ml,\beta}_{j}
}
{
\widetilde{H}^{kim,\alpha\beta}_{jln,\chi\delta} 
\mathcal{O}^{ml,(\beta\times f)}_{j}
}
F^{imk,\alpha\delta}_{lnj,\beta\chi}  
=
(\Lambda^{ml,\beta}_{j})^{*}
\widetilde{\zeta}^{kim,\alpha\beta}_{jl}
F^{imk,\alpha\delta}_{lnj,\beta\chi}  
\text{, \ \ if }m\text{ is q-type},
\label{Fvert1}
\end{align}
which is the equivalence between two $F$-moves with the fermion parity on the first two vertical fusion states changed. On the other hand, the projective unitary condition of dual $H$ move also induce an additional condition for $F$-move:

\begin{equation}
\underset{n\chi\delta}{\sum}
\frac{d_{n}}{n_{n}} 
F^{im'k,\alpha'\delta}_{lnj,\beta'\chi}
(F^{imk,\alpha\delta}_{lnj,\beta\chi})^{*}
=\frac{d_{j}d_{k}n_{m}}{n_{j}n_{k}d_{m}} 
\left\{\begin{array}{l}
\delta_{mm'}\delta_{\alpha\alpha'}\delta_{\beta\beta'}
\text{, \ \ if }m\text{ is m-type
}\\ 
\frac{1}{n_{m}} (\delta_{mm'}\delta_{\alpha\alpha'}\delta_{\beta\beta'}+
\Omega^{kim,\alpha\beta}_{jl}
\delta_{mm'}\delta_{(\alpha\times f)\alpha'}\delta_{(\beta\times f)\beta'})
\text{, \ \ if }m\text{ is q-type}
\end{array}\right.
\label{projcond3}
\end{equation}
where 
\begin{align}
\Omega^{kim,\alpha\beta}_{jl}:=
(\Lambda^{ml,\beta}_{j})^*
\widetilde{\zeta}^{kim,\alpha\beta}_{jl}
\label{omega1}
\end{align}
is the combination of two phase factors, and it also satisfies $(\Omega^{kim,\alpha\beta}_{jl})^*=\Omega^{kim,(\alpha\times f)(\beta\times f)}_{jl}.$

Inversely, if we sum over the states $\{m,\alpha,\beta\}$, the dual-$H$ move also satisfies:
\begin{equation}
\underset{m\alpha\beta}{\sum} 
(\widetilde{H}^{kim,\alpha\beta}_{jln',\chi'\delta'})^{*}
\widetilde{H}^{kim,\alpha\beta}_{jln,\chi\delta} 
=\left\{\begin{array}{l}
\delta_{nn'}\delta_{\chi\chi'}\delta_{\delta\delta'}
\text{, \ \ if }n\text{ is m-type
}\\ 
\frac{1}{n_{n}} (\delta_{nn'}\delta_{\chi\chi'}\delta_{\delta\delta'}+
\widetilde{\zeta}^{ki}_{jln,\chi\delta}
\delta_{nn'}\delta_{(\chi\times f)\chi'}\delta_{(\delta\times f)\delta'})
\text{, \ \ if }n\text{ is q-type}
\end{array}\right.
\end{equation}
If $m$ is q-type, and in the fermion parity-even sector for example, this projective unitary condition for dual $H$-move can be viewed as the following projective map:
\begin{align}
\frac{1}{2}\psi_{\text{fix}}\begin{pmatrix} \includegraphics[scale=.40]{dy1} 
\end{pmatrix}+
\frac{(\widetilde{\zeta}^{ki}_{jln,\chi\delta})^*}{2}\psi_{\text{fix}}\begin{pmatrix} \includegraphics[scale=.40]{dy4fff} 
\end{pmatrix}
\rightarrow
\psi_{\text{fix}}\begin{pmatrix} \includegraphics[scale=.40]{dy1} 
\end{pmatrix}
,
\end{align}
\begin{align}
\frac{\widetilde{\zeta}^{ki}_{jln,\chi\delta}}{2}
\psi_{\text{fix}}\begin{pmatrix} \includegraphics[scale=.40]{dy1} 
\end{pmatrix}+
\frac{1}{2}\psi_{\text{fix}}\begin{pmatrix} \includegraphics[scale=.40]{dy4fff} 
\end{pmatrix}
\rightarrow
\psi_{\text{fix}}\begin{pmatrix} \includegraphics[scale=.40]{dy4fff} 
\end{pmatrix}
,
\end{align}
In terms of matrix form, we have:
\begin{equation}
\widetilde{P}'
=
\left(
\begin{array}{cc}
\frac{1}{2} &  \frac{(\widetilde{\zeta}^{ki}_{jln,\chi\delta})^*}{2}\\ 
\frac{\widetilde{\zeta}^{ki}_{jln,\chi\delta}}{2}& \frac{1}{2}%
\end{array}%
\right),
\end{equation}
which also satisfies $(\widetilde{P}')^2=\widetilde{P}'$.  Relation in Eq.(\ref{hyfo}) again induces the following equivalence relation on $F$-move:
\begin{align}
(F^{imk,\alpha(\delta\times f)}_{lnj,\beta(\chi\times f)}  )^*
=
\frac{
(\widetilde{H}^{kim,\alpha\beta}_{jln,(\chi\times f)(\delta\times f)} 
Y^{kl}_{n,\delta})^*
}
{
(\widetilde{H}^{kim,\alpha\beta}_{jln,\chi\delta} 
Y^{kl}_{n,(\delta\cdot f)})^*
}
(F^{imk,\alpha\delta}_{lnj,\beta\chi}  )^*
=
(\widetilde{\Lambda}^{kl}_{n,\delta})^*
\widetilde{\zeta}^{ki}_{jln,\chi\delta}
(F^{imk,\alpha\delta}_{lnj,\beta\chi}  )^*
\text{, \ \ if }n\text{ is q-type}
.
\label{Fvert2}
\end{align}
which is the equivalence between two $F$-moves with the fermion parity on the second two vertical fusion states changed. We also have another condition for $F$-move:

\begin{equation}
\underset{m\alpha\beta}{\sum}
\frac{d_{m}}{n_{m}} 
(F^{imk,\alpha\delta'}_{ln'j,\beta\chi'})^{*}
F^{imk,\alpha\delta}_{lnj,\beta\chi}
=\frac{d_{j}d_{k}n_{n}}{n_{j}n_{k}d_{n}} 
\left\{\begin{array}{l}
\delta_{nn'}\delta_{\chi\chi'}\delta_{\delta\delta'}
\text{, \ \ if }n\text{ is m-type
}\\ 
\frac{1}{n_{n}} (\delta_{nn'}\delta_{\chi\chi'}\delta_{\delta\delta'}+
\Omega^{ki}_{jln,\chi\delta}
\delta_{nn'}\delta_{(\chi\times f)\chi'}\delta_{(\delta\times f)\delta'})
\text{, \ \ if }n\text{ is q-type}
\end{array}\right.
\label{projcond4}
\end{equation}
where 
\begin{align}
\Omega^{ki}_{jln,\chi\delta}:=
(\widetilde{\Lambda}^{kl}_{n,\delta})^*
\widetilde{\zeta}^{ki}_{jln,\chi\delta},
\label{omega2}
\end{align}
and it also satisfies $(\Omega^{ki}_{jln,\chi\delta})^*=\Omega^{ki}_{jln,(\chi\times f)(\delta\times f)}.$

Similarly, we can also derive $H$-move from $F$, $Y$ and $O$-moves:
\begin{align}
H^{kim,\alpha\beta}_{jln,\chi\delta}=
n_{i} 
Y^{kl}_{n,\delta}
(F^{kmi,\alpha\chi}_{jnl,\beta\delta})^*
O^{km,\alpha}_{i}.
\label{hhyfo}
\end{align}

When $m$ is q-type, there is also such an equivalence relation:
\begin{align}
H^{kim,(\alpha\times f)(\beta\times f)}_{jln,\chi\delta}
=
{\zeta}^{kim,\alpha\beta}_{jl}
H^{kim,\alpha\beta}_{jln,\chi\delta} 
\text{, \ \ if }m\text{ is q-type}
,
\label{hequi1}
\end{align}
and when $n$ is q-type, we have another equivalence relation:
\begin{align}
(H^{kim,\alpha\beta}_{jln,(\chi\times f)(\delta\times f)})^*
=
{\zeta}^{ki}_{jln,\chi\delta}
(H^{kim,\alpha\beta}_{jln,\chi\delta} )^*
\text{, \ \ if }n\text{ is q-type}
.
\label{hequi2}
\end{align}
From Eq.(\ref{hyfo}) and Eq.(\ref{hhyfo}), the phase factor between equivalent dual $H$-moves ${\zeta}^{kim,\alpha\beta}_{jl}$ is related to the phase factor between equivalent $H$-moves $\widetilde{\zeta}^{kim,\alpha\beta}_{jl}$ by:
\begin{align}
\zeta^{ikm,\alpha\beta}_{lj}
=
\Lambda^{im,\alpha}_{k}
\Lambda^{ml,\beta}_{j}
(\widetilde{\zeta}^{kim,\alpha\beta}_{jl})^*
,
\end{align}
which can be proven from Eq.(\ref{Fvert1}),  Eq.(\ref{Fvert2}) and Eq.(\ref{hhyfo}). And ${\zeta}^{ki}_{jln,\chi\delta}=\widetilde{\zeta}^{ki}_{jln,\chi\delta}$. The projectively-unitary conditions of $H$-moves are:
\begin{equation}
\underset{n\chi\delta}{\sum} 
H^{kim',\alpha'\beta'}_{jln,\chi\delta} 
(H^{kim,\alpha\beta}_{jln,\chi\delta})^{*}
=\left\{\begin{array}{l}
\delta_{mm'}\delta_{\alpha\alpha'}\delta_{\beta\beta'}
\text{, \ \ if }m\text{ is m-type
}\\ 
\frac{1}{n_{m}} (\delta_{mm'}\delta_{\alpha\alpha'}\delta_{\beta\beta'}+
\zeta^{kim,\alpha\beta}_{jl}
\delta_{mm'}\delta_{(\alpha\times f)\alpha'}\delta_{(\beta\times f)\beta'})
\text{, \ \ if }m\text{ is q-type}
\end{array}\right.
\end{equation}
\begin{equation}
\underset{m\alpha\beta}{\sum} 
(H^{kim,\alpha\beta}_{jln',\chi'\delta'})^{*}
H^{kim,\alpha\beta}_{jln,\chi\delta} 
=\left\{\begin{array}{l}
\delta_{nn'}\delta_{\chi\chi'}\delta_{\delta\delta'}
\text{, \ \ if }n\text{ is m-type
}\\ 
\frac{1}{n_{n}} (\delta_{nn'}\delta_{\chi\chi'}\delta_{\delta\delta'}+
\zeta^{ki}_{jln,\chi\delta}
\delta_{nn'}\delta_{(\chi\times f)\chi'}\delta_{(\delta\times f)\delta'})
\text{, \ \ if }n\text{ is q-type}
\end{array}\right.
\end{equation}
which will give exactly the same conditions for $F$-moves Eq.(\ref{projcond3}) and Eq.(\ref{projcond4}). 




\subsection{Relations among the phase factors}
\label{relationphase}

The four phase factors $\Xi^{ijm,\alpha\beta}_{kl}$, $\Xi^{ij}_{kln,\chi\delta}$, $\Omega^{kim,\alpha\beta}_{jl}$ and $\Omega^{ki}_{jln,\chi\delta}$ are not independent. Consistency between the fermionic Pentagon equation in Eq.(\ref{fpenta}) and equivalence relatioins in Eq.(\ref{Fequiup}), Eq.(\ref{Fequidown}), Eq.(\ref{Fvert1}) and Eq.(\ref{Fvert2}) give rise to many relations among the phase factors. We will only show the following two relations here:
\begin{align}
\Omega^{kim,\alpha\beta}_{jl}
=
\Xi^{is}_{tkm,\eta\alpha}
(\Xi^{stm,\eta\beta}_{lj})^*
,
\label{pr01}
\end{align}
\begin{align}
\Omega^{ki}_{jln,\chi\delta}
=
\Xi^{si}_{jtn,\delta\eta}
(\Xi^{sk}_{ltn,\chi\eta})^*,
\label{pr02}
\end{align}
where strings $s,t$ and fusion state $\eta$ can be arbitrarily chosen as long as fusion rules are satisfied in the above two equations.

Now we show how to derive the first relation in Eq.(\ref{pr01}). We divide the summation over $t$ in Eq.(\ref{fpenta}) into two parts: the summation over $t$ strings that are m-type, and the summation over $t$ string sthat are q-type. Then we relabel $\eta$ and $\psi$ by $\eta\times f$ and $\psi\times f$ in the summation that $t$ strings are q-type:

\begin{align}
\underset{\epsilon}{\sum}
F^{ijm,\alpha\epsilon}_{qps,\phi\gamma}
F^{mkn,\beta\chi}_{lpq,\delta\epsilon}
&=
(-1)^{s(\alpha)s(\delta)}
\underset{\{\text{m-type }t\}\eta\psi\kappa}{\sum}
F^{ijm,\alpha\beta}_{knt,\eta\psi}
F^{itn,\psi\chi}_{lps,\kappa\gamma}
F^{jkt,\eta\kappa}_{lsq,\delta\phi}
\nonumber\\
&
\ \ \ \ \ 
+
(-1)^{s(\alpha)s(\delta)}
\underset{\{\text{q-type }t\}(\eta\times f)(\psi\times f)(\kappa\times f)}{\sum}
F^{ijm,\alpha\beta}_{knt,(\eta\times f)(\psi\times f)}
F^{itn,(\psi\times f)\chi}_{lps,(\kappa\times f)\gamma}
F^{jkt,(\eta\times f)(\kappa\times f)}_{lsq,\delta\phi}
\nonumber\\
&
=
(-1)^{s(\alpha)s(\delta)}
\underset{\{\text{m-type }t\}\eta\psi\kappa}{\sum}
F^{ijm,\alpha\beta}_{knt,\eta\psi}
F^{itn,\psi\chi}_{lps,\kappa\gamma}
F^{jkt,\eta\kappa}_{lsq,\delta\phi}
\nonumber\\
&
\ \ \ \ \ 
+
(-1)^{s(\alpha)s(\delta)}
\underset{\{\text{q-type }t\}\eta\psi\kappa}{\sum}
(\Xi^{ij}_{knt,\eta\psi})^*
\Omega^{nit,\psi\kappa}_{sl}
\Xi^{jkt,\eta\kappa}_{ls}
F^{ijm,\alpha\beta}_{knt,\eta\psi}
F^{itn,\psi\chi}_{lps,\kappa\gamma}
F^{jkt,\eta\kappa}_{lsq,\delta\phi}
,
\label{23move11}
\end{align}
where we note that for the summation over q-type $t$ strings, we can only change the fermion parity for even number of fusion states for a single $F$-move (as the $F$-move should preserve fermion-parity), and only the state $\kappa$ which is summed over can compensate the fermion-parity change in $F^{itn,\psi\chi}_{lps,\kappa\gamma}$ and $F^{jkt,\eta\kappa}_{lsq,\delta\phi}$. So that here $\kappa$ must also be replaced by $\kappa\times f$. The summation $\underset{(\eta\times f)(\psi\times f)(\kappa\times f)}{\sum}$ is actually equivalent to the summation $\underset{\eta\psi\kappa}{\sum}$ (only up to changing the summation order). Comparing Eq.(\ref{23move11}) with Eq.(\ref{fpenta}), we obtain 
\begin{align}
\underset{\{\text{q-type }t\}\eta\psi\kappa}{\sum}
F^{ijm,\alpha\beta}_{knt,\eta\psi}
F^{itn,\psi\chi}_{lps,\kappa\gamma}
F^{jkt,\eta\kappa}_{lsq,\delta\phi},
=
\underset{\{\text{q-type }t\}\eta\psi\kappa}{\sum}
(\Xi^{ij}_{knt,\eta\psi})^*
\Omega^{nit,\psi\kappa}_{sl}
\Xi^{jkt,\eta\kappa}_{ls}
F^{ijm,\alpha\beta}_{knt,\eta\psi}
F^{itn,\psi\chi}_{lps,\kappa\gamma}
F^{jkt,\eta\kappa}_{lsq,\delta\phi}.
\end{align}
We see that Eq.(\ref{pr01}) is a simple solution to the above equation (up to a relabelling). 

Then we derive Eq.(\ref{pr02}). When string $s$ is q-type, we relabel $\phi$ and $\gamma$ by $\phi\times f$ and $\gamma\times f$. And in order to conserve the fermion-parity for a single $F$-move, we also need to replace $\kappa$ by $\kappa\times f$:
\begin{equation}
\underset{\epsilon}{\sum}
F^{ijm,\alpha\epsilon}_{qps,(\phi\times f)(\gamma\times f)}
F^{mkn,\beta\chi}_{lpq,\delta\epsilon}
=(-1)^{s(\alpha)s(\delta)}
\underset{t\eta\psi\kappa}{\sum}
F^{ijm,\alpha\beta}_{knt,\eta\psi}
F^{itn,\psi\chi}_{lps,(\kappa\times f)(\gamma\times f)}
F^{jkt,\eta(\kappa\times f)}_{lsq,\delta(\phi\times f)}.
\end{equation}
\begin{equation}
\underset{\epsilon}{\sum}
(\Xi^{ij}_{qps,\phi\gamma})^*
F^{ijm,\alpha\epsilon}_{qps,\phi\gamma}
F^{mkn,\beta\chi}_{lpq,\delta\epsilon}
=(-1)^{s(\alpha)s(\delta)}
\underset{t\eta\psi\kappa}{\sum}
(\Xi^{it}_{lps,\kappa\gamma})^*
(\Omega^{tj}_{qls,\kappa\phi})^*
F^{ijm,\alpha\beta}_{knt,\eta\psi}
F^{itn,\psi\chi}_{lps,\kappa\gamma}
F^{jkt,\eta\kappa}_{lsq,\delta\phi}.
\end{equation}
We see that Eq.(\ref{pr02}) is a simple solution to the above equation (up to a relabelling).

Further, from Eq.(\ref{omega1}) and Eq.(\ref{pr01}), we have
\begin{align}
(\Lambda^{tl,\kappa}_{s})^*
\widetilde{\zeta}^{nit,\psi\kappa}_{sl}
=
(\Xi^{jkt,\eta\kappa}_{ls})^*
\Xi^{ij}_{knt,\eta\psi}
,
\label{lett}
\end{align}
where strings $j,k$ and fusion states $\eta$ on right-hand side of the above equation can be chosen arbitrarily as long as fusion rules are satisfied. Here we can choose the values of $\Lambda^{ij,\alpha\beta}_{k}$ and $\widetilde{\zeta}^{kim,\alpha\beta}_{jl}$ arbitrarily as long as Eq.(\ref{lett}) is satisfied. And there always exists a gauge such that all $\Lambda^{ij,\alpha\beta}_{k}=1$. In such a gauge, $\widetilde{\zeta}^{kim,\alpha\beta}_{jl}$ is determined by $\widetilde{\zeta}^{nit,\psi\kappa}_{sl}
=
(\Xi^{jkt,\eta\kappa}_{ls})^*
\Xi^{ij}_{knt,\eta\psi}$.

Also, Eq.(\ref{dualhh2}) reduces to Eq.(\ref{yy1}) if string $k$ is identified with $i$ and $l$ is identified with $j$, which implies
\begin{align}
\widetilde{\Lambda}^{ij}_{n,\chi\delta}
=
\widetilde{\zeta}^{ii}_{jjn,\chi\delta}
.
\end{align}
Then from Eq.(\ref{omega2}) and Eq.(\ref{pr02}), we have
\begin{align}
(\widetilde{\Lambda}^{kl}_{n,\delta})^*
\widetilde{\zeta}^{ki}_{jln,\chi\delta}
=
\Xi^{si}_{jtn,\delta\eta}
(\Xi^{sk}_{ltn,\chi\eta})^*,
\end{align}
where strings $s,t$ and fusion states $\eta$ on right-hand side of the above equation can be chosen arbitrarily as long as fusion rules are satisfied. We note that if in certain example or under certain gauge we always have $\Xi^{si}_{jtn,\delta\eta}
(\Xi^{sk}_{ltn,\chi\eta})^*=1$, we can then choose all $\widetilde{\Lambda}^{ij,\alpha\beta}_{k}=1$ and all $\widetilde{\zeta}^{ki}_{jln,\chi\delta}=1$.

\subsection{Summary}
\label{Summary}

We collect all conditions and list them below:

\begin{equation}
N^{ij}_{k}=B^{ij}_{k}+F^{ij}_{k}.
\end{equation}

\begin{equation}
\underset{m}{\sum} \frac{N^{ij}_{m}N^{mk}_{l}}{n_{m}}
=\underset{n}{\sum} \frac{N^{in}_{l}N^{jk}_{n}}{n_{n}}.
\end{equation}

\begin{equation}
\underset{m}{\sum} \frac{B^{ij}_{m}F^{mk}_{l}+F^{ij}_{m}B^{mk}_{l}}{n_{m}}
=\underset{n}{\sum} \frac{B^{in}_{l}F^{jk}_{n}+F^{in}_{l}B^{jk}_{n}}{n_{n}}.
\label{F4}
\end{equation}

\begin{equation}
F^{ijm,\alpha\beta}_{kln,\chi\delta}=0 \text{ when }
N^{ij}_{m}<1 \text{ or } N^{mk}_{l}<1 \text{ or }
N^{jk}_{n}<1 \text{ or } N^{in}_{l}<1 \text{, or } s(\alpha)+s(\beta)+s(\chi)+s(\delta)=1 \text{ mod } 2.
\label{F5}
\end{equation}

\begin{equation}
\underset{ij}{\sum}(B^{ij}_{k})^{2}+(F^{ij}_{k})^{2}\geq 1.
\label{S5}
\end{equation}

\begin{equation}
\underset{n\chi\delta}{\sum} F^{ijm',\alpha'\beta'}_{kln,\chi\delta} (F^{ijm,\alpha\beta}_{kln,\chi\delta})^{*}=
\left\{\begin{array}{l}
\delta_{mm'}\delta_{\alpha\alpha'}\delta_{\beta\beta'}
\text{, \ \ if }m\text{ is m-type
}\\ 
\frac{1}{n_{m}} (\delta_{mm'}\delta_{\alpha\alpha'}\delta_{\beta\beta'}+
\Xi^{ijm,\alpha\beta}_{kl}
\delta_{mm'}\delta_{(\alpha\times f)\alpha'}\delta_{(\beta\times f)\beta'})
\text{, \ \ if }m\text{ is q-type}
\end{array}\right.
\label{w1}
\end{equation}

\begin{equation}
\underset{m\alpha\beta}{\sum} (F^{ijm,\alpha\beta}_{kln',\chi'\delta'})^{*} F^{ijm,\alpha\beta}_{kln,\chi\delta} =
\left\{\begin{array}{l}
\delta_{nn'}\delta_{\chi\chi'}\delta_{\delta\delta'}
\text{, \ \ if }n\text{ is m-type
}\\ 
\frac{1}{n_{n}} (\delta_{nn'}\delta_{\chi\chi'}\delta_{\delta\delta'}
+
\Xi^{ij}_{kln,\chi\delta}
\delta_{nn'}\delta_{(\chi\times f)\chi'}\delta_{(\delta\times f)\delta'})
\text{, \ \ if }n\text{ is q-type}
\end{array}\right.
\label{w2}
\end{equation}

\begin{equation}
\underset{n\chi\delta}{\sum}
\frac{d_{n}}{n_{n}} 
F^{im'k,\alpha'\delta}_{lnj,\beta'\chi}
(F^{imk,\alpha\delta}_{lnj,\beta\chi})^{*}
=\frac{d_{j}d_{k}n_{m}}{n_{j}n_{k}d_{m}} 
\left\{\begin{array}{l}
\delta_{mm'}\delta_{\alpha\alpha'}\delta_{\beta\beta'}
\text{, \ \ if }m\text{ is m-type
}\\ 
\frac{1}{n_{m}} (\delta_{mm'}\delta_{\alpha\alpha'}\delta_{\beta\beta'}+
\Xi^{is}_{tkm,\eta\alpha}
(\Xi^{stm,\eta\beta}_{lj})^*
\delta_{mm'}\delta_{(\alpha\times f)\alpha'}\delta_{(\beta\times f)\beta'})
\text{, \ \ if }m\text{ is q-type}
\end{array}\right.
\label{w3}
\end{equation}

\begin{equation}
\underset{m\alpha\beta}{\sum}
\frac{d_{m}}{n_{m}} 
(F^{imk,\alpha\delta'}_{ln'j,\beta\chi'})^{*}
F^{imk,\alpha\delta}_{lnj,\beta\chi}
=\frac{d_{j}d_{k}n_{n}}{n_{j}n_{k}d_{n}} 
\left\{\begin{array}{l}
\delta_{nn'}\delta_{\chi\chi'}\delta_{\delta\delta'}
\text{, \ \ if }n\text{ is m-type
}\\ 
\frac{1}{n_{n}} (\delta_{nn'}\delta_{\chi\chi'}\delta_{\delta\delta'}+
\Xi^{si}_{jtn,\delta\eta}
(\Xi^{sk}_{ltn,\chi\eta})^*
\delta_{nn'}\delta_{(\chi\times f)\chi'}\delta_{(\delta\times f)\delta'})
\text{, \ \ if }n\text{ is q-type}
\end{array}\right.
\label{w4}
\end{equation}

(In Eq.(\ref{w3}) and Eq.(\ref{w4}), strings $s,t$ and fusion state $\eta$ can be arbitrarily chosen as long as fusion rules are satisfied.)

\begin{align}
(\Xi^{ijm,\alpha\beta}_{kl})^*=\Xi^{ijm,(\alpha\times f)(\beta\times f)}_{kl}, \ \ \
(\Xi^{ij}_{kln,\chi\delta})^*=\Xi^{ij}_{kln,(\chi\times f)(\delta\times f)},
\end{align}


\begin{equation}
\underset{t\eta\phi\kappa}{\sum} F^{ijm,\alpha\beta}_{knt,\eta\psi} F^{itn,\psi\chi}_{lps,\kappa\gamma} F^{jkt,\eta\kappa}_{lsq,\delta\phi} 
=(-1)^{s(\alpha)s(\delta)}\underset{\epsilon}{\sum} F^{mkn,\beta\chi}_{lpq,\delta\epsilon} F^{ijm,\alpha\epsilon}_{qps,\phi\gamma}.
\label{fpentas}
\end{equation}

\begin{equation}
\label{QD}
\underset{ij}{\sum} 
\frac{N^{ij}_{k}d_{i}d_{j}}{n_{i}n_{j}}
=d_{k}D^{2}, 
\text{ where }
D^{2}=\underset{i}{\sum} 
\frac{d^{2}_{i}}{n_{i}}.
\end{equation}

\subsection{Hamiltonian for general 2D non-chiral topological orders}

We construct the parent Hamiltonian that realizes the fixed-point wavefunctions satisfying all algebraic conditions listed in section \ref{Summary} as the gapped ground state. The Hamiltonian is constructed on a 2D lattice, and let us consider a honeycomb lattice for example. It is a Hamiltonian that contains three terms:
\begin{align}
\hat{H}=-\sum_{v} \hat{Q}_{v} -
\sum_{e} \hat{D}_{l}-
 \sum_{p} \hat{B}_{p},
 \label{H}
\end{align}
where $\sum_{v}$ sums over all vertices, $\sum_{l}$ sums over all links, and $\sum_{p}$ sums over all plaquettes, as shown in Fig. \ref{Hlatt}.


\begin{figure}[h]
\begin{center}
\includegraphics[scale=0.25]{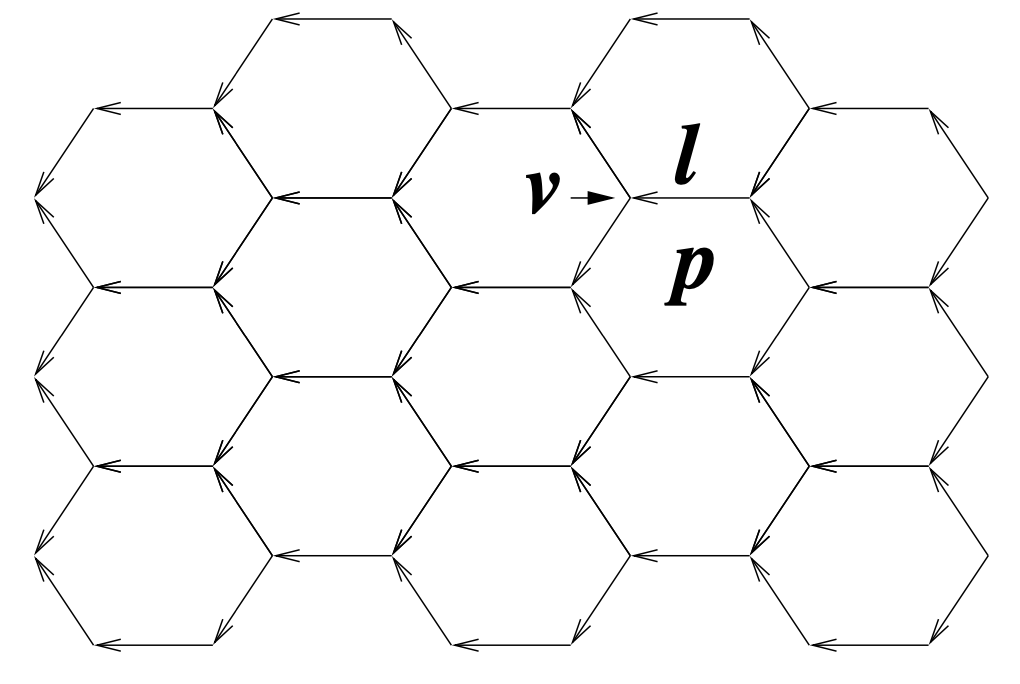}
\end{center}
\caption{
A honeycomb lattice. The vertices are labeled by $v$,
hexagons by $p$, and links by $l$.
}
\label{Hlatt}
\end{figure}

The vertex term is defined in the same way as in Ref. \onlinecite{Gu2015}, which encodes all string fusion rules. Let the Hilbert space on a patch $G$ be $V_{G}$. We expand the Hilbert space by adding an auxiliary qubit to each vertex $v$:
\begin{align}
 V^{ex}_G=V_G \otimes (\otimes_{\nu} V_{qubit}),
\end{align}
where $V_{qubit}$ is a two dimensional Hilbert space of qubit $| I_{v} \rangle$, $I_{v}=0,1$. Then in the expanded Hilbert space $V^{ex}_G$, $\hat{Q}_{v}$ acts on each vertex $v$ and the three links connected to $v$ as:
\begin{align}
\hat{Q}_{v}
\left |
\vcenter{\hbox{\includegraphics[scale=.35]{ingoing}}}
\right\rangle
\otimes
| I_{\alpha} \rangle
=
\left\{\begin{array}{l}
\left |
\vcenter{\hbox{\includegraphics[scale=.35]{ingoing}}}
\right \rangle
\otimes
| I_{\alpha} \rangle
\text{, \ \ if }N^{ij}_{k}>0, I_{\alpha}=s(\alpha),
\\ 
0
\text{,  \ \ otherwise}.
\end{array}\right.
\label{Qv}
\end{align}
We see that $\hat{Q}_{v}$ is a projector satisfying $\hat{Q}^2_{v}=\hat{Q}_{v}$. Equivalently, we can express $\hat{Q}_{v}$ as
\begin{align}
\hat{Q}_{v}
=
\sum_{ijk\alpha}
\left|\Psi_\text{fix}\begin{pmatrix}
\vcenter{\hbox{\includegraphics[scale=.35]{ingoing}}}
\end{pmatrix}\right \rangle
\left\langle \Psi_\text{fix}\begin{pmatrix}
\vcenter{\hbox{\includegraphics[scale=.35]{ingoing}}}
\end{pmatrix}\right |,
\end{align}
where the states of fermionic ground state fixed-point wavefunctions $\Psi_\text{fix}$ automatically satisfy $N^{ij}_{k}>0, I_{\alpha}=s(\alpha)$, and are assigned with ordered Majorana numbers on vertices.

The link term $\hat{D}_{l}$ is needed when there are q-type strings involved. $\hat{D}_{l}$ projects the following states into vacuum if the corresponding inner strings are q-type:
\begin{align}
\hat{D}_{l}
\begin{pmatrix}
\left|\psi_\text{fix}\begin{pmatrix}
\vcenter{\hbox{\includegraphics[scale=.35]{dh4fff}}}
\end{pmatrix}\right \rangle
-
\widetilde{\zeta}^{kim,\alpha\beta}_{jl}
\left|\psi_\text{fix}\begin{pmatrix}
\vcenter{\hbox{\includegraphics[scale=.35]{dh1}}}
\end{pmatrix}\right \rangle
\end{pmatrix}
=
0
\text{,\ \ if } m \text{ is q-type},
\label{dl1}
\end{align}
\begin{align}
\hat{D}_{l}
\begin{pmatrix}
\left|\psi_\text{fix}\begin{pmatrix}
\vcenter{\hbox{\includegraphics[scale=.35]{dy4fff}}}
\end{pmatrix}\right \rangle
-
\widetilde{\zeta}^{ki}_{jln,\chi\delta}
\left|\psi_\text{fix}\begin{pmatrix}
\vcenter{\hbox{\includegraphics[scale=.35]{dy1}}}
\end{pmatrix}\right \rangle
\end{pmatrix}
=
0
\text{,\ \ if } n \text{ is q-type},
\end{align}
\begin{align}
\hat{D}_{l}
\begin{pmatrix}
\left|\psi_\text{fix}\begin{pmatrix}
\vcenter{\hbox{\includegraphics[scale=.35]{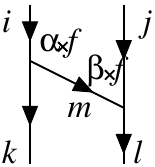}}}
\end{pmatrix}\right \rangle
-
\zeta^{kim,\alpha\beta}_{jl}
\left|\psi_\text{fix}\begin{pmatrix}
\vcenter{\hbox{\includegraphics[scale=.35]{H2}}}
\end{pmatrix}\right \rangle
\end{pmatrix}
=
0
\text{,\ \ if } m \text{ is q-type},
\end{align}
where the states if bosonic part ground state fixed-point wavefunction $\psi_\text{fix}$ contain no Majorana number. Equivalently, we can express $\hat{D}_{l}$ as 
\begin{align}
\hat{D}_{l}
=&
\underset{\{\text{q-type }m\}ijkl\alpha\beta}{\sum}
\frac{1}{2}
\begin{pmatrix}
(\widetilde{\zeta}^{kim,\alpha\beta}_{jl})^*
(-1)^{s(\beta)}
\theta_{\underline{\alpha}}
\theta_{\underline{\beta}}
\left|\Psi_\text{fix}\begin{pmatrix}
\vcenter{\hbox{\includegraphics[scale=.35]{dh4fff}}}
\end{pmatrix}\right \rangle
+
\left|\Psi_\text{fix}\begin{pmatrix}
\vcenter{\hbox{\includegraphics[scale=.35]{dh1}}}
\end{pmatrix}\right \rangle
\end{pmatrix}
\left\langle \Psi_\text{fix}\begin{pmatrix}
\vcenter{\hbox{\includegraphics[scale=.35]{dh1}}}
\end{pmatrix}\right |
\nonumber\\&
+\underset{\{\text{q-type }n\}ijkl\chi\delta}{\sum}
\frac{1}{2}
\begin{pmatrix}
(\widetilde{\zeta}^{ki}_{jln,\chi\delta})^*
(-1)^{s(\chi)}
\theta_{\underline{\delta}}
\theta_{\underline{\chi}}
\left|\Psi_\text{fix}\begin{pmatrix}
\vcenter{\hbox{\includegraphics[scale=.35]{dy4fff}}}
\end{pmatrix}\right \rangle
+
\left|\Psi_\text{fix}\begin{pmatrix}
\vcenter{\hbox{\includegraphics[scale=.35]{dy1}}}
\end{pmatrix}\right \rangle
\end{pmatrix}
\left\langle \Psi_\text{fix}\begin{pmatrix}
\vcenter{\hbox{\includegraphics[scale=.35]{dy1}}}
\end{pmatrix}\right |
\nonumber\\&
+\underset{\{\text{q-type }m\}ijkl\alpha\beta}{\sum}
\frac{1}{2}
\begin{pmatrix}
(\zeta^{kim,\alpha\beta}_{jl})^*
(-1)^{s(\alpha)}
\theta_{\underline{\beta}}
\theta_{\underline{\alpha}}
\left|\Psi_\text{fix}\begin{pmatrix}
\vcenter{\hbox{\includegraphics[scale=.35]{h3ff}}}
\end{pmatrix}\right \rangle
+
\left|\Psi_\text{fix}\begin{pmatrix}
\vcenter{\hbox{\includegraphics[scale=.35]{H2}}}
\end{pmatrix}\right \rangle
\end{pmatrix}
\left\langle \Psi_\text{fix}\begin{pmatrix}
\vcenter{\hbox{\includegraphics[scale=.35]{H2}}}
\end{pmatrix}\right |,
\end{align}
where $\{\text{q-type }m\}$ in the summation means that we only sum over $m$ strings that are q-type. The attached Grassmann numbers for the first part of the $\hat{D}_{l}$ operator for example is derived from the equivalence relation in Eq.(\ref{dualhe1}):
\begin{align}
\left|\Psi_\text{fix}\begin{pmatrix}
\vcenter{\hbox{\includegraphics[scale=.35]{dh1}}}
\end{pmatrix}\right \rangle
=
(\widetilde{\zeta}^{kim,\alpha\beta}_{jl})^*
\theta^{s(\beta)}_{\underline{\beta}}
\theta^{s(\alpha)}_{\underline{\alpha}}
\theta^{s(\alpha\times f)}_{\underline{\alpha}}
\theta^{s(\beta\times f)}_{\underline{\beta}}
\left|\Psi_\text{fix}\begin{pmatrix}
\vcenter{\hbox{\includegraphics[scale=.35]{dh4fff}}}
\end{pmatrix}\right \rangle
=
(\widetilde{\zeta}^{kim,\alpha\beta}_{jl})^*
(-1)^{s(\beta)}
\theta_{\underline{\alpha}}
\theta_{\underline{\beta}}
\left|\Psi_\text{fix}\begin{pmatrix}
\vcenter{\hbox{\includegraphics[scale=.35]{dh4fff}}}
\end{pmatrix}\right \rangle.
\end{align}
And it is easy to see that $\hat{D}_{l}$ as a projector satisfies $\hat{D}^2_{l}=\hat{D}_{l}$. 



The plaquette term $\hat{B}_{p}$ is also defined similarly as in Ref. \onlinecite{Gu2015}. It acts on the six vertices $\alpha,\beta,\gamma,\lambda,\mu,\nu$ and six inner links $a,b,c,d,e,f$ of a hexagon $p$, and the six outer links $i,j,k,l,m,n$ connected to the hexagon (the outer links are fixed). The Majorana number valued matrix element $ \mathcal{B}^{a\alpha,b\beta,c\gamma,d\lambda,e\mu,f\nu } _{
a'\alpha',b'\beta',c'\gamma',d'\lambda',e'\mu',f'\nu' } (i,j,k,l,m,n)$ is defined as:
\begin{align}
\mathcal{B}^{a\alpha,b\beta,c\gamma,d\lambda,e\mu,f\nu } _{
a'\alpha',b'\beta',c'\gamma',d'\lambda',e'\mu',f'\nu' } (i,j,k,l,m,n)
=
\left\langle \Psi_\text{fix}\begin{pmatrix}
\vcenter{\hbox{\includegraphics[scale=.25]{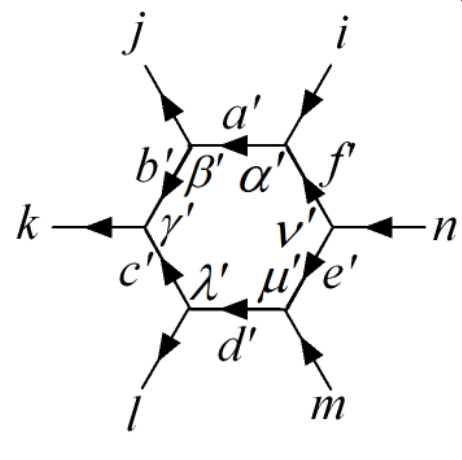}}}
\end{pmatrix}\right |
\hat{B}_{p}
\left|\Psi_\text{fix}\begin{pmatrix}
\vcenter{\hbox{\includegraphics[scale=.25]{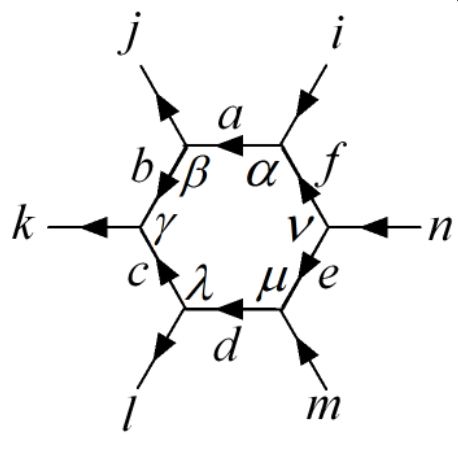}}}
\end{pmatrix}\right \rangle.
\end{align}
The matrix $\mathcal{B}=\mathcal{U}^{\dagger}_{p} I\mathcal{U}_p$, where 
\begin{align}
(\mathcal{U}_p)^{a\alpha,b\beta,c\gamma,d\lambda,e\mu,f\nu}_{trs,\chi\eta\varphi\epsilon}(i,j,k,l,m,n)
=
(\mathcal{H}^{jit,\chi\delta}_{fba,\alpha\beta})^{*}
\mathcal{F}^{tfb,\delta\gamma}_{ckr,\kappa\eta}
\widetilde{\mathcal{H}}^{rfc,\kappa\lambda}_{dls,\rho\varphi}
(\mathcal{F}^{fen,\nu'\epsilon'}_{msd,\mu\rho})^{*}
{\widetilde{\mathcal{O}_{2}}}^{fe,\nu\nu'\epsilon'}_{nms,\epsilon},
\end{align}
where the involved $F$-moves, $H$-moves, dual $H$-moves and $\widetilde{O}$-moves should satisfy the equivalence relations in Eq.(\ref{Fequiup}), Eq.(\ref{Fequidown}), Eq.(\ref{eo}), Eq.(\ref{dualhequi1}), Eq.(\ref{dualhequi2}), Eq.(\ref{hequi1}) and Eq.(\ref{hequi2}) when certain strings are q-type.

Then we argue that our constructed Hamiltonian in Eq.(\ref{H}) is a commuting-projector Hamiltonian. First, in Ref. \onlinecite{Gu2015}, it has been shown that $ \hat{Q}_{v}$ commutes with $\hat{B}_{p}$. Next, the link term $\hat{D}_{l}$ automatically commutes with $\hat{Q}_{v}$ as long as the states that $ \hat{D}_{l}$ projects onto satisfy all string fusion rules, which is exactly the case. Then the link term $\hat{D}_{l}$ automatically commutes with $ \hat{B}_{p}$ as along as the involved $F$-moves, $H$-moves, dual $H$-moves and $\widetilde{O}$-moves satisfy their corresponding equivalence relations when certain strings are q-type, which is also the case.

\section{Topological Invariant Partition Function}
\label{pf}

\subsection{Partition function and spin structure}

\begin{figure}[h]
\begin{center}
\includegraphics[width=2.1in]
{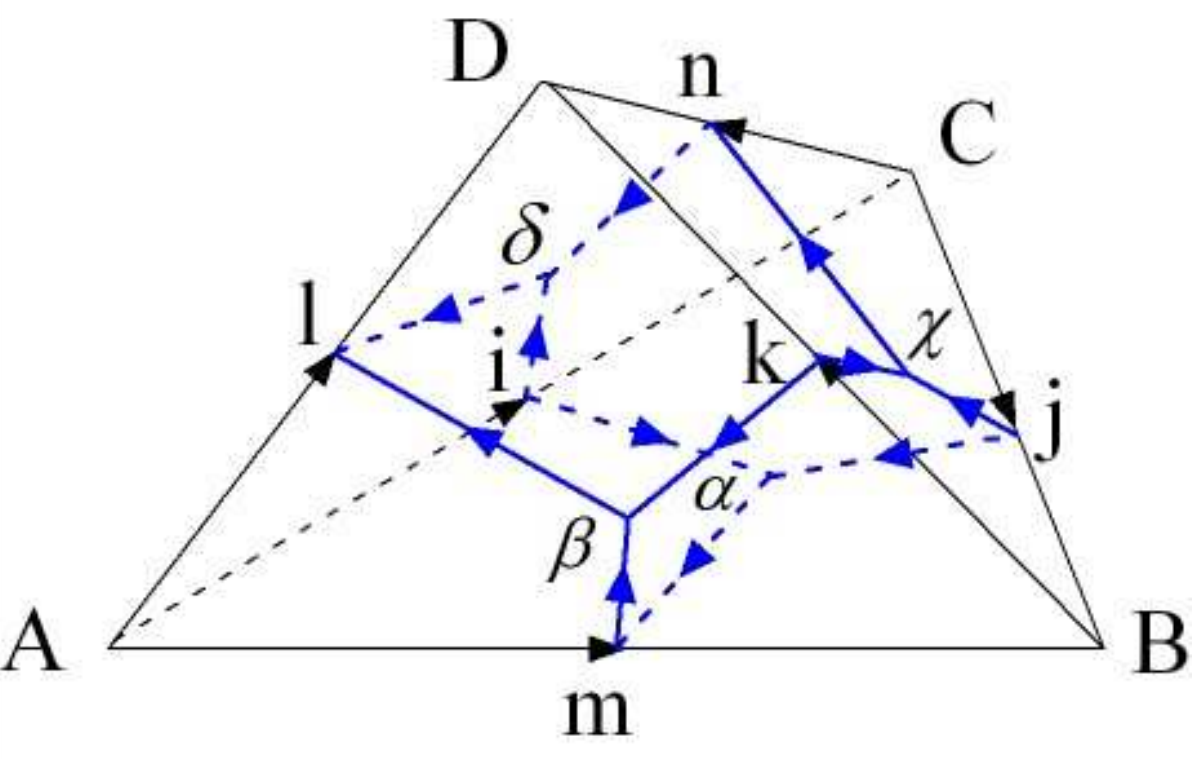}
\end{center}
\caption{The graphical representation of the $G$-symbol is actually a
dual representation of the $F$-symbol.} \label{G1}
\end{figure}

Based on the above algebriac relations, we can construct the fellowing topological invariant partition function\cite{Gu2014} for an arbitrary triangulation of 3D spin manifold $\mathcal{M}$:
\begin{equation}
\mathcal{Z}=
\frac{1}{D^{2N_{\nu}}}
\sum_{ijklmn...\alpha\beta\chi\delta...}
\prod_{\text{link}} 
\frac{d_{i}}{n_{i}}
\int
\prod_{\text{face}}
d\theta^{s(\alpha)}_{\underline{\alpha}}
d\overline{\theta}^{s(\alpha)}_{\underline{\alpha}}
\prod_{E} (-1)^{s(\alpha)}
\prod_{\text{tetrahedron}}
(\mathcal{G}^{ijm,\alpha\beta}_{kln,\chi\delta})^{\sigma_{ijmkln}},
\label{partition}
\end{equation}
where $D^{2}=\sum_{i} d^{2}_{i}/n_{i}$ is the total quantum dimension and $N_{\nu}$ is the total number of vertices for a given triangulation. We evaluate the Grassmann integral on all interior faces, where we choose that $d\theta$ always comes before $d\overline{\theta}$.  
$\mathcal{G}^{ijm}_{kln}$ is the ordering-independent Grassmann valued $G$-symbol and $\sigma_{ijmkln}=\pm$ is the orientation of the tetrahedron.:
\begin{equation}
(\mathcal{G}^{ijm,\alpha\beta}_{kln,\chi\delta})^+ =
\theta^{s(\alpha)}_{\underline{\alpha}}
\theta^{s(\beta)}_{\underline{\beta}}
\overline{\theta}^{s(\delta)}_{\underline{\delta}}
\overline{\theta}^{s(\chi)}_{\underline{\chi}}
G^{ijm,\alpha\beta}_{kln,\chi\delta},
\end{equation}
\begin{equation}
(\mathcal{G}^{ijm,\alpha\beta}_{kln,\chi\delta})^- =
\theta^{s(\chi)}_{\underline{\chi}}
\theta^{s(\delta)}_{\underline{\delta}}
\overline{\theta}^{s(\beta)}_{\underline{\beta}}
\overline{\theta}^{s(\alpha)}_{\underline{\alpha}}
(G^{ijm,\alpha\beta}_{kln,\chi\delta})^*.
\end{equation}
The $G$-symbol is actually the dual representation of the original $F$-symbol, as shown Fig. \ref{G1}, and $G^{ijm,\alpha\beta}_{kln,\chi\delta}$ is related to $F$-symbol via: 
\begin{equation}
G^{ijm,\alpha\beta}_{kln,\chi\delta}
=
\sqrt{\frac{n_{n}n_{m}}{d_{n}d_{m}}}
F^{ijm,\alpha\beta}_{kln,\chi\delta}.
\label{gf}
\end{equation}

Specifically, $\prod_{E} (-1)^{s(\alpha)}$ is the spin structure term. We include this spin structure term such that the partition function is invariant under all Pachner moves\cite{Gu2014}, i.e. retriangulations. Mathematically, the fermionic partition function can only be defined on a spin manifold, i.e. a manifold that admits spin structures. It is known that an oriented manifold $\mathcal{M}$ admits spin structures if and only if its second Stiefel-Whitney class $[\omega^2]\in H^2(\mathcal{M},\mathbb{Z}_{2})$ vanishes. We denote the Poincare dual of $\omega^2$ to be $\omega_1$ in 2+1D, which is a set of some 1-simplices. Therefore, the requirement that $\omega^2$ vanishes (being a coboundary) is equivalent to $\omega_1$ being the boundary of some surface $E$: $\partial E=\omega_1$. Different choices of $E$ correspond to different admitted choices of spin structures $\eta$, where $E$ is the Poincare dual of the 1-cochain $\eta\in C^1(\mathcal{M},\mathbb{Z}_{2})$. In Ref. \onlinecite{Gu2014}, the spin structure term is expressed as $\prod_{\omega_{1}}(-1)^{m(i)}$, where $m(i)$ is a $\mathbb{Z}_2$ function defined on link $i$ satisfying $s(\alpha)=m(i)+m(j)+m(k)$ (mod 2). And $\omega_{1}$ are certain links given as
\begin{align}
\omega_{1}
&=\text{
\{all 1-simplices\}+\{(02) in any $+$ tetrahedron (0123)\}+\{(13) in any $-$ tetrahedron (0123)\}}
\nonumber\\&
=\text{
\{all 1-simplices\}+\{(02) in any 2-simplex\}+\{(03) in any 3-simplex\}},
\end{align}
where we have relabelled the vertices $A,B,C,D$ in Fig. \ref{G1} by $0,1,2,3$, and the two expressions of $\omega_1$ are equivalent as shown in Ref. \onlinecite{Nat2017}. And both expressions are further equivalent to our spin structure term $\prod_{E} (-1)^{s(\alpha)}$. 
It is known that all oriented 3D manifolds admit spin structures.  The $E$ surfaces for all eight time ordered 2-3 moves are listed in Ref. \onlinecite{Wang2018}.

\subsection{2-3 moves}

In 2+1D, the first type of Pachner move is the 2-3 move. There are in total eight 2-3 moves that can be induced by a time ordering. The standard 2-3 move is given by
\begin{align}
&\ \ \ \ \ \ 
\underset{\epsilon}{\sum}
\int
d\theta^{s(\epsilon)}_{\underline{\epsilon}}
d\overline{\theta}^{s(\epsilon)}_{\underline{\epsilon}}
(\mathcal{G}^{ijm,\alpha\epsilon}_{qps,\phi\gamma})^-
(\mathcal{G}^{mkn,\beta\chi}_{lpq,\delta\epsilon})^-
\nonumber\\ 
&\ \ 
=
\underset{t\eta\phi\kappa}{\sum}
\int
d\theta^{s(\eta)}_{\underline{\eta}}
d\overline{\theta}^{s(\eta)}_{\underline{\eta}}
d\theta^{s(\phi)}_{\underline{\phi}}
d\overline{\theta}^{s(\phi)}_{\underline{\phi}}
d\theta^{s(\kappa)}_{\underline{\kappa}}
d\overline{\theta}^{s(\kappa)}_{\underline{\kappa}}
\frac{d_{t}}{n_{t}} 
(\mathcal{G}^{ijm,\alpha\beta}_{knt,\eta\phi})^-
(\mathcal{G}^{itn,\phi\chi}_{lps,\kappa\gamma})^-
(\mathcal{G}^{jkt,\eta\kappa}_{lsq,\delta\phi})^-,
\end{align}
where the spin structure term is trivial for the standrad 2-3 move, i.e. $\prod_{E} (-1)^{s(\alpha)}=1$. After integrating out the Grassmann numbers and comparing the rest Grassmann numbers on both sides, this equation is reduced to
\begin{equation}
\underset{\epsilon}{\sum}
G^{ijm,\alpha\epsilon}_{qps,\phi\gamma}
G^{mkn,\beta\chi}_{lpq,\delta\epsilon}
=(-1)^{s(\alpha)s(\delta)}
\underset{t\eta\psi\kappa}{\sum}
\frac{d_{t}}{n_{t}} 
G^{ijm,\alpha\beta}_{knt,\eta\psi}
G^{itn,\psi\chi}_{lps,\kappa\gamma}
G^{jkt,\eta\kappa}_{lsq,\delta\phi}.
\label{23move1}
\end{equation}
which is exactly the same as Eq.(\ref{fpentas}),
as shown graphically in Fig. \ref{pentagon}.

\begin{figure}[h]
\begin{center}
\includegraphics[width=3in]
{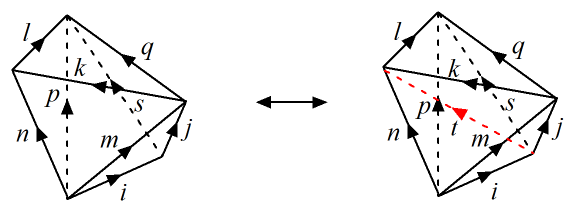}
\end{center}
\caption{Graphical representation of the standard 2-3 move for $G$-symbol, or the fermionic Pentagon relation for $F$-symbol.}\label{pentagon}
\end{figure}

\begin{figure}[h]
\begin{center}
\includegraphics[width=6in]
{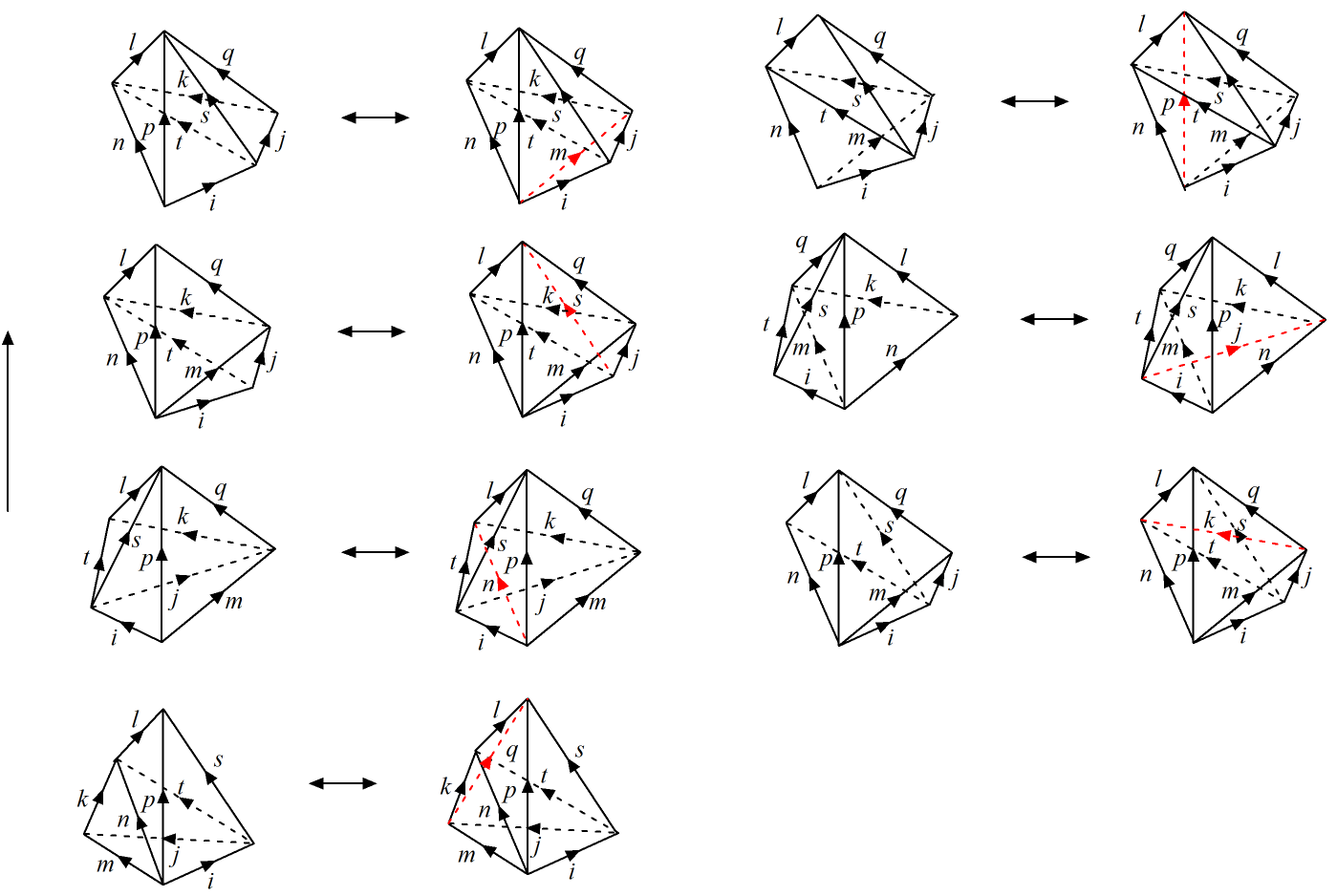}
\end{center}
\caption{All possible time ordered 2-3 move for $G$-symbol.}\label{T23}
\end{figure}

The other seven 2-3 moves induced by time-ordering are:
\begin{equation}
\underset{\kappa}{\sum}
G^{itn,\psi\chi}_{lps,\kappa\gamma}
G^{jkt,\eta\kappa}_{lsq,\delta\phi}
=
\underset{m\alpha\beta\epsilon}{\sum}
(-1)^{s(\alpha)s(\delta)}
\frac{d_{m}}{n_{m}}
(G^{ijm,\alpha\beta}_{knt,\eta\psi})^*
G^{ijm,\alpha\epsilon}_{qps,\phi\gamma}
G^{mkn,\beta\chi}_{lpq,\delta\epsilon},
\label{23move2}
\end{equation}
\begin{equation}
\underset{\beta}{\sum}
(G^{ijm,\alpha\beta}_{knt,\eta\psi})^*
G^{mkn,\beta\chi}_{lpq,\delta\epsilon}
=
(-1)^{s(\alpha)s(\delta)}
\underset{s\phi\gamma\kappa}{\sum}
\frac{d_{s}}{n_{s}}
(G^{ijm,\alpha\epsilon}_{qps,\phi\gamma})^*
G^{itn,\psi\chi}_{lps,\kappa\gamma}
G^{jkt,\eta\kappa}_{lsq,\delta\phi},
\label{23move3}
\end{equation}
\begin{equation}
\underset{\phi}{\sum}
(G^{ijm,\alpha\epsilon}_{qps,\phi\gamma})^*
G^{jkt,\eta\kappa}_{lsq,\delta\phi}
=
(-1)^{s(\alpha)s(\delta)}
\underset{n\psi\beta\chi}{\sum}
\frac{d_{n}}{n_{n}}
(G^{ijm,\alpha\beta}_{knt,\eta\psi})^*
(G^{itn,\psi\chi}_{lps,\kappa\gamma})^*
G^{mkn,\beta\chi}_{lpq,\delta\epsilon},
\label{23move4}
\end{equation}
\begin{equation}
\underset{\psi}{\sum}
G^{ijm,\alpha\beta}_{knt,\eta\psi}
G^{itn,\psi\chi}_{lps,\kappa\gamma}
=
\underset{q\epsilon\delta\phi}{\sum}
(-1)^{s(\alpha)s(\delta)}
\frac{d_{q}}{n_{q}} 
G^{ijm,\alpha\epsilon}_{qps,\phi\gamma}
G^{mkn,\beta\chi}_{lpq,\delta\epsilon}
(G^{jkt,\eta\kappa}_{lsq,\delta\phi})^*
,
\label{23move5}
\end{equation}
\begin{equation}
\underset{\eta}{\sum}
G^{ijm,\alpha\beta}_{knt,\eta\psi}
G^{jkt,\eta\kappa}_{lsq,\delta\phi}
=
(-1)^{s(\alpha)s(\delta)}
\underset{p\gamma\epsilon\chi}{\sum}
\frac{d_{p}}{n_{p}} 
G^{ijm,\alpha\epsilon}_{qps,\phi\gamma}
G^{mkn,\beta\chi}_{lpq,\delta\epsilon}
(G^{itn,\psi\chi}_{lps,\kappa\gamma})^*
,
\label{23move6}
\end{equation}
\begin{equation}
\underset{\chi}{\sum}
G^{mkn,\beta\chi}_{lpq,\delta\epsilon}
(G^{itn,\psi\chi}_{lps,\kappa\gamma})^*
=
\underset{j\alpha\eta\phi}{\sum}
(-1)^{s(\alpha)s(\delta)}
\frac{d_{j}}{n_{j}}
G^{ijm,\alpha\beta}_{knt,\eta\psi}
G^{jkt,\eta\kappa}_{lsq,\delta\phi}
(G^{ijm,\alpha\epsilon}_{qps,\phi\gamma})^*
,
\label{23move7}
\end{equation}
\begin{equation}
\underset{\gamma}{\sum}
(G^{ijm,\alpha\epsilon}_{qps,\phi\gamma})^*
G^{itn,\psi\chi}_{lps,\kappa\gamma}
=
\underset{k\beta\delta\eta}{\sum}
(-1)^{s(\alpha)s(\delta)}
\frac{d_{k}}{n_{k}}
(G^{ijm,\alpha\beta}_{knt,\eta\psi})^*
G^{mkn,\beta\chi}_{lpq,\delta\epsilon}
(G^{jkt,\eta\kappa}_{lsq,\delta\phi})^*
,
\label{23move8}
\end{equation}
as shown graphically in Fig. \ref{T23}. Below we will show how to derive these seven 2-3 moves.

\subsection{The additional relations among projective phase factors}
\label{phaserelation}

Since the $G$-move and $F$-move are related through Eq.(\ref{gf}), we can rewrite the four projective unitary conditions in Eq.(\ref{fprojc1}), Eq.(\ref{fprojc2}), Eq.(\ref{projcond3}) and Eq.(\ref{projcond4}) in terms of $G$-move as

\begin{equation}
\underset{n\chi\delta}{\sum}
\frac{d_{n}d_{m}}{n_{n}n_{m}}
G^{ijm',\alpha'\beta'}_{kln,\chi\delta}
(G^{ijm,\alpha\beta}_{kln,\chi\delta})^{*}=
\left\{\begin{array}{l}
\delta_{mm'}\delta_{\alpha\alpha'}\delta_{\beta\beta'}
\text{, \ \ if }m\text{ is m-type
}\\ 
\frac{1}{n_{m}} (\delta_{mm'}\delta_{\alpha\alpha'}\delta_{\beta\beta'}+
\Xi^{ijm,\alpha\beta}_{kl}
\delta_{mm'}\delta_{(\alpha\times f)\alpha'}\delta_{(\beta\times f)\beta'})
\text{, \ \ if }m\text{ is q-type}
\end{array}\right.
\label{gp1}
\end{equation}
\begin{equation}
\underset{m\alpha\beta}{\sum}
\frac{d_{n}d_{m}}{n_{n}n_{m}}
(G^{ijm,\alpha\beta}_{kln',\chi'\delta'})^{*}
G^{ijm,\alpha\beta}_{kln,\chi\delta} =
\left\{\begin{array}{l}
\delta_{nn'}\delta_{\chi\chi'}\delta_{\delta\delta'}
\text{, \ \ if }n\text{ is m-type
}\\ 
\frac{1}{n_{n}} (\delta_{nn'}\delta_{\chi\chi'}\delta_{\delta\delta'}
+
\Xi^{ij}_{kln,\chi\delta}
\delta_{nn'}\delta_{(\chi\times f)\chi'}\delta_{(\delta\times f)\delta'})
\text{, \ \ if }n\text{ is q-type}
\end{array}\right.
\label{gp2}
\end{equation}
\begin{equation}
\underset{n\chi\delta}{\sum}
\frac{d_{n}d_{m}}{n_{n}n_{m}}
G^{im'k,\alpha'\delta}_{lnj,\beta'\chi}
(G^{imk,\alpha\delta}_{lnj,\beta\chi})^{*}
=
\left\{\begin{array}{l}
\delta_{mm'}\delta_{\alpha\alpha'}\delta_{\beta\beta'}
\text{, \ \ if }m\text{ is m-type
}\\ 
\frac{1}{n_{m}} (\delta_{mm'}\delta_{\alpha\alpha'}\delta_{\beta\beta'}+
\Omega^{kim,\alpha\beta}_{jl}
\delta_{mm'}\delta_{(\alpha\times f)\alpha'}\delta_{(\beta\times f)\beta'})
\text{, \ \ if }m\text{ is q-type}
\end{array}\right.
\label{gp3}
\end{equation}
\begin{equation}
\underset{m\alpha\beta}{\sum}
\frac{d_{n}d_{m}}{n_{n}n_{m}} 
(G^{imk,\alpha\delta'}_{ln'j,\beta\chi'})^{*}
G^{imk,\alpha\delta}_{lnj,\beta\chi}
=
\left\{\begin{array}{l}
\delta_{nn'}\delta_{\chi\chi'}\delta_{\delta\delta'}
\text{, \ \ if }n\text{ is m-type
}\\ 
\frac{1}{n_{n}} (\delta_{nn'}\delta_{\chi\chi'}\delta_{\delta\delta'}+
\Omega^{ki}_{jln,\chi\delta}
\delta_{nn'}\delta_{(\chi\times f)\chi'}\delta_{(\delta\times f)\delta'})
\text{, \ \ if }n\text{ is q-type}
\end{array}\right.
\label{gp4}
\end{equation}

\end{widetext}

Consistency between the fermionic Pentagon equation in Eq.(\ref{23move1}) and four projective unitary conditions in Eq.(\ref{gp1})-Eq.(\ref{gp4}) (all in terms of $G$-move) can induce many relations among the phase factors. Here we only focus on the relations that are required to fully construct the fermionic partition function in Eq.(\ref{partition}). 

The above four projective unitary conditions induce the following four equivalence relations for $G$-move:
\begin{equation}
G^{ijm,(\alpha\times f)(\beta\times f)}_{kln,\chi\delta}
=
\Xi^{ijm,\alpha\beta}_{kl}
G^{ijm,\alpha\beta}_{kln,\chi\delta}
\text{, \ \ if }m\text{ is q-type}
\label{geq1}
\end{equation}
\begin{equation}
(G^{ijm,\alpha\beta}_{kln,(\chi\times f)(\delta\times f)})^*
=
\Xi^{ij}_{kln,\chi\delta}
(G^{ijm,\alpha\beta}_{kln,\chi\delta})^*
\text{, \ \ if }n\text{ is q-type}
\label{geq2}
\end{equation}
\begin{equation}
G^{imk,(\alpha\times f)\delta}_{lnj,(\beta\times f)\chi}
=
\Omega^{kim,\alpha\beta}_{jl}
G^{imk,\alpha\delta}_{lnj,\beta\chi}
\text{, \ \ if }m\text{ is q-type}
\label{geq3}
\end{equation}
\begin{equation}
(G^{imk,\alpha(\delta\times f)}_{lnj,\beta(\chi\times f)})^*
=
\Omega^{ki}_{jln,\chi\delta}
(G^{imk,\alpha\delta}_{lnj,\beta\chi})^*
\text{, \ \ if }n\text{ is q-type}
\label{geq4}
\end{equation}
In addition to the four phase factors $\Xi^{ijm,\alpha\beta}_{kl}$, $\Xi^{ij}_{kln,\chi\delta}$, $\Omega^{kim,\alpha\beta}_{jl}$ and $\Omega^{ki}_{jln,\chi\delta}$ we defined above, we need to define a new phase factor $\Delta^{mji,\alpha\delta}_{nl}$ to construct the topological invariant partition function, as the following 
\begin{equation}
G^{ijm,(\alpha\times f)\beta}_{kln,\chi(\delta\times f)}
=
\Delta^{mji,\alpha\delta}_{nl}
G^{ijm,\alpha\beta}_{kln,\chi\delta}
\text{, \ \ if }i\text{ is q-type,}
\label{geq5}
\end{equation}
which corresponds to the changing of fermion parity on two diagonal fusion states $\alpha$ and $\delta$, and the phase factor $\Delta^{mji,\alpha\delta}_{nl}$ can be explicitly constructed through a sequence of $F$-move and $O$-moves, as introduced in Appendix \ref{Amove}.

To derive the rest seven 2-3 moves induced by time-ordering, i.e., to fully establish the topological invariance of partition function, the following four relations on phase factors are required (See full details in Appendix \ref{check23}):
\begin{align}
\Xi^{ij}_{knt,\eta\psi}
=
\Omega^{nit,\psi\kappa}_{sl}
\Xi^{jkt,\eta\kappa}_{ls},
\label{pr1}
\end{align}

\begin{align}
\Xi^{ijm,\alpha\epsilon}_{qp}=\Xi^{ijm,\alpha\beta}_{kn}
(\Delta^{nkm,\beta\epsilon}_{qp})^*
,
\label{pr2}
\end{align}
\begin{align}
(\Xi^{it}_{lps,\kappa\gamma})^*=(\Xi^{ij}_{qps,\phi\gamma})^*
\Omega^{tj}_{qls,\kappa\phi}
,
\label{pr3}
\end{align}
\begin{align}
(\Omega^{mi}_{sqp,\epsilon\gamma})^*=\Omega^{nm}_{qlp,\chi\epsilon}
(\Omega^{ni}_{slp,\chi\gamma})^*
,
\label{pr6}
\end{align}

All above relations can be obtained as simple solutions by comparing the fermionic Pentagon equation in Eq.(\ref{23move1}) with the equivalence relations in Eq.(\ref{geq1})-Eq.(\ref{geq5}). Eq. (\ref{pr1}) and Eq. (\ref{pr3}) are derived in section \ref{relationphase}. We can obtain the rest relations in Eq.(\ref{pr2}) and Eq.(\ref{pr6}) in a similar manner.

\begin{widetext}

\subsection{1-4 moves}

The second type of Pachner move is the 1-4 move. There are three different 1-4 moves induced by a global time ordering:
\begin{equation}
G^{ijm,\alpha\epsilon}_{qps,\phi\gamma}
=
(-1)^{s(\alpha)s(\delta)}
\frac{1}{D^2}
\underset{ntkl}{\sum}
\underset{\beta\eta\psi\chi\kappa\delta}{\sum}
\frac{d_{n}d_{t}d_{k}d_{l}}
{n_{n}n_{t}n_{k}n_{l}} 
G^{ijm,\alpha\beta}_{knt,\eta\psi}
G^{itn,\psi\chi}_{lps,\kappa\gamma}
(G^{mkn,\beta\chi}_{lpq,\delta\epsilon})^*
G^{jkt,\eta\kappa}_{lsq,\delta\phi},
\label{14move1}
\end{equation}
\begin{equation}
G^{mkn,\beta\chi}_{lpq,\delta\epsilon}
=
(-1)^{s(\alpha)s(\delta)}
\frac{1}{D^2}
\underset{ijts}{\sum}
\underset{\alpha\eta\psi\phi\kappa\gamma}{\sum}
\frac{d_{i}d_{j}d_{t}d_{s}}
{n_{i}n_{j}n_{t}n_{s}} 
G^{ijm,\alpha\beta}_{knt,\eta\psi}
(G^{ijm,\alpha\epsilon}_{qps,\phi\gamma})^*
G^{itn,\psi\chi}_{lps,\kappa\gamma}
G^{jkt,\eta\kappa}_{lsq,\delta\phi},
\label{14move2}
\end{equation}
\begin{equation}
G^{itn,\psi\chi}_{lps,\kappa\gamma}
=
(-1)^{s(\alpha)s(\delta)}
\frac{1}{D^2}
\underset{mjkq}{\sum}
\underset{\alpha\beta\eta\epsilon\phi\delta}{\sum}
\frac{d_{m}d_{j}d_{k}d_{q}}
{n_{m}n_{j}n_{k}n_{q}} 
(G^{ijm,\alpha\beta}_{knt,\eta\psi})^*
G^{ijm,\alpha\epsilon}_{qps,\phi\gamma}
G^{mkn,\beta\chi}_{lpq,\delta\epsilon}
(G^{jkt,\eta\kappa}_{lsq,\delta\phi})^*.
\label{14move3}
\end{equation}

Combining all the 2-3 moves in Eq.(\ref{23move1})-Eq.(\ref{23move8}) and all the 1-4 moves in Eq.(\ref{14move1})-Eq.(\ref{14move3}), the following relations can be derived:
\begin{equation}
\sum_{kjm\alpha\beta\chi}
\frac{d_{k}d_{j}d_{m}}
{n_{k}n_{j}n_{m}} 
(G^{ijm,\alpha\beta}_{kln,\chi\delta})^*
G^{ijm,\alpha\beta}_{kln,\chi\delta}
=
D^2,
\label{r1p}
\end{equation}
\begin{equation}
\sum_{ijn\alpha\chi\delta}
\frac{d_{i}d_{j}d_{n}}
{n_{i}n_{j}n_{n}} 
(G^{ijm,\alpha\beta}_{kln,\chi\delta})^*
G^{ijm,\alpha\beta}_{kln,\chi\delta}
=
D^2,
\label{r2p}
\end{equation}
\begin{equation}
\sum_{ilm\alpha\beta\delta}
\frac{d_{i}d_{l}d_{m}}
{n_{i}n_{l}n_{m}} 
G^{ijm,\alpha\beta}_{kln,\chi\delta}
(G^{ijm,\alpha\beta}_{kln,\chi\delta})^*
=
D^2,
\label{r3p}
\end{equation}
\begin{equation}
\sum_{kln\beta\chi\delta}
\frac{d_{k}d_{l}d_{n}}
{n_{k}n_{l}n_{n}} 
G^{ijm,\alpha\beta}_{kln,\chi\delta}
(G^{ijm,\alpha\beta}_{kln,\chi\delta})^*
=
D^2,
\label{r4p}
\end{equation}
For example, we show how to derive Eq.(\ref{r1p}) by comparing Eq.(\ref{23move1}) with Eq.(\ref{14move1}). We multiply by $(G^{mkn,\beta\chi}_{lpq,\delta\epsilon})^*$ and sum over $l,k,n,\beta,\chi,\delta$ on both sides of Eq.(\ref{23move1}):
\begin{equation}
\underset{\epsilon,lkn\beta\chi\delta}{\sum}
G^{ijm,\alpha\epsilon}_{qps,\phi\gamma}
(G^{mkn,\beta\chi}_{lpq,\delta\epsilon})^*
G^{mkn,\beta\chi}_{lpq,\delta\epsilon}
=(-1)^{s(\alpha)s(\delta)}
\underset{t\eta\psi\kappa,lkn\beta\chi\delta}{\sum}
\frac{d_{t}}{n_{t}} 
G^{ijm,\alpha\beta}_{knt,\eta\psi}
G^{itn,\psi\chi}_{lps,\kappa\gamma}
(G^{mkn,\beta\chi}_{lpq,\delta\epsilon})^*
G^{jkt,\eta\kappa}_{lsq,\delta\phi}.
\label{r1pd}
\end{equation}
We see that the difference between Eq.(\ref{23move1}) and Eq.(\ref{14move1}) can be exactly compensated by Eq.(\ref{r1p}) up to a relabelling on indices.

Eq.(\ref{r1p})-Eq.(\ref{r4p}) together with the projective unitary conditions in Eq.(\ref{gp1})-Eq.(\ref{gp4}) further imply:
\begin{equation}
\underset{ij}{\sum} 
\frac{N^{ij}_{k}d_{i}d_{j}}{n_{i}n_{j}}
=d_{k}D^{2}.
\label{qdr}
\end{equation}
which is exactly Eq.(\ref{QD}). For example, we show how to derive Eq.(\ref{qdr}) from Eq.(\ref{r1p}) and Eq.(\ref{gp2}) in two cases below:

(1) If $n$ is m-type, replacing 
$
\underset{m\alpha\beta}{\sum}
\frac{d_{m}d_{n}}{n_{m}}
(G^{ijm,\alpha\beta}_{kln,\chi\delta})^{*}
G^{ijm,\alpha\beta}_{kln,\chi\delta} =
1
$ 
into Eq.(\ref{r1p}), we obtain
\begin{equation}
\sum_{kj\chi}
\frac{d_{k}d_{j}}
{n_{k}n_{j}d_{n}} 
=
D^2,
\end{equation}
where $\sum_{\chi}$ only counts the number of possible states of $\chi$ and can be replaced by $N^{jk}_{n}$. We find it is exactly Eq.(\ref{qdr}) up to a relabelling on indices.

(2) If $n$ is q-type, replacing 
$
\underset{m\alpha\beta}{\sum}
\frac{d_{m}d_{n}}{n_{m}n_{n}}
(G^{ijm,\alpha\beta}_{kln,\chi\delta})^{*}
G^{ijm,\alpha\beta}_{kln,\chi\delta} =
\frac{1}{n_{n}} 
$ 
into Eq.(\ref{r1p}), it becomes the same equation as above. So that we can again obtain Eq.(\ref{qdr}).


\section{Examples}
\label{ex}

In this section, we derive all equivalence relations and $F$-moves for all following examples, as listed in Appendix \ref{equirela}. We write down the explicit expressions of the phase factors $\Xi^{ijm,\alpha\beta}_{kl}$, $\Xi^{ij}_{kln,\chi\delta}$, $\Omega^{kim,\alpha\beta}_{jl}=\Xi^{is}_{tkm,\eta\alpha}
(\Xi^{stm,\eta\beta}_{lj})^*$ and $\Omega^{ki}_{jln,\chi\delta}=\Xi^{si}_{jtn,\delta\eta}
(\Xi^{sk}_{ltn,\chi\eta})^*$ in Appendix \ref{fcm}. We note that we didn't choose the gauge such that all $\Lambda^{ij,\alpha\beta}_{k}=1$ as illustrated in section \ref{relationphase}. But we choose the gauge such that all $\Xi^{si}_{jtn,\delta\eta}
(\Xi^{sk}_{ltn,\chi\eta})^*=1$, and then all $\widetilde{\Lambda}^{ij,\alpha\beta}_{k}=1$ and all $\widetilde{\zeta}^{ki}_{jln,\chi\delta}=1$. We verify that all $F$-moves in each example exactly satisfy the corresponding four projective unitary condition, as well as the fermionic Pentagon equation.

\subsection{Fermionic topological order $SO(3)_6/\psi$}

In the fermionic topological phase $SO(3)_6/\psi$, we have two strings $\{1,s\}$, where $1$ is the vacuum string, and $s$ is an m-type string.  The quantum dimensions are given by
\begin{align}
d_1=1, \ \ \ 
d_{s}=1+\sqrt{2}.
\end{align}
 
The fusion rules are given by 
\begin{align}
1\times s=s\times 1=s, \ \ \ 
s\times s=1+\mathbb{C}^{1|1}s,
\end{align}
where we use the notation $\mathbb{C}^{B^{ab}_c|F^{ab}_c}$ to denote the number of bosonic and fermionic fusion state for a given fusion space $V^{ab}_{c}$. The fusion rules written as fusion tensors are
\begin{equation}
B^{11}_{1}=B^{1s}_{s}=B^{s1}_{s}=B^{ss}_{1}=B^{ss}_{s}=F^{ss}_{s}=1,
\end{equation}
and all other fusion tensors are zero.

Since the fermionic theory $SO(3)_6/\psi$ only contains m-type strings, and the dimension of endomorphism for m-type strings $n_{i}=1$. 
We list all $F$-moves of $SO(3)_6/\psi$ in Appendix \ref{example1}.

\subsection{Majorana toric code}

In the Majorana toric code, we have two string types $\{1,\sigma\}$, where $1$ is the vacuum string, $\sigma$ is a q-type Majorana string.

The quantum dimensions are given by
\begin{align}
d_1=1, \ \ \ 
d_{\sigma}=\sqrt{2}.
\end{align}
 
The fusion rules are given by 
\begin{align}
1\times \sigma=\sigma\times 1=\mathbb{C}^{1|1}\sigma, \ \ \ 
\sigma\times \sigma=\mathbb{C}^{1|1}1,
\end{align}
written in fusion tensors as
\begin{equation}
B^{11}_{1}=B^{1\sigma}_{\sigma}=B^{\sigma 1}_{\sigma}=B^{\sigma\sigma}_{1}=F^{1\sigma}_{\sigma}=F^{\sigma1}_{\sigma}=F^{\sigma\sigma}_{1}=1,
\end{equation}
and all other fusion tensors are zero.

The four projective unitary conditions for Majorana toric code are:

\begin{equation}
\underset{n\chi\delta}{\sum} F^{ijm',\alpha'\beta'}_{kln,\chi\delta} (F^{ijm,\alpha\beta}_{kln,\chi\delta})^{*}=
\left\{\begin{array}{l}
\delta_{mm'}\delta_{\alpha\alpha'}\delta_{\beta\beta'}
\text{, \ \ if }m\text{ is m-type
}\\ 
\frac{1}{2} (\delta_{mm'}\delta_{\alpha\alpha'}\delta_{\beta\beta'}+
\delta_{mm'}\delta_{(\alpha\times f)\alpha'}\delta_{(\beta\times f)\beta'})
\text{, \ \ if }m\text{ is q-type}
\end{array}\right.
\end{equation}

\begin{equation*}
\underset{m\alpha\beta}{\sum} (F^{ijm,\alpha\beta}_{kln',\chi'\delta'})^{*} F^{ijm,\alpha\beta}_{kln,\chi\delta} =
\left\{\begin{array}{l}
\delta_{nn'}\delta_{\chi\chi'}\delta_{\delta\delta'}
\text{, \ \ if }n\text{ is m-type
}\\ 
\frac{1}{2} (\delta_{nn'}\delta_{\chi\chi'}\delta_{\delta\delta'}
+
\Theta_{fi}^{\Gamma_{\delta}}
\delta_{nn'}\delta_{(\chi\times f)\chi'}\delta_{(\delta\times f)\delta'})
\text{, \ \ if }n\text{ is q-type}
\end{array}\right.
\end{equation*}
\begin{equation}
\text{ \\where }
\Theta_{fi}=
\left\{\begin{array}{l}
1
\text{, \ \ if }i\text{ is m-type
}\\ 
i
\text{, \ \ if }i\text{ is q-type}
\end{array}\right.
\text{\ \ and } \Gamma_{\delta}=
\left\{\begin{array}{l}
1
\text{, \ \ if } s(\delta)=0
\\ 
*
\text{, \ \ if } s(\delta)=1
\end{array}\right.
\end{equation}


\begin{equation*}
\underset{n\chi\delta}{\sum}
\frac{d_{n}}{n_{n}} 
F^{im'k,\alpha'\delta}_{lnj,\beta'\chi}
(F^{imk,\alpha\delta}_{lnj,\beta\chi})^{*}
=\frac{d_{j}d_{k}n_{m}}{n_{j}n_{k}d_{m}} 
\left\{\begin{array}{l}
\delta_{mm'}\delta_{\alpha\alpha'}\delta_{\beta\beta'}
\text{, \ \ if }m\text{ is m-type
}\\ 
\frac{1}{2} (\delta_{mm'}\delta_{\alpha\alpha'}\delta_{\beta\beta'}+
\Theta_{fi}^{\Gamma_{\alpha}}
\delta_{mm'}\delta_{(\alpha\times f)\alpha'}\delta_{(\beta\times f)\beta'})
\text{, \ \ if }m\text{ is q-type}
\end{array}\right.
\end{equation*}

\begin{equation}
\underset{m\alpha\beta}{\sum}
\frac{d_{m}}{n_{m}} 
(F^{imk,\alpha\delta'}_{ln'j,\beta\chi'})^{*}
F^{imk,\alpha\delta}_{lnj,\beta\chi}
=\frac{d_{j}d_{k}n_{n}}{n_{j}n_{k}d_{n}} 
\left\{\begin{array}{l}
\delta_{nn'}\delta_{\chi\chi'}\delta_{\delta\delta'}
\text{, \ \ if }n\text{ is m-type
}\\ 
\frac{1}{2} (\delta_{nn'}\delta_{\chi\chi'}\delta_{\delta\delta'}+
\delta_{nn'}\delta_{(\chi\times f)\chi'}\delta_{(\delta\times f)\delta'})
\text{, \ \ if }n\text{ is q-type}
\end{array}\right.
\label{w6}
\end{equation}

We list all $F$-moves of Majorana toric code in Appendix \ref{example2}. We have checked numerically that the $F$-moves satisfy all above projective unitary conditions.



\subsection{Fermionic topological order $\frac{1}{2}E_6/\psi$}

In the fermionic topological phase $\frac{1}{2}E_6/\psi$, we have two string types $\{1,x\}$, where $1$ is the vacuum string, and $x$ is a q-type Majorana string. 

The quantum dimensions are given by
\begin{align}
d_1=1, \ \ \ 
d_x=1+\sqrt{3}.
\end{align}
 
The fusion rules are given by 
\begin{align}
1\times x=x\times 1=\mathbb{C}^{1|1}x, \ \ \ 
x\times x=\mathbb{C}^{1|1}1+\mathbb{C}^{2|2}x,
\end{align}
written in fusion tensors as
\begin{align}
B^{11}_{1}=B^{1x}_{x}=B^{x1}_{x}=B^{xx}_{1}=F^{1x}_{x}=F^{x1}_{x}=F^{xx}_{1}=1, 
\ \ \ \ \ \ 
B^{xx}_{x}=F^{xx}_{x}=2,
\end{align}
where all other fusion tensors vanish.  We note that non-trivially $B^{xx}_{x}=2$, i.e. we have two bosonic fusion states if we fuse two $x$ and again obtain $x$. We denote the two bosonic fusion states as "1" and "2" respectively.

By fermion condensation, invoking the $F$-symbols in the bosonic $\frac{1}{2}E_6$ theory in a certain gauge in Eq.(\ref{bf1})-Eq.(\ref{bf6}) in Appendix \ref{fc}, the equivalence relations Eq.(\ref{feq1a}), Eq.(\ref{feq2a}), Eq.(\ref{oeq1a}), Eq.(\ref{heq1a}) and Eq.(\ref{heq2a}) have the forms:
\begin{equation}
F^{ijm,(\alpha\times f)(\beta\times f)}_{kln,\chi\delta}
=
\left\{\begin{array}{l}
F^{ijm,\alpha\beta}_{kln,\chi\delta}
\text{, \ \ if }m\text{ is q-type and } B^{mk}_{l}=1
\\
(-\sigma_{y})_{B(\beta\times f)B(\beta)}
F^{ijm,\alpha\beta}_{kln,\chi\delta}
\text{, \ \ if }m\text{ is q-type and } B^{mk}_{l}=2
\end{array}\right.
\label{e6f1}
\end{equation}
\begin{equation}
(F^{ijm,\alpha\beta}_{kln,(\chi\times f)(\delta\times f)})^*
=
\left\{\begin{array}{l}
\Theta_{fi}^{\Gamma_{\delta}}
(F^{ijm,\alpha\beta}_{kln,\chi\delta})^*
\text{, \ \ if }n\text{ is q-type and } B^{in}_{l}=1
\\ 
(\sigma_{x})_{B(\delta\times f)B(\delta)}
(F^{ijm,\alpha\beta}_{kln,\chi\delta})^*
\text{, \ \ if }n\text{ is q-type and } B^{in}_{l}=2
\end{array}\right.
\label{e6f2}
\end{equation}

\begin{equation}
O^{ij,(\alpha\times f)(\beta\times f)}_{k}
=
\left\{\begin{array}{l}
O^{ij,\alpha\beta}_{k}   
\text{, \ \ if }k\text{ is q-type and } B^{ij}_{k}=1
\\ 
(-\sigma_{y})_{B(\alpha\times f)B(\alpha)}
(-\sigma_{y})_{B(\beta\times f)B(\beta)}
O^{ij,\alpha\beta}_{k}   
\text{, \ \ if }k\text{ is q-type and } B^{ij}_{k}=2
\end{array}\right.
\label{oeq1a}
\end{equation}

Our notation of Pauli matrices appear whenever a changing of fermion-parity alters the bosonic states in any equivalence relation. If a Pauli matrix $\sigma_{i}$, where $i=x,y$,  corresponds to a fusion state $\alpha$, then the rows represent $B(\alpha\times f)$ is 1 or 2, and columns of $\sigma_{i}$ represent the values of $B(\alpha)$ is 1 or 2. And we use $(\sigma_{i})_{B(\alpha\times f)B(\alpha)}$ to represent the entry of $\sigma_{i}$ in row $B(\alpha\times f)$ and column $B(\alpha)$, which is simply a phase factor. For example, the notation $(-\sigma_{y})_{B(\beta\times f)B(\beta)}$ in Eq.(\ref{e6f1}) represents the phase factor:
\begin{equation}
(-\sigma_{y})_{B(\beta\times f)B(\beta)}
=
\left\{\begin{array}{l}
i  
\text{, \ \ if }B(\beta\times f)=1, B(\beta)=2
\\ 
-i  
\text{, \ \ if }B(\beta\times f)=2, B(\beta)=1
\\
0
\text{, \ \ otherwise}
\end{array}\right.
\end{equation}
And the notation $(\sigma_{x})_{B(\delta\times f)B(\delta)}$ in Eq.(\ref{e6f2}) represents:
\begin{equation}
(\sigma_{x})_{B(\delta\times f)B(\delta)}
=
\left\{\begin{array}{l}
1
\text{, \ \ if }B(\delta\times f)=1, B(\delta)=2
\\ 
1
\text{, \ \ if }B(\delta\times f)=2, B(\delta)=1
\\
0
\text{, \ \ otherwise}
\end{array}\right.
\end{equation}

Taking the gauge choice Eq.(\ref{Ogauge}) on the involved $O$-move in Eq.(\ref{hyfo}), 
\begin{equation}
O^{ml,(\beta\times f)}_{j}
=
\left\{\begin{array}{l}
O^{ml,\beta}_{j}
\text{, \ \ if }j\text{ is q-type and } B^{ml}_{j}=1
\\ 
(-\sigma_{y})_{B(\beta\times f)B(\beta)}
(-\sigma_{y})_{B(\beta\times f)B(\beta)}
O^{ml,\beta}_{j}
\text{, \ \ if }j\text{ is q-type and } B^{ml}_{j}=2
\end{array}\right.
\label{e3o}
\end{equation}

\begin{equation}
\widetilde{H}^{kim,(\alpha\times f)(\beta\times f)}_{jln,\chi\delta}
=
\left\{\begin{array}{l}
\Theta_{fi}^{\Gamma_{\alpha}}
\widetilde{H}^{kim,\alpha\beta}_{jln,\chi\delta}
\text{,\ \ if }m\text{ is q-type,  } B^{im}_{k}=1, B^{ml}_{j}=1
\\
\Theta_{fi}^{\Gamma_{\alpha}}
(-\sigma_{y})_{B(\beta\times f)B(\beta)}
\widetilde{H}^{kim,\alpha\beta}_{jln,\chi\delta}
\text{,\ \ if }m\text{ is q-type,  } B^{im}_{k}=1, B^{ml}_{j}=2
\\ 
(\sigma_{x})_{B(\alpha\times f)B(\alpha)}
\widetilde{H}^{kim,\alpha\beta}_{jln,\chi\delta}
\text{,\ \ if }m\text{ is q-type,  } B^{im}_{k}=2, B^{ml}_{j}=1
\\
(-\sigma_{y})_{B(\beta\times f)B(\beta)}
(\sigma_{x})_{B(\alpha\times f)B(\alpha)}
\widetilde{H}^{kim,\alpha\beta}_{jln,\chi\delta}
\text{,\ \ if }m\text{ is q-type,  } B^{im}_{k}=2, B^{ml}_{j}=2
\end{array}\right.
\label{e3h3}
\end{equation}

\begin{equation}
(\widetilde{H}^{kim,\alpha\beta}_{jln,(\chi\times f)(\delta\times f)})^*
=
(\widetilde{H}^{kim,\alpha\beta}_{jln,\chi\delta})^*
\text{,\ \ if }n\text{ is q-type,  }.
\end{equation}

By Eq.(\ref{hyfo}), we replace the dual $H$-move in Eq.(\ref{e3h3}) by $F$-move, which will also bring a phase factor from equivalent $O$-moves in Eq.(\ref{e3o}). We should also note that when two such matrices multiply together, we are not doing matrix multiplication, but we should multiply by each entry.  For example, $(-\sigma_{y})_{B(\beta\times f)B(\beta)}(-\sigma_{y})_{B(\beta\times f)B(\beta)}=(-\sigma_{x})_{B(\beta\times f)B(\beta)}$.

We obtain the four projective unitary conditions for fermonic topological order $\frac{1}{2}E_6/\psi$:

\begin{equation}
\underset{n\chi\delta}{\sum} 
F^{ijm',\alpha'\beta'}_{kln,\chi\delta}
(F^{ijm,\alpha \beta}_{kln,\chi\delta})^*=
\left\{\begin{array}{l}
\delta_{mm'}\delta_{\alpha\alpha'}\delta_{\beta\beta'}
\text{, \ \ if }m\text{ is m-type}
\\ 
\frac{1}{2} (
\delta_{mm'}\delta_{\alpha\alpha'}\delta_{\beta\beta'}+
\delta_{mm'}\delta_{(\alpha\times f)\alpha'}
\delta_{(\beta\times f)\beta'})
\text{, \ \ if }m\text{ is q-type and } B^{mk}_{l}=1
\\
\frac{1}{2} (
\delta_{mm'}\delta_{\alpha\alpha'}\delta_{\beta\beta'}+
(-\sigma_{y})_{B(\beta\times f)B(\beta)}
\delta_{mm'}\delta_{(\alpha\times f)\alpha'}
\delta_{(\beta\times f)\beta'})
\text{, \ \ if }m\text{ is q-type and } B^{mk}_{l}=2
\end{array}\right.
\end{equation}

\begin{equation}
\underset{m\alpha\beta}{\sum} 
(F^{ijm,\alpha\beta}_{kln',\chi' \delta'})^*
F^{ijm,\alpha\beta}_{kln,\chi \delta} =
\left\{\begin{array}{l}
\delta_{nn'}\delta_{\chi\chi'}\delta_{\delta\delta'}
\text{, \ \ if }n\text{ is m-type
}
\\ 
\frac{1}{2}(\delta_{nn'}\delta_{\chi\chi'}\delta_{\delta\delta'}
+
\Theta_{fi}^{\Gamma_{\delta}}
\delta_{nn'}\delta_{(\chi\times f)\chi'}\delta_{(\delta\times f)\delta'})
\text{, \ \ if }n\text{ is q-type and } B^{in}_{l}=1
\\
\frac{1}{2} (
\delta_{nn'}\delta_{\chi\chi'}\delta_{\delta\delta'}
+
(\sigma_{x})_{B(\delta\times f)B(\delta)}
\delta_{nn'}\delta_{(\chi\times f)\chi'}\delta_{(\delta\times f)\delta'})
\text{, \ \ if }n\text{ is q-type and } B^{in}_{l}=2
\end{array}\right.
\end{equation}
\begin{equation*}
\text{ \\ where }
\Theta_{fi}=
\left\{\begin{array}{l}
1
\text{, \ \ if }i\text{ is m-type
}\\ 
i
\text{, \ \ if }i\text{ is q-type}
\end{array}\right.
\text{ and }
\Gamma_{\delta}=
\left\{\begin{array}{l}
1
\text{, \ \ if }s(\delta)=0
\\ 
*
\text{, \ \ if }s(\delta)=1
\end{array}\right.,
\end{equation*}

\begin{equation}
\underset{n\chi\delta}{\sum}
\frac{d_{n}}{n_{n}} 
F^{im'k,\alpha'\delta}_{lnj,\beta'\chi}
(F^{imk,\alpha\delta}_{lnj,\beta\chi})^*
=\frac{d_{j}d_{k}n_{m}}{n_{j}n_{k}d_{m}} 
\left\{\begin{array}{l}
\delta_{mm'}\delta_{\alpha\alpha'}\delta_{\beta\beta'}
\text{, \ \ if }m\text{ is m-type
}
\\ 
\frac{1}{2}(\delta_{mm'}\delta_{\alpha\alpha'}\delta_{\beta\beta'}+
\Theta_{fi}^{\Gamma_{\alpha}}
\delta_{mm'}
\delta_{(\alpha\times f)\alpha'}\delta_{(\beta\times f)\beta'}),
\\
\text{\ \ \ \ \ \ \ \ \ \ \ \ \ \ \ \ \ \ \ \ \ \ \ \ \ \ \ \ \ \ \ \ \ \ \ \ \ \ \ \ \ \ \ \ \ \ \ \ \ \ \ \ \ \ 
 if }m\text{ is q-type,  } B^{im}_{k}=1, B^{ml}_{j}=1
\\ 
\frac{1}{2}(\delta_{mm'}\delta_{\alpha\alpha'}\delta_{\beta\beta'}+
\Theta_{fi}^{\Gamma_{\alpha}}
(\sigma_{y})_{B(\beta\times f)B(\beta)}
\delta_{mm'}
\delta_{(\alpha\times f)\alpha'}\delta_{(\beta\times f)\beta'}),
\\
\text{\ \ \ \ \ \ \ \ \ \ \ \ \ \ \ \ \ \ \ \ \ \ \ \ \ \ \ \ \ \ \ \ \ \ \ \ \ \ \ \ \ \ \ \ \ \ \ \ \ \ \ \ \ \ 
 if }m\text{ is q-type,  } B^{im}_{k}=1, B^{ml}_{j}=2
\\
\frac{1}{2} (
\delta_{mm'}\delta_{\alpha\alpha'}\delta_{\beta\beta'}+
(\sigma_{x})_{B(\alpha\times f)B(\alpha)}
\delta_{mm'}
\delta_{(\alpha\times f)\alpha'}\delta_{(\beta\times f)\beta'}),
\\
\text{\ \ \ \ \ \ \ \ \ \ \ \ \ \ \ \ \ \ \ \ \ \ \ \ \ \ \ \ \ \ \ \ \ \ \ \ \ \ \ \ \ \ \ \ \ \ \ \ \ \ \ \ \ \ 
 if }m\text{ is q-type, } B^{im}_{k}=2, B^{ml}_{j}=1
 \\
\frac{1}{2} (
\delta_{mm'}\delta_{\alpha\alpha'}\delta_{\beta\beta'}+
(\sigma_{y})_{B(\beta\times f)B(\beta)}
(\sigma_{x})_{B(\alpha\times f)B(\alpha)}
\delta_{mm'}
\delta_{(\alpha\times f)\alpha'}
\delta_{(\beta\times f)\beta'}),
\\
\text{\ \ \ \ \ \ \ \ \ \ \ \ \ \ \ \ \ \ \ \ \ \ \ \ \ \ \ \ \ \ \ \ \ \ \ \ \ \ \ \ \ \ \ \ \ \ \ \ \ \ \ \ \ \ 
 if }m\text{ is q-type, } B^{im}_{k}=2, B^{ml}_{j}=2
\end{array}\right.
\end{equation}


\begin{equation}
\underset{m\alpha\beta}{\sum}
\frac{d_{m}}{n_{m}} 
(F^{imk,\alpha\delta'}_{ln'j,\beta\chi'})^*
F^{imk,\alpha\delta}_{lnj,\beta\chi}
=\frac{d_{j}d_{k}n_{n}}{n_{j}n_{k}d_{n}} 
\left\{\begin{array}{l}
\delta_{nn'}\delta_{\chi\chi'}\delta_{\delta\delta'}
\text{, \ \ if }n\text{ is m-type
}\\ 
\frac{1}{2} (\delta_{nn'}\delta_{\chi\chi'}\delta_{\delta\delta'}+
\delta_{nn'}\delta_{(\chi\times f)\chi'}
\delta_{(\delta\times f)\delta'})
\text{, \ \ if }n\text{ is q-type}
\end{array}\right.
\end{equation}

We list all $F$-moves of $\frac{1}{2}E_6/\psi$ in Appendix \ref{example3}. We have checked numerically that the $F$-moves satisfies all above projective unitary conditions.

\subsection{Fermionic topological order $\mathrm{TY}_{\Z_{2N}}^{t,\varkappa}/\psi_{N}$}

$\mathrm{TY}_{\Z_{2N}}^{t,\varkappa}/\psi_{N}$ is the Tambara-Yamagami category after condensing the fermion $\psi_{N}$, with symmetric non-degenerate bicharacter of type $t$ ($1\le t \le 2N-1$ and $\operatorname{gcd}(t,2N)=1$):
\begin{align}
\chi_t(a,b)=e^{2\pi itab/(2N)},
\end{align}
and $\varkappa$ is the Frobenius-Schur indicator.  The fusion category $\mathrm{TY}_{\Z_{2N}}^{t,\varkappa}/\psi_{N}$ is a generalization of the Majorana toric code, i.e., the Majorana toric code is the special case $N=1$ in $\mathrm{TY}_{\Z_{2N}}^{t,\varkappa}/\psi_{N}$.

$\mathrm{TY}_{\Z_{2N}}^{t,\varkappa}/\psi_{N}$ has string types $\Z_{N}\cup \{\s\}$, where $\sigma$ is a q-type string, $\Z_{N}=\{0,1,\cdots,N-1\}$ are labels of m-type strings.

The quantum dimensions are given by
\begin{align}
d_i=1, \ \ \ 
d_\s=\sqrt{2N}.
\end{align}
 
The fusion rules are given by 
\begin{align}
a\times b=\mathbb{C}^{(1-\floor*{\frac{a+b}{N}})|  \floor*{\frac{a+b}{N}}}[a+b]_{N}, \ \ \ 
a\times \sigma=\sigma\times a=\mathbb{C}^{1|1}\sigma, \ \ \ 
\sigma\times \sigma=\sum_{a\in\Z_{N}}\mathbb{C}^{1|1}a,
\end{align}
where we define $[a+b]_{N}=(a+b)$ (mod $N$), and $\floor*{\frac{a+b}{N}}$ means taking the integer part of $\frac{a+b}{N}$. In fusion tensors,
\begin{align}
B^{ab}_{[a+b]_{N}}=1, F^{ab}_{[a+b]_{N}}=0, \text{if }a+b< N,
\nonumber\\
B^{ab}_{[a+b]_{N}}=0, F^{ab}_{[a+b]_{N}}=1, \text{if }a+b\geq N,
\nonumber\\
B^{1\sigma}_{\sigma}=B^{\sigma1}_{\sigma}=B^{\sigma\sigma}_{1}=F^{1\sigma}_{\sigma}=F^{\sigma 1}_{\sigma}=F^{\sigma\sigma}_{1}=1, 
\end{align}
where all other fusion tensors vanish.  

We only consider $N$ to be odd here. The four projective unitary conditions for $\mathrm{TY}_{\Z_{2N}}^{t,\varkappa}/\psi_{N}$  are:

\begin{equation}
\underset{n\chi\delta}{\sum} F^{ijm',\alpha'\beta'}_{kln,\chi\delta} (F^{ijm,\alpha\beta}_{kln,\chi\delta})^{*}=
\left\{\begin{array}{l}
\delta_{mm'}\delta_{\alpha\alpha'}\delta_{\beta\beta'}
\text{, \ \ if }m\text{ is m-type
}
\\
\frac{1}{2} (\delta_{mm'}\delta_{\alpha\alpha'}\delta_{\beta\beta'}+
\delta_{mm'}\delta_{(\alpha\times f)\alpha'}\delta_{(\beta\times f)\beta'})
\text{, \ \ if }m,k\text{ are q-type, and } l \text{ is m-type}
\\
\frac{1}{2} (\delta_{mm'}\delta_{\alpha\alpha'}\delta_{\beta\beta'}+
(-1)^k
\delta_{mm'}\delta_{(\alpha\times f)\alpha'}\delta_{(\beta\times f)\beta'})
\text{, \ \ if }m,l\text{ are q-type, and } k \text{ is m-type}
\end{array}\right.
\end{equation}
where we note that in the third case, $m,l\text{ are q-type and } k \text{ is m-type}$, which means that $m,l$ are the $\s$ string and $k\in \Z_{N}$. So that In the phase factor $(-1)^k$, $k$ just takes value in $\Z_{N}$.

\begin{equation*}
\underset{m\alpha\beta}{\sum} (F^{ijm,\alpha\beta}_{kln',\chi'\delta'})^{*} F^{ijm,\alpha\beta}_{kln,\chi\delta} =
\left\{\begin{array}{l}
\delta_{nn'}\delta_{\chi\chi'}\delta_{\delta\delta'}
\text{, \ \ if }n\text{ is m-type
}
\\ 
\frac{1}{2} (\delta_{nn'}\delta_{\chi\chi'}\delta_{\delta\delta'}
+
(-1)^i
\delta_{nn'}\delta_{(\chi\times f)\chi'}\delta_{(\delta\times f)\delta'})
\text{, \ \ if }n,l\text{ are q-type, and } i \text{ is m-type}
\\ 
\frac{1}{2} (\delta_{nn'}\delta_{\chi\chi'}\delta_{\delta\delta'}
+
(-1)^l
\Theta_{fi}^{\Gamma_{\delta}}
\delta_{nn'}\delta_{(\chi\times f)\chi'}\delta_{(\delta\times f)\delta'})
\text{, \ \ if }n,i\text{ are q-type, and } l \text{ is m-type}
\end{array}\right.
\end{equation*}
\begin{equation}
\text{ \\where }
\Theta_{fi}=
\left\{\begin{array}{l}
1
\text{, \ \ if }i\text{ is m-type
}\\ 
i
\text{, \ \ if }i\text{ is q-type}
\end{array}\right.
\text{\ \ and } \Gamma_{\delta}=
\left\{\begin{array}{l}
1
\text{, \ \ if } s(\delta)=0
\\ 
*
\text{, \ \ if } s(\delta)=1
\end{array}\right.
\end{equation}
where similarly in the second case $i\in \Z_{N}$,  while in the third case $l\in \Z_{N}$.

\begin{equation}
\underset{n\chi\delta}{\sum}
\frac{d_{n}}{n_{n}} 
F^{im'k,\alpha'\delta}_{lnj,\beta'\chi}
(F^{imk,\alpha\delta}_{lnj,\beta\chi})^*
=\frac{d_{j}d_{k}n_{m}}{n_{j}n_{k}d_{m}} 
\left\{\begin{array}{l}
\delta_{mm'}\delta_{\alpha\alpha'}\delta_{\beta\beta'}
\text{, \ \ if }m\text{ is m-type
}
\\ 
\frac{1}{2}(\delta_{mm'}\delta_{\alpha\alpha'}\delta_{\beta\beta'}+
(-1)^i
\delta_{mm'}
\delta_{(\alpha\times f)\alpha'}\delta_{(\beta\times f)\beta'}),
\\
\text{\ \ \ \ \ \ \ \ \ \ \ \ \ \ \ \ \ \ \ \ \ \ \ \ \ \ \ \ \ \ \ \ \ \ \ \ \ \ \ \ \ \ \ \ \ \ \ \ \ \ \ \ \ \ 
 if }m,k,l\text{ are q-type,  and } i,j \text{ are m-type}
\\ 
\frac{1}{2}(\delta_{mm'}\delta_{\alpha\alpha'}\delta_{\beta\beta'}+
(-1)^{i+l}
\delta_{mm'}
\delta_{(\alpha\times f)\alpha'}\delta_{(\beta\times f)\beta'}),
\\
\text{\ \ \ \ \ \ \ \ \ \ \ \ \ \ \ \ \ \ \ \ \ \ \ \ \ \ \ \ \ \ \ \ \ \ \ \ \ \ \ \ \ \ \ \ \ \ \ \ \ \ \ \ \ \ 
 if }m,j,k \text{ are q-type,  and } i,l \text{ are m-type}
\\
\frac{1}{2} (
\delta_{mm'}\delta_{\alpha\alpha'}\delta_{\beta\beta'}+
(-1)^k
\Theta_{fi}^{\Gamma_{\alpha}}
\delta_{mm'}
\delta_{(\alpha\times f)\alpha'}\delta_{(\beta\times f)\beta'}),
\\
\text{\ \ \ \ \ \ \ \ \ \ \ \ \ \ \ \ \ \ \ \ \ \ \ \ \ \ \ \ \ \ \ \ \ \ \ \ \ \ \ \ \ \ \ \ \ \ \ \ \ \ \ \ \ \ 
 if }m,i,l \text{ are q-type,  and } j,k \text{ are m-type}
 \\
\frac{1}{2} (
\delta_{mm'}\delta_{\alpha\alpha'}\delta_{\beta\beta'}+
(-1)^{k+l}
\Theta_{fi}^{\Gamma_{\alpha}}
\delta_{mm'}
\delta_{(\alpha\times f)\alpha'}
\delta_{(\beta\times f)\beta'}),
\\
\text{\ \ \ \ \ \ \ \ \ \ \ \ \ \ \ \ \ \ \ \ \ \ \ \ \ \ \ \ \ \ \ \ \ \ \ \ \ \ \ \ \ \ \ \ \ \ \ \ \ \ \ \ \ \ 
if }m,i,j \text{ are q-type,  and } k,l \text{ are m-type}
\end{array}\right.
\end{equation}


\begin{equation}
\underset{m\alpha\beta}{\sum}
\frac{d_{m}}{n_{m}} 
(F^{imk,\alpha\delta'}_{ln'j,\beta\chi'})^*
F^{imk,\alpha\delta}_{lnj,\beta\chi}
=\frac{d_{j}d_{k}n_{n}}{n_{j}n_{k}d_{n}} 
\left\{\begin{array}{l}
\delta_{nn'}\delta_{\chi\chi'}\delta_{\delta\delta'}
\text{, \ \ if }n\text{ is m-type
}\\ 
\frac{1}{2} (\delta_{nn'}\delta_{\chi\chi'}\delta_{\delta\delta'}+
\delta_{nn'}\delta_{(\chi\times f)\chi'}
\delta_{(\delta\times f)\delta'})
\text{, \ \ if }n\text{ is q-type}
\end{array}\right.
\end{equation}

We list all $F$-moves of $\mathrm{TY}_{\Z_{2N}}^{t,\varkappa}/\psi_{N}$ in Appendix \ref{example4}.

\end{widetext}

\section{Conclusion and discussions}
\label{con}

In conclusion, we obtain a hopefully complete classification of all 2D non-chiral fermionic topological orders characterized by a set of tensors $(N^{ij}_{k},F^{ij}_{k},F^{ijm,\alpha\beta}_{kln,\chi\delta},n_{i},d_{i})$, which satisfy a set of nonlinear algebraic equations parameterized by phase factors $\Xi^{ijm,\alpha\beta}_{kl}$, $\Xi^{ij}_{kln,\chi\delta}$, $\Omega^{kim,\alpha\beta}_{jl}$ and $\Omega^{ki}_{jln,\chi\delta}$.
By considering the consistency between the fermionic Pentagon equation and the four projective unitary conditions, we get more relations for the phase factors $\Xi^{ijm,\alpha\beta}_{kl}$, $\Xi^{ij}_{kln,\chi\delta}$, $\Omega^{kim,\alpha\beta}_{jl}$,$\Omega^{ki}_{jln,\chi\delta}$, $\Delta^{mji,\alpha\delta}_{nl}$, from which we can define a topological invariant partition function for arbitrary 3-manifold with a spin structure. Finally, we also discussed four examples which satisfy all algebraic conditions.


For future study, it would also be very interesting to generalize the construction in Ref.\onlinecite{Cheng2017} and Ref.\onlinecite{Levin2016} for 2D non-chiral fermionic symmetry-enriched topological (fSET) phases,  
including those anomalous 2D fermionic SET states\cite{wangc13b,bonderson13,chen14a,metlitski15,Nfset2021a,Nfset2021b} which can only exist on the surface of some 3D  fermionic symmetry-protected (fSPT) phases. We believe that the q-type strings, or called Majorana-type strings, are very likely to characterize the anomaly of 3D fSPT phases with Kitaev-chain decoration\cite{Wang2020}. Moreover, it will also be very interesting to understand the generic algebraic structure\cite{TianSET,fset2021a,fset2021b} of fSET phases from equivalence class of symmetric fLU transformations.  

\section{acknowledgements}
We are grateful to Tian Lan for enlightening discussions. We also thank Zhenghan Wang for insightful discussions on the $\frac{1}{2}E_6$ example. This work was supported by Research Grant Council of Hong Kong(GRF 14306420,
ANR/RGC Joint Research Scheme no. A-CUHK402/18).

\appendix

\section{Super Fusion Category}
\label{AppendixA}

This section is a review on some basic concepts about super fusion categories\cite{brundan2017,usher2018,fc2019}. In the point of view of category theory, the string types are the simple objects in a super fusion category, or more precisely, a super pivotal category $\mathcal{S}$, where the "pivotal" structure is covered by the $H$-move we defined in the fermionic string-net model.  The super pivotal category only covers 2D fermoinic topological orders that can be obtained from fermion condensation. We believe that our approach from fixed-point wavefunction realizes more general fermonic topological than super pivotal category.

The number $n_{i}$ we introduced in the general fermionic string-net model is actually the dimension of endomorphism of the string:
\begin{equation}
n_{i}:= \text{dim End}(i).
\end{equation}

Explicitly, the string types are further divided into m-type strings and q-type strings:

(1) A string $i$ is an m-type string if
\begin{equation}
\text{End}(i)\cong\mathbb{C},
\end{equation}
where End$(i)$ is the endomorphism algebra of string $i$ (maps from string $i$ to itself), and the dimension of it is dim(End$(i))=1$. It means that the map from an m-type string to itself is one dimensional.

(2) A string $i$ is a q-type string if
\begin{equation}
\text{End}(i)\cong\mathbb{C}l_1,
\end{equation}
where dim End$(i)=2$. The first complex Clifford algebra is $\mathbb{C}l_1=\{1,\gamma\}$, where $1$ is the parity-even generator and $\gamma$ is the parity-odd generator.

\subsection{The modified fusion space}

In fermionic case, the fusion space $V^{ij}_{k}$ is different from the super vector space $\Delta^{ij}_{k}$ appearing in the fusion rule (while in bosonic case, $V^{ij}_{k}\cong \Delta^{ij}_{k}$):

(1) The super vector space $\Delta^{ij}_{k}$ is defined as the space of fusion coefficients in the string fusion rule:
\begin{equation}
i\otimes j\cong\underset{k}{\oplus} \Delta^{ij}_{k}\cdot k,
\end{equation}
where the fusion outcome is generally a composite object, which can be written in the from of multiplying an object with a super vector space.

(2) The fusion space $V^{ij}_{k}$ is defined as the vector space of morphisms from $k$ to $i\otimes j$:
\begin{equation}
V^{ij}_{k}:=\text{mor}(k\rightarrow i\otimes j)\cong \Delta^{ij}_{k}\otimes \text{mor}(k\rightarrow k)\cong\Delta^{ij}_{k}\otimes \text{End}(k),
\end{equation}
and therefore the dimension of the super vector space $\Delta^{ij}_{k}$ is given by
\begin{equation}
\eta^{ij}_{k}:=\text{dim}(\Delta^{ij}_{k})=\frac{N^{ij}_{k}}{\text{dim End}(k)},
\end{equation}
where $N^{ij}_{k}=$ dim$(V^{ij}_{k})$ is the dimension of the fusion space.


\subsection{The modified tensor product}

\begin{figure}[h]
\centering
\includegraphics[
width=0.15\textwidth]{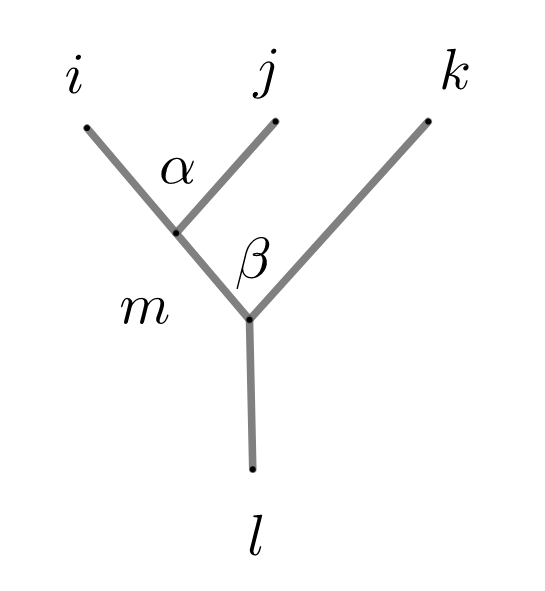}
\caption{The fusion space of fusing three strings $i,j,k$ into $l$. }
\label{Fig1}
\end{figure}

In larger fusion spaces involving more strings, the tensor product should also be modified. For example, we consider the fusion space $V^{ijk}_{l}$ in Fig. \ref{Fig1}. In bosonic case, we have
\begin{equation}
V^{ijk}_{l}\cong \underset{m}{\oplus} V^{ij}_{m} \otimes V^{mk}_{l}.
\end{equation}

However, in fermionic case, the two fusion spaces on two sides are not isomorphic. Explicitly,
\begin{eqnarray}
V^{ijk}_{l}&:=&\text{mor}(l\rightarrow i\otimes j\otimes k)
\nonumber\\
&\cong& \underset{m}{\oplus} \Delta^{ij}_{m}\otimes \text{mor}(l\rightarrow m\otimes k)
\nonumber\\
&\cong& \underset{m}{\oplus} \Delta^{ij}_{m}\otimes \Delta^{mk}_{l}\otimes\text{End}(l),
\end{eqnarray}
and
\begin{eqnarray}
\underset{m}{\oplus} V^{ij}_{m} \otimes V^{mk}_{l}&\cong&
\underset{m}{\oplus} (\Delta^{ij}_{m}\otimes \text{End}(m)) \otimes (\Delta^{mk}_{l}\otimes \text{End}(l))
\nonumber\\
&\cong& \underset{m}{\oplus} \Delta^{ij}_{m}\otimes \Delta^{mk}_{l}\otimes\text{End}(m)\otimes\text{End}(l).
\end{eqnarray}

Therefore in fermionic case the fusion space $V^{ijk}_{l}$ should be decomposed as
\begin{equation}
V^{ijk}_{l}\cong \underset{m}{\oplus} V^{ij}_{m} \otimes_{\text{End}(m)} V^{mk}_{l}
:=\underset{m}{\oplus} V^{ij}_{m} \otimes V^{mk}_{l}\backslash\text{End}(m),
\end{equation}
where $\otimes_{\text{End}(m)}$ is the relative tensor product, which is just the original tensor product modulo out the equivalence relations induced by End$(m)$.

The support dimension of $V^{ijk}_{l}$ is
\begin{equation}
\text{dim}(V^{ijk}_{l})=\underset{m}{\sum} \frac{N^{ij}_{m}N^{mk}_{l}}{\text{dim End}(m)}.
\end{equation}

\subsection{$F$-move}

We define the $F$-move as
\begin{equation}
F^{ijk}_{l}: \underset{n}{\oplus} V^{in}_{l} \otimes_{\text{End}(n)} V^{jk}_{n}\rightarrow
\underset{m}{\oplus} V^{ij}_{m} \otimes_{\text{End}(m)} V^{mk}_{l},
\end{equation}
where the support dimensions on two sides are equal:
\begin{equation}
\underset{m}{\sum} \frac{N^{ij}_{m}N^{mk}_{l}}{\text{dim End}(m)}
=\underset{n}{\sum} \frac{N^{in}_{l}N^{jk}_{n}}{\text{dim End}(n)}.
\label{F2}
\end{equation}

\subsection{Quantum dimensions}

We define the quantum dimension $d_{i}$ of a string $i$ as the largest eigenvalue of the fusion matrix $%
\widehat{\eta}_{i}$, where $\widehat{\eta}_{i}=(\eta^{ij}_{k}:j,k\in \mathcal{S})$ and $\eta^{ij}_{k}=$ dim$(\Delta^{ij}_{k})$. We define the vector $\left\vert\omega\right\rangle=\underset{i}{\sum}d_{i}\left\vert i \right\rangle$ as the common eigenvector of $\widehat{\eta}_{i}$, such that
\begin{equation}
\widehat{\eta}_{i}\left\vert\omega\right\rangle=d_{i}\left\vert \omega \right\rangle,
\end{equation}
where $\left\vert i \right\rangle$ is the Dirac notation of string type $i$. Specially, if $i$ is a q-type string, we have $\left\vert i \rangle \langle i \vert j \right\rangle=\frac{1}{2}\delta_{ij}\left\vert i \right\rangle$, where $\left\vert i \rangle\langle i \right\vert$ is a projective operator: when it acts on $\left\vert i \right\rangle$, it only maps to half of the initial state due to the equivalence relations generated by End$(i)$:
\begin{align}
 \includegraphics[scale=.45]{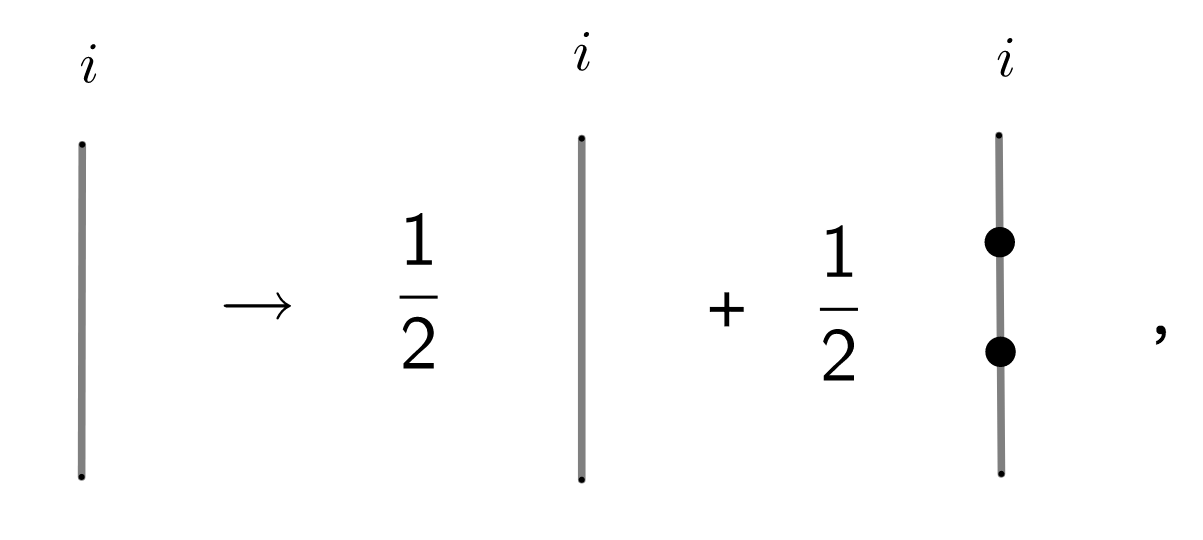},
\end{align}
where we consider the simplest strings in parity-even sector with up to two fermions. For strings with more fermions, we can also divide all the configurations into two sets, and pair one element from each set by the above equivalence relation. This can be concluded as
\begin{equation}
\left\langle i \vert i \right\rangle=\frac{1}{\text{dim End}(i)}.
\end{equation}

Remark: The string state $i$ here is not the fixed point wavefunction state as above.

The total quantum dimension is defined as the inner product of the common eigenvector:
\begin{equation}
D^{2}=\left\langle\omega\vert\omega\right\rangle=\underset{i}{\sum}d_{i}^{2}\left\langle i \vert i \right\rangle=\underset{i}{\sum}\frac{d_{i}^{2}}{\text{dim End}(i)}.
\end{equation}

From relation in Eq.(\ref{F2}), we can derive a general formula for quantum dimensions:
\begin{equation*}
\underset{ml}{\sum} \frac{N^{ij}_{m}N^{mk}_{l}}{\text{dim End}(m)\text{dim End}(l)} \left\vert\omega\right\rangle
=\underset{nl}{\sum} \frac{N^{in}_{l}N^{jk}_{n}}{\text{dim End}(n)\text{dim End}(l)}\left\vert\omega\right\rangle,
\end{equation*}
\begin{equation*}
\underset{ml}{\sum} \frac{N^{ij}_{m}\eta^{mk}_{l}}{\text{dim End}(m)}\left\vert\omega\right\rangle
=\underset{nl}{\sum} \eta^{in}_{l}\eta^{jk}_{n}\left\vert\omega\right\rangle,
\end{equation*}
\begin{equation*}
\underset{m}{\sum} \frac{N^{ij}_{m}\widehat{\eta}_{m}}{\text{dim End}(m)}\left\vert\omega\right\rangle
= \widehat{\eta}_{i}\widehat{\eta}_{j}\left\vert\omega\right\rangle,
\end{equation*}
\begin{equation}
d_{i}d_{j}=\underset{m}{\sum} \frac{N^{ij}_{m}d_{m}}{\text{dim End}(m)}.
\end{equation}

\section{Fermion Condensation}
\label{fc}

Anyon condensation is a systematic approach to construct new topological orders from old ones~\cite{bais2009,eliens2014,kong2014,burnell2018}. In the fermionic system, there is an analogous fermion condensation to obtain fermionic topological orders from a bosonic one~\cite{wan2017,fc2019,lou2021}.

In this appendix, we will discuss one version of fermion condensation that produces a super fusion category (discussed in Appendix~\ref{AppendixA}) from a fusion category if there is an object that is promoted to a fermion in the Drinfeld center of the fusion category. We will use this scheme to produce several super fusion categories. We have checked that they all satisfy the conditions summarized in Section~\ref{Summary}, although the solutions of which are assumed to be more general than super fusion categories.

\subsection{Fermion Condensation Scheme}
\label{fca}

The easiest way to understand fermion condensation is from the string diagram. We assume that there is a special object $y$ in the fusion category $\mathcal C$ that is promoted to a fermion $(y,\beta_y)$ ($\beta_y(x)$ is the half braiding of $y$ with respect to $x$) in the Drinfeld center $\mathcal Z(\mathcal C)$ of $\mathcal C$. We will denote the fermion string $y$ by red color in the string diagram.

In the construction of super fusion category from a fusion category $\mathcal C$, we need only the half braiding of $y$ with other objects. The string diagram for other objects are still planer. We assume that $y$ in $\mathcal C$ is lifted to a fermion $(y,\beta_y)$ in the Drinfeld center $\mathcal Z(\mathcal C)$. Therefore, a self-twist of $y$ will give us a minus fermion sign. The braiding of $y$ should satisfy the naturality condition:
\begin{align}\label{eq:half-braiding}
\vcenter{\hbox{\includegraphics[scale=1]{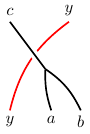}}}\ 
=
\vcenter{\hbox{\includegraphics[scale=1]{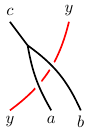}}}\ 
,\quad
\mathrm{for}\ 
\forall a, b, c.
\end{align}
It implies that we can move the red $y$ string freely under any vertex (morphism) of the diagram.

Since $y^2=y\otimes y =1$, the fusion of $y$ with the simple objects in $\mathcal C$ gives us an involution. Since $y$ is condensed, $a$ and $y\otimes a\in \mathcal C$ should be identified in the super fusion category $\mathcal C/y$. So the simple objects of $\mathcal C/y$ consist of the orbits of this action. We will denote the representative object of the orbit of $a$ as $[a]$, which can be viewed as an object in $\mathcal C/y$. If $y\otimes a\cong a\in \mathcal C$, then $[a]$ is a q-type object in $\mathcal C/y$. Otherwise, $[a]$ is m-type in $\mathcal C/y$.

On the other hand, the hom space of $\mathcal C/y$ is defined to be
\begin{align}
\mathrm{Hom}_{\mathcal C/y}([a],[b]) := \mathrm{Hom}_{\mathcal C}([a],[b]) \oplus \mathrm{Hom}_{\mathcal C}([a],y\otimes [b]),
\end{align}
which is a direct sum of bosonic and fermionic fusion spaces. On the right-hand side of the equation, $[a]$ is understood as the representative object of the $y$-fusion-action orbit in $\mathcal C$. Graphically, the fermionic fusion space (with a black dot on the vertex) of $\mathrm{Hom}_{\mathcal C/y}([a],[b])$ in $\mathcal C/y$ is defined to be the bosonic fusion space $\mathrm{Hom}_{\mathcal C}([a],y\otimes [b])$ in $\mathcal C$ as, for example,
\begin{align}\label{dot_string}
\vcenter{\hbox{\includegraphics[scale=1]{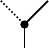}}}:=
\vcenter{\hbox{\includegraphics[scale=1]{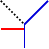}}},
\end{align}
where the red string $y$ is fused from left to the vertex in $\mathcal C$ by convention. The condensed red string of $y$ should be understood as under all other strings in the diagram. It is paired up with another red string at the left infinity, such that the total digram is fermion even. The examples we consider in this paper all have trivial Frobenius-Schur indicator for the fermion string: $\varkappa_y=1$. For simplicity, we assume that the straight $y$ string from leftmost to a vertex can be regularized arbitrarily near the vertex:
\begin{align}
\vcenter{\hbox{\includegraphics[scale=1]{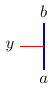}}}
=
\vcenter{\hbox{\includegraphics[scale=1]{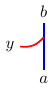}}}
=
\vcenter{\hbox{\includegraphics[scale=1]{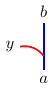}}}.
\end{align}
So the notion of horizontal red line makes sense in the string diagram.

In such way, we construct both objects and morphisms of the super fusion category $\mathcal C/y$ from the fusion category $\mathcal C$. This procedure is called fermion condensation. Every string diagram of $\mathcal C/y$ should be understood as a digram of $\mathcal C$ with black dots replaced by red strings using the rule \eq{dot_string}. In particular, the $F$ move of $\mathcal C/y$ can be derived from the $F$ move of the fusion category $\mathcal C$.


\subsection{$F$-move in fermion condensation}

For the fermion condensation part, we use a notation of $F$-move differ from Eq.(\ref{F1}) in our fixed-point wavefunction approach. In the fermionic theory, the $F$-move is denoted as
\begin{align}
\begin{matrix} \includegraphics[scale=.45]{F1g} \end{matrix}
=
\sum_{n\chi\delta}
[F^{ijk}_{l}]^{m,\alpha\beta}_{n,\chi\delta}
\begin{matrix} \includegraphics[scale=.45]{F2g1} \end{matrix}.
\end{align}
While in the bosonic theory before fermion condensation, a bosonic $F$-move is denoted in blue color:
\begin{align}
\begin{matrix} \includegraphics[scale=.45]{F1g} \end{matrix}
=
\sum_{n\chi\delta}
[\textcolor{blue}{F^{ijk}_{l}}]^{m,\alpha\beta}_{n,\chi\delta}
\begin{matrix} \includegraphics[scale=.45]{F2g1} \end{matrix}.
\end{align}

So that if we take the notation $F^{ijk}_{l}$ or $\textcolor{blue}{F^{ijk}_{l}}$, it is in general a matrix.

\subsection{From bosonic to fermionic pentagon equation}
\label{sec:bfpentagon}

In the former section, we have discussed how to obtain a new category $\mathcal C/y$ from a fusion category $\mathcal C$. We have to show that the $\mathcal C/y$ is indeed a super fusion category. In particular, the new $F$ move should satisfy the super (fermionic) pentagon equation, which was first proposed in Ref.~\onlinecite{Gu2015}. Compared to the bosonic counterpart, there is an additional fermion sign in the super pentagon equation.


The super pentagon equation is derived by replacing all the fermionic vertex Hom states (represented by black dots) by red strings of fermions going into left infinity under all other strings. For instance, a diagram with four dots is understood as
\begin{align}
\begin{matrix} \includegraphics[scale=.45]{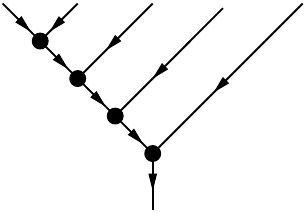} 
\end{matrix}
:=
 \begin{matrix} \includegraphics[scale=.45]{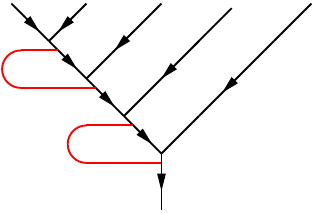} 
\end{matrix}
,
\end{align}
where the right-hand diagram is a bosonic one for $\mathcal C$. The four red fermion strings are paired up from top to bottom.

The string diagrams of a fermionic pentagon equation of $\mathcal C/y$ can be also expressed as diagrams of $\mathcal C$. There are possibly red fermion strings going from a vertex to the left, if the fusion space of the super fusion category is fermionic. If we switch the order (height) of two vertices which are both fermionic, there is a fermion sign
\begin{align}\label{fsign}
\begin{matrix} \includegraphics[scale=.45]{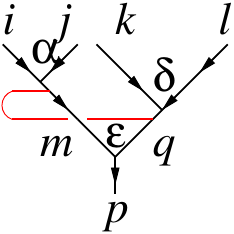} 
\end{matrix}
=
\begin{matrix} \includegraphics[scale=.45]{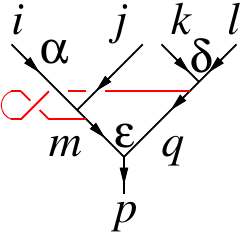} 
\end{matrix}
=
(-1)^{s(\alpha)s(\delta)}
\begin{matrix} \includegraphics[scale=.45]{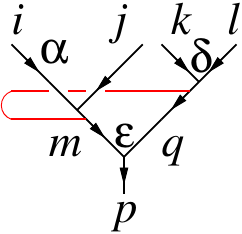} 
\end{matrix}
.
\end{align}
In general, this fermion sign of the above move is $(-1)^{s(\alpha)s(\delta)}$. Now the super pentagon equation of the super fusion category is in fact a hexagon equation:
\begin{align}
\includegraphics[]{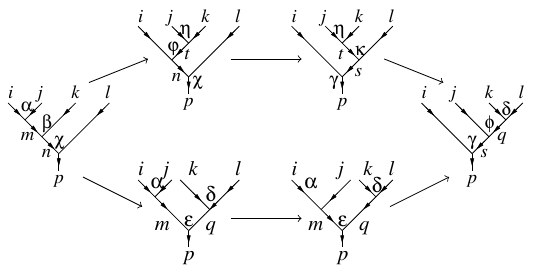}
\end{align}
where the bottom move is the fermion sign move shown in \eq{fsign}. Therefore, we show that the condensed theory $\mathcal/y$ satisfies the super pentagon equation \eq{fpenta} which is first derived in Ref.~\onlinecite{Gu2015}.


\subsection{Equivalence Relations}
\label{equirela}

In doing fermion condensation, we set all transformations related to adding and removing vertices, e.g.  $O$-move and $Y$-move, to be normalized as 1,which will cause no harm as the number of vertices is invariant for the initial and final state in any fermion condensation step.

Now we derive the equivalence relations in our fixed-point wavefunction approach (So that we need to put back the notation $\psi_{\text{fix}}$). From our fermion condensation convention,  if $m$ is q-type, the two equivalent states in Eq.(\ref{ef1}) and Eq.(\ref{ef2}) are related by:
\begin{align}
\psi_{\text{fix}} \begin{pmatrix} \includegraphics[scale=.45]{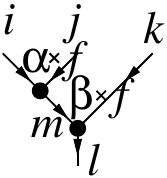} 
\end{pmatrix}
&:=\psi_{\text{fix}} \begin{pmatrix} \includegraphics[scale=.45]{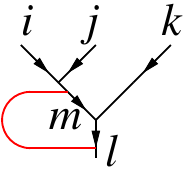} 
\end{pmatrix}
\nonumber\\
&=
\textcolor{blue}{(F^{ymk}_{\textcolor{red}{l}})_{\beta}}
\psi_{\text{fix}}\begin{pmatrix} \includegraphics[scale=.45]{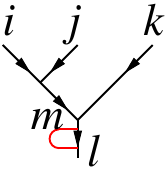} 
\end{pmatrix}
\nonumber\\
&:=
\textcolor{blue}{(F^{y
mk}_{\textcolor{red}{l}})_{\beta}}
\psi_{\text{fix}}\begin{pmatrix} \includegraphics[scale=.45]{F1g} 
\end{pmatrix},
\label{fc1}
\end{align}
\begin{align}
\psi_{\text{fix}} \begin{pmatrix} \includegraphics[scale=.45]{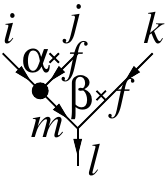} 
\end{pmatrix}
&:=\psi_{\text{fix}} \begin{pmatrix} \includegraphics[scale=.45]{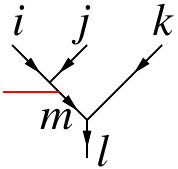} 
\end{pmatrix}
\nonumber\\
&=\textcolor{blue}{(F^{ymk}_{l})_{\beta}}
\psi_{\text{fix}}\begin{pmatrix} \includegraphics[scale=.45]{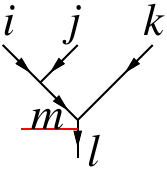} 
\end{pmatrix}
\nonumber\\
&:=\textcolor{blue}{(F^{ymk}_{l})_{\beta}}
\psi_{\text{fix}}\begin{pmatrix} \includegraphics[scale=.45]{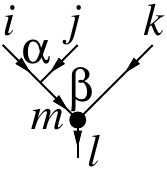} 
\end{pmatrix},
\label{fc2}
\end{align}
where $\color{blue}{y}$ denotes for the fermion string in the bosonic theory. 

We introduce some of our notations here:

(1) The index in the bottom right corner, e.g. $\textcolor{blue}{\beta}$ in $\textcolor{blue}{(F^{y
mk}_{\textcolor{red}{l}})_{\beta}}$, refers to that the $F$-move in the uncondensed bosonic theory acts on the vertex with fusion state $\beta$. So that any bosonic $F$-move in such notation is just a phase factor, as the inner states are all fixed as long as the state $\beta$ (including the two incoming strings and one outgoing string that define $\beta$) is fixed.

(2) Some of the strings are denoted in red color, e.g. string $\textcolor{red}{l}$ in $\textcolor{blue}{(F^{y
\textcolor{red}{m}k}_{\textcolor{red}{l}})_{\beta}}$.  A string is denoted in red color if it satisfies the following property:

\textit{When it is a trivial string in the fermionic theory, it is a fermion string in the bosonic theory before fermion condensation.}

\noindent This property is important as it may cause difference on the phase factors in fermion-parity even and odd sectors.


We obtain the relation between the $F$-moves on two equivalent states:
\begin{equation}
F^{ijm,(\alpha\times f)(\beta\times f)}_{kln,\chi\delta}
=
\left\{\begin{array}{l}
\textcolor{blue}{(F^{y
mk}_{\textcolor{red}{l}})_{\beta}}
F^{ijm,\alpha\beta}_{kln,\chi\delta}
\text{, \ \ parity-even}
\\ 
\textcolor{blue}{(F^{ymk}_{l})_{\beta}}
F^{ijm,\alpha\beta}_{kln,\chi\delta}
\text{, \ \ parity-odd}
\end{array}\right.
\label{feq1a}
\end{equation}

Similarly, if $n$ is q-type, the two equivalent states in Eq.(\ref{ef3}) and Eq.(\ref{ef4}) are related by:
\begin{align}
\psi_{\text{fix}} \begin{pmatrix} \includegraphics[scale=.45]{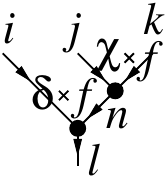} 
\end{pmatrix}
&:=\psi_{\text{fix}} \begin{pmatrix} \includegraphics[scale=.45]{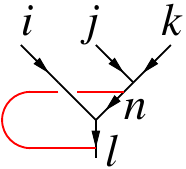} 
\end{pmatrix}
\nonumber\\
&=\Theta_{fi}
\psi_{\text{fix}}\begin{pmatrix} \includegraphics[scale=.45]{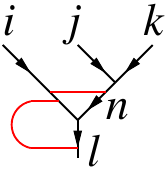} 
\end{pmatrix}
\nonumber\\
&=\Theta_{fi}
\textcolor{blue}{(F^{yin}_{\textcolor{red}{l}})_{\delta}}
\psi_{\text{fix}}\begin{pmatrix} \includegraphics[scale=.45]{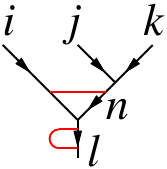} 
\end{pmatrix}
\nonumber\\
&=\Theta_{fi}\
\textcolor{blue}{(F^{yin}_{\textcolor{red}{l}}F^{iyn}_{l})_{\delta}}
\psi_{\text{fix}}\begin{pmatrix} \includegraphics[scale=.45]{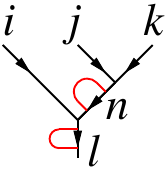} 
\end{pmatrix}
\nonumber\\
&:=
\Theta_{fi}\
\textcolor{blue}{(F^{yin}_{\textcolor{red}{l}}F^{iyn}_{l})_{\delta}}
\psi_{\text{fix}}\begin{pmatrix} \includegraphics[scale=.45]{F2g1} 
\end{pmatrix},
\label{fc3}
\end{align}
\begin{align}
\psi_{\text{fix}} \begin{pmatrix} \includegraphics[scale=.45]{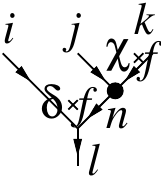} 
\end{pmatrix}
&:=\psi_{\text{fix}} \begin{pmatrix} \includegraphics[scale=.45]{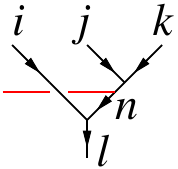} 
\end{pmatrix}
\nonumber\\
&=\Theta_{fi}\
\textcolor{blue}{(F^{yin}_{l}F^{iyn}_{\textcolor{red}{l}})_{\delta}}
\psi_{\text{fix}}\begin{pmatrix} \includegraphics[scale=.45]{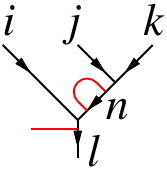} 
\end{pmatrix}
\nonumber\\
&:=
\Theta_{fi}\
\textcolor{blue}{(F^{yin}_{l}F^{iyn}_{\textcolor{red}{l}})_{\delta}}
\psi_{\text{fix}}\begin{pmatrix} \includegraphics[scale=.45]{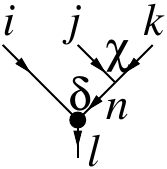} 
\end{pmatrix},
\label{fc4}
\end{align}
where $\Theta_{fi}$ is the half-braiding phase between the string $i$ and the fermion string also in the uncondensed bosnic theory. When $i$ is m-type, the half-braiding is trivial. When $i$ is q-type, $\Theta_{fi}=\pm i$, as proved in Section \ref{fca}, and we can always choose a gauge $\Theta_{fi}= i$. Therefore, we have the expresssion:
\begin{equation}
\Theta_{fi}=
\left\{\begin{array}{l}
1
\text{, \ \ if }i\text{ is m-type
}\\ 
i
\text{, \ \ if }i\text{ is q-type}
\end{array}\right.
\end{equation}

We obtain the relation between inverse $F$-moves on two equivalent states:

\begin{equation}
(F^{ijm,\alpha\beta}_{kln,(\chi\times f)(\delta\times f)})^*
=
\left\{\begin{array}{l}
\Theta_{fi}\
\textcolor{blue}{(F^{yin}_{\textcolor{red}{l}}F^{iyn}_{l})_{\delta}}
(F^{ijm,\alpha\beta}_{kln,\chi\delta})^*,
\\
\text{\ \ \ \ \ \ \ \ \ \ \ \ \ \ \ \ \ \ \ \ \ \ \ \ \ \ \ \ \ \ \ \ \ \ \ \ \ \
parity-even}
\\ 
\Theta_{fi}\
\textcolor{blue}{(F^{yin}_{l}F^{iyn}_{\textcolor{red}{l}})_{\delta}}
(F^{ijm,\alpha\beta}_{kln,\chi\delta})^*,
\\
\text{\ \ \ \ \ \ \ \ \ \ \ \ \ \ \ \ \ \ \ \ \ \ \ \ \ \ \ \ \ \ \ \ \ \ \ \ \ \
parity-odd}
\end{array}\right.
\label{feq2a}
\end{equation}


If $k$ is q-type, The two equivalent $O$-moves in Eq.(\ref{eo1}) are related by:
\begin{align}
\psi_{\text{fix}} \begin{pmatrix} \includegraphics[scale=.45]{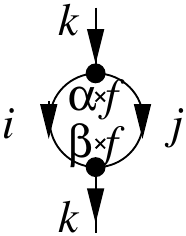} 
\end{pmatrix}
&:=\psi_{\text{fix}} \begin{pmatrix} \includegraphics[scale=.45]{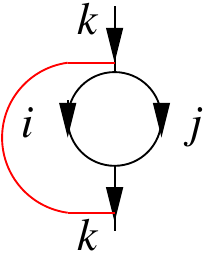} 
\end{pmatrix}
\nonumber\\
&=
\textcolor{blue}{(\widetilde{F}^{yij}_{\textcolor{red}{k}})_{\alpha}}
\psi_{\text{fix}}\begin{pmatrix} \includegraphics[scale=.45]{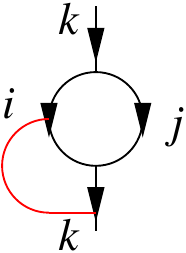} 
\end{pmatrix}
\nonumber\\
&=
\textcolor{blue}{(\widetilde{F}^{yij}_{\textcolor{red}{k}})_{\alpha}}
\textcolor{blue}{(F^{yij}_{\textcolor{red}{k}})_{\beta}}
\psi_{\text{fix}}\begin{pmatrix} \includegraphics[scale=.45]{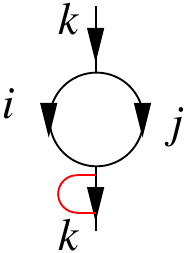} 
\end{pmatrix}
\nonumber\\
&:=
\textcolor{blue}{(\widetilde{F}^{yij}_{\textcolor{red}{k}})_{\alpha}}
\textcolor{blue}{(F^{yij}_{\textcolor{red}{k}})_{\beta}}
\psi_{\text{fix}}\begin{pmatrix} \includegraphics[scale=.45]{iOip} 
\end{pmatrix},
\label{oc1}
\end{align}

We obtain the relation between $O$-moves on two equivalent states:

\begin{equation}
O^{ij,(\alpha\times f)(\beta\times f)}_{k}
=
\textcolor{blue}{(\widetilde{F}^{yij}_{\textcolor{red}{k}})_{\alpha}}
\textcolor{blue}{(F^{yij}_{\textcolor{red}{k}})_{\beta}}
O^{ij,\alpha\beta}_{k} .
\label{oeq1a}
\end{equation}

If $k$ is q-type, The two equivalent $Y$-moves in Eq.(\ref{ey1}) are related by: related by:
\begin{align}
\psi_{\text{fix}} \begin{pmatrix} \includegraphics[scale=.45]{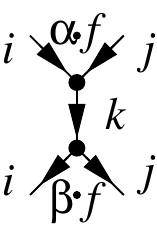} 
\end{pmatrix}
&:=\psi_{\text{fix}} \begin{pmatrix} \includegraphics[scale=.55]{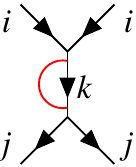} 
\end{pmatrix}
\nonumber\\
&:=
\psi_{\text{fix}}\begin{pmatrix} \includegraphics[scale=.45]{Y} 
\end{pmatrix},
\end{align}

For the dual $H$-move, if $m$ is q-type, the two equivalent states in Eq.(\ref{eh1}) and Eq.(\ref{eh2}) are related by:
\begin{align}
\psi_{\text{fix}} \begin{pmatrix} \includegraphics[scale=.45]{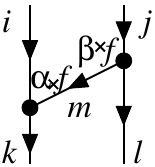} 
\end{pmatrix}
&:=\psi_{\text{fix}} \begin{pmatrix} \includegraphics[scale=.45]{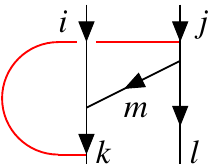} 
\end{pmatrix}
\nonumber\\
&=\Theta_{fi}
\psi_{\text{fix}}\begin{pmatrix} \includegraphics[scale=.45]{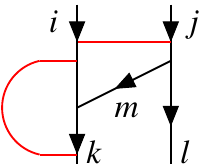} 
\end{pmatrix}
\nonumber\\
&=\Theta_{fi}
\textcolor{blue}{(F^{yim}
_{\textcolor{red}{k}})_{\alpha}}
\psi_{\text{fix}}\begin{pmatrix} \includegraphics[scale=.45]{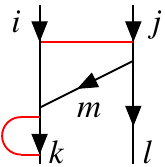} 
\end{pmatrix}
\nonumber\\
&=\Theta_{fi}\
\textcolor{blue}{(
F^{yim}
_{\textcolor{red}{k}}
F^{iym}_{k})_{\alpha}}
\psi_{\text{fix}}\begin{pmatrix} \includegraphics[scale=.45]{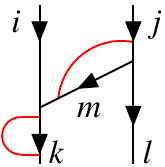} 
\end{pmatrix}
\nonumber\\
&=\Theta_{fi}\
\textcolor{blue}{(F^{yim}
_{\textcolor{red}{k}}
F^{iym}_{k})_{\alpha}}
\textcolor{blue}{(\widetilde{F}^{yml}_{\textcolor{red}{j}})_{\beta}}
\psi_{\text{fix}}\begin{pmatrix} \includegraphics[scale=.45]{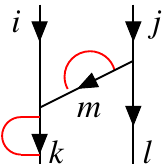} 
\end{pmatrix}
\nonumber\\
&:=
\Theta_{fi}\
\textcolor{blue}{(F^{yim}
_{\textcolor{red}{k}}
F^{iym}_{k})_{\alpha}}
\textcolor{blue}{(\widetilde{F}^{yml}_{\textcolor{red}{j}})_{\beta}}
\psi_{\text{fix}}\begin{pmatrix} \includegraphics[scale=.45]{dh1} 
\end{pmatrix},
\label{fc5}
\end{align}
\begin{align}
\psi_{\text{fix}} \begin{pmatrix} \includegraphics[scale=.45]{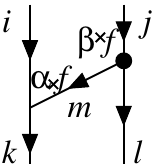} 
\end{pmatrix}
&:=\psi_{\text{fix}} \begin{pmatrix} \includegraphics[scale=.45]{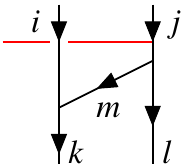} 
\end{pmatrix}
\nonumber\\
&=\Theta_{fi}\
\textcolor{blue}{(F^{yim}_{k}F^{iym}_{\textcolor{red}{k}})_{\alpha}}
\textcolor{blue}{(\widetilde{F}^{yml}_{\textcolor{red}{j}})_{\beta}}
\psi_{\text{fix}}\begin{pmatrix} \includegraphics[scale=.45]{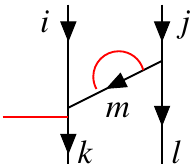} 
\end{pmatrix}
\nonumber\\
&:=
\Theta_{fi}
\textcolor{blue}{(F^{yim}_{k}F^{iym}_{\textcolor{red}{k}})_{\alpha}}
\textcolor{blue}{(\widetilde{F}^{yml}_{\textcolor{red}{j}})_{\beta}}
\psi_{\text{fix}}\begin{pmatrix} \includegraphics[scale=.45]{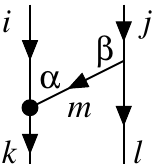} 
\end{pmatrix},
\label{fc6}
\end{align}

We obtain the relation between $H$-moves on two equivalent states:

\begin{equation}
\widetilde{H}^{kim,(\alpha\times f)(\beta\times f)}_{jln,\chi\delta}
=
\left\{\begin{array}{l}
\Theta_{fi}\
\textcolor{blue}{(F^{yim}
_{\textcolor{red}{k}}
F^{iym}_{k})_{\alpha}}
\textcolor{blue}{(\widetilde{F}^{yml}_{\textcolor{red}{j}})_{\beta}}
\widetilde{H}^{kim,\alpha\beta}_{jln,\chi\delta},
\\
\text{\ \ \ \ \ \ \ \ \ \ \ \ \ \ \ \ \ \ \ \ \ \ \ \ \ \ \ \ \ \ \ \ \ \ \ \ \ \ \ \ \ \ \ \ \ 
parity-even}
\\ 
\Theta_{fi}
\textcolor{blue}{(F^{yim}_{k}F^{iym}_{\textcolor{red}{k}})_{\alpha}}
\textcolor{blue}{(\widetilde{F}^{yml}_{\textcolor{red}{j}})_{\beta}}
\widetilde{H}^{kim,\alpha\beta}_{jln,\chi\delta},
\\
\text{\ \ \ \ \ \ \ \ \ \ \ \ \ \ \ \ \ \ \ \ \ \ \ \ \ \ \ \ \ \ \ \ \ \ \ \ \ \ \ \ \ \ \ \ \ 
parity-odd}
\end{array}\right.
\label{heq1a}
\end{equation}


If $n$ is q-type, the two equivalent states in Eq.(\ref{eh3}) and Eq.(\ref{eh4}) are related by:
\begin{align}
\psi_{\text{fix}} \begin{pmatrix} \includegraphics[scale=.45]{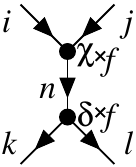} 
\end{pmatrix}
&:=\psi_{\text{fix}} \begin{pmatrix} \includegraphics[scale=.45]{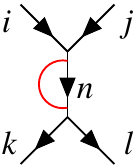} 
\end{pmatrix}
\nonumber\\
&:=
\psi_{\text{fix}}\begin{pmatrix} \includegraphics[scale=.45]{dy1} 
\end{pmatrix},
\label{fc7}
\end{align}
\begin{align}
\psi_{\text{fix}} \begin{pmatrix} \includegraphics[scale=.45]{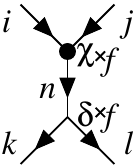} 
\end{pmatrix}
&:=\psi_{\text{fix}} \begin{pmatrix} \includegraphics[scale=.45]{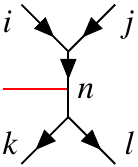} 
\end{pmatrix}
\nonumber\\
&:=
\psi_{\text{fix}}\begin{pmatrix} \includegraphics[scale=.45]{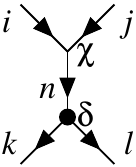} 
\end{pmatrix},
\label{fc8}
\end{align}

We obtain the relation between inverse $H$-moves on two equivalent states:
\begin{equation}
(\widetilde{H}^{kim,\alpha\beta}_{jln,(\chi\times f)(\delta\times f)})^*
=
(\widetilde{H}^{kim,\alpha\beta}_{jln,\chi\delta})^*
\label{heq2a}
\end{equation}

In the following several subsections of this Appendix~\ref{fc}, we will present several examples of super fusion categories by fermion condensation. They all satisfy the conditions summarized in Section~\ref{Summary}, although the solutions of which are supposed to be more general than super fusion categories.

\begin{widetext}

\subsection{Phase factors from fermion condensation}
\label{fcm}

We can obtain explicit forms of the five phase factors in Eq.(\ref{geq1})-Eq.(\ref{geq5}) from fermion condensation:
\begin{align}
\Xi^{ijm,\alpha\beta}_{kl}=\textcolor{blue}{(F^{ymk}_{l})_{\beta}},
\label{phase1}
\end{align}
\begin{align}
\Xi^{ij}_{kln,\chi\delta}=\Theta_{fi}\
\textcolor{blue}{(F^{yin}_{l}F^{iyn}_{l})_{\delta}},
\label{phase2}
\end{align}
\begin{align}
\Omega^{kim,\alpha\beta}_{jl}=\Theta_{fi}
\textcolor{blue}{(F^{yim}_{k}F^{iym}_{k})_{\alpha}}
\textcolor{blue}{(F^{yml}_{j})_{\beta}^*},
\label{phase3}
\end{align}
\begin{align}
\Omega^{ki}_{jln,\chi\delta}=1,
\label{phase4}
\end{align}
\begin{align}
\Delta^{mji,\alpha\delta}_{nl}=\textcolor{blue}{(F^{yin}_{l})_{\delta}^*}
\textcolor{blue}{(F^{yij}_{m})_{\alpha}},
\label{phase5}
\end{align}
where the $\color{blue}{F}$-symbols in blue color are the $F$-symbols in the bosonic theory before fermion condensation. With the above forms of phase factors, we can check that the relations among phase factors in Eq.(\ref{pr1})-Eq.(\ref{pr6}) are satisfied.

We note that the expression for each phase factor can be differ for fermion-parity even and odd sector. Here we do not distinguish parity-even and odd sector for simplicity. But we should keep in mind that the phase factor in parity-even and odd sector can be different, as explicitly shown in Appendix \ref{equirela}.

The five relations among phase factors are obtained from the following relations for $F$-move, $O$-move, $H$-move and a special sequence of moves in Appendix \ref{Amove} from fermion condensation:
\begin{equation}
F^{ijm,(\alpha\times f)(\beta\times f)}_{kln,\chi\delta}
=
\textcolor{blue}{(F^{ymk}_{l})_{\beta}}
F^{ijm,\alpha\beta}_{kln,\chi\delta}
\text{, \ \ if }m\text{ is q-type;}
\label{feq1}
\end{equation}
\begin{equation}
(F^{ijm,\alpha\beta}_{kln,(\chi\times f)(\delta\times f)})^*
=
\Theta_{fi}\
\textcolor{blue}{(F^{yin}_{l}F^{iyn}_{l})_{\delta}}
(F^{ijm,\alpha\beta}_{kln,\chi\delta})^*
\text{, \ \ if }n\text{ is q-type;}
\label{feq2}
\end{equation}

\begin{equation}
O^{ij,(\alpha\times f)(\beta\times f)}_{k}
=
\textcolor{blue}{(\widetilde{F}^{yij}_{k})_{\alpha}}
\textcolor{blue}{(F^{yij}_{k})_{\beta}}
O^{ij,\alpha\beta}_{k}   
\text{, \ \ if }k\text{ is q-type;}
\label{oeq1}
\end{equation}

\begin{equation}
Y^{ij}_{k,(\alpha\cdot f)(\beta\cdot f)}
=
Y^{ij}_{k,\alpha\beta}   
\text{, \ \ if }k\text{ is q-type;}
\label{yeq1}
\end{equation}

\begin{equation}
\widetilde{H}^{imk,(\alpha\times f)\delta}_{lnj,(\beta\times f)\chi}
=
\Theta_{fi}
\textcolor{blue}{(F^{yim}_{k}F^{iym}_{k})_{\alpha}}
\textcolor{blue}{(\widetilde{F}^{yml}_{j})_{\beta}}
\widetilde{H}^{imk,\alpha\delta}_{lnj,\beta\chi}
\text{, \ \ if }m\text{ is q-type;}
\label{heq1}
\end{equation}

\begin{equation}
(\widetilde{H}^{imk,\alpha(\delta\times f)}_{lnj,\beta(\chi\times f)})^*
=
(\widetilde{H}^{imk,\alpha\delta}_{lnj,\beta\chi})^*
\text{, \ \ if }n\text{ is q-type;}
\label{heq2}
\end{equation}

\begin{equation}
F^{ijm,(\alpha\times f)\beta}_{kln,\chi(\delta\times f)}
=
\textcolor{blue}{(F^{yin}_{l})_{\delta}^*}
\textcolor{blue}{(F^{yij}_{m})_{\alpha}}
F^{ijm,\alpha\beta}_{kln,\chi\delta}
\text{, \ \ if }i\text{ is q-type.}
\label{feq5}
\end{equation}
where we derive these relations in Appendix \ref{equirela} and Appendix \ref{Amove}.

We take the gauge in Eq.(\ref{Ogauge}) on the relation in Eq.(\ref{oeq1}),
\begin{equation}
O^{ij,(\alpha\times f)}_{k}
=
\textcolor{blue}{(\widetilde{F}^{yij}_{k})_{\alpha,\underline{\alpha}}}
\textcolor{blue}{(F^{yij}_{k})_{\alpha,\underline{\beta}}}
O^{ij,\alpha}_{k}   
\text{, \ \ if }k\text{ is q-type,}
\label{oeq1g}
\end{equation}
where we note that the bosonic $F$-move $\textcolor{blue}{(\widetilde{F}^{yij}_{k})_{\alpha,\underline{\alpha}}}$ and $\textcolor{blue}{(F^{yij}_{k})_{\alpha,\underline{\beta}}}$ acts on different vertices, but the two vertices are in the same space spanned by $\alpha$ and $\alpha\times f$.  In order not to cause confusion, we in addition label the vertices $\underline{\alpha}$ and $\underline{\beta}$ in Eq. (\ref{oeq1g}).
We take the gauge in Eq.(\ref{Ygauge}) on Eq. (\ref{yeq1}),
\begin{equation}
Y^{ij}_{k,(\alpha\cdot f)}
=
Y^{ij}_{k,\alpha}   
\text{, \ \ if }k\text{ is q-type,}
\label{yeq1g}
\end{equation}
Combining Eq.(\ref{heq1}), Eq.(\ref{heq2}), Eq.(\ref{oeq1g}), Eq.(\ref{yeq1g}) and a relation among $F$-move, $O$-move and $H$-move in Eq.(\ref{hyfo}), we obtain two more relations between $F$-moves:
\begin{equation}
F^{imk,(\alpha\times f)\delta}_{lnj,(\beta\times f)\chi}
=
\Theta_{fi}
\textcolor{blue}{(F^{yim}_{k}F^{iym}_{k})_{\alpha}}
\textcolor{blue}{(F^{yml}_{j})_{\beta}^*}
F^{imk,\alpha\delta}_{lnj,\beta\chi}
\text{, \ \ if }m\text{ is q-type;}
\label{feq3}
\end{equation}
\begin{equation}
(F^{imk,\alpha(\delta\times f)}_{lnj,\beta(\chi\times f)})^*
=
(F^{imk,\alpha\delta}_{lnj,\beta\chi})^*
\text{, \ \ if }n\text{ is q-type.}
\label{feq4}
\end{equation}

Therefore, relations in Eq.(\ref{feq1}), Eq.(\ref{feq2}), Eq.(\ref{feq3}),Eq.(\ref{feq4}) and Eq.(\ref{feq5}) exactly give the forms of the five phase factors in Eq.(\ref{phase1})-Eq.(\ref{phase5}).

\subsection{Fermionic topological order $SO(3)_6/\psi$ from $SO(3)_6$}
\label{example1}

\subsubsection{$SO(3)_6$ string-net model}

The full data for the bosonic $SO(3)_6$ theory are listed in, for example, Ref~.\onlinecite{fid2013}. There are 4 string types or simple objects $1,s,\tilde s,\psi$ in $SO(3)_6$. Some important fusion rules are $s\times \psi=\tilde s, \psi\times\psi=1, s\times s= 1+s+\tilde s$, from which we can derive all other fusion rules. The quantum dimensions of the simple objects are $d_1=d_\psi=1$, $d_s=d_{\tilde s}=1+\sqrt{2}$. The object $\psi$ is a fermion (in together with half-braiding) in modular tensor category $\mathcal{Z}(SO(3)_6)$.  So we can try to condense the fermion $\psi$ to obtain a super fusion category $SO(3)_6/\psi$.

\subsubsection{Fermionic topological order $SO(3)_6/\psi$}

Since the nontrivial simple objects $s$ and $\tilde s$ are changed into each other under the fusion of $y$: $s\times \psi=\tilde s$, they become the same simple object (which is also denoted as $s$) after the fermion condensation. So the simple objects of $SO(3)_6/\psi$ are $1$ and $s$. The quantum dimensions of them are $d_1=1$ and $d_s=1+\sqrt{2}$.

If we use dashed line and solid line to indicate the simple objects $1$ and $s$, the fusion rules of $SO(3)_6/\psi$ are
\NewFigName{Fig_SO36_fusion_}
\begin{align}
\vcenter{\hbox{\includegraphics[scale=1]{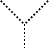}}}\ ,\ \ 
\vcenter{\hbox{\includegraphics[scale=1]{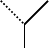}}}\ ,\ \ 
\vcenter{\hbox{\includegraphics[scale=1]{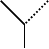}}}\ ,\ \ 
\vcenter{\hbox{\includegraphics[scale=1]{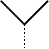}}}\ ,\ \ 
\vcenter{\hbox{\includegraphics[scale=1]{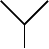}}}\ ,\ \ 
\vcenter{\hbox{\includegraphics[scale=1]{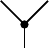}}}\ ,\ \ 
\end{align}
where the last three fusion diagram of $SO(3)_6/\psi$ come from the fusion rule $s\times s=1+s+\psi\times s$ in $SO(3)_6$. Only the last one is fermionic with a black dot which should be understood as a red string of $\psi$ going out of the vertex.

From the $F$ moves of $SO(3)_6$, we can derive the following (trivial) $F$ moves of $SO(3)_6/\psi$:
\NewFigName{Fig_SO(3)6_F_}
\begin{align}
\vcenter{\hbox{\includegraphics[scale=1]{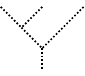}}}
&=
F^{111}_1
\vcenter{\hbox{\includegraphics[scale=1]{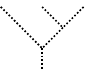}}},
\quad
F^{111}_1 = 1,
\end{align}
\begin{align}
\vcenter{\hbox{\includegraphics[scale=1]{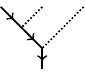}}}
&=
F^{s11}_s
\vcenter{\hbox{\includegraphics[scale=1]{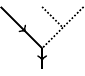}}},
\quad
F^{s11}_s = 1,
\end{align}
\begin{align}
\vcenter{\hbox{\includegraphics[scale=1]{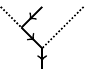}}}
&=
F^{1s1}_s
\vcenter{\hbox{\includegraphics[scale=1]{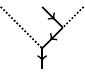}}},
\quad
F^{1s1}_s = 1,
\end{align}
\begin{align}
\vcenter{\hbox{\includegraphics[scale=1]{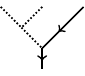}}}
&=
F^{11s}_s
\vcenter{\hbox{\includegraphics[scale=1]{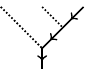}}}
\quad
F^{11s}_s = 1,
\end{align}
\begin{align}
\vcenter{\hbox{\includegraphics[scale=1]{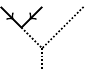}}}
&=
F^{ss1}_1
\vcenter{\hbox{\includegraphics[scale=1]{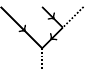}}},
\quad
F^{ss1}_1 = 1,
\end{align}
\begin{align}
\vcenter{\hbox{\includegraphics[scale=1]{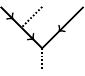}}}
&=
F^{s1s}_1
\vcenter{\hbox{\includegraphics[scale=1]{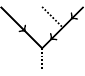}}},
\quad
F^{s1s}_1 = 1,
\end{align}
\begin{align}
\vcenter{\hbox{\includegraphics[scale=1]{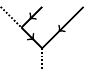}}}
&=
F^{1ss}_1
\vcenter{\hbox{\includegraphics[scale=1]{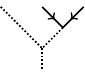}}}
\quad
F^{1ss}_1 = 1,
\end{align}
\begin{align}\nonumber
\begin{pmatrix}
\vcenter{\hbox{\includegraphics[scale=1]{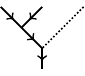}}}\ ,\ 
\vcenter{\hbox{\includegraphics[scale=1]{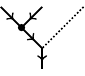}}}
\end{pmatrix}^T
&=
F^{ss1}_s
\begin{pmatrix}
\vcenter{\hbox{\includegraphics[scale=1]{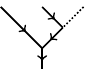}}}\ ,\ 
\vcenter{\hbox{\includegraphics[scale=1]{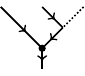}}}
\end{pmatrix}^T,\\
F^{ss1}_s &= 
\begin{pmatrix}
1&0\\
0&1\\
\end{pmatrix},
\end{align}
\begin{align}\nonumber
\begin{pmatrix}
\vcenter{\hbox{\includegraphics[scale=1]{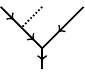}}}\ ,\ 
\vcenter{\hbox{\includegraphics[scale=1]{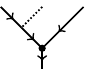}}}
\end{pmatrix}^T
&=
F^{s1s}_s
\begin{pmatrix}
\vcenter{\hbox{\includegraphics[scale=1]{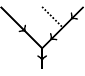}}}\ ,\ 
\vcenter{\hbox{\includegraphics[scale=1]{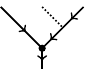}}}
\end{pmatrix}^T,\\
F^{s1s}_s &= 
\begin{pmatrix}
1&0\\
0&1\\
\end{pmatrix},
\end{align}
\begin{align}\nonumber
\begin{pmatrix}
\vcenter{\hbox{\includegraphics[scale=1]{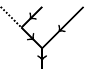}}}\ ,\ 
\vcenter{\hbox{\includegraphics[scale=1]{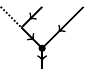}}}
\end{pmatrix}^T
&=
F^{1ss}_s
\begin{pmatrix}
\vcenter{\hbox{\includegraphics[scale=1]{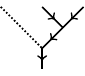}}}\ ,\ 
\vcenter{\hbox{\includegraphics[scale=1]{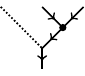}}}
\end{pmatrix}^T,\\
F^{1ss}_s &= 
\begin{pmatrix}
1&0\\
0&1\\
\end{pmatrix}.
\end{align}
The nontrivial $F$ moves of $SO(3)_6/\psi$ are
\begin{align}\nonumber
\begin{pmatrix}
\vcenter{\hbox{\includegraphics[scale=1]{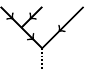}}}\ ,\ 
\vcenter{\hbox{\includegraphics[scale=1]{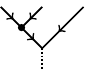}}}
\end{pmatrix}^T
&=
F^{sss}_1
\begin{pmatrix}
\vcenter{\hbox{\includegraphics[scale=1]{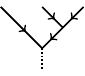}}}\ ,\ 
\vcenter{\hbox{\includegraphics[scale=1]{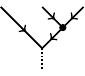}}}
\end{pmatrix}^T,\\
F^{sss}_1 &= 
\begin{pmatrix}
1&0\\
0&-1\\
\end{pmatrix},
\end{align}
\begin{align}\nonumber
&\begin{pmatrix}
\vcenter{\hbox{\includegraphics[scale=1]{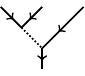}}}\ ,\ 
\vcenter{\hbox{\includegraphics[scale=1]{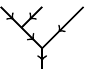}}}\ ,\ 
\vcenter{\hbox{\includegraphics[scale=1]{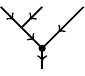}}}\ ,\ 
\vcenter{\hbox{\includegraphics[scale=1]{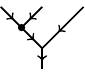}}}\ ,\ 
\vcenter{\hbox{\includegraphics[scale=1]{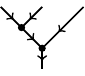}}}
\end{pmatrix}^T
\\\nonumber
&=
F^{sss}_s
\begin{pmatrix}
\vcenter{\hbox{\includegraphics[scale=1]{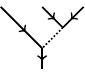}}}\ ,\ 
\vcenter{\hbox{\includegraphics[scale=1]{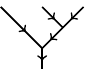}}}\ ,\ 
\vcenter{\hbox{\includegraphics[scale=1]{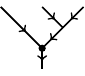}}}\ ,\ 
\vcenter{\hbox{\includegraphics[scale=1]{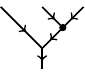}}}\ ,\ 
\vcenter{\hbox{\includegraphics[scale=1]{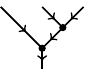}}}
\end{pmatrix}^T,\\
&\ \ \ \ \ \ \ \ \ \ \ \ \ \ \ \ \ \ \ \ 
F^{sss}_s 
= 
\begin{pmatrix}
\frac{1}{1+\sqrt{2}}&\frac{1}{\sqrt{1+\sqrt{2}}}&0&0&\frac{1}{\sqrt{1+\sqrt{2}}}\\
\frac{1}{\sqrt{1+\sqrt{2}}}&\frac{1}{2+\sqrt{2}}&0&0&-\frac{1}{\sqrt{2}}\\
0&0&-\frac{1}{\sqrt{2}}&-\frac{1}{\sqrt{2}}&0\\
0&0&-\frac{1}{\sqrt{2}}&\frac{1}{\sqrt{2}}&0\\
-\frac{1}{\sqrt{1+\sqrt{2}}}&\frac{1}{\sqrt{2}}&0&0&-\frac{1}{2+\sqrt{2}}
\end{pmatrix}.
\end{align}
One can check that they all satisfy the super pentagon equation and other conditions summarized in Section~\ref{Summary}.

\subsection{Majorana toric code from Ising string-net model}

\subsubsection{Ising string-net model}

Ising string-net model is one of the simplest models with non-Abelian fusion rules. The Ising fusion category has three simple objects $\{1,\psi,\s\}$ with quantum dimensions $d_1=d_\psi=1$, $d_\s=\sqrt{2}$. The fusion rules of them are: $\psi\times\psi=1$, $\psi\times\s=\s\times\psi=\s$, $\s\times\s=1+\psi$. If we use dashed, red and blue lines to represent $1$, $\psi$ and $\s$, the trivalent vertices of fusion rules are
\begin{align}
\vcenter{\hbox{\includegraphics[scale=1]{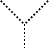}}}\ ,\ \ 
\vcenter{\hbox{\includegraphics[scale=1]{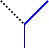}}}\ ,\ \ 
\vcenter{\hbox{\includegraphics[scale=1]{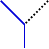}}}\ ,\ \ 
\vcenter{\hbox{\includegraphics[scale=1]{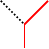}}}\ ,\ \ 
\vcenter{\hbox{\includegraphics[scale=1]{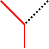}}}\ ,\ \ 
\vcenter{\hbox{\includegraphics[scale=1]{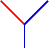}}}\ ,\ \ 
\vcenter{\hbox{\includegraphics[scale=1]{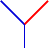}}}\ ,\ \ 
\vcenter{\hbox{\includegraphics[scale=1]{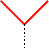}}}\ ,\ \ 
\vcenter{\hbox{\includegraphics[scale=1]{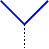}}}\ ,\ \ 
\vcenter{\hbox{\includegraphics[scale=1]{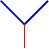}}}.
\end{align}
The $F$ symbols of the Ising fusion category can be found in, for example, Ref.~\onlinecite{Kitaev2006}.

\subsubsection{Majorana toric code model}
\label{example2}

Now we want to condense the fermion $\psi$ (in together with half-braiding) in the Drinfeld center of Ising fusion category to obtain a super fusion category. After the condensation, we have two simple objects 0 and 1, which come from 1 and $\s$ in the Ising model. The quantum dimension of them are $d_0=1$, $d_1=\sqrt{2}$. The fusion rules of them are
\begin{align}
\vcenter{\hbox{\includegraphics[scale=1]{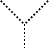}}}\ ,\ \ 
\vcenter{\hbox{\includegraphics[scale=1]{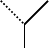}}}\ ,\ \ 
\vcenter{\hbox{\includegraphics[scale=1]{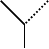}}}\ ,\ \ 
\vcenter{\hbox{\includegraphics[scale=1]{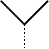}}}\ ,\ \ 
\vcenter{\hbox{\includegraphics[scale=1]{Fig_Maj_Fusion_5}}}:=
\vcenter{\hbox{\includegraphics[scale=1]{Fig_Maj_Fusion_6}}}\ ,\ \ 
\vcenter{\hbox{\includegraphics[scale=1]{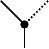}}}:=
\vcenter{\hbox{\includegraphics[scale=1]{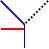}}}\ ,\ \ 
\vcenter{\hbox{\includegraphics[scale=1]{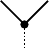}}}:=
\vcenter{\hbox{\includegraphics[scale=1]{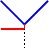}}},
\end{align}
with fusion coefficient $N_{00}^0=1$, $N_{01}^1=N_{10}^1=N_{11}^0=2$. The black dot in the diagram means that the fermion is condensed at this vertex, which can be understood as a fermion string going out of the vertex from the bosonic string diagram. Since we have fusion rule $\psi\times\s=\s$ in the Ising model, the nontrivial object 1 in the super fusion category is a q-type one. The fusion rule $\s\times\s=1+\psi$ in the Ising model becomes the fusion of $1$ and $1$ into $0$ with one bosonic channel and one fermionic channel in the super fusion category.


The $F$ moves for the super fusion category can be divided into two kinds according to whether or not the $F$ symbol is of full rank. The full rank unitary $F$ moves are:
\NewFigName{Fig_Maj_F_}
\begin{align}
\vcenter{\hbox{\includegraphics[scale=1]{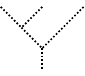}}}
&=
F^{000}_0
\vcenter{\hbox{\includegraphics[scale=1]{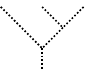}}},
\quad
F^{000}_0 = 1,
\end{align}
\begin{align}\nonumber
\begin{pmatrix}
\vcenter{\hbox{\includegraphics[scale=1]{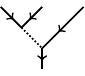}}}\ ,\ 
\vcenter{\hbox{\includegraphics[scale=1]{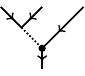}}}\ ,\ 
\vcenter{\hbox{\includegraphics[scale=1]{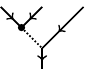}}}\ ,\ 
\vcenter{\hbox{\includegraphics[scale=1]{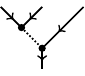}}}
\end{pmatrix}^T
&=
F^{111}_1
\begin{pmatrix}
\vcenter{\hbox{\includegraphics[scale=1]{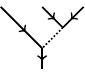}}}\ ,\ 
\vcenter{\hbox{\includegraphics[scale=1]{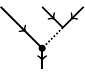}}}\ ,\ 
\vcenter{\hbox{\includegraphics[scale=1]{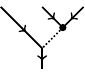}}}\ ,\ 
\vcenter{\hbox{\includegraphics[scale=1]{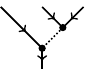}}}
\end{pmatrix}^T,\\
F^{111}_1 &= \frac{1}{\sqrt{2}}
\begin{pmatrix}
1&0&0&-i\\
0&1&-i&0\\
0&1&i&0\\
1&0&0&i
\end{pmatrix}.
\end{align}
The projective $F$ moves are
\begin{align}\nonumber
\begin{pmatrix}
\vcenter{\hbox{\includegraphics[scale=1]{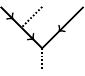}}}\ ,\ 
\vcenter{\hbox{\includegraphics[scale=1]{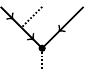}}}\ ,\ 
\vcenter{\hbox{\includegraphics[scale=1]{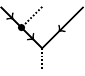}}}\ ,\ 
\vcenter{\hbox{\includegraphics[scale=1]{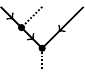}}}
\end{pmatrix}^T
&=
F^{101}_0
\begin{pmatrix}
\vcenter{\hbox{\includegraphics[scale=1]{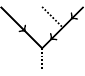}}}\ ,\ 
\vcenter{\hbox{\includegraphics[scale=1]{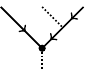}}}\ ,\ 
\vcenter{\hbox{\includegraphics[scale=1]{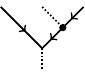}}}\ ,\ 
\vcenter{\hbox{\includegraphics[scale=1]{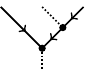}}}
\end{pmatrix}^T,\\
F^{101}_0 &= \frac{1}{2}
\begin{pmatrix}
1&0&0&-i\\
0&1&i&0\\
0&1&i&0\\
1&0&0&-i
\end{pmatrix},
\end{align}
\begin{align}\nonumber
\begin{pmatrix}
\vcenter{\hbox{\includegraphics[scale=1]{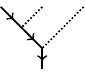}}}\ ,\ 
\vcenter{\hbox{\includegraphics[scale=1]{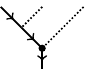}}}\ ,\ 
\vcenter{\hbox{\includegraphics[scale=1]{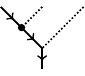}}}\ ,\ 
\vcenter{\hbox{\includegraphics[scale=1]{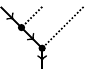}}}
\end{pmatrix}^T
&=
F^{100}_1
\begin{pmatrix}
\vcenter{\hbox{\includegraphics[scale=1]{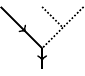}}}\ ,\ 
\vcenter{\hbox{\includegraphics[scale=1]{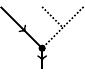}}}
\end{pmatrix}^T,\\
F^{100}_1 &= {\frac{1}{\sqrt{2}}}
\begin{pmatrix}
1&0\\
0&1\\
0&1\\
1&0\\
\end{pmatrix}
\end{align}
\begin{align}\nonumber
\begin{pmatrix}
\vcenter{\hbox{\includegraphics[scale=1]{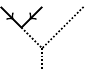}}}\ ,\ 
\vcenter{\hbox{\includegraphics[scale=1]{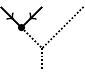}}}
\end{pmatrix}^T
&=
F^{110}_0
\begin{pmatrix}
\vcenter{\hbox{\includegraphics[scale=1]{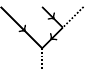}}}\ ,\ 
\vcenter{\hbox{\includegraphics[scale=1]{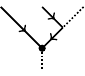}}}\ ,\ 
\vcenter{\hbox{\includegraphics[scale=1]{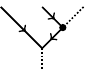}}}\ ,\ 
\vcenter{\hbox{\includegraphics[scale=1]{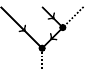}}}
\end{pmatrix}^T,\\
F^{110}_0 &= {\frac{1}{\sqrt{2}}}
\begin{pmatrix}
1&0&0&-i\\
0&1&i&0\\
\end{pmatrix},
\end{align}
\begin{align}\nonumber
\begin{pmatrix}
\vcenter{\hbox{\includegraphics[scale=1]{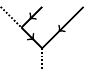}}}\ ,\ 
\vcenter{\hbox{\includegraphics[scale=1]{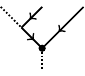}}}\ ,\ 
\vcenter{\hbox{\includegraphics[scale=1]{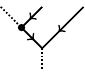}}}\ ,\ 
\vcenter{\hbox{\includegraphics[scale=1]{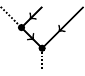}}}
\end{pmatrix}^T
&=
F^{011}_0
\begin{pmatrix}
\vcenter{\hbox{\includegraphics[scale=1]{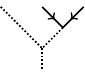}}}\ ,\ 
\vcenter{\hbox{\includegraphics[scale=1]{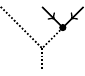}}}
\end{pmatrix}^T,\\
F^{011}_0 &= {\frac{1}{\sqrt{2}}}
\begin{pmatrix}
1&0\\
0&1\\
0&1\\
1&0\\
\end{pmatrix},
\end{align}
\begin{align}\nonumber
\begin{pmatrix}
\vcenter{\hbox{\includegraphics[scale=1]{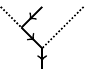}}}\ ,\ 
\vcenter{\hbox{\includegraphics[scale=1]{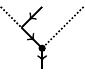}}}\ ,\ 
\vcenter{\hbox{\includegraphics[scale=1]{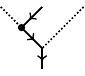}}}\ ,\ 
\vcenter{\hbox{\includegraphics[scale=1]{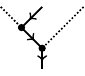}}}
\end{pmatrix}^T
&=
F^{010}_1
\begin{pmatrix}
\vcenter{\hbox{\includegraphics[scale=1]{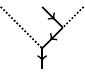}}}\ ,\ 
\vcenter{\hbox{\includegraphics[scale=1]{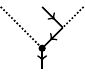}}}\ ,\ 
\vcenter{\hbox{\includegraphics[scale=1]{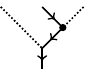}}}\ ,\ 
\vcenter{\hbox{\includegraphics[scale=1]{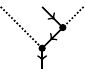}}}
\end{pmatrix}^T,\\
F^{010}_1 &= \frac{1}{2}
\begin{pmatrix}
1&0&0&1\\
0&1&1&0\\
0&1&1&0\\
1&0&0&1
\end{pmatrix},
\end{align}
\begin{align}\nonumber
\begin{pmatrix}
\vcenter{\hbox{\includegraphics[scale=1]{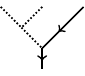}}}\ ,\ 
\vcenter{\hbox{\includegraphics[scale=1]{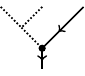}}}
\end{pmatrix}^T
&=
F^{001}_1
\begin{pmatrix}
\vcenter{\hbox{\includegraphics[scale=1]{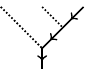}}}\ ,\ 
\vcenter{\hbox{\includegraphics[scale=1]{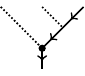}}}\ ,\ 
\vcenter{\hbox{\includegraphics[scale=1]{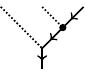}}}\ ,\ 
\vcenter{\hbox{\includegraphics[scale=1]{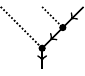}}}
\end{pmatrix}^T,\\
F^{001}_1 &= {\frac{1}{\sqrt{2}}}
\begin{pmatrix}
1&0&0&1\\
0&1&1&0\\
\end{pmatrix}.
\end{align}
Since there are always two independent basis states (with different fermion parities) on both sides of the above equations, all the above $F$ matrices have rank 2.

One can check that the $F$ moves satisfy the conditions summarized in Section~\ref{Summary}.

\subsection{Fermionic topological order $\left(\frac{1}{2}E_6\right)/y$ from $\frac{1}{2}E_6$}
\label{e3}

\subsubsection{Unitary fusion category $\frac{1}{2}E_6$}

The unitary fusion category $\frac{1}{2}E_6$ does not admit a braiding structure. It has three simple objects: $1$, $x$, $y$. The fusion rules are given by: $x\times x=1+2x+y$, $x\times y=y\times x=x$ and $y\times y=1$. If we represent $1$, $x$ and $y$ by dotted, blue and red strings respectively, the fusion configurations can be shown as:
\begin{align}\label{halfE6fusion}
\vcenter{\hbox{\includegraphics[scale=1]{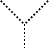}}}\ ,\ 
\vcenter{\hbox{\includegraphics[scale=1]{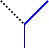}}}\ ,\ 
\vcenter{\hbox{\includegraphics[scale=1]{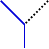}}}\ ,\ 
\vcenter{\hbox{\includegraphics[scale=1]{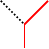}}}\ ,\ 
\vcenter{\hbox{\includegraphics[scale=1]{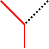}}}\ ,\ 
\vcenter{\hbox{\includegraphics[scale=1]{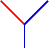}}}\ ,\ 
\vcenter{\hbox{\includegraphics[scale=1]{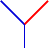}}}\ ,\ 
\vcenter{\hbox{\includegraphics[scale=1]{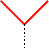}}}\ ,\ 
\vcenter{\hbox{\includegraphics[scale=1]{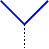}}}\ ,\ 
\vcenter{\hbox{\includegraphics[scale=1]{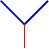}}}\ ,\ 
\vcenter{\hbox{\includegraphics[scale=1]{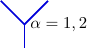}}}.
\end{align}
The last fusion of $x$ and $x$ to $y$ has two fusion channels. So we use a vertex index $\al=1,2$. All the simple objects are self-dual. The quantum dimensions can be calculated from the fusion rules as: $d_1=d_y=1$, $d_x=1+\sqrt{3}$.

The $F$ matrices in this tensor category are summarized as ($\sigma^i$'s are Pauli matrices)\cite{hong2008exotic}
\begin{align}
\textcolor{blue}{F^{abc}_d}&=1,\quad\text{ if $a=1$ or $b=1$ or $c=1$},\label{bf1} \\
\textcolor{blue}{F^{yyy}_y}&=\textcolor{blue}{F^{xyy}_x}=\textcolor{blue}{F^{yyx}_x}=\textcolor{blue}{F^{xyx}_1}=\textcolor{blue}{F^{yxx}_1}=\textcolor{blue}{F^{yxx}_y}=\textcolor{blue}{F^{xxy}_1}=\textcolor{blue}{F^{xxy}_y}=1,\label{bf2} \\
\textcolor{blue}{F^{xyx}_y}&=\textcolor{blue}{F^{yxy}_x}=-1,\label{bf3} \\
\textcolor{blue}{F^{yxx}_x}&=-\sigma^y,\quad \textcolor{blue}{F^{xyx}_x}=\sigma^z,\quad \textcolor{blue}{F^{xxy}_x}=\sigma^x,\label{bf4} \\ 
\textcolor{blue}{F^{xxx}_1}&=\frac{e^{7\pi i/12}}{2\sqrt{2}}
\begin{pmatrix}
1&i\\
1&-i
\end{pmatrix},\quad 
\textcolor{blue}{F^{xxx}_y}=\frac{e^{7\pi i/12}}{2\sqrt{2}}
\begin{pmatrix}
i&1 \label{bf5}\\
-i&1
\end{pmatrix},\\
\textcolor{blue}{F^{xxx}_x}&=
\begin{pmatrix}
\frac{\sqrt{3}-1}{2} & \frac{\sqrt{3}-1}{2} &
\frac{\sqrt{\sqrt{3}-1}}{2}e^{\pi i/6} & \frac{\sqrt{\sqrt{3}-1}}{2}e^{2\pi i/3} & \frac{\sqrt{\sqrt{3}-1}}{2}e^{2\pi i/3} & \frac{\sqrt{\sqrt{3}-1}}{2}e^{\pi i/6}\\
\frac{\sqrt{3}-1}{2} & \frac{1-\sqrt{3}}{2} 
& \frac{\sqrt{\sqrt{3}-1}}{2}e^{\pi i/6} & \frac{\sqrt{\sqrt{3}-1}}{2}e^{2\pi i/3} & -\frac{\sqrt{\sqrt{3}-1}}{2}e^{2\pi i/3} & -\frac{\sqrt{\sqrt{3}-1}}{2}e^{\pi i/6}\\
-\frac{1}{\sqrt{2(1+\sqrt{3})}} & -\frac{1}{\sqrt{2(1+\sqrt{3})}} & -\frac{1}{2}(e^{\pi i/6}-1) & \frac{1}{2}e^{5\pi i/6} & \frac{1}{2}(e^{-\pi i/3}+i) & \frac{1}{2}e^{\pi i/3}\\
-\frac{1}{\sqrt{2(1+\sqrt{3})}} & -\frac{1}{\sqrt{2(1+\sqrt{3})}} & \frac{1}{2}e^{\pi i/3} & \frac{1}{2}(e^{-\pi i/3}+i) & \frac{1}{2}e^{5\pi i/6} & -\frac{1}{2}(e^{\pi i/6}-1)\\
-\frac{1}{\sqrt{2(1+\sqrt{3})}} & \frac{1}{\sqrt{2(1+\sqrt{3})}} & -\frac{1}{2}(e^{\pi i/6}-1) & \frac{1}{2}e^{5\pi i/6} & -\frac{1}{2}(e^{-\pi i/3}+i) & -\frac{1}{2}e^{\pi i/3}\\
\frac{1}{\sqrt{2(1+\sqrt{3})}} & -\frac{1}{\sqrt{2(1+\sqrt{3})}} & -\frac{1}{2}e^{\pi i/3} & -\frac{1}{2}(e^{-\pi i/3}+i) & \frac{1}{2}e^{5\pi i/6} & -\frac{1}{2}(e^{\pi i/6}-1)
\end{pmatrix}.
\label{bf6}
\end{align}
\NewFigName{Fig_halfE6_F_}
The bases for $\textcolor{blue}{F^{xxx}_1}$ (similar for $\textcolor{blue}{F^{xxx}_y}$) are ordered as $\left(
\vcenter{\hbox{\includegraphics[scale=1]{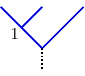}}}\ ,\ 
\vcenter{\hbox{\includegraphics[scale=1]{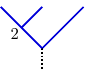}}}
\right)^T$ and $\left(
\vcenter{\hbox{\includegraphics[scale=1]{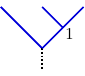}}}\ ,\ 
\vcenter{\hbox{\includegraphics[scale=1]{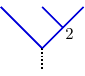}}}
\right)^T$ for left and right fusion spaces respectively. The left and right bases for $F^{xxx}_x$ are $\left(
\vcenter{\hbox{\includegraphics[scale=1]{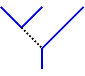}}}\ ,\ 
\vcenter{\hbox{\includegraphics[scale=1]{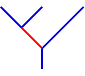}}}\ ,\ 
\vcenter{\hbox{\includegraphics[scale=1]{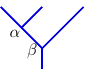}}}
\right)^T$ and $\left(
\vcenter{\hbox{\includegraphics[scale=1]{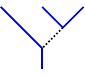}}}\ ,\ 
\vcenter{\hbox{\includegraphics[scale=1]{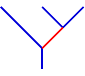}}}\ ,\ 
\vcenter{\hbox{\includegraphics[scale=1]{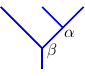}}}
\right)^T$ respectively. Note that the last diagram with $\al$ and $\be$ is in fact a vector of four bases with the usual tensor product order, i.e., $(\al,\beta)=(1,1),(1,2),(2,1),(2,2)$. One can check that the above $F$ moves satisfy (bosonic) pentagon equations.

To condense the fermion $y$, we first have to show that $y$ is indeed lifted to a fermion in the Drinfeld center of the fusion category $\frac{1}{2}E_6$. In fact, by solving the naturality condition in \eq{eq:half-braiding}, we have the following half-braiding of $y$:
\begin{align}
\beta_y(1)=1,
\quad \beta_y(y)=-v^{yy}_1\circ v^1_{yy},
\quad \beta_y(x)=i v^{xy}_x\circ v^x_{yx},
\end{align}
where $v^c_{ab}$ is the basis of morphism in $\textrm{Hom}(a\times b,c)$, and $v_c^{ba}$ is the dual. Graphically, the half-braiding induces the relations: (we use the convention that the $y$ string in $\beta_y(a)$ is the under-crossing line)
\NewFigName{Fig_halfE6_ey_}
\begin{align}
\beta_y(y)&=
\vcenter{\hbox{\includegraphics[scale=1]{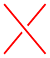}}}
=-
\vcenter{\hbox{\includegraphics[scale=1]{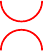}}}
,\\
\beta_y(x)&=
\vcenter{\hbox{\includegraphics[scale=1]{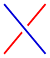}}}
=i
\vcenter{\hbox{\includegraphics[scale=1]{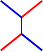}}}
.
\end{align}
The first half-braiding equation implies that $(y,\beta_y)$ is indeed a fermion in $\mathcal Z(\frac{1}{2}E_6)$. We can condense this fermion to obtain a super fusion category.

\subsubsection{Fermionic topological order $\left(\frac{1}{2}E_6\right)/\psi$}
\label{example3}

After the fermion condensation, the simple objects are 0 and 1 from the objects $1$ and $x$ in the fusion category $\frac{1}{2}E_6$. The quantum dimensions are $d_0=1$ and $d_1=1+\sqrt{3}$. Using the fusion rules listed in \eq{halfE6fusion}, we can obtain the fusion rules of the super fusion category $\left(\frac{1}{2}E_6\right)/\psi$ as
\NewFigName{Fig_FhalfE6_Fusion_}
\begin{align}
\vcenter{\hbox{\includegraphics[scale=1]{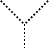}}}\ ,\ 
\vcenter{\hbox{\includegraphics[scale=1]{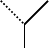}}}\ ,\ 
\vcenter{\hbox{\includegraphics[scale=1]{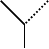}}}\ ,\ 
\vcenter{\hbox{\includegraphics[scale=1]{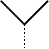}}}\ ,\ 
\vcenter{\hbox{\includegraphics[scale=1]{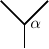}}}\ ,\ 
\vcenter{\hbox{\includegraphics[scale=1]{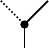}}}\ :=\ 
\vcenter{\hbox{\includegraphics[scale=1]{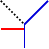}}}\ ,\ 
\vcenter{\hbox{\includegraphics[scale=1]{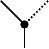}}}\ :=\ 
\vcenter{\hbox{\includegraphics[scale=1]{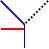}}}\ ,\ 
\vcenter{\hbox{\includegraphics[scale=1]{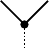}}}\ :=\ 
\vcenter{\hbox{\includegraphics[scale=1]{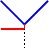}}}\ ,\ 
\vcenter{\hbox{\includegraphics[scale=1]{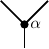}}}\ :=\ 
\vcenter{\hbox{\includegraphics[scale=1]{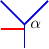}}}.
\end{align}

We can also derive the fermionic $F$ moves. The trivial vacuum $F$ matrix is
\begin{align}
\vcenter{\hbox{\includegraphics[scale=1]{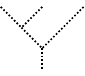}}}\ 
&=
F^{000}_0
\vcenter{\hbox{\includegraphics[scale=1]{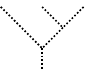}}}\ ,\ 
\quad
F^{000}_0 = 1.
\end{align}
The projective $F$ moves with two outgoing strings are
\NewFigName{Fig_FhalfE6_F_}
\begin{align}\nonumber
\begin{pmatrix}
\vcenter{\hbox{\includegraphics[scale=1]{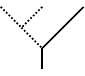}}}\ ,\ 
\vcenter{\hbox{\includegraphics[scale=1]{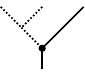}}}
\end{pmatrix}^T
&=
F^{001}_1
\begin{pmatrix}
\vcenter{\hbox{\includegraphics[scale=1]{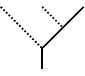}}}\ ,\ 
\vcenter{\hbox{\includegraphics[scale=1]{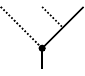}}}\ ,\ 
\vcenter{\hbox{\includegraphics[scale=1]{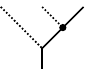}}}\ ,\ 
\vcenter{\hbox{\includegraphics[scale=1]{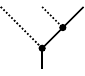}}}
\end{pmatrix}^T,\\
F^{001}_1 &= 
\frac{1}{\sqrt{2}}
\begin{pmatrix}
1&0&0&1\\
0&1&1&0\\
\end{pmatrix}.
\end{align}
\begin{align}\nonumber
\begin{pmatrix}
\vcenter{\hbox{\includegraphics[scale=1]{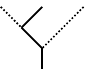}}}\ ,\ 
\vcenter{\hbox{\includegraphics[scale=1]{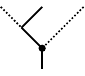}}}\ ,\ 
\vcenter{\hbox{\includegraphics[scale=1]{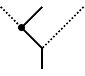}}}\ ,\ 
\vcenter{\hbox{\includegraphics[scale=1]{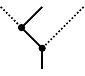}}}
\end{pmatrix}^T
&=
F^{010}_1
\begin{pmatrix}
\vcenter{\hbox{\includegraphics[scale=1]{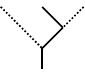}}}\ ,\ 
\vcenter{\hbox{\includegraphics[scale=1]{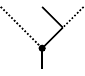}}}\ ,\ 
\vcenter{\hbox{\includegraphics[scale=1]{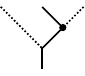}}}\ ,\ 
\vcenter{\hbox{\includegraphics[scale=1]{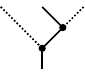}}}
\end{pmatrix}^T,\\
F^{010}_1 &= \frac{1}{2}
\begin{pmatrix}
1&0&0&1\\
0&1&1&0\\
0&1&1&0\\
1&0&0&1
\end{pmatrix},
\end{align}
\begin{align}\nonumber
\begin{pmatrix}
\vcenter{\hbox{\includegraphics[scale=1]{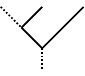}}}\ ,\ 
\vcenter{\hbox{\includegraphics[scale=1]{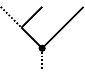}}}\ ,\ 
\vcenter{\hbox{\includegraphics[scale=1]{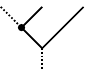}}}\ ,\ 
\vcenter{\hbox{\includegraphics[scale=1]{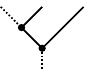}}}
\end{pmatrix}^T
&=
F^{011}_0
\begin{pmatrix}
\vcenter{\hbox{\includegraphics[scale=1]{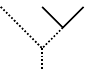}}}\ ,\ 
\vcenter{\hbox{\includegraphics[scale=1]{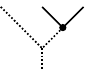}}}
\end{pmatrix}^T,\\
F^{011}_0 &=
\frac{1}{\sqrt{2}}
\begin{pmatrix}
1&0\\
0&1\\
0&1\\
1&0\\
\end{pmatrix},
\end{align}
\begin{align}\nonumber
\begin{pmatrix}
\vcenter{\hbox{\includegraphics[scale=1]{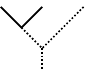}}}\ ,\ 
\vcenter{\hbox{\includegraphics[scale=1]{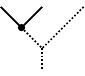}}}
\end{pmatrix}^T
&=
F^{100}_1
\begin{pmatrix}
\vcenter{\hbox{\includegraphics[scale=1]{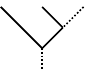}}}\ ,\ 
\vcenter{\hbox{\includegraphics[scale=1]{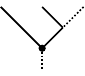}}}\ ,\ 
\vcenter{\hbox{\includegraphics[scale=1]{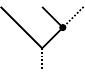}}}\ ,\ 
\vcenter{\hbox{\includegraphics[scale=1]{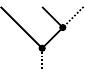}}}
\end{pmatrix}^T,\\
F^{110}_0 &= \frac{1}{\sqrt{2}}
\begin{pmatrix}
1&0&0&-i\\
0&1&i&0\\
\end{pmatrix},
\end{align}
\begin{align}\nonumber
\begin{pmatrix}
\vcenter{\hbox{\includegraphics[scale=1]{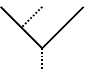}}}\ ,\ 
\vcenter{\hbox{\includegraphics[scale=1]{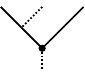}}}\ ,\ 
\vcenter{\hbox{\includegraphics[scale=1]{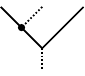}}}\ ,\ 
\vcenter{\hbox{\includegraphics[scale=1]{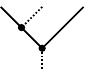}}}
\end{pmatrix}^T
&=
F^{101}_0
\begin{pmatrix}
\vcenter{\hbox{\includegraphics[scale=1]{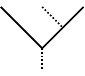}}}\ ,\ 
\vcenter{\hbox{\includegraphics[scale=1]{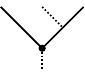}}}\ ,\ 
\vcenter{\hbox{\includegraphics[scale=1]{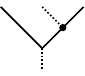}}}\ ,\ 
\vcenter{\hbox{\includegraphics[scale=1]{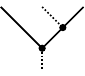}}}
\end{pmatrix}^T,\\
F^{101}_0 &= \frac{1}{2}
\begin{pmatrix}
1&0&0&-i\\
0&1&i&0\\
0&1&i&0\\
1&0&0&-i
\end{pmatrix},
\end{align}
\begin{align}\nonumber
\begin{pmatrix}
\vcenter{\hbox{\includegraphics[scale=1]{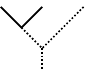}}}\ ,\ 
\vcenter{\hbox{\includegraphics[scale=1]{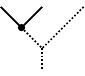}}}
\end{pmatrix}^T
&=
F^{110}_0
\begin{pmatrix}
\vcenter{\hbox{\includegraphics[scale=1]{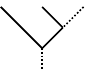}}}\ ,\ 
\vcenter{\hbox{\includegraphics[scale=1]{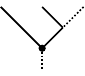}}}\ ,\ 
\vcenter{\hbox{\includegraphics[scale=1]{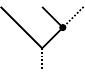}}}\ ,\ 
\vcenter{\hbox{\includegraphics[scale=1]{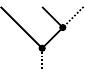}}}
\end{pmatrix}^T,
\\
F^{110}_0 &= \frac{1}{\sqrt{2}}
\begin{pmatrix}
1&0&0&-i\\
0&1&i&0\\
\end{pmatrix}.
\end{align}
They all have rank 2. 
The projective $F$ moves with three outgoing strings are all $8\times 8$ matrices with rank 4:
\begin{align}\nonumber
\begin{pmatrix}
\vcenter{\hbox{\includegraphics[scale=1]{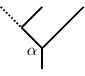}}}\ ,\ 
\vcenter{\hbox{\includegraphics[scale=1]{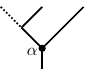}}}\ ,\ 
\vcenter{\hbox{\includegraphics[scale=1]{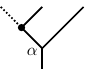}}}\ ,\ 
\vcenter{\hbox{\includegraphics[scale=1]{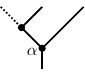}}}
\end{pmatrix}^T
&=
F^{011}_1
\begin{pmatrix}
\vcenter{\hbox{\includegraphics[scale=1]{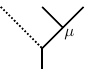}}}\ ,\ 
\vcenter{\hbox{\includegraphics[scale=1]{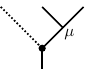}}}\ ,\ 
\vcenter{\hbox{\includegraphics[scale=1]{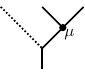}}}\ ,\ 
\vcenter{\hbox{\includegraphics[scale=1]{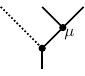}}}
\end{pmatrix}^T,\\
F^{011}_1 &= \frac{1}{2}
\begin{pmatrix}
\s^0 & 0 & 0 & \s^0 \\
0 & \s^0 & \s^0 & 0 \\
0 & -\s^y & -\s^y & 0 \\
-\s^y & 0 & 0 & -\s^y \\
\end{pmatrix},
\end{align}
\begin{align}\nonumber
\begin{pmatrix}
\vcenter{\hbox{\includegraphics[scale=1]{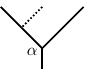}}}\ ,\ 
\vcenter{\hbox{\includegraphics[scale=1]{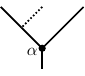}}}\ ,\ 
\vcenter{\hbox{\includegraphics[scale=1]{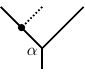}}}\ ,\ 
\vcenter{\hbox{\includegraphics[scale=1]{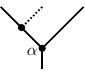}}}
\end{pmatrix}^T
&=
F^{101}_1
\begin{pmatrix}
\vcenter{\hbox{\includegraphics[scale=1]{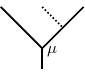}}}\ ,\ 
\vcenter{\hbox{\includegraphics[scale=1]{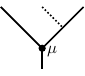}}}\ ,\ 
\vcenter{\hbox{\includegraphics[scale=1]{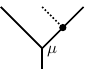}}}\ ,\ 
\vcenter{\hbox{\includegraphics[scale=1]{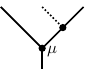}}}
\end{pmatrix}^T,\\
F^{101}_1 &= \frac{1}{2}
\begin{pmatrix}
\s^0 & 0 & 0 & \s^x \\
0 & \s^0 & \s^x & 0 \\
0 & -\s^y & i\s^z & 0 \\
-\s^y & 0 & 0 & i\s^z \\
\end{pmatrix},
\end{align}
\begin{align}\nonumber
\begin{pmatrix}
\vcenter{\hbox{\includegraphics[scale=1]{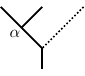}}}\ ,\ 
\vcenter{\hbox{\includegraphics[scale=1]{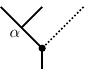}}}\ ,\ 
\vcenter{\hbox{\includegraphics[scale=1]{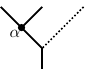}}}\ ,\ 
\vcenter{\hbox{\includegraphics[scale=1]{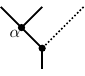}}}
\end{pmatrix}^T
&=
F^{110}_1
\begin{pmatrix}
\vcenter{\hbox{\includegraphics[scale=1]{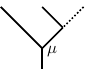}}}\ ,\ 
\vcenter{\hbox{\includegraphics[scale=1]{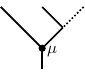}}}\ ,\ 
\vcenter{\hbox{\includegraphics[scale=1]{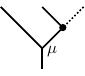}}}\ ,\ 
\vcenter{\hbox{\includegraphics[scale=1]{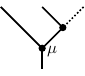}}}
\end{pmatrix}^T,\\
F^{110}_1 &= \frac{1}{2}
\begin{pmatrix}
\s^0 & 0 & 0 & \s^x \\
0 & \s^0 & \s^x & 0 \\
0 & \s^0 & \s^x & 0 \\
\s^0 & 0 & 0 & \s^x \\
\end{pmatrix},
\end{align}
\begin{align}\nonumber
\begin{pmatrix}
\vcenter{\hbox{\includegraphics[scale=1]{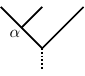}}}\ ,\ 
\vcenter{\hbox{\includegraphics[scale=1]{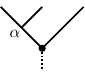}}}\ ,\ 
\vcenter{\hbox{\includegraphics[scale=1]{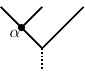}}}\ ,\ 
\vcenter{\hbox{\includegraphics[scale=1]{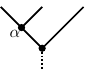}}}\ ,\ 
\end{pmatrix}^T
&=
F^{111}_0
\begin{pmatrix}
\vcenter{\hbox{\includegraphics[scale=1]{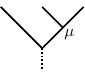}}}\ ,\ 
\vcenter{\hbox{\includegraphics[scale=1]{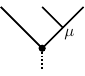}}}\ ,\ 
\vcenter{\hbox{\includegraphics[scale=1]{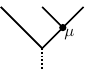}}}\ ,\ 
\vcenter{\hbox{\includegraphics[scale=1]{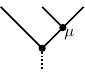}}}
\end{pmatrix}^T,\\
F^{111}_0 = \frac{1}{2}
\begin{pmatrix}
F^{xxx}_1 & 0 & 0 & -iF^{xxx}_1\\
0 & F^{xxx}_y & iF^{xxx}_y & 0 \\
0 & F^{xxx}_y & iF^{xxx}_y & 0 \\
F^{xxx}_1 & 0 & 0 & -iF^{xxx}_1 \\
\end{pmatrix}
&= \frac{e^{7\pi i/12}}{2\sqrt{2}}
\begin{pmatrix}
1&i  & 0&0 & 0&0 & -i&1  \\
1&-i & 0&0 & 0&0 & -i&-1 \\
0&0 & i&1  & -1&i & 0&0  \\
0&0 & -i&1 & 1&i  & 0&0  \\
0&0 & i&1  & -1&i & 0&0  \\
0&0 & -i&1 & 1&i  & 0&0  \\
1&i  & 0&0 & 0&0 & -i&1  \\
1&-i & 0&0 & 0&0 & -i&-1 \\
\end{pmatrix}.
\end{align}
The most complicated $F$ move is $F^{111}_1$. 
Let us first denote
\begin{align}
\tilde F:=F^{xxx}_x =
\begin{pmatrix}
\tilde F_{1,1} & \tilde F_{1,y} & \tilde F_{1,x} \\
\tilde F_{y,1} & \tilde F_{y,y} & \tilde F_{y,x} \\
\tilde F_{x,1} & \tilde F_{x,y} & \tilde F_{x,x} \\
\end{pmatrix}
\end{align}
to be one of the $F$ matrix in the bosonic $\frac{1}{2}E_6$ category. $\tilde F_{1,x}$ and $\tilde F_{y,x}$ ($\tilde F_{x,1}$ and $\tilde F_{x,y}$) are matrices of size $1\times 4$ ($4\times 1$). $\tilde F_{x,x}$ is a matrix of size $4\times 4$.
Then the $F^{111}_1$ of the super fusion category $\left(\frac{1}{2}E_6\right)/\psi$ is
\begin{align}
&\quad\begin{pmatrix}
\vcenter{\hbox{\includegraphics[scale=1]{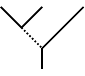}}}\ ,\ 
\vcenter{\hbox{\includegraphics[scale=1]{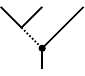}}}\ ,\ 
\vcenter{\hbox{\includegraphics[scale=1]{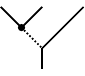}}}\ ,\ 
\vcenter{\hbox{\includegraphics[scale=1]{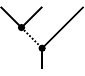}}}\ ,\ 
\vcenter{\hbox{\includegraphics[scale=1]{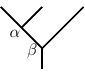}}}\ ,\ 
\vcenter{\hbox{\includegraphics[scale=1]{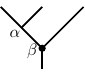}}}\ ,\ 
\vcenter{\hbox{\includegraphics[scale=1]{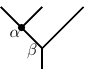}}}\ ,\ 
\vcenter{\hbox{\includegraphics[scale=1]{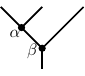}}}
\end{pmatrix}^T\\\nonumber
&=
F^{111}_1
\begin{pmatrix}
\vcenter{\hbox{\includegraphics[scale=1]{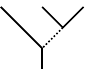}}}\ ,\ 
\vcenter{\hbox{\includegraphics[scale=1]{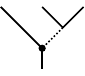}}}\ ,\ 
\vcenter{\hbox{\includegraphics[scale=1]{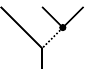}}}\ ,\ 
\vcenter{\hbox{\includegraphics[scale=1]{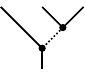}}}\ ,\ 
\vcenter{\hbox{\includegraphics[scale=1]{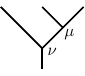}}}\ ,\ 
\vcenter{\hbox{\includegraphics[scale=1]{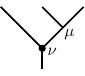}}}\ ,\ 
\vcenter{\hbox{\includegraphics[scale=1]{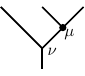}}}\ ,\ 
\vcenter{\hbox{\includegraphics[scale=1]{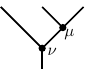}}}
\end{pmatrix}^T,
\end{align}
where the $F^{111}_1$ matrix is of size $20\times 20$ and defined by
\begin{align}\nonumber
&F^{111}_1 \nonumber\\
=&
\scalebox{0.9}{$
\begin{pmatrix}
\s^0\otimes\tilde F_{1,1} & \s^x\otimes(-i\tilde F_{1,y}) & \s^0\otimes(\frac{1}{\sqrt{2}}\tilde F_{1,x}) & \s^x\otimes(\frac{1}{\sqrt{2}}\tilde F_{1,x}  (\s^0 \otimes \s^x)) \\
\s^x\otimes\tilde F_{y,1} & \s^0\otimes(-i\tilde F_{y,y}) & \s^x\otimes(\frac{1}{\sqrt{2}}\tilde F_{y,x}) & \s^0\otimes(\frac{1}{\sqrt{2}}\tilde F_{y,x}  (\s^0 \otimes \s^x)) \\
\s^0\otimes (\frac{1}{\sqrt{2}}\tilde F_{x,1}) & \s^x\otimes(-\frac{i}{\sqrt{2}}\tilde F_{x,y}) & \s^0\otimes(\frac{1}{2}\tilde F_{x,x}) & \s^x\otimes(\frac{1}{2}\tilde F_{x,x}  (\s^0 \otimes \s^x)) \\
\s^x\otimes((\s^0 \otimes -\s^y) (\frac{1}{\sqrt{2}}\tilde F_{x,1})) & \s^0\otimes((\s^0 \otimes -\s^y)(-\frac{i}{\sqrt{2}}\tilde F_{x,y})) & \s^x\otimes((\s^0 \otimes -\s^y)\frac{1}{2}\tilde F_{x,x}) & \s^0\otimes((\s^0 \otimes -\s^y)\frac{1}{2}\tilde F_{x,x}  (\s^0 \otimes \s^x)) \\
\end{pmatrix}
$}\\
=&
\scalebox{0.7}{$
\begin{pmatrix}
\tilde F_{1,1} & 0 & 0 & -i\tilde F_{1,y} & \frac{1}{\sqrt{2}}\tilde F_{1,x} & 0 & 0 & \frac{1}{\sqrt{2}}\tilde F_{1,x}  (\s^0 \otimes \s^x) \\
0 & \tilde F_{1,1} & -i\tilde F_{1,y} & 0 & 0 & \frac{1}{\sqrt{2}}\tilde F_{1,x} & \frac{1}{\sqrt{2}}\tilde F_{1,x}  (\s^0 \otimes \s^x) & 0\\
0 & \tilde F_{y,1} & -i\tilde F_{y,y} & 0 & 0 & \frac{1}{\sqrt{2}}\tilde F_{y,x} & \frac{1}{\sqrt{2}}\tilde F_{y,x}  (\s^0 \otimes \s^x) & 0\\
\tilde F_{y,1} & 0 & 0 & -i\tilde F_{y,y} & \frac{1}{\sqrt{2}}\tilde F_{y,x} & 0 & 0 & \frac{1}{\sqrt{2}}\tilde F_{y,x}  (\s^0 \otimes \s^x) \\
\frac{1}{\sqrt{2}}\tilde F_{x,1} & 0 & 0 & -\frac{i}{\sqrt{2}}\tilde F_{x,y} & \frac{1}{2}\tilde F_{x,x} & 0 & 0 & \frac{1}{2}\tilde F_{x,x}  (\s^0 \otimes \s^x) \\
0 & \frac{1}{\sqrt{2}}\tilde F_{x,1} & -\frac{i}{\sqrt{2}}\tilde F_{x,y} & 0 & 0 & \frac{1}{2}\tilde F_{x,x} & \frac{1}{2}\tilde F_{x,x}  (\s^0 \otimes \s^x) & 0\\
0 & \frac{1}{\sqrt{2}}(\s^0 \otimes -\s^y) \tilde F_{x,1} & \frac{i}{\sqrt{2}}(\s^0 \otimes \s^y) \tilde F_{x,y} & 0 & 0 & \frac{1}{2}(\s^0 \otimes -\s^y) \tilde F_{x,x} & \frac{1}{2}(\s^0 \otimes -\s^y) \tilde F_{x,x}  (\s^0 \otimes \s^x) & 0\\
-\frac{1}{\sqrt{2}}\begin{pmatrix}\s^y&0\\0&\s^y\end{pmatrix} \tilde F_{x,1} & 0 & 0 & \frac{i}{\sqrt{2}}(\s^0 \otimes \s^y)\tilde F_{x,y} & \frac{1}{2}(\s^0 \otimes -\s^y)\tilde F_{x,x} & 0 & 0 & \frac{1}{2}(\s^0 \otimes -\s^y)\tilde F_{x,x}  (\s^0 \otimes \s^x) \\
\end{pmatrix}
$}
\end{align}
One can show that $F^{111}_1$ has rank 12.

We have checked that the above $F$ matrices satisfy the consistent equations such as the super pentagon equation.

\subsection{Fermionic topological order from Tambara-Yamagami category for $\Z_{2N}$}

\subsubsection{Unitary fusion category $\mathrm{TY}_{\Z_{2N}}^{t,\varkappa}$}

The fusion category $\mathrm{TY}_{\Z_{2N}}^{t,\varkappa}$ is the Tambara-Yamagami category~\cite{tambara1998} for $\Z_{2N}$ with symmetric non-degenerate bicharacter of type $t$ ($1\le t \le 2N-1$ and $\operatorname{gcd}(t,2N)=1$) defined as
\begin{align}\label{bicha}
\chi_t(a,b)=e^{2\pi itab/(2N)}.
\end{align}
The simple objects are labelled by $\Z_{2N}\cup \{\s\}$ where $\Z_{2N}=\{0,1,\cdots,2N-1\}$ is the cyclic group of order $2N$. $\s$ is an additional object. The quantum dimensions of them are $d_i=1$ ($\forall i\in\Z_{2N}$) and $d_\s=\sqrt{2N}$. The fusion rule for objects in $\Z_{2N}$ is simply the addition modulo $2N$: $a\times b=[a+b]_{2N}$. The $\s$ object can absorb all $\Z_{2N}$ objects: $a\times \s=\s\times a=\s$. The fusion of $\s$ with itself is $\s\times \s=\sum_{a\in\Z_{2N}}a$. If we use red strings and blue string to denote the simple objects in $\Z_{2N}$ and $\s$, then the fusion rules can be represented as
\NewFigName{Fig_TY_Fusion_}
\begin{align}
\vcenter{\hbox{\includegraphics[scale=1]{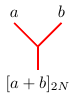}}}\ ,\ 
\vcenter{\hbox{\includegraphics[scale=1]{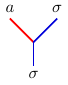}}}\ ,\ 
\vcenter{\hbox{\includegraphics[scale=1]{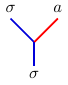}}}\ ,\ 
\vcenter{\hbox{\includegraphics[scale=1]{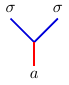}}}\ .
\end{align}

The nontrivial $F$ matrices for the fusion category $\mathrm{TY}_{\Z_{2N}}^{t,\varkappa}$ are related to the bicharacter $\chi_t$ defined in \eq{bicha} as:
\NewFigName{Fig_TY_F_}
\begin{align}
\vcenter{\hbox{\includegraphics[scale=1]{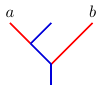}}}
&=
F^{a\s b}_\s
\vcenter{\hbox{\includegraphics[scale=1]{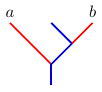}}}\ ,\ 
\quad
F^{a\s b}_\s=\chi_t(a,b),\\
\vcenter{\hbox{\includegraphics[scale=1]{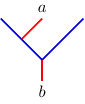}}}
&=
F^{\s a\s}_b
\vcenter{\hbox{\includegraphics[scale=1]{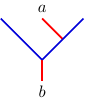}}}\ ,\ 
\quad
F^{\s a\s}_b=\chi_t(a,b),\\
\vcenter{\hbox{\includegraphics[scale=1]{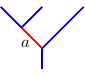}}}
&=
\sum_b (F^{\s\s\s}_\s)_{ab}
\vcenter{\hbox{\includegraphics[scale=1]{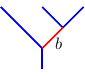}}}\ ,\ 
\quad
(F^{\s\s\s}_\s)_{ab}=\frac{\varkappa}{\sqrt{2N}}\chi_t(a,b)^{-1},
\end{align}
where $\varkappa=\pm 1$ is the Frobenius-Schur indicator of simple object $\s$.

\subsubsection{Fermionic topological order $\mathrm{TY}_{\Z_{2N}}^{t,\varkappa}/\psi_N$ ($N$ odd)}
\label{example4}

To perform fermion condensation in the category $\mathrm{TY}_{\Z_{2N}}^{t,\varkappa}$, we have to find a fermion in the Drinfeld center of the category. Let us try to find the half-braiding of the object $N\in\Z_{2N}$. Direct calculations of the naturality condition \eq{eq:half-braiding} for the half-braiding of $N$ give the results:
\NewFigName{Fig_TY_halfbraiding_}
\begin{align}
\beta_N(a)&=
\vcenter{\hbox{\includegraphics[scale=1]{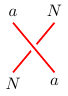}}}
=(-1)^a
\vcenter{\hbox{\includegraphics[scale=1]{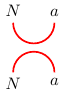}}}\ 
,\\
\beta_N(\s)&=
\vcenter{\hbox{\includegraphics[scale=1]{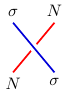}}}
= i
\vcenter{\hbox{\includegraphics[scale=1]{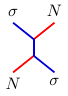}}}\ 
.
\end{align}
Now let us assume that $N$ is an odd integer. Then the object $(N,\beta_N)$ in $\mathcal Z\left(\mathrm{TY}_{\Z_{2N}}^{t,\varkappa}\right)$ has twist $\theta_{(N,\beta_N)}=\beta_N(N)=-1$ and is a fermion. Therefore, we can condense the fermion $\psi_N=N$ in $\mathrm{TY}_{\Z_{2N}}^{t,\varkappa}$.

After the fermion condensation, the object $a$ and $[a+N]_{2N}$ in $\Z_{2N}$ of $\mathrm{TY}_{\Z_{2N}}^{t,\varkappa}$ are identified. So the simple objects in the super fusion category $\mathrm{TY}_{\Z_{2N}}^{t,\varkappa}/\psi_N$ are $0,1,\cdots,N-1,\s$. The fusion rules of them can be represented as
\NewFigName{Fig_FTY_Fusion1_}
\begin{align}
\vcenter{\hbox{\includegraphics[scale=1]{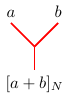}}}
\text{if }a+b< N,
\vcenter{\hbox{\includegraphics[scale=1]{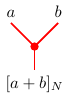}}}
\text{if }a+b\geq N,
\vcenter{\hbox{\includegraphics[scale=1]{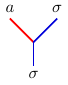}}}\ ,\ 
\vcenter{\hbox{\includegraphics[scale=1]{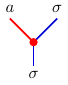}}}\ ,\ 
\vcenter{\hbox{\includegraphics[scale=1]{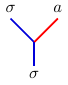}}}\ ,\ 
\vcenter{\hbox{\includegraphics[scale=1]{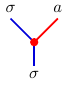}}}\ ,\ 
\vcenter{\hbox{\includegraphics[scale=1]{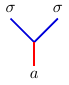}}}\ ,\ 
\vcenter{\hbox{\includegraphics[scale=1]{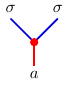}}}\ ,\ 
\end{align}
where $a,b\in\Z_{N}$ are objects from $\Z_{2N}$, and $\s$ is the same one in the fusion category. Every vertex of the above fusion rules can be either bosonic or fermionic.

From the $F$ matrices of the original fusion category, we can obtain the $F$ matrices of the super fusion category as
\NewFigName{Fig_FTY_FF_}
\begin{align}
\vcenter{\hbox{\includegraphics[scale=1]{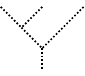}}}\ 
&=
F^{000}_0
\vcenter{\hbox{\includegraphics[scale=1]{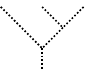}}}\ ,\ 
\quad
F^{000}_0 = 1,
\end{align}
\begin{align}
\vcenter{\hbox{\includegraphics[scale=1]{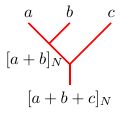}}}\ 
&=
F^{abc}_{[a+b+c]_N}
\vcenter{\hbox{\includegraphics[scale=1]{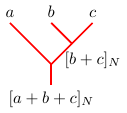}}}\ ,\ 
\quad
F^{abc}_{[a+b+c]_N}= (-1)^{a\floor*{\frac{b+c}{N}}}.
\end{align}
If $a+b<N$, the $F^{ab\s}_\s$-type and $F^{\s ab}_\s$-type $F$ moves are
\begin{align}\nonumber
\begin{pmatrix}
\vcenter{\hbox{\includegraphics[scale=1]{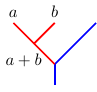}}}\ 
,
\vcenter{\hbox{\includegraphics[scale=1]{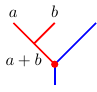}}}\ 
\end{pmatrix}^T
&=
F^{ab\s}_{\s}
\begin{pmatrix}
\vcenter{\hbox{\includegraphics[scale=1]{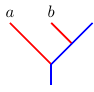}}}\ 
,
\vcenter{\hbox{\includegraphics[scale=1]{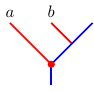}}}\ 
,
\vcenter{\hbox{\includegraphics[scale=1]{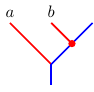}}}\ 
,
\vcenter{\hbox{\includegraphics[scale=1]{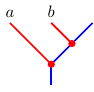}}}\ 
\end{pmatrix}^T,
\\
F^{ab\s}_{\s}
&=
\frac{1}{2}
\begin{pmatrix}
1&0&0&(-1)^a\\
0&1&(-1)^a&0\\
\end{pmatrix},
\end{align}
\begin{align}\nonumber
\begin{pmatrix}
\vcenter{\hbox{\includegraphics[scale=1]{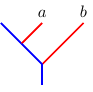}}}\ 
,
\vcenter{\hbox{\includegraphics[scale=1]{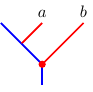}}}\ 
,
\vcenter{\hbox{\includegraphics[scale=1]{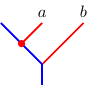}}}\ 
,
\vcenter{\hbox{\includegraphics[scale=1]{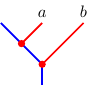}}}\ 
\end{pmatrix}^T
&=
F^{\s ab}_{\s}
\begin{pmatrix}
\vcenter{\hbox{\includegraphics[scale=1]{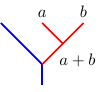}}}\ 
,
\vcenter{\hbox{\includegraphics[scale=1]{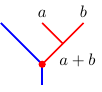}}}\ 
\end{pmatrix}^T,
\\
F^{\s ab}_{\s}
&=
\frac{1}{2}
\begin{pmatrix}
1&0\\
0&1\\
0&(-1)^b\\
(-1)^b&0\\
\end{pmatrix}.
\end{align}

On the other hand, if $a+b\geq N$, the $F$ moves are
\begin{align}\nonumber
\begin{pmatrix}
\vcenter{\hbox{\includegraphics[scale=1]{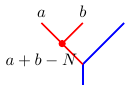}}}\ 
,
\vcenter{\hbox{\includegraphics[scale=1]{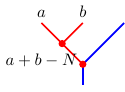}}}\ 
\end{pmatrix}^T
&=
F^{ab\s}_{\s}
\begin{pmatrix}
\vcenter{\hbox{\includegraphics[scale=1]{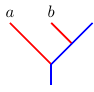}}}\ 
,
\vcenter{\hbox{\includegraphics[scale=1]{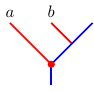}}}\ 
,
\vcenter{\hbox{\includegraphics[scale=1]{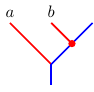}}}\ 
,
\vcenter{\hbox{\includegraphics[scale=1]{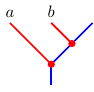}}}\ 
\end{pmatrix}^T,
\\
F^{ab\s}_{\s}
&=
\frac{1}{2}
\begin{pmatrix}
0&1&(-1)^a&0\\
1&0&0&(-1)^a\\
\end{pmatrix},
\end{align}
\begin{align}\nonumber
\begin{pmatrix}
\vcenter{\hbox{\includegraphics[scale=1]{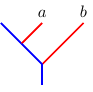}}}\ 
,
\vcenter{\hbox{\includegraphics[scale=1]{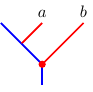}}}\ 
,
\vcenter{\hbox{\includegraphics[scale=1]{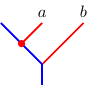}}}\ 
,
\vcenter{\hbox{\includegraphics[scale=1]{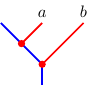}}}\ 
\end{pmatrix}^T
&=
F^{\s ab}_{\s}
\begin{pmatrix}
\vcenter{\hbox{\includegraphics[scale=1]{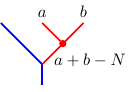}}}\ 
,
\vcenter{\hbox{\includegraphics[scale=1]{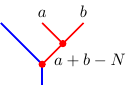}}}\ 
\end{pmatrix}^T,
\\
F^{\s ab}_{\s}
&=
\frac{1}{2}
\begin{pmatrix}
0&(-1)^{a+b}i\\
(-1)^{a+b}i&1\\
(-1)^{a}i&0\\
0&(-1)^{a}i\\
\end{pmatrix}.
\end{align}
They are all projective $F$ matrices with rank 2.
The $F^{a\s b}_\s$-type $F$ move is
\begin{align}\nonumber
\begin{pmatrix}
\!\!
\vcenter{\hbox{\includegraphics[scale=1]{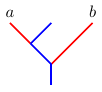}}}\ 
,
\vcenter{\hbox{\includegraphics[scale=1]{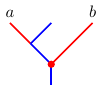}}}\ 
,
\vcenter{\hbox{\includegraphics[scale=1]{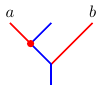}}}\ 
,
\vcenter{\hbox{\includegraphics[scale=1]{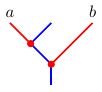}}}\ 
\!\!
\end{pmatrix}^T
&=
F^{a\s b}_{\s}
\begin{pmatrix}
\!\!
\vcenter{\hbox{\includegraphics[scale=1]{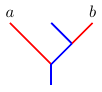}}}\ 
,
\vcenter{\hbox{\includegraphics[scale=1]{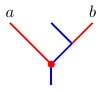}}}\ 
,
\vcenter{\hbox{\includegraphics[scale=1]{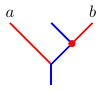}}}\ 
,
\vcenter{\hbox{\includegraphics[scale=1]{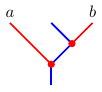}}}\ 
\!\!
\end{pmatrix}^T,\\
F^{a\s b}_{\s}
&=
\frac{\chi_{t}(a,b)}{2}
\begin{pmatrix}
1&0&0&(-1)^a\\
0&1&(-1)^a&0\\
0&(-1)^b&(-1)^{a+b}&0\\
(-1)^b&0&0&(-1)^{a+b}
\end{pmatrix},
\end{align}
which also has rank 2. For the $F^{\s\s a}_b$-type and $F^{a\s\s}_b$-type $F$ moves, we need to compare $a$ and $b$. If $b\geq a$, we have
\begin{align}\nonumber
\begin{pmatrix}
\vcenter{\hbox{\includegraphics[scale=1]{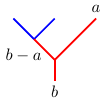}}}\ 
,
\vcenter{\hbox{\includegraphics[scale=1]{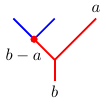}}}\ 
\end{pmatrix}^T
&=
F^{\s\s a}_{b}
\begin{pmatrix}
\vcenter{\hbox{\includegraphics[scale=1]{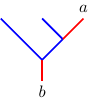}}}\ 
,
\vcenter{\hbox{\includegraphics[scale=1]{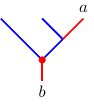}}}\ 
,
\vcenter{\hbox{\includegraphics[scale=1]{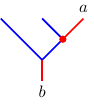}}}\ 
,
\vcenter{\hbox{\includegraphics[scale=1]{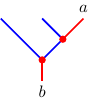}}}\ 
\end{pmatrix}^T,
\\
F^{\s\s a}_{b}
&=
\frac{1}{2}
\begin{pmatrix}
1&0&0&(-1)^{b+1}i\\
0&1&(-1)^{b}i&0\\
\end{pmatrix},
\end{align}
\begin{align}\nonumber
\begin{pmatrix}
\vcenter{\hbox{\includegraphics[scale=1]{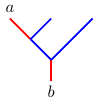}}}\ 
,
\vcenter{\hbox{\includegraphics[scale=1]{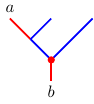}}}\ 
,
\vcenter{\hbox{\includegraphics[scale=1]{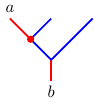}}}\ 
,
\vcenter{\hbox{\includegraphics[scale=1]{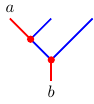}}}\ 
\end{pmatrix}^T
&=
F^{a\s\s}_{b}
\begin{pmatrix}
\vcenter{\hbox{\includegraphics[scale=1]{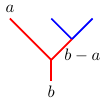}}}\ 
,
\vcenter{\hbox{\includegraphics[scale=1]{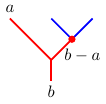}}}\ 
\end{pmatrix}^T,
\\
F^{a\s\s}_{b}
&=
\frac{1}{2}
\begin{pmatrix}
1&0\\
0&(-1)^a\\
0&(-1)^a\\
1&0\\
\end{pmatrix}.
\end{align}
If $b< a$, on the other hand, we have
\begin{align}\nonumber
\begin{pmatrix}
\vcenter{\hbox{\includegraphics[scale=1]{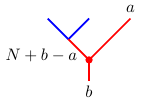}}}\ 
,
\vcenter{\hbox{\includegraphics[scale=1]{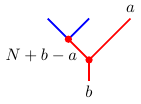}}}\ 
\end{pmatrix}^T
&=
F^{\s\s a}_{b}
\begin{pmatrix}
\vcenter{\hbox{\includegraphics[scale=1]{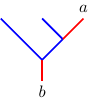}}}\ 
,
\vcenter{\hbox{\includegraphics[scale=1]{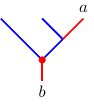}}}\ 
,
\vcenter{\hbox{\includegraphics[scale=1]{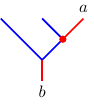}}}\ 
,
\vcenter{\hbox{\includegraphics[scale=1]{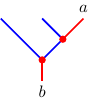}}}\ 
\end{pmatrix}^T,
\\
F^{\s\s a}_{b}
&=
\frac{1}{2}
\begin{pmatrix}
0&1&(-1)^{b}i&0\\
1&0&0&(-1)^{b+1}i\\
\end{pmatrix},
\end{align}
\begin{align}\nonumber
\begin{pmatrix}
\vcenter{\hbox{\includegraphics[scale=1]{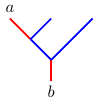}}}\ 
,
\vcenter{\hbox{\includegraphics[scale=1]{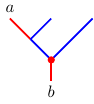}}}\ 
,
\vcenter{\hbox{\includegraphics[scale=1]{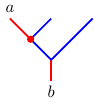}}}\ 
,
\vcenter{\hbox{\includegraphics[scale=1]{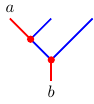}}}\ 
\end{pmatrix}^T
&=
F^{a\s\s}_{b}
\begin{pmatrix}
\vcenter{\hbox{\includegraphics[scale=1]{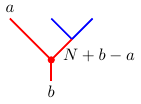}}}\ 
,
\vcenter{\hbox{\includegraphics[scale=1]{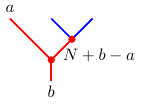}}}\ 
\end{pmatrix}^T,
\\
F^{a\s\s}_{b}
&=
\frac{1}{2}
\begin{pmatrix}
0&(-1)^a\\
1&0\\
1&0\\
0&(-1)^a\\
\end{pmatrix}.
\end{align}
Finally, the $F$ moves of types $F^{\s a\s}_b$ and $F^{\s\s\s}_\s$ are
\begin{align}\nonumber
\begin{pmatrix}
\vcenter{\hbox{\includegraphics[scale=1]{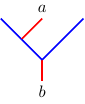}}}\ 
,
\vcenter{\hbox{\includegraphics[scale=1]{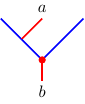}}}\ 
,
\vcenter{\hbox{\includegraphics[scale=1]{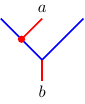}}}\ 
,
\vcenter{\hbox{\includegraphics[scale=1]{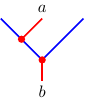}}}\ 
\end{pmatrix}^T
&=
F^{\s a\s}_{b}
\begin{pmatrix}
\vcenter{\hbox{\includegraphics[scale=1]{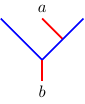}}}\ 
,
\vcenter{\hbox{\includegraphics[scale=1]{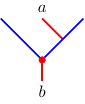}}}\ 
,
\vcenter{\hbox{\includegraphics[scale=1]{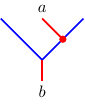}}}\ 
,
\vcenter{\hbox{\includegraphics[scale=1]{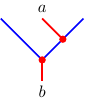}}}\ 
\end{pmatrix}^T,\\
F^{\s a\s}_{b}
&=
 \frac{\chi_{t}(a,b)}{2}
\begin{pmatrix}
1&0&0&(-1)^{b+1}i\\
0&(-1)^a&(-1)^{a+b}i&0\\
0&(-1)^a&(-1)^{a+b}i&0\\
1&0&0&(-1)^{b+1}i
\end{pmatrix},
\end{align}
\begin{align}\nonumber
\begin{pmatrix}
\vcenter{\hbox{\includegraphics[scale=1]{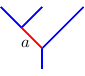}}}\ 
,
\vcenter{\hbox{\includegraphics[scale=1]{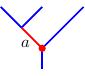}}}\ 
,
\vcenter{\hbox{\includegraphics[scale=1]{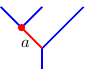}}}\ 
,
\vcenter{\hbox{\includegraphics[scale=1]{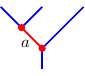}}}\ 
\end{pmatrix}^T
&=
\sum_{b\in \mathbb{Z}_{N}}
(F^{\s\s\s}_{\s})_{ab}
\begin{pmatrix}
\vcenter{\hbox{\includegraphics[scale=1]{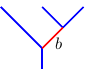}}}\ 
,
\vcenter{\hbox{\includegraphics[scale=1]{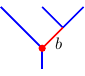}}}\ 
,
\vcenter{\hbox{\includegraphics[scale=1]{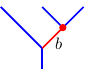}}}\ 
,
\vcenter{\hbox{\includegraphics[scale=1]{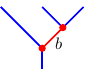}}}\ 
\end{pmatrix}^T,\\
(F^{\s\s\s}_{\s})_{ab}
&=
 \frac{\varkappa}{\sqrt{2N}}
 \chi_{t}^{-1}(a,b)
\begin{pmatrix}
1&0&0&(-1)^{a+b+1}i\\
0&1&(-1)^{a+b+1}i&0\\
0&(-1)^b&(-1)^{a}i&0\\
(-1)^b&0&0&(-1)^{a}i
\end{pmatrix}.
\end{align}

We have checked that the above $F$ matrices satisfy fermionic pentagon equations.

\section{Equivalence relation for diagonal fusion states}
\label{Amove}

To obtain relation in Eq.(\ref{pr2}), we need to consider the following sequence of $F$-move and $O$-moves:
\begin{align}
\Psi_\text{fix}\begin{pmatrix} \includegraphics[scale=.35]{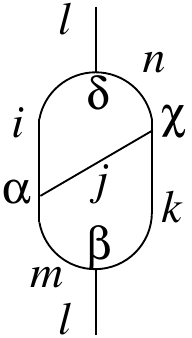} \end{pmatrix}
\simeq
\frac{n_{n}}{n_{k}}
\mathcal{F}^{ijm,\alpha\beta}_{kln,\chi\delta}
\mathcal{O}^{jk,\chi}_{n}
\mathcal{O}^{in,\delta}_{l}
(\mathcal{O}^{mk,\beta}_{l})^{-1}
(\mathcal{O}^{jn,\chi}_{k})^{-1}
\Psi_\text{fix}
\begin{pmatrix} \includegraphics[scale=.35]{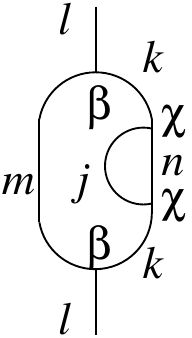} \end{pmatrix}
,
\label{A1}
\end{align}
where the Majorana numbers on the gauged $O$-move is still assigned from top to bottom, e.g., $\mathcal{O}^{jn,\chi}_{k}=\theta^{s(\chi)}_{\underline{\chi}}\theta^{s(\chi)}_{\underline{\chi'}}O^{jn,\chi}_{k}$.  The above equation is derived by
\begin{align}
\Psi_\text{fix}\begin{pmatrix} \includegraphics[scale=.35]{A1} \end{pmatrix}
&\simeq
\sum_{n'\chi'\delta'}
\mathcal{F}^{ijm,\alpha\beta}_{kln',\chi'\delta'}
\Psi_\text{fix}
\begin{pmatrix} \includegraphics[scale=.35]{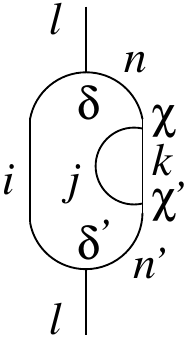} \end{pmatrix}
\nonumber\\
&\simeq
\mathcal{F}^{ijm,\alpha\beta}_{kln,\chi\delta}
\Psi_\text{fix}
\begin{pmatrix} \includegraphics[scale=.35]{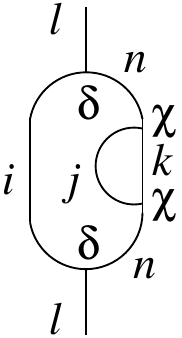} \end{pmatrix}
+
(n_n-1)
\mathcal{F}^{ijm,\alpha\beta}_{kln,(\chi\times f)(\delta\times f)}
\Psi_\text{fix}
\begin{pmatrix} \includegraphics[scale=.35]{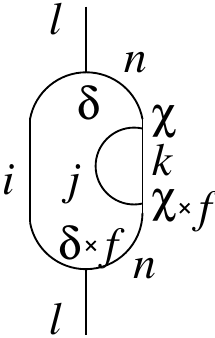} \end{pmatrix}
\nonumber\\
&\simeq
\mathcal{F}^{ijm,\alpha\beta}_{kln,\chi\delta}
\mathcal{O}^{jk,\chi}_{n}
\mathcal{O}^{in,\delta}_{l}
\Psi_\text{fix}
\begin{pmatrix} \includegraphics[scale=.35]{iline} \end{pmatrix}
+
(n_n-1)
\mathcal{F}^{ijm,\alpha\beta}_{kln,(\chi\times f)(\delta\times f)}
\widetilde{O_1}^{jk,\chi(\chi\times f)(\delta\times f)}_{nil,\delta}
\mathcal{O}^{in,\delta}_{l}
\Psi_\text{fix}
\begin{pmatrix} \includegraphics[scale=.35]{iline} \end{pmatrix}
\nonumber\\
&\simeq
n_n
\mathcal{F}^{ijm,\alpha\beta}_{kln,\chi\delta}
\mathcal{O}^{jk,\chi}_{n}
\mathcal{O}^{in,\delta}_{l}
\Psi_\text{fix}
\begin{pmatrix} \includegraphics[scale=.35]{iline} \end{pmatrix}
\nonumber\\
&\simeq
\frac{n_{n}}{n_{k}}
\mathcal{F}^{ijm,\alpha\beta}_{kln,\chi\delta}
\mathcal{O}^{jk,\chi}_{n}
\mathcal{O}^{in,\delta}_{l}
(\mathcal{O}^{mk,\beta}_{l})^{-1}
(\mathcal{O}^{jn,\chi}_{k})^{-1}
\Psi_\text{fix}
\begin{pmatrix} \includegraphics[scale=.35]{A2} \end{pmatrix}
,
\end{align}
where in the second line the factor $(n_n-1)$ means that the term $\mathcal{F}^{ijm,\alpha\beta}_{kln,(\chi\times f)(\delta\times f)}$ only exist when $n_n=2$. In the final line we have in addition a factor $\frac{1}{n_k}$, as when $k$ is q-type, the inverse $O$-move $(\mathcal{O}^{jn,\chi}_{k})^{-1}$ can only map to one-half of the state (only one of the two equivalent states).

From Eq.(\ref{A1}), we can obtain the relation in Eq.(\ref{geq5}) that is needed to derive all 2-3 moves in the fermionic partition function:
\begin{equation}
F^{ijm,(\alpha\times f)\beta}_{kln,\chi(\delta\times f)}
=
\Delta^{mji,\alpha\delta}_{nl}
F^{ijm,\alpha\beta}_{kln,\chi\delta}
\text{, \ \ if }i\text{ is q-type},
\end{equation}
where the Majorana numbers are removed when considering equivalence relations.

From fermion condensation, if $i$ is q-type, we have the following equivalent relations for fermion parity-even and odd sector respectively:
\begin{align}
\psi_\text{fix}\begin{pmatrix} \includegraphics[scale=.35]{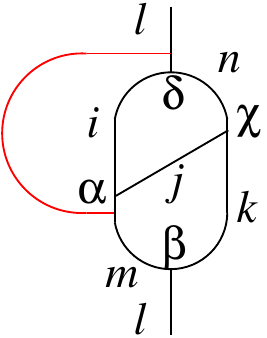} \end{pmatrix}
=
\textcolor{blue}{(\widetilde{F}^{yin}_{\textcolor{red}{l}})_{\delta}}
\psi_\text{fix}
\begin{pmatrix} \includegraphics[scale=.35]{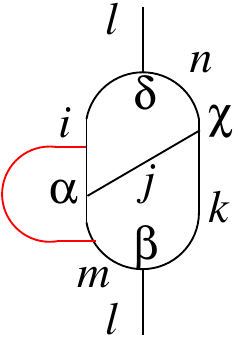} \end{pmatrix}
=
\textcolor{blue}{(\widetilde{F}^{yin}_{\textcolor{red}{l}})_{\delta}}
\textcolor{blue}{(F^{y\textcolor{red}{i}j}_{\textcolor{red}{m}})_{\alpha}}
\psi_\text{fix}
\begin{pmatrix} \includegraphics[scale=.35]{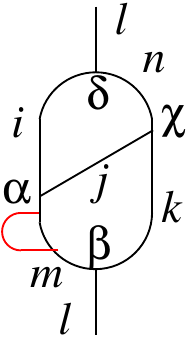} \end{pmatrix}
,
\end{align}
\begin{align}
\psi_\text{fix}\begin{pmatrix} \includegraphics[scale=.35]{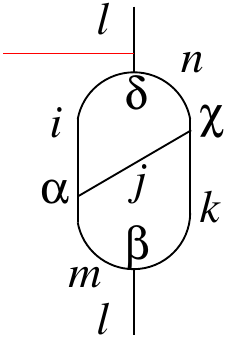} \end{pmatrix}
=
\textcolor{blue}{(\widetilde{F}^{yin}_{\textcolor{red}{l}})_{\delta}}
\psi_\text{fix}
\begin{pmatrix} \includegraphics[scale=.35]{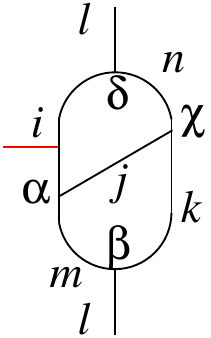} \end{pmatrix}
=
\textcolor{blue}{(\widetilde{F}^{yin}_{\textcolor{red}{l}})_{\delta}}
\textcolor{blue}{(F^{y\textcolor{red}{i}j}_{m})_{\alpha}}
\psi_\text{fix}
\begin{pmatrix} \includegraphics[scale=.35]{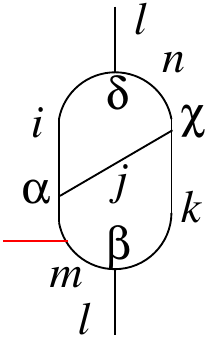} \end{pmatrix}
.
\end{align}
When the fermion parity on $\alpha$ and $\delta$ are changed, the following $O$-move will also induce a phase factor:
\begin{equation}
O^{in,(\delta\times f)}_{l}
=
\left\{\begin{array}{l}
\textcolor{blue}{(\widetilde{F}^{yin}_{\textcolor{red}{l}})_{\delta}}
\textcolor{blue}{(F^{y\textcolor{red}{i}n}_{\textcolor{red}{l}})_{\delta}}
O^{in,\delta}_{l}
\text{, \ \ parity-even}
\\ 
\textcolor{blue}{(\widetilde{F}^{yin}_{\textcolor{red}{l}})_{\delta}}
\textcolor{blue}{(F^{y\textcolor{red}{i}n}_{l})_{\delta}}
O^{in,\delta}_{l}
\text{, \ \ parity-odd}
\end{array}\right.
\end{equation}

Therefore, we obtain another equivalence relation for $F$-move if string $i$ is q-type:
\begin{equation}
F^{ijm,(\alpha\times f)\beta}_{kln,\chi(\delta\times f)}
=
\left\{\begin{array}{l}
\textcolor{blue}{(F^{y\textcolor{red}{i}n}_{\textcolor{red}{l}})_{\delta}^*}
\textcolor{blue}{(F^{y\textcolor{red}{i}j}_{\textcolor{red}{m}})_{\alpha}}
F^{ijm,\alpha\beta}_{kln,\chi\delta}
\text{, \ \ parity-even}
\\ 
\textcolor{blue}{(F^{y\textcolor{red}{i}n}_{l})_{\delta}^*}
\textcolor{blue}{(F^{y\textcolor{red}{i}j}_{m})_{\alpha}}
F^{ijm,\alpha\beta}_{kln,\chi\delta}
\text{, \ \ parity-odd}
\end{array}\right.
\label{feqaa}
\end{equation}
which corresponds to the fermion parity change on two diagonal fusion states $\alpha$ and $\delta$.

\section{Check the consistency between the 2-3 moves and the projective unitary conditions}
\label{check23}

\subsection{Obtain the second 2-3 move}

To obtain the second 2-3 move Eq.(\ref{23move2}), we consider two cases:

(1) Let the string $t'$ be m-type. We multiply by $(G^{ijm,\alpha\beta}_{knt',\eta'\psi'})^*$ and sum over $m,\alpha,\beta$ on both sides of the standard 2-3 move Eq.(\ref{23move1}):
\begin{align}
\underset{\epsilon,m\alpha\beta}{\sum}
(-1)^{s(\alpha)s(\delta)}
(G^{ijm,\alpha\beta}_{knt',\eta'\psi'})^*
G^{ijm,\alpha\epsilon}_{qps,\phi\gamma}
G^{mkn,\beta\chi}_{lpq,\delta\epsilon}
=
\underset{t\eta\psi\kappa,m\alpha\beta}{\sum}
\frac{d_{t}}{n_{t}}
(G^{ijm,\alpha\beta}_{knt',\eta'\psi'})^*
G^{ijm,\alpha\beta}_{knt,\eta\psi}
G^{itn,\psi\chi}_{lps,\kappa\gamma}
G^{jkt,\eta\kappa}_{lsq,\delta\phi},
\end{align}
where
\begin{equation}
\underset{m\alpha\beta}{\sum}
\frac{d_{t}d_{m}}{n_{t}n_{m}}
(G^{ijm,\alpha\beta}_{knt',\eta'\psi'})^{*}
G^{ijm,\alpha\beta}_{knt,\eta\psi} =
\delta_{tt'}\delta_{\eta\eta'}\delta_{\psi\psi'}
\text{, \ \ if }t'\text{ is m-type.}
\end{equation}

We can get Eq.(\ref{23move2}) straightforwardly.

(2) Let the string $t'$ be q-type. We multiply by $\frac{1}{2}[(G^{ijm,\alpha\beta}_{knt',\eta'\psi'})^*+(\Xi^{ij}_{knt',\eta'\psi'})^*G^{ijm,\alpha\beta}_{knt',(\eta'\times f)(\psi'\times f)})^*]$ and sum over $m,\alpha,\beta$ on both sides of Eq.(\ref{23move1}):
\begin{align}
&\ \ \ \ 
\underset{\epsilon,m\alpha\beta}{\sum}
(-1)^{s(\alpha)s(\delta)}
\frac{1}{2}
[(G^{ijm,\alpha\beta}_{knt',\eta'\psi'})^*
G^{ijm,\alpha\epsilon}_{qps,\phi\gamma}
G^{mkn,\beta\chi}_{lpq,\delta\epsilon}
+
(\Xi^{ij}_{knt',\eta'\psi'})^*
(G^{ijm,\alpha\beta}_{knt',(\eta'\times f)(\psi'\times f)})^*
G^{ijm,\alpha\epsilon}_{qps,\phi\gamma}
G^{mkn,\beta\chi}_{lpq,\delta\epsilon}
]
\nonumber\\
&=
\underset{t\eta\psi\kappa,m\alpha\beta}{\sum}
\frac{d_{t}}{n_{t}}
\frac{1}{2}
[(G^{ijm,\alpha\beta}_{knt',\eta'\psi'})^*
G^{ijm,\alpha\beta}_{knt,\eta\psi}
G^{itn,\psi\chi}_{lps,\kappa\gamma}
G^{jkt,\eta\kappa}_{lsq,\delta\phi}
+
(\Xi^{ij}_{knt',\eta'\psi'})^*
(G^{ijm,\alpha\beta}_{knt',(\eta'\times f)(\psi'\times f)})^*
G^{ijm,\alpha\beta}_{knt,\eta\psi}
G^{itn,\psi\chi}_{lps,\kappa\gamma}
G^{jkt,\eta\kappa}_{lsq,\delta\phi}
],
\end{align}
where the second projective unitary condition for $G$-move in Eq.(\ref{gp2}) is applied:
\begin{equation}
\underset{m\alpha\beta}{\sum}
\frac{d_{t}d_{m}}{n_{t}n_{m}}
(G^{ijm,\alpha\beta}_{knt',\eta'\psi'})^{*}
G^{ijm,\alpha\beta}_{knt,\eta\psi} =
\frac{1}{2} (\delta_{tt'}\delta_{\eta\eta'}\delta_{\psi\psi'}
+
(\Xi^{ij}_{knt,\eta'\psi'})^*
\delta_{tt'}\delta_{\eta(\eta'\times f)}\delta_{\psi(\psi'\times f)})
\text{, \ \ if }t'\text{ is q-type.}
\end{equation}
where we note that the phase factor here $(\Xi^{ij}_{knt,\eta'\psi'})^*=(\Xi^{ij}_{knt,(\eta\times f)(\psi\times f)})^*=\Xi^{ij}_{knt,\eta\psi}$. Then 
\begin{align}
&\ \ \ \ \ \ 
\underset{m\alpha\beta\epsilon}{\sum}
(-1)^{s(\alpha)s(\delta)}
\frac{d_{m}}{n_{m}}
\frac{1}{2}
[
(G^{ijm,\alpha\beta}_{knt,\eta'\psi'})^*
G^{ijm,\alpha\epsilon}_{qps,\phi\gamma}
G^{mkn,\beta\chi}_{lpq,\delta\epsilon}
+
(\Xi^{ij}_{knt,\eta'\psi'})^*
(G^{ijm,\alpha\beta}_{knt,(\eta'\times f)(\psi'\times f)})^*
G^{ijm,\alpha\epsilon}_{qps,\phi\gamma}
G^{mkn,\beta\chi}_{lpq,\delta\epsilon}
]
\nonumber\\ 
&\ \ 
=
\underset{\kappa}{\sum}
\frac{1}{2}
[
G^{itn,\psi'\chi}_{lps,\kappa\gamma}
G^{jkt,\eta'\kappa}_{lsq,\delta\phi}
+
(\Xi^{ij}_{knt,\eta'\psi'})^*
G^{itn,(\psi'\times f)\chi}_{lps,\kappa\gamma}
G^{jkt',(\eta'\times f)\kappa}_{lsq,\delta\phi}
],
\label{splitting}
\end{align}
where the two terms on the left hand side are actually equal, as well as the two terms on the right hand side:
\begin{align}
(G^{ijm,\alpha\beta}_{knt,\eta'\psi'})^*
&=
(\Xi^{ij}_{knt,\eta'\psi'})^*
(G^{ijm,\alpha\beta}_{knt,(\eta'\times f)(\psi'\times f)})^*,
\nonumber\\
G^{itn,\psi'\chi}_{lps,\kappa\gamma}
G^{jkt,\eta'\kappa}_{lsq,\delta\phi}
&=
(\Xi^{ij}_{knt,\eta'\psi'})^*
G^{itn,(\psi'\times f)\chi}_{lps,(\kappa\times f)\gamma}
G^{jkt,(\eta'\times f)(\kappa\times f)}_{lsq,\delta\phi},
\label{r1}
\end{align}
where as we require each $G$-move preserve fermion parity, the term $G^{itn,(\psi'\times f)\chi}_{lps,\kappa\gamma}
G^{jkt,(\eta'\times f)\kappa}_{lsq,\delta\phi}$ actually varies the fermion parity on $\kappa$ and should be written as $G^{itn,(\psi'\times f)\chi}_{lps,(\kappa\times f)\gamma}
G^{jkt,(\eta'\times f)(\kappa\times f)}_{lsq,\delta\phi}$. The first equality in Eq.(\ref{r1}) is exactly the complex conjugate of the second equivalence relation on $G$-move Eq.(\ref{geq2}). The second equality in Eq.(\ref{r1}) is satisfies straightforwardly by the relation among phase factors in Eq.(\ref{pr1}):
\begin{align}
(\Xi^{ij}_{knt,\eta'\psi'})^*
=
(\Omega^{nit,\psi'\kappa}_{sl})^*
(\Xi^{jkt,\eta'\kappa}_{ls})^*,
\label{p1a}
\end{align}
where $G^{itn,\psi'\chi}_{lps,\kappa\gamma}=(\Omega^{nit,\psi'\kappa}_{sl})^*G^{itn,(\psi'\times f)\chi}_{lps,(\kappa\times f)\gamma}$, and $G^{jkt,\eta'\kappa}_{lsq,\delta\phi}=(\Xi^{jkt,\eta'\kappa}_{ls})^*G^{jkt,(\eta'\times f)(\kappa\times f)}_{lsq,\delta\phi}$. We can again check this relation from fermion condensation: $\Xi^{ij}_{knt,\eta'\psi'}=\Theta_{fi}\textcolor{blue}{(F^{yit}_{n}F^{iyt}_{n})_{\psi'}}$, $\Omega^{nit,\psi'\kappa}_{sl}=\Theta_{fi}
\textcolor{blue}{(F^{yit}_{n}F^{iyt}_{n})_{\psi'}}
\textcolor{blue}{(F^{ytl}_{s})_{\kappa}^*}$
and $\Xi^{jkt,\eta'\kappa}_{ls}=\textcolor{blue}{(F^{ytl}_{s})_{\kappa}}$, we find Eq.(\ref{p1a}) is satisfied straightforwardly.

Physically, from the point of view of fixed-point wavefunctions, there is a splitting on the two channels. Suppose that $\eta$ and $\psi$ are two bosonic fusion states. In terms of $F$-moves, the splitting of Eq.(\ref{splitting}) is graphically represented as:
\begin{align}
\frac{1}{2}
\psi_\text{fix}\begin{pmatrix} \includegraphics[scale=.45]{pent4g} \end{pmatrix}
&\simeq
\sum_{\kappa}
\frac{1}{2}
F^{itn,\psi'\chi}_{lps,\kappa\gamma}
F^{jkt,\eta'\kappa}_{lsq,\delta\phi}
\psi_\text{fix}
\begin{pmatrix} \includegraphics[scale=.45]{pent3g} \end{pmatrix}
\nonumber\\
&\simeq
\underset{m\alpha\beta\epsilon}{\sum}
(-1)^{s(\alpha)s(\delta)}
\frac{1}{2}
(F^{ijm,\alpha\beta}_{knt,\eta'\psi'})^*
F^{ijm,\alpha\epsilon}_{qps,\phi\gamma}
F^{mkn,\beta\chi}_{lpq,\delta\epsilon}
\psi_\text{fix}
\begin{pmatrix} \includegraphics[scale=.45]{pent3g} \end{pmatrix},
\label{splitting1}
\end{align}
\begin{align}
\frac{1}{2}
(\Xi^{ij}_{knt,\eta'\psi'})^*
\psi_\text{fix}\begin{pmatrix} \includegraphics[scale=.45]{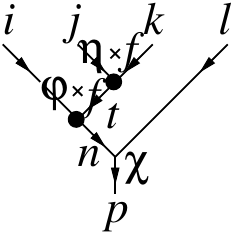} \end{pmatrix}
&=
\sum_{\kappa}
\frac{1}{2}
(\Xi^{ij}_{knt,\eta'\psi'})^*
F^{itn,(\psi'\times f)\chi}_{lps,(\kappa\times f)\gamma}
F^{jkt,(\eta'\times f)(\kappa\times f)}_{lsq,\delta\phi}
\psi_\text{fix}
\begin{pmatrix} \includegraphics[scale=.45]{pent3g} \end{pmatrix}
\nonumber\\
&=
\underset{m\alpha\beta\epsilon}{\sum}
(-1)^{s(\alpha)s(\delta)}
\frac{1}{2}
(\Xi^{ij}_{knt,\eta'\psi'})^*
(F^{ijm,\alpha\beta}_{knt,(\eta'\times f)(\psi'\times f)})^*
F^{ijm,\alpha\epsilon}_{qps,\phi\gamma}
F^{mkn,\beta\chi}_{lpq,\delta\epsilon}
\psi_\text{fix}
\begin{pmatrix} \includegraphics[scale=.45]{pent3g} \end{pmatrix}.
\label{splitting2}
\end{align}

Therefore, Eq.(\ref{splitting}) again implies: 
\begin{equation}
\underset{m\alpha\beta\epsilon}{\sum}
(-1)^{s(\alpha)s(\delta)}
\frac{d_{m}}{n_{m}}
(G^{ijm,\alpha\beta}_{knt,\eta\psi})^*
G^{ijm,\alpha\epsilon}_{qps,\phi\gamma}
G^{mkn,\beta\chi}_{lpq,\delta\epsilon}
=
\underset{\kappa}{\sum}
G^{itn,\psi\chi}_{lps,\kappa\gamma}
G^{jkt,\eta\kappa}_{lsq,\delta\phi}.
\label{23move2'}
\end{equation}

\subsection{Obtain the third 2-3 move}

To obtain the third 2-3 move Eq.(\ref{23move3}), we consider two cases:

(1) Let the string $m'$ be m-type. We multiply by $(G^{ijm',\alpha'\epsilon'}_{qps,\phi\gamma})^*$ and sum over $s,\phi,\gamma$ on both sides of the second 2-3 move Eq.(\ref{23move2}). We can obtain Eq.(\ref{23move3}) by applying 
\begin{equation}
\underset{s\phi\gamma}{\sum}
\frac{d_{s}d_{m}}{n_{s}n_{m}}
(G^{ijm',\alpha'\epsilon'}_{qps,\phi\gamma})^{*}
G^{ijm,\alpha\epsilon}_{qps,\phi\gamma} =
\delta_{mm'}\delta_{\alpha\alpha'}\delta_{\epsilon\epsilon'}
\text{, \ \ if }m'\text{ is m-type.}
\end{equation}

(2) Let the string $m'$ be q-type. We multiply by $\frac{1}{2}[(G^{ijm',\alpha'\epsilon'}_{qps,\phi\gamma})^*+\Xi^{ijm',\alpha'\epsilon'}_{qp}(G^{ijm',(\alpha'\times f)(\epsilon'\times f)}_{qps,\phi\gamma})^*]$ and sum over $s,\phi,\gamma$ on both sides of Eq.(\ref{23move2}):

\begin{align}
&\ \ \ \ \ \ 
\underset{s\phi\kappa\gamma}{\sum}
\frac{d_{s}}{n_{s}}
\frac{1}{2}
(
G^{ijm',\alpha'\epsilon'}_{qps,\phi\gamma})^*
G^{itn,\psi\chi}_{lps,\kappa\gamma}
G^{jkt,\eta\kappa}_{lsq,\delta\phi}
+
\Xi^{ijm',\alpha'\epsilon'}_{qp}
(G^{ijm',(\alpha'\times f)(\epsilon'\times f)}_{qps,\phi\gamma})^*
G^{itn,\psi\chi}_{lps,\kappa\gamma}
G^{jkt,\eta\kappa}_{lsq,\delta\phi}
)
\nonumber\\ 
&\ \ 
=
(-1)^{s(\alpha)s(\delta)}
\underset{\beta}{\sum}
\frac{1}{2}(
(G^{ijm,\alpha'\beta}_{knt,\eta\psi})^*
G^{mkn,\beta\chi}_{lpq,\delta\epsilon'}
+
\Xi^{ijm',\alpha'\epsilon'}_{qp}
(G^{ijm,(\alpha'\times f)\beta}_{knt,\eta\psi})^*
G^{mkn,\beta\chi}_{lpq,\delta(\epsilon'\times f)}
).
\end{align}

Splitting the two channels, we again get Eq.(\ref{23move3}) by applying
\begin{equation}
\underset{s\phi\gamma}{\sum}
\frac{d_{s}d_{m}}{n_{s}n_{m}}
G^{ijm,\alpha\epsilon}_{qps,\phi\gamma}
(G^{ijm',\alpha'\epsilon'}_{qps,\phi\gamma})^* =
\frac{1}{2} (\delta_{mm'}\delta_{\alpha\alpha'}\delta_{\epsilon\epsilon'}
+
\Xi^{ijm',\alpha'\epsilon'}_{qp}
\delta_{mm'}\delta_{\alpha(\alpha'\times f)}\delta_{\epsilon(\epsilon'\times f)})
\text{, \ \ if }m'\text{ is q-type.}
\end{equation}

We obtain Eq.(\ref{23move3}) from Eq.(\ref{23move2}) by requiring Eq.(\ref{gp1}) and the relation in Eq.(\ref{pr2}): 
\begin{equation}
\Xi^{ijm,\alpha'\epsilon'}_{qp}=\Xi^{ijm,\alpha'\beta}_{kn}(\Delta^{nkm,\beta\epsilon'}_{qp})^*,
\label{p2a}
\end{equation} 
as from which we have
\begin{align}
(G^{ijm,\alpha'\epsilon'}_{qps,\phi\gamma})^{*}
&=
\Xi^{ijm,\alpha'\epsilon'}_{qp}
(G^{ijm,(\alpha'\times f)(\epsilon'\times f)}_{qps,\phi\gamma})^{*},
\nonumber\\
(G^{ijm,\alpha'\beta}_{knt,\eta\psi})^*
G^{mkn,\beta\chi}_{lpq,\delta\epsilon'}
&=
\Xi^{ijm,\alpha'\epsilon'}_{qp}
(G^{ijm,(\alpha'\times f)(\beta\times f)}_{knt,\eta\psi})^*
G^{mkn,(\beta\times f)\chi}_{lpq,\delta(\epsilon'\times f)}.
\label{r2}
\end{align}
We can again check Eq.(\ref{p2a}) from fermion condensation: $\Xi^{ijm,\alpha'\epsilon'}_{qp}=\textcolor{blue}{(F^{ymq}_{p})_{\epsilon'}}$, $\Xi^{ijm,\alpha'\beta}_{kn}=\textcolor{blue}{(F^{ymk}_{n})_{\beta}}$ and $\Delta^{nkm,\beta\epsilon'}_{qp}=\textcolor{blue}{(F^{ymq}_{p})_{\epsilon'}^*}
\textcolor{blue}{(F^{ymk}_{n})_{\beta}}$.

\subsection{Obtain the rest five 2-3 moves}

Similarly, we obtain Eq.(\ref{23move4}) from Eq.(\ref{23move3}) by requiring Eq.(\ref{gp2}) and $(\Xi^{it}_{lps,\kappa'\gamma'})^*=(\Xi^{ij}_{qps,\phi\gamma'})^*
\Omega^{tj}_{qls,\kappa'\phi}$, where
\begin{align}
(G^{itn,\psi\chi}_{lps,\kappa'\gamma'})^*
&=
(\Xi^{it}_{lps,\kappa'\gamma'})^*
(G^{itn,\psi\chi}_{lps,(\kappa'\times f)(\gamma'\times f)})^*,
\nonumber\\
(G^{ijm,\alpha\epsilon}_{qps,\phi\gamma'})^*
G^{jkt,\eta\kappa'}_{lsq,\delta\phi}
&=
(\Xi^{ij}_{qps,\phi\gamma'})^*
\Omega^{tj}_{qls,\kappa'\phi}
(G^{ijm,\alpha\epsilon}_{qps,(\phi\times f)(\gamma'\times f)})^*
G^{jkt,\eta(\kappa'\times f)}_{lsq,\delta(\phi\times f)}.
\label{r3}
\end{align}

We obtain Eq.(\ref{23move5}) from Eq.(\ref{23move1}) by requiring Eq.(\ref{gp1}) and $\Xi^{jkt,\eta'\kappa'}_{ls}=\Xi^{ij}_{knt,\eta'\psi}
(\Omega^{nit,\psi\kappa'}_{sl})^*$ (the same relation as Eq.(\ref{p1a})), where
\begin{align}
(G^{jkt,\eta'\kappa'}_{lsq,\delta\phi})^*
&=
\Xi^{jkt,\eta'\kappa'}_{ls}
(G^{jkt,(\eta'\times f)(\kappa\times f)}_{lsq,\delta\phi})^*,
\nonumber\\
G^{ijm,\alpha\beta}_{knt,\eta'\psi}
G^{itn,\psi\chi}_{lps,\kappa'\gamma}
&=
\Xi^{ij}_{knt,\eta'\psi}
(\Omega^{nit,\psi\kappa'}_{sl})^*
G^{ijm,\alpha\beta}_{knt,(\eta'\times f)(\psi\times f)}
G^{itn,(\psi\times f)\chi}_{lps,(\kappa'\times f)\gamma}.
\label{r4}
\end{align}

We obtain Eq.(\ref{23move6}) from Eq.(\ref{23move1}) by requiring Eq.(\ref{gp3}) and $\Omega^{nit,\psi'\kappa'}_{sl}=\Xi^{ij}_{knt,\eta\psi'}
(\Xi^{jkt,\eta\kappa'}_{ls})^*$ (the same relation as Eq.(\ref{p1a})), where
\begin{align}
(G^{itn,\psi'\chi}_{lps,\kappa'\gamma})^*
&=
\Omega^{nit,\psi'\kappa'}_{sl}
(G^{itn,(\psi'\times f)\chi}_{lps,(\kappa'\times f)\gamma})^*,
\nonumber\\
G^{ijm,\alpha\beta}_{knt,\eta\psi'}
G^{jkt,\eta\kappa'}_{lsq,\delta\phi}
&=
\Xi^{ij}_{knt,\eta\psi'}
(\Xi^{jkt,\eta\kappa'}_{ls})^*
G^{ijm,\alpha\beta}_{knt,(\eta\times f)(\psi'\times f)}
G^{jkt,(\eta\times f)(\kappa'\times f)}_{lsq,\delta\phi}.
\label{r5}
\end{align}

We obtain Eq.(\ref{23move7}) from Eq.(\ref{23move6}) by requiring Eq.(\ref{gp4}) and $(\Omega^{mi}_{sqp,\epsilon'\gamma'})^*=\Omega^{nm}_{qlp,\chi\epsilon'}
(\Omega^{ni}_{slp,\chi\gamma'})^*$, where
\begin{align}
(G^{ijm,\alpha\epsilon‘}_{qps,\phi\gamma’})^*
&=
(\Omega^{mi}_{sqp,\epsilon'\gamma'})^*
(G^{ijm,\alpha(\epsilon‘\times f)}_{qps,\phi(\gamma'\times f)})^*,
\nonumber\\
G^{mkn,\beta\chi}_{lpq,\delta\epsilon'}
(G^{itn,\psi\chi}_{lps,\kappa\gamma'})^*
&=
\Omega^{nm}_{qlp,\chi\epsilon'}
(\Omega^{ni}_{slp,\chi\gamma'})^*
G^{mkn,\beta(\chi\times f)}_{lpq,\delta(\epsilon‘\times f)}
(G^{itn,\psi(\chi\times f)}_{lps,\kappa(\gamma‘\times f)})^*.
\label{r6}
\end{align}

We obtain Eq.(\ref{23move8}) from Eq.(\ref{23move3}) by requiring Eq.(\ref{gp4}) and $(\Omega^{tj}_{qls,\kappa'\phi'})^*=(\Xi^{ij}_{qps,\phi'\gamma})^*
\Xi^{it}_{lps,\kappa'\gamma}$, where
\begin{align}
(G^{jkt,\eta\kappa'}_{lsq,\delta\phi'})^*
&=
(\Omega^{tj}_{qls,\kappa'\phi'})^*
(G^{jkt,\eta(\kappa'\times f)}_{lsq,\delta(\phi'\times f)})^*,
\nonumber\\
(G^{ijm,\alpha\epsilon}_{qps,\phi'\gamma})^*
G^{itn,\psi\chi}_{lps,\kappa'\gamma}
&=
(\Xi^{ij}_{qps,\phi'\gamma})^*
\Xi^{it}_{lps,\kappa'\gamma}
(G^{ijm,\alpha\epsilon}_{qps,(\phi'\times f)(\gamma\times f)})^*
G^{itn,\psi\chi}_{lps,(\kappa'\times f)(\gamma\times f)}
.
\label{r7}
\end{align}

\end{widetext}

\bibliography{fSN.bib}

\begin{thebibliography}{45}%
\makeatletter
\providecommand \@ifxundefined [1]{%
 \@ifx{#1\undefined}
}%
\providecommand \@ifnum [1]{%
 \ifnum #1\expandafter \@firstoftwo
 \else \expandafter \@secondoftwo
 \fi
}%
\providecommand \@ifx [1]{%
 \ifx #1\expandafter \@firstoftwo
 \else \expandafter \@secondoftwo
 \fi
}%
\providecommand \natexlab [1]{#1}%
\providecommand \enquote  [1]{``#1''}%
\providecommand \bibnamefont  [1]{#1}%
\providecommand \bibfnamefont [1]{#1}%
\providecommand \citenamefont [1]{#1}%
\providecommand \href@noop [0]{\@secondoftwo}%
\providecommand \href [0]{\begingroup \@sanitize@url \@href}%
\providecommand \@href[1]{\@@startlink{#1}\@@href}%
\providecommand \@@href[1]{\endgroup#1\@@endlink}%
\providecommand \@sanitize@url [0]{\catcode `\\12\catcode `\$12\catcode
  `\&12\catcode `\#12\catcode `\^12\catcode `\_12\catcode `\%12\relax}%
\providecommand \@@startlink[1]{}%
\providecommand \@@endlink[0]{}%
\providecommand \url  [0]{\begingroup\@sanitize@url \@url }%
\providecommand \@url [1]{\endgroup\@href {#1}{\urlprefix }}%
\providecommand \urlprefix  [0]{URL }%
\providecommand \Eprint [0]{\href }%
\providecommand \doibase [0]{http://dx.doi.org/}%
\providecommand \selectlanguage [0]{\@gobble}%
\providecommand \bibinfo  [0]{\@secondoftwo}%
\providecommand \bibfield  [0]{\@secondoftwo}%
\providecommand \translation [1]{[#1]}%
\providecommand \BibitemOpen [0]{}%
\providecommand \bibitemStop [0]{}%
\providecommand \bibitemNoStop [0]{.\EOS\space}%
\providecommand \EOS [0]{\spacefactor3000\relax}%
\providecommand \BibitemShut  [1]{\csname bibitem#1\endcsname}%
\let\auto@bib@innerbib\@empty
\bibitem [{\citenamefont {Tsui}\ \emph {et~al.}(1982)\citenamefont {Tsui},
  \citenamefont {Stormer},\ and\ \citenamefont {Gossard}}]{Tsui1982}%
  \BibitemOpen
  \bibfield  {author} {\bibinfo {author} {\bibfnamefont {D.~C.}\ \bibnamefont
  {Tsui}}, \bibinfo {author} {\bibfnamefont {H.~L.}\ \bibnamefont {Stormer}}, \
  and\ \bibinfo {author} {\bibfnamefont {A.~C.}\ \bibnamefont {Gossard}},\
  }\href {\doibase 10.1103/PhysRevLett.48.1559} {\bibfield  {journal} {\bibinfo
   {journal} {Phys. Rev. Lett.}\ }\textbf {\bibinfo {volume} {48}},\ \bibinfo
  {pages} {1559} (\bibinfo {year} {1982})}\BibitemShut {NoStop}%
\bibitem [{\citenamefont {Wen}(1990)}]{Wen90}%
  \BibitemOpen
  \bibfield  {author} {\bibinfo {author} {\bibfnamefont {X.-G.}\ \bibnamefont
  {Wen}},\ }\href@noop {} {\bibfield  {journal} {\bibinfo  {journal}
  {International Journal of Modern Physics B}\ }\textbf {\bibinfo {volume}
  {4}},\ \bibinfo {pages} {239} (\bibinfo {year} {1990})}\BibitemShut {NoStop}%
\bibitem [{\citenamefont {Laughlin}(1983)}]{Laughlin1983}%
  \BibitemOpen
  \bibfield  {author} {\bibinfo {author} {\bibfnamefont {R.~B.}\ \bibnamefont
  {Laughlin}},\ }\href {\doibase 10.1103/PhysRevLett.50.1395} {\bibfield
  {journal} {\bibinfo  {journal} {Phys. Rev. Lett.}\ }\textbf {\bibinfo
  {volume} {50}},\ \bibinfo {pages} {1395} (\bibinfo {year}
  {1983})}\BibitemShut {NoStop}%
\bibitem [{\citenamefont {Kitaev}(2006)}]{Kitaev2006}%
  \BibitemOpen
  \bibfield  {author} {\bibinfo {author} {\bibfnamefont {A.}~\bibnamefont
  {Kitaev}},\ }\href {\doibase 10.1016/j.aop.2005.10.005} {\bibfield  {journal}
  {\bibinfo  {journal} {Annals of Physics}\ }\textbf {\bibinfo {volume}
  {321}},\ \bibinfo {pages} {2–111} (\bibinfo {year} {2006})}\BibitemShut
  {NoStop}%
\bibitem [{\citenamefont {Chen}\ \emph {et~al.}(2010)\citenamefont {Chen},
  \citenamefont {Gu},\ and\ \citenamefont {Wen}}]{Chen2010}%
  \BibitemOpen
  \bibfield  {author} {\bibinfo {author} {\bibfnamefont {X.}~\bibnamefont
  {Chen}}, \bibinfo {author} {\bibfnamefont {Z.-C.}\ \bibnamefont {Gu}}, \ and\
  \bibinfo {author} {\bibfnamefont {X.-G.}\ \bibnamefont {Wen}},\ }\href
  {\doibase 10.1103/PhysRevB.82.155138} {\bibfield  {journal} {\bibinfo
  {journal} {Phys. Rev. B}\ }\textbf {\bibinfo {volume} {82}},\ \bibinfo
  {pages} {155138} (\bibinfo {year} {2010})}\BibitemShut {NoStop}%
\bibitem [{\citenamefont {Kitaev}(2003)}]{Kit03}%
  \BibitemOpen
  \bibfield  {author} {\bibinfo {author} {\bibfnamefont {A.}~\bibnamefont
  {Kitaev}},\ }\href {\doibase http://dx.doi.org/10.1016/S0003-4916(02)00018-0}
  {\bibfield  {journal} {\bibinfo  {journal} {Annals of Physics}\ }\textbf
  {\bibinfo {volume} {303}},\ \bibinfo {pages} {2 } (\bibinfo {year}
  {2003})}\BibitemShut {NoStop}%
\bibitem [{\citenamefont {Levin}\ and\ \citenamefont {Wen}(2005)}]{Levin2005}%
  \BibitemOpen
  \bibfield  {author} {\bibinfo {author} {\bibfnamefont {M.~A.}\ \bibnamefont
  {Levin}}\ and\ \bibinfo {author} {\bibfnamefont {X.-G.}\ \bibnamefont
  {Wen}},\ }\href {\doibase 10.1103/PhysRevB.71.045110} {\bibfield  {journal}
  {\bibinfo  {journal} {Phys. Rev. B}\ }\textbf {\bibinfo {volume} {71}},\
  \bibinfo {pages} {045110} (\bibinfo {year} {2005})}\BibitemShut {NoStop}%
\bibitem [{\citenamefont {Freedman}\ \emph {et~al.}(2004)\citenamefont
  {Freedman}, \citenamefont {Nayak}, \citenamefont {Shtengel}, \citenamefont
  {Walker},\ and\ \citenamefont {Wang}}]{FNSWW04}%
  \BibitemOpen
  \bibfield  {author} {\bibinfo {author} {\bibfnamefont {M.}~\bibnamefont
  {Freedman}}, \bibinfo {author} {\bibfnamefont {C.}~\bibnamefont {Nayak}},
  \bibinfo {author} {\bibfnamefont {K.}~\bibnamefont {Shtengel}}, \bibinfo
  {author} {\bibfnamefont {K.}~\bibnamefont {Walker}}, \ and\ \bibinfo {author}
  {\bibfnamefont {Z.}~\bibnamefont {Wang}},\ }\href {\doibase
  http://dx.doi.org/10.1016/j.aop.2004.01.006} {\bibfield  {journal} {\bibinfo
  {journal} {Annals of Physics}\ }\textbf {\bibinfo {volume} {310}},\ \bibinfo
  {pages} {428 } (\bibinfo {year} {2004})}\BibitemShut {NoStop}%
\bibitem [{\citenamefont {Gu}\ \emph {et~al.}(2014)\citenamefont {Gu},
  \citenamefont {Wang},\ and\ \citenamefont {Wen}}]{fto}%
  \BibitemOpen
  \bibfield  {author} {\bibinfo {author} {\bibfnamefont {Z.-C.}\ \bibnamefont
  {Gu}}, \bibinfo {author} {\bibfnamefont {Z.}~\bibnamefont {Wang}}, \ and\
  \bibinfo {author} {\bibfnamefont {X.-G.}\ \bibnamefont {Wen}},\ }\href
  {\doibase 10.1103/PhysRevB.90.085140} {\bibfield  {journal} {\bibinfo
  {journal} {Phys. Rev. B}\ }\textbf {\bibinfo {volume} {90}},\ \bibinfo
  {pages} {085140} (\bibinfo {year} {2014})}\BibitemShut {NoStop}%
\bibitem [{\citenamefont {Gu}\ \emph {et~al.}(2015)\citenamefont {Gu},
  \citenamefont {Wang},\ and\ \citenamefont {Wen}}]{Gu2015}%
  \BibitemOpen
  \bibfield  {author} {\bibinfo {author} {\bibfnamefont {Z.-C.}\ \bibnamefont
  {Gu}}, \bibinfo {author} {\bibfnamefont {Z.}~\bibnamefont {Wang}}, \ and\
  \bibinfo {author} {\bibfnamefont {X.-G.}\ \bibnamefont {Wen}},\ }\href
  {\doibase 10.1103/PhysRevB.91.125149} {\bibfield  {journal} {\bibinfo
  {journal} {Phys. Rev. B}\ }\textbf {\bibinfo {volume} {91}},\ \bibinfo
  {pages} {125149} (\bibinfo {year} {2015})}\BibitemShut {NoStop}%
\bibitem [{\citenamefont {Lan}\ \emph {et~al.}(2016)\citenamefont {Lan},
  \citenamefont {Kong},\ and\ \citenamefont {Wen}}]{Lan2016}%
  \BibitemOpen
  \bibfield  {author} {\bibinfo {author} {\bibfnamefont {T.}~\bibnamefont
  {Lan}}, \bibinfo {author} {\bibfnamefont {L.}~\bibnamefont {Kong}}, \ and\
  \bibinfo {author} {\bibfnamefont {X.-G.}\ \bibnamefont {Wen}},\ }\href
  {\doibase 10.1103/PhysRevB.94.155113} {\bibfield  {journal} {\bibinfo
  {journal} {Phys. Rev. B}\ }\textbf {\bibinfo {volume} {94}},\ \bibinfo
  {pages} {155113} (\bibinfo {year} {2016})}\BibitemShut {NoStop}%
\bibitem [{\citenamefont {Aasen}\ \emph {et~al.}(2019)\citenamefont {Aasen},
  \citenamefont {Lake},\ and\ \citenamefont {Walker}}]{fc2019}%
  \BibitemOpen
  \bibfield  {author} {\bibinfo {author} {\bibfnamefont {D.}~\bibnamefont
  {Aasen}}, \bibinfo {author} {\bibfnamefont {E.}~\bibnamefont {Lake}}, \ and\
  \bibinfo {author} {\bibfnamefont {K.}~\bibnamefont {Walker}},\ }\href
  {\doibase 10.1063/1.5045669} {\bibfield  {journal} {\bibinfo  {journal}
  {Journal of Mathematical Physics}\ }\textbf {\bibinfo {volume} {60}},\
  \bibinfo {pages} {121901} (\bibinfo {year} {2019})}\BibitemShut {NoStop}%
\bibitem [{\citenamefont {Wan}\ and\ \citenamefont {Wang}(2017)}]{wan2017}%
  \BibitemOpen
  \bibfield  {author} {\bibinfo {author} {\bibfnamefont {Y.}~\bibnamefont
  {Wan}}\ and\ \bibinfo {author} {\bibfnamefont {C.}~\bibnamefont {Wang}},\
  }\href@noop {} {\bibfield  {journal} {\bibinfo  {journal} {Journal of High
  Energy Physics}\ }\textbf {\bibinfo {volume} {2017}},\ \bibinfo {pages} {172}
  (\bibinfo {year} {2017})}\BibitemShut {NoStop}%
\bibitem [{\citenamefont {Lou}\ \emph {et~al.}(2021)\citenamefont {Lou},
  \citenamefont {Shen}, \citenamefont {Chen},\ and\ \citenamefont
  {Hung}}]{lou2021}%
  \BibitemOpen
  \bibfield  {author} {\bibinfo {author} {\bibfnamefont {J.}~\bibnamefont
  {Lou}}, \bibinfo {author} {\bibfnamefont {C.}~\bibnamefont {Shen}}, \bibinfo
  {author} {\bibfnamefont {C.}~\bibnamefont {Chen}}, \ and\ \bibinfo {author}
  {\bibfnamefont {L.-Y.}\ \bibnamefont {Hung}},\ }\href@noop {} {\bibfield
  {journal} {\bibinfo  {journal} {Journal of High Energy Physics}\ }\textbf
  {\bibinfo {volume} {2021}},\ \bibinfo {pages} {171} (\bibinfo {year}
  {2021})}\BibitemShut {NoStop}%
\bibitem [{\citenamefont {Hastings}\ and\ \citenamefont
  {Wen}(2005)}]{Hastings05}%
  \BibitemOpen
  \bibfield  {author} {\bibinfo {author} {\bibfnamefont {M.~B.}\ \bibnamefont
  {Hastings}}\ and\ \bibinfo {author} {\bibfnamefont {X.-G.}\ \bibnamefont
  {Wen}},\ }\href {\doibase 10.1103/PhysRevB.72.045141} {\bibfield  {journal}
  {\bibinfo  {journal} {Phys. Rev. B}\ }\textbf {\bibinfo {volume} {72}},\
  \bibinfo {pages} {045141} (\bibinfo {year} {2005})}\BibitemShut {NoStop}%
\bibitem [{\citenamefont {Bravyi}\ \emph {et~al.}(2006)\citenamefont {Bravyi},
  \citenamefont {Hastings},\ and\ \citenamefont {Verstraete}}]{Bravyi06}%
  \BibitemOpen
  \bibfield  {author} {\bibinfo {author} {\bibfnamefont {S.}~\bibnamefont
  {Bravyi}}, \bibinfo {author} {\bibfnamefont {M.~B.}\ \bibnamefont
  {Hastings}}, \ and\ \bibinfo {author} {\bibfnamefont {F.}~\bibnamefont
  {Verstraete}},\ }\href {\doibase 10.1103/PhysRevLett.97.050401} {\bibfield
  {journal} {\bibinfo  {journal} {Phys. Rev. Lett.}\ }\textbf {\bibinfo
  {volume} {97}},\ \bibinfo {pages} {050401} (\bibinfo {year}
  {2006})}\BibitemShut {NoStop}%
\bibitem [{\citenamefont {Bravyi}\ \emph {et~al.}(2010)\citenamefont {Bravyi},
  \citenamefont {Hastings},\ and\ \citenamefont {Michalakis}}]{Bravyi10}%
  \BibitemOpen
  \bibfield  {author} {\bibinfo {author} {\bibfnamefont {S.}~\bibnamefont
  {Bravyi}}, \bibinfo {author} {\bibfnamefont {M.~B.}\ \bibnamefont
  {Hastings}}, \ and\ \bibinfo {author} {\bibfnamefont {S.}~\bibnamefont
  {Michalakis}},\ }\href {https://doi.org/10.1063/1.3490195} {\bibfield
  {journal} {\bibinfo  {journal} {J. Math. Phys.}\ }\textbf {\bibinfo {volume}
  {51}},\ \bibinfo {pages} {093512} (\bibinfo {year} {2010})}\BibitemShut
  {NoStop}%
\bibitem [{\citenamefont {Zeng}\ and\ \citenamefont {Wen}(2015)}]{Zeng2015}%
  \BibitemOpen
  \bibfield  {author} {\bibinfo {author} {\bibfnamefont {B.}~\bibnamefont
  {Zeng}}\ and\ \bibinfo {author} {\bibfnamefont {X.-G.}\ \bibnamefont {Wen}},\
  }\href {\doibase 10.1103/PhysRevB.91.125121} {\bibfield  {journal} {\bibinfo
  {journal} {Phys. Rev. B}\ }\textbf {\bibinfo {volume} {91}},\ \bibinfo
  {pages} {125121} (\bibinfo {year} {2015})}\BibitemShut {NoStop}%
\bibitem [{\citenamefont {Haah}(2011)}]{FC1}%
  \BibitemOpen
  \bibfield  {author} {\bibinfo {author} {\bibfnamefont {J.}~\bibnamefont
  {Haah}},\ }\href {\doibase 10.1103/PhysRevA.83.042330} {\bibfield  {journal}
  {\bibinfo  {journal} {Phys. Rev. A}\ }\textbf {\bibinfo {volume} {83}},\
  \bibinfo {pages} {042330} (\bibinfo {year} {2011})}\BibitemShut {NoStop}%
\bibitem [{\citenamefont {Yoshida}(2013)}]{FC2}%
  \BibitemOpen
  \bibfield  {author} {\bibinfo {author} {\bibfnamefont {B.}~\bibnamefont
  {Yoshida}},\ }\href {\doibase 10.1103/PhysRevB.88.125122} {\bibfield
  {journal} {\bibinfo  {journal} {Phys. Rev. B}\ }\textbf {\bibinfo {volume}
  {88}},\ \bibinfo {pages} {125122} (\bibinfo {year} {2013})}\BibitemShut
  {NoStop}%
\bibitem [{\citenamefont {Vijay}\ \emph {et~al.}(2015)\citenamefont {Vijay},
  \citenamefont {Haah},\ and\ \citenamefont {Fu}}]{FC3}%
  \BibitemOpen
  \bibfield  {author} {\bibinfo {author} {\bibfnamefont {S.}~\bibnamefont
  {Vijay}}, \bibinfo {author} {\bibfnamefont {J.}~\bibnamefont {Haah}}, \ and\
  \bibinfo {author} {\bibfnamefont {L.}~\bibnamefont {Fu}},\ }\href {\doibase
  10.1103/PhysRevB.92.235136} {\bibfield  {journal} {\bibinfo  {journal} {Phys.
  Rev. B}\ }\textbf {\bibinfo {volume} {92}},\ \bibinfo {pages} {235136}
  (\bibinfo {year} {2015})}\BibitemShut {NoStop}%
\bibitem [{\citenamefont {Gu}\ and\ \citenamefont {Wen}(2014)}]{Gu2014}%
  \BibitemOpen
  \bibfield  {author} {\bibinfo {author} {\bibfnamefont {Z.-C.}\ \bibnamefont
  {Gu}}\ and\ \bibinfo {author} {\bibfnamefont {X.-G.}\ \bibnamefont {Wen}},\
  }\href {\doibase 10.1103/PhysRevB.90.115141} {\bibfield  {journal} {\bibinfo
  {journal} {Phys. Rev. B}\ }\textbf {\bibinfo {volume} {90}},\ \bibinfo
  {pages} {115141} (\bibinfo {year} {2014})}\BibitemShut {NoStop}%
\bibitem [{\citenamefont {Tantivasadakarn}(2017)}]{Nat2017}%
  \BibitemOpen
  \bibfield  {author} {\bibinfo {author} {\bibfnamefont {N.}~\bibnamefont
  {Tantivasadakarn}},\ }\href {\doibase 10.1103/PhysRevB.96.195101} {\bibfield
  {journal} {\bibinfo  {journal} {Phys. Rev. B}\ }\textbf {\bibinfo {volume}
  {96}},\ \bibinfo {pages} {195101} (\bibinfo {year} {2017})}\BibitemShut
  {NoStop}%
\bibitem [{\citenamefont {Wang}\ and\ \citenamefont {Gu}(2018)}]{Wang2018}%
  \BibitemOpen
  \bibfield  {author} {\bibinfo {author} {\bibfnamefont {Q.-R.}\ \bibnamefont
  {Wang}}\ and\ \bibinfo {author} {\bibfnamefont {Z.-C.}\ \bibnamefont {Gu}},\
  }\href {\doibase 10.1103/PhysRevX.8.011055} {\bibfield  {journal} {\bibinfo
  {journal} {Phys. Rev. X}\ }\textbf {\bibinfo {volume} {8}},\ \bibinfo {pages}
  {011055} (\bibinfo {year} {2018})}\BibitemShut {NoStop}%
\bibitem [{\citenamefont {Cheng}\ \emph {et~al.}(2017)\citenamefont {Cheng},
  \citenamefont {Gu}, \citenamefont {Jiang},\ and\ \citenamefont
  {Qi}}]{Cheng2017}%
  \BibitemOpen
  \bibfield  {author} {\bibinfo {author} {\bibfnamefont {M.}~\bibnamefont
  {Cheng}}, \bibinfo {author} {\bibfnamefont {Z.-C.}\ \bibnamefont {Gu}},
  \bibinfo {author} {\bibfnamefont {S.}~\bibnamefont {Jiang}}, \ and\ \bibinfo
  {author} {\bibfnamefont {Y.}~\bibnamefont {Qi}},\ }\href {\doibase
  10.1103/PhysRevB.96.115107} {\bibfield  {journal} {\bibinfo  {journal} {Phys.
  Rev. B}\ }\textbf {\bibinfo {volume} {96}},\ \bibinfo {pages} {115107}
  (\bibinfo {year} {2017})}\BibitemShut {NoStop}%
\bibitem [{\citenamefont {Heinrich}\ \emph {et~al.}(2016)\citenamefont
  {Heinrich}, \citenamefont {Burnell}, \citenamefont {Fidkowski},\ and\
  \citenamefont {Levin}}]{Levin2016}%
  \BibitemOpen
  \bibfield  {author} {\bibinfo {author} {\bibfnamefont {C.}~\bibnamefont
  {Heinrich}}, \bibinfo {author} {\bibfnamefont {F.}~\bibnamefont {Burnell}},
  \bibinfo {author} {\bibfnamefont {L.}~\bibnamefont {Fidkowski}}, \ and\
  \bibinfo {author} {\bibfnamefont {M.}~\bibnamefont {Levin}},\ }\href
  {\doibase 10.1103/PhysRevB.94.235136} {\bibfield  {journal} {\bibinfo
  {journal} {Phys. Rev. B}\ }\textbf {\bibinfo {volume} {94}},\ \bibinfo
  {pages} {235136} (\bibinfo {year} {2016})}\BibitemShut {NoStop}%
\bibitem [{\citenamefont {Wang}\ \emph {et~al.}(2013)\citenamefont {Wang},
  \citenamefont {Potter},\ and\ \citenamefont {Senthil}}]{wangc13b}%
  \BibitemOpen
  \bibfield  {author} {\bibinfo {author} {\bibfnamefont {C.}~\bibnamefont
  {Wang}}, \bibinfo {author} {\bibfnamefont {A.~C.}\ \bibnamefont {Potter}}, \
  and\ \bibinfo {author} {\bibfnamefont {T.}~\bibnamefont {Senthil}},\ }\href
  {\doibase 10.1103/PhysRevB.88.115137} {\bibfield  {journal} {\bibinfo
  {journal} {Phys. Rev. B}\ }\textbf {\bibinfo {volume} {88}},\ \bibinfo
  {pages} {115137} (\bibinfo {year} {2013})}\BibitemShut {NoStop}%
\bibitem [{\citenamefont {Bonderson}\ \emph {et~al.}(2013)\citenamefont
  {Bonderson}, \citenamefont {Nayak},\ and\ \citenamefont {Qi}}]{bonderson13}%
  \BibitemOpen
  \bibfield  {author} {\bibinfo {author} {\bibfnamefont {P.}~\bibnamefont
  {Bonderson}}, \bibinfo {author} {\bibfnamefont {C.}~\bibnamefont {Nayak}}, \
  and\ \bibinfo {author} {\bibfnamefont {X.-L.}\ \bibnamefont {Qi}},\ }\href
  {http://stacks.iop.org/1742-5468/2013/i=09/a=P09016} {\bibfield  {journal}
  {\bibinfo  {journal} {Journal of Statistical Mechanics: Theory and
  Experiment}\ }\textbf {\bibinfo {volume} {2013}},\ \bibinfo {pages} {P09016}
  (\bibinfo {year} {2013})}\BibitemShut {NoStop}%
\bibitem [{\citenamefont {Chen}\ \emph {et~al.}(2014)\citenamefont {Chen},
  \citenamefont {Fidkowski},\ and\ \citenamefont {Vishwanath}}]{chen14a}%
  \BibitemOpen
  \bibfield  {author} {\bibinfo {author} {\bibfnamefont {X.}~\bibnamefont
  {Chen}}, \bibinfo {author} {\bibfnamefont {L.}~\bibnamefont {Fidkowski}}, \
  and\ \bibinfo {author} {\bibfnamefont {A.}~\bibnamefont {Vishwanath}},\
  }\href {\doibase 10.1103/PhysRevB.89.165132} {\bibfield  {journal} {\bibinfo
  {journal} {Phys. Rev. B}\ }\textbf {\bibinfo {volume} {89}},\ \bibinfo
  {pages} {165132} (\bibinfo {year} {2014})}\BibitemShut {NoStop}%
\bibitem [{\citenamefont {Metlitski}\ \emph {et~al.}(2015)\citenamefont
  {Metlitski}, \citenamefont {Kane},\ and\ \citenamefont
  {Fisher}}]{metlitski15}%
  \BibitemOpen
  \bibfield  {author} {\bibinfo {author} {\bibfnamefont {M.~A.}\ \bibnamefont
  {Metlitski}}, \bibinfo {author} {\bibfnamefont {C.~L.}\ \bibnamefont {Kane}},
  \ and\ \bibinfo {author} {\bibfnamefont {M.~P.~A.}\ \bibnamefont {Fisher}},\
  }\href {\doibase 10.1103/PhysRevB.92.125111} {\bibfield  {journal} {\bibinfo
  {journal} {Phys. Rev. B}\ }\textbf {\bibinfo {volume} {92}},\ \bibinfo
  {pages} {125111} (\bibinfo {year} {2015})}\BibitemShut {NoStop}%
\bibitem [{\citenamefont {{Tata}}\ \emph {et~al.}(2021)\citenamefont {{Tata}},
  \citenamefont {{Kobayashi}}, \citenamefont {{Bulmash}},\ and\ \citenamefont
  {{Barkeshli}}}]{Nfset2021a}%
  \BibitemOpen
  \bibfield  {author} {\bibinfo {author} {\bibfnamefont {S.}~\bibnamefont
  {{Tata}}}, \bibinfo {author} {\bibfnamefont {R.}~\bibnamefont {{Kobayashi}}},
  \bibinfo {author} {\bibfnamefont {D.}~\bibnamefont {{Bulmash}}}, \ and\
  \bibinfo {author} {\bibfnamefont {M.}~\bibnamefont {{Barkeshli}}},\
  }\href@noop {} {\bibfield  {journal} {\bibinfo  {journal} {arXiv e-prints}\ }
  (\bibinfo {year} {2021})},\ \Eprint {http://arxiv.org/abs/2104.14567}
  {arXiv:2104.14567} \BibitemShut {NoStop}%
\bibitem [{\citenamefont {Bulmash}\ and\ \citenamefont
  {Barkeshli}(2022{\natexlab{a}})}]{Nfset2021b}%
  \BibitemOpen
  \bibfield  {author} {\bibinfo {author} {\bibfnamefont {D.}~\bibnamefont
  {Bulmash}}\ and\ \bibinfo {author} {\bibfnamefont {M.}~\bibnamefont
  {Barkeshli}},\ }\href {\doibase 10.1103/PhysRevB.105.155126} {\bibfield
  {journal} {\bibinfo  {journal} {Phys. Rev. B}\ }\textbf {\bibinfo {volume}
  {105}},\ \bibinfo {pages} {155126} (\bibinfo {year}
  {2022}{\natexlab{a}})}\BibitemShut {NoStop}%
\bibitem [{\citenamefont {Wang}\ and\ \citenamefont {Gu}(2020)}]{Wang2020}%
  \BibitemOpen
  \bibfield  {author} {\bibinfo {author} {\bibfnamefont {Q.-R.}\ \bibnamefont
  {Wang}}\ and\ \bibinfo {author} {\bibfnamefont {Z.-C.}\ \bibnamefont {Gu}},\
  }\href {\doibase 10.1103/PhysRevX.10.031055} {\bibfield  {journal} {\bibinfo
  {journal} {Phys. Rev. X}\ }\textbf {\bibinfo {volume} {10}},\ \bibinfo
  {pages} {031055} (\bibinfo {year} {2020})}\BibitemShut {NoStop}%
\bibitem [{\citenamefont {Lan}\ \emph {et~al.}(2017)\citenamefont {Lan},
  \citenamefont {Kong},\ and\ \citenamefont {Wen}}]{TianSET}%
  \BibitemOpen
  \bibfield  {author} {\bibinfo {author} {\bibfnamefont {T.}~\bibnamefont
  {Lan}}, \bibinfo {author} {\bibfnamefont {L.}~\bibnamefont {Kong}}, \ and\
  \bibinfo {author} {\bibfnamefont {X.-G.}\ \bibnamefont {Wen}},\ }\href
  {\doibase 10.1103/PhysRevB.95.235140} {\bibfield  {journal} {\bibinfo
  {journal} {Phys. Rev. B}\ }\textbf {\bibinfo {volume} {95}},\ \bibinfo
  {pages} {235140} (\bibinfo {year} {2017})}\BibitemShut {NoStop}%
\bibitem [{\citenamefont {{Aasen}}\ \emph {et~al.}(2021)\citenamefont
  {{Aasen}}, \citenamefont {{Bonderson}},\ and\ \citenamefont
  {{Knapp}}}]{fset2021a}%
  \BibitemOpen
  \bibfield  {author} {\bibinfo {author} {\bibfnamefont {D.}~\bibnamefont
  {{Aasen}}}, \bibinfo {author} {\bibfnamefont {P.}~\bibnamefont
  {{Bonderson}}}, \ and\ \bibinfo {author} {\bibfnamefont {C.}~\bibnamefont
  {{Knapp}}},\ }\href@noop {} {\bibfield  {journal} {\bibinfo  {journal} {arXiv
  e-prints}\ } (\bibinfo {year} {2021})},\ \Eprint
  {http://arxiv.org/abs/2109.10911} {arXiv:2109.10911} \BibitemShut {NoStop}%
\bibitem [{\citenamefont {Bulmash}\ and\ \citenamefont
  {Barkeshli}(2022{\natexlab{b}})}]{fset2021b}%
  \BibitemOpen
  \bibfield  {author} {\bibinfo {author} {\bibfnamefont {D.}~\bibnamefont
  {Bulmash}}\ and\ \bibinfo {author} {\bibfnamefont {M.}~\bibnamefont
  {Barkeshli}},\ }\href {\doibase 10.1103/PhysRevB.105.125114} {\bibfield
  {journal} {\bibinfo  {journal} {Phys. Rev. B}\ }\textbf {\bibinfo {volume}
  {105}},\ \bibinfo {pages} {125114} (\bibinfo {year}
  {2022}{\natexlab{b}})}\BibitemShut {NoStop}%
\bibitem [{\citenamefont {Brundan}\ and\ \citenamefont
  {Ellis}(2017)}]{brundan2017}%
  \BibitemOpen
  \bibfield  {author} {\bibinfo {author} {\bibfnamefont {J.}~\bibnamefont
  {Brundan}}\ and\ \bibinfo {author} {\bibfnamefont {A.~P.}\ \bibnamefont
  {Ellis}},\ }\href@noop {} {\bibfield  {journal} {\bibinfo  {journal}
  {Communications in Mathematical Physics}\ }\textbf {\bibinfo {volume}
  {351}},\ \bibinfo {pages} {1045} (\bibinfo {year} {2017})}\BibitemShut
  {NoStop}%
\bibitem [{\citenamefont {Usher}(2018)}]{usher2018}%
  \BibitemOpen
  \bibfield  {author} {\bibinfo {author} {\bibfnamefont {R.}~\bibnamefont
  {Usher}},\ }\href@noop {} {\bibfield  {journal} {\bibinfo  {journal} {Journal
  of Algebra}\ }\textbf {\bibinfo {volume} {503}},\ \bibinfo {pages} {453}
  (\bibinfo {year} {2018})}\BibitemShut {NoStop}%
\bibitem [{\citenamefont {Bais}\ and\ \citenamefont
  {Slingerland}(2009)}]{bais2009}%
  \BibitemOpen
  \bibfield  {author} {\bibinfo {author} {\bibfnamefont {F.~A.}\ \bibnamefont
  {Bais}}\ and\ \bibinfo {author} {\bibfnamefont {J.~K.}\ \bibnamefont
  {Slingerland}},\ }\href {\doibase 10.1103/PhysRevB.79.045316} {\bibfield
  {journal} {\bibinfo  {journal} {Phys. Rev. B}\ }\textbf {\bibinfo {volume}
  {79}},\ \bibinfo {pages} {045316} (\bibinfo {year} {2009})}\BibitemShut
  {NoStop}%
\bibitem [{\citenamefont {Eli\"ens}\ \emph {et~al.}(2014)\citenamefont
  {Eli\"ens}, \citenamefont {Romers},\ and\ \citenamefont {Bais}}]{eliens2014}%
  \BibitemOpen
  \bibfield  {author} {\bibinfo {author} {\bibfnamefont {I.~S.}\ \bibnamefont
  {Eli\"ens}}, \bibinfo {author} {\bibfnamefont {J.~C.}\ \bibnamefont
  {Romers}}, \ and\ \bibinfo {author} {\bibfnamefont {F.~A.}\ \bibnamefont
  {Bais}},\ }\href {\doibase 10.1103/PhysRevB.90.195130} {\bibfield  {journal}
  {\bibinfo  {journal} {Phys. Rev. B}\ }\textbf {\bibinfo {volume} {90}},\
  \bibinfo {pages} {195130} (\bibinfo {year} {2014})}\BibitemShut {NoStop}%
\bibitem [{\citenamefont {Kong}(2014)}]{kong2014}%
  \BibitemOpen
  \bibfield  {author} {\bibinfo {author} {\bibfnamefont {L.}~\bibnamefont
  {Kong}},\ }\href@noop {} {\bibfield  {journal} {\bibinfo  {journal} {Nuclear
  Physics B}\ }\textbf {\bibinfo {volume} {886}},\ \bibinfo {pages} {436}
  (\bibinfo {year} {2014})}\BibitemShut {NoStop}%
\bibitem [{\citenamefont {Burnell}(2018)}]{burnell2018}%
  \BibitemOpen
  \bibfield  {author} {\bibinfo {author} {\bibfnamefont {F.~J.}\ \bibnamefont
  {Burnell}},\ }\href@noop {} {\bibfield  {journal} {\bibinfo  {journal}
  {Annual Review of Condensed Matter Physics}\ }\textbf {\bibinfo {volume}
  {9}},\ \bibinfo {pages} {307} (\bibinfo {year} {2018})}\BibitemShut {NoStop}%
\bibitem [{\citenamefont {Fidkowski}\ \emph {et~al.}(2013)\citenamefont
  {Fidkowski}, \citenamefont {Chen},\ and\ \citenamefont
  {Vishwanath}}]{fid2013}%
  \BibitemOpen
  \bibfield  {author} {\bibinfo {author} {\bibfnamefont {L.}~\bibnamefont
  {Fidkowski}}, \bibinfo {author} {\bibfnamefont {X.}~\bibnamefont {Chen}}, \
  and\ \bibinfo {author} {\bibfnamefont {A.}~\bibnamefont {Vishwanath}},\
  }\href {\doibase 10.1103/PhysRevX.3.041016} {\bibfield  {journal} {\bibinfo
  {journal} {Phys. Rev. X}\ }\textbf {\bibinfo {volume} {3}},\ \bibinfo {pages}
  {041016} (\bibinfo {year} {2013})}\BibitemShut {NoStop}%
\bibitem [{\citenamefont {Hong}\ \emph {et~al.}(2008)\citenamefont {Hong},
  \citenamefont {Rowell},\ and\ \citenamefont {Wang}}]{hong2008exotic}%
  \BibitemOpen
  \bibfield  {author} {\bibinfo {author} {\bibfnamefont {S.-M.}\ \bibnamefont
  {Hong}}, \bibinfo {author} {\bibfnamefont {E.}~\bibnamefont {Rowell}}, \ and\
  \bibinfo {author} {\bibfnamefont {Z.}~\bibnamefont {Wang}},\ }\href@noop {}
  {\bibfield  {journal} {\bibinfo  {journal} {Communications in Contemporary
  Mathematics}\ }\textbf {\bibinfo {volume} {10}},\ \bibinfo {pages} {1049}
  (\bibinfo {year} {2008})}\BibitemShut {NoStop}%
\bibitem [{\citenamefont {Tambara}\ and\ \citenamefont
  {Yamagami}(1998)}]{tambara1998}%
  \BibitemOpen
  \bibfield  {author} {\bibinfo {author} {\bibfnamefont {D.}~\bibnamefont
  {Tambara}}\ and\ \bibinfo {author} {\bibfnamefont {S.}~\bibnamefont
  {Yamagami}},\ }\href@noop {} {\bibfield  {journal} {\bibinfo  {journal}
  {Journal of Algebra}\ }\textbf {\bibinfo {volume} {209}},\ \bibinfo {pages}
  {692} (\bibinfo {year} {1998})}\BibitemShut {NoStop}%
\end{thebibliography}%

\end{document}